%% file: AFTER-REVIEW-arxiv-v4.tex
\documentclass[3p,times,11pt]{elsarticle}

\usepackage{float}
\usepackage{amsmath}
\usepackage[normalem]{ulem}
\usepackage{times}
\usepackage{amsmath}
\usepackage{txfonts}
\usepackage{graphicx}
\usepackage{comment}
\usepackage{color}
\usepackage{fancyhdr}
\usepackage{subfigure}
\usepackage{setspace}
\usepackage{array}
\usepackage{xspace}
\usepackage{upgreek}
\usepackage{multirow}
\usepackage{textcomp}
\usepackage{tabularx}
\usepackage{scrextend}
\usepackage{hyperref}
\usepackage{adjustbox}
\usepackage{tablefootnote}
\usepackage{makecell}

\journal{Physics Reports}

\bibpunct{[}{]}{,}{n}{}{;}

\graphicspath{{./implementation-fixed-target/}}

\usepackage{lineno}

\newcount\savefnused
\newcount\savefndone

\newcommand{\savefootnote}[2][\empty]%
{\ifx\empty#1\footnotemark\else\footnotemark[#1]\fi
 \global\advance\savefnused by 1
 \expandafter\xdef\csname savefnmark\the\savefnused\endcsname{\thefootnote}%
 \expandafter\xdef\csname savefntext\the\savefnused\endcsname{#2}%
}
\newcommand{\flushfootnote}{\loop\ifnum\savefndone<\savefnused
  \global\advance\savefndone by 1
  \footnotetext[\csname savefnmark\the\savefndone\endcsname]%
    {\csname savefntext\the\savefndone\endcsname}%
  \global\expandafter\let\csname savefnmark\the\savefndone\endcsname\relax
  \global\expandafter\let\csname savefntext\the\savefndone\endcsname\relax
\repeat}

\def\ycms   {\mbox{$y_{\rm c.m.s.}$}\xspace}
\def\ylab   {\mbox{$y_{\rm lab.}$}\xspace}
\def\etalab   {\mbox{$\eta_{\rm lab.}$}\xspace}

\def\pp   {$pp$\xspace}

\def\pA   {$pA$\xspace}

\def\dAu  {$d$Au}

\def\AA   {$AA$}

\def\PbPb {PbPb\xspace}
\def\PbA {Pb$A$\xspace}
\def\PbXe {PbXe\xspace}
\def\PbAr {PbAr\xspace}
\def\PbW {PbW\xspace}
\def\pPb {\ensuremath{p \rm Pb}\xspace}
\def\pCa {\ensuremath{p \rm Ca}\xspace}
\def\pCu {\ensuremath{p \rm Cu}\xspace}
\def\pXe {\ensuremath{p \rm Xe}\xspace}

\def\Pbp {Pb$p$\xspace}
\def\pA {$pA$\xspace}
\newcommand{\chic}{\mbox{$\chi_c$}}

\def\sqrts {\mbox{$\sqrt{s}$}\xspace}
\def\sqrtsNN {\mbox{$\sqrt{s_{NN}}$}\xspace}

\newcommand{\xF}{\mbox{$x_{F}$}}
\newcommand{\kT}{\mbox{$k_{T}$}}
\newcommand{\pT}{p_{T}}
\newcommand{\An}{\ensuremath{A_N}\xspace}
\newcommand{\ie}{{\it i.e.}}
\newcommand{\eg}{{\it e.g.}}

\renewcommand{\L}{\mathcal L}
\def\jpsi    {\mbox{$J/\psi$}}

\newcommand{\cms}{{\rm c.m.s.}}
\newcommand{\lab}{{\rm lab.}}

\newcommand{\cf}[1]{{Fig.~\ref{#1}}}
\newcommand{\ct}[1]{{Table~\ref{#1}}}

\def\L {\mathcal L}

\usepackage{hyperref}
\hypersetup{
  pdftitle={A Fixed-Target Programme at the LHC},%
  pdfauthor={AFTER Study Group (Contact: J.P. Lansberg)},%
  pdfsubject={},%
  pdfkeywords={},%
  pdfstartview={},%
  bookmarksopen=true, breaklinks=true, debug=true, %
  colorlinks=true, linkcolor=blue, citecolor=blue, urlcolor=blue
}

\newcolumntype{C}{>{\centering\arraybackslash}X}

\makeatletter
\let\@fnsymbol\@alph
\makeatother

\input{symbols-def.tex}
\begin{document}

\begin{frontmatter}

\title{A Fixed-Target Programme at the LHC: \\Physics Case and Projected Performances for Heavy-Ion, Hadron, Spin and Astroparticle Studies}

\date{\today}

\input{authors.tex}

\begin{abstract}
We review the context, the motivations and the expected performances of a comprehensive and ambitious fixed-target program using the multi-TeV proton and ion LHC beams. We also provide a detailed account of the different possible technical implementations ranging from an internal wire target to a full dedicated beam line extracted with a bent crystal. The possibilities offered by the use of the ALICE and LHCb detectors in the fixed-target mode are also reviewed. 
\end{abstract}

\end{frontmatter}

\tableofcontents

\newpage

\input{introduction.tex}

\input{motivation/motivation.tex}

\input{implementation-fixed-target/implementation-fixed-target.tex}

\input{detector/detector.tex}

\section{Physics Projections}
\label{sec:physics}

In this section, we review the projected performances for each of the 3 main topics. The assumptions considered for the generation of pseudo-data for LHCb and ALICE will be given. Figures-of-Merits (FoM) will refer to AFTER@LHCb, AFTER@ALICE$_\mu$ or AFTER@ALICE$_{\rm CB}$ for pseudo-data generated in the acceptance of LHCb, the ALICE Muon Spectrometer and the ALICE Central Barrel, respectively.

\input{physics-high-x/physics-high-x.tex}

\input{physics-spin/physics-spin.tex}

\input{physics-heavy-ion-collisions/physics-heavy-ion-collisions.tex}

\input{summary/summary.tex}

\input{acknowledgements.tex}

\newpage

\section{Appendices}

\input{Appendix_Implementation_B.tex}

\input{Appendix_Implementation_Abis.tex}

\input{Appendix_Implementation_C.tex}

\input{Appendix_Implementation_D.tex}
\input{Appendix_Detector_ALICE.tex}
\input{Appendix_Detector_LHCB.tex}

\newpage

\bibliographystyle{utphys}

\addcontentsline{toc}{section}{References}
\bibliography{references}

\typeout{get arXiv to do 4 passes: Label(s) may have changed. Rerun}
\end{document}

%% file: symbols-def.tex
\def\RpXe {\ensuremath{R_{p\text{Xe}}}\xspace}
\def\RpA {\ensuremath{R_{pA}}\xspace}

\def\RPbXe {\ensuremath{R_{\text{PbXe}}}\xspace}

\def\AFTERALICECB {\mbox{AFTER@ALICE$_{\rm CB}$}\xspace}
\def\AFTERALICEMU {\mbox{AFTER@ALICE$_\mu$}\xspace}
\def\AFTERALICE {\mbox{AFTER@ALICE}\xspace}
\def\AFTERLHCb {\mbox{AFTER@LHCb}\xspace}
\def\AFTER {\mbox{AFTER@LHC}\xspace}

\def\herschel {\mbox{\textsc{HeRSCheL}}\xspace}

 \def\Pmu         {\ensuremath{\mu}\xspace}

 \def\Ppi         {\ensuremath{\pi}\xspace}

 \def\Pchi        {\ensuremath{\chi}\xspace}

 \def\PLambda      {\ensuremath{\Lambda}\xspace}

 \def\PB      {\ensuremath{\mathrm{B}}\xspace}                 
                  
 \def\PD      {\ensuremath{\mathrm{D}}\xspace}

 \def\PK      {\ensuremath{\mathrm{K}}\xspace}

 \def\Pi      {\ensuremath{\mathrm{i}}\xspace}

\def\mumu       {{\ensuremath{\Pmu^+\Pmu^-}}\xspace}

\def\pion   {{\ensuremath{\Ppi}}\xspace}
\def\piz    {{\ensuremath{\pion^0}}\xspace}

\def\kaon    {{\ensuremath{\PK}}\xspace}
  \def\Kbar    {{\kern 0.2em\overline{\kern -0.2em \PK}{}}\xspace}

\def\KS      {{\ensuremath{\kaon^0_{\mathrm{ \scriptscriptstyle S}}}}\xspace}

  \def\Dbar    {{\kern 0.2em\overline{\kern -0.2em \PD}{}}\xspace}

\def\Bbar    {{\ensuremath{\kern 0.18em\overline{\kern -0.18em \PB}{}}}\xspace}

\def\jpsi     {{\ensuremath{J/\psi}}\xspace}
\def\Jpsi {\jpsi}
\def\psitwos  {{\ensuremath{\psi{(2S)}}}\xspace}

\def\psip {\psitwos}

  \def\Y#1S{\ensuremath{\Upsilon{(#1S)}}\xspace}%
\def\OneS  {{\Y1S}}

\def\ups  {{\ensuremath{\Upsilon(nS)}}}

\def\chic  {{\ensuremath{\Pchi_{c}}}\xspace}

\def\Lbar        {{\ensuremath{\kern 0.1em\overline{\kern -0.1em\PLambda}}}\xspace}

\def\BF         {{\ensuremath{\mathcal{B}}}\xspace}

\def\BR         {\BF}

\def\to                 {\ensuremath{\rightarrow}\xspace}

\newcommand{\lqcd}{{\ensuremath{\Lambda_{\mathrm{QCD}}}}\xspace}

\def\AT#1     {\ensuremath{A_{\mathrm{T}}^{#1}}\xspace}           %

\def\C#1      {\ensuremath{\mathcal{C}_{#1}}\xspace}                       %
\def\Cp#1     {\ensuremath{\mathcal{C}_{#1}^{'}}\xspace}                    %
\def\Ceff#1   {\ensuremath{\mathcal{C}_{#1}^{\mathrm{(eff)}}}\xspace}        %
\def\Cpeff#1  {\ensuremath{\mathcal{C}_{#1}^{'\mathrm{(eff)}}}\xspace}       %
\def\Ope#1    {\ensuremath{\mathcal{O}_{#1}}\xspace}                       %
\def\Opep#1   {\ensuremath{\mathcal{O}_{#1}^{'}}\xspace}                    %

\newcommand{\tev}{\ensuremath{\mathrm{\,Te\kern -0.1em V}}\xspace}
\newcommand{\gev}{\ensuremath{\mathrm{\,Ge\kern -0.1em V}}\xspace}
\newcommand{\mev}{\ensuremath{\mathrm{\,Me\kern -0.1em V}}\xspace}
\newcommand{\kev}{\ensuremath{\mathrm{\,ke\kern -0.1em V}}\xspace}
\newcommand{\ev}{\ensuremath{\mathrm{\,e\kern -0.1em V}}\xspace}
\newcommand{\gevc}{\ensuremath{{\mathrm{\,Ge\kern -0.1em V\!/}c}}\xspace}
\newcommand{\mevc}{\ensuremath{{\mathrm{\,Me\kern -0.1em V\!/}c}}\xspace}
\newcommand{\gevcc}{\ensuremath{{\mathrm{\,Ge\kern -0.1em V\!/}c^2}}\xspace}
\newcommand{\gevgevcccc}{\ensuremath{{\mathrm{\,Ge\kern -0.1em V^2\!/}c^4}}\xspace}
\newcommand{\mevcc}{\ensuremath{{\mathrm{\,Me\kern -0.1em V\!/}c^2}}\xspace}

\def\m    {\ensuremath{\mathrm{ \,m}}\xspace}

\def\mm   {\ensuremath{\mathrm{ \,mm}}\xspace}

\def\mum  {\ensuremath{{\,\mu\mathrm{m}}}\xspace}

\def\mub{\ensuremath{{\mathrm{ \,\mu b}}}\xspace}

\def\mhz  {\ensuremath{{\mathrm{ \,MHz}}}\xspace}

\def\gsim{{~\raise.15em\hbox{$>$}\kern-.85em
          \lower.35em\hbox{$\sim$}~}\xspace}
\def\lsim{{~\raise.15em\hbox{$<$}\kern-.85em
          \lower.35em\hbox{$\sim$}~}\xspace}

\def\pt         {\mbox{$p_{T}$}\xspace}

\newcommand{\etc}{\mbox{\itshape etc.}\xspace}

%% file: authors.tex
\author[IJCLab]{C.~Hadjidakis\fnref{Editor}}
\address[IJCLab]{Universit\'e Paris-Saclay, CNRS, IJCLab, 91405 Orsay, France}

\author[WUT]{D.~Kiko\l a\fnref{Editor}}
\address[WUT]{Faculty of Physics, Warsaw University of Technology, ul. Koszykowa 75, 00-662 Warsaw, Poland}

\author[IJCLab]{J.P.~Lansberg\corref{JPL}\fnref{Editor}}

\author[IJCLab]{L.~Massacrier\fnref{Editor}}

\fntext[Editor]{Editor}
\fntext[SE]{Section editor}

\author[Pavia,UAH]{M.G.~Echevarria\fnref{SE}}
\address[Pavia]{Istituto Nazionale di Fisica Nucleare, Sezione di Pavia, via Bassi 6, 27100 Pavia, Italy}
\address[UAH]{Dpto. de F\'isica y Matem\'aticas, Universidad de Alcal\'a, 28805 Alcal\'a de Henares (Madrid), Spain}

\author[Krakow]{A.~Kusina\fnref{SE}}
\address[Krakow]{Institute of Nuclear Physics Polish Academy of Sciences, PL-31342 Krakow, Poland}

\author[LPSC]{I.~Schienbein\fnref{SE}}
\address[LPSC]{Laboratoire de Physique Subatomique et de Cosmologie, Universit\'e Grenoble Alpes, CNRS/IN2P3, 
53 Avenue des Martyrs, F-38026 Grenoble, France}

\author[IST,LIP,CeFEMA]{J.~Seixas\fnref{SE}}
\address[IST]{Dep. Fisica, Instituto Superior Tecnico, Av. Rovisco Pais 1, 1049-001 Lisboa, Portugal}
\address[LIP]{LIP, Av. Prof. Gama Pinto, 2, 1649-003 Lisboa, Portugal}
\address[CeFEMA]{Centro de F\'isica e Engenharia de Materiais Avan\c cados, Av. Rovisco Pais 1, 1049-001 Lisboa, Portugal}

\author[LPTHE]{H.S.~Shao\fnref{SE}}
\address[LPTHE]{LPTHE, UMR 7589, Sorbonne University\'e et CNRS, 4 place Jussieu, 75252 Paris Cedex 05, France}

\author[JLAB,Pavia,Pavia-U]{A.~Signori\fnref{SE}}
\address[JLAB]{Theory Center, Thomas Jefferson National Accelerator Facility, 12000 Jefferson Avenue, Newport News, VA 23606, USA}
\address[Pavia-U]{Dipartimento di Fisica, Universit\`a di Pavia, via Bassi 6, I-27100 Pavia, Italy}

\author[Utrecht,Prague]{B.~Trzeciak\fnref{SE}}
\address[Utrecht]{Institute for Subatomic Physics, Utrecht University, Utrecht, The Netherlands}
\address[Prague]{Faculty of Nuclear Sciences and Physical Engineering, Czech Technical University in Prague, Prague, Czech Republic}

\author[SLAC]{S.J.~Brodsky} 
\address[SLAC]{SLAC National Accelerator Laboratory, Stanford University, Menlo Park, CA 94025, USA}

\author[Roma]{G.~Cavoto} 
\address[Roma]{``Sapienza" Universit\`a di Roma, Dipartimento di Fisica \&
INFN, Sez. di Roma, P.le A. Moro 2, 00185 Roma, Italy}

\author[LANL]{C.~Da~Silva}
\address[LANL]{P-25, Los Alamos National Laboratory, Los Alamos, NM 87545, USA}

\author[Turin]{F.~Donato}
\address[Turin]{Turin University, Department of Physics, and INFN, Sezione of Turin, Turin, Italy}

\author[USC,LLR]{E.G.~Ferreiro}
\address[USC]{Dept. de F{\'\i}sica de Part{\'\i}culas \& IGFAE, Universidade de Santiago de Compostela, 15782 Santiago de Compostela, Spain}
\address[LLR]{Laboratoire Leprince-Ringuet, Ecole polytechnique, CNRS/IN2P3,  Palaiseau, France}

\author[IJCLab]{I.~H\v{r}ivn\'{a}\v{c}ov\'{a}}

\author[LANL]{A.~Klein}

\author[INR]{A.~Kurepin}
\address[INR]{Institute for Nuclear Research, Russian Academy of Sciences, 117312 Moscow, Russia}                    

\author[CPHT]{C.~Lorc\'e}
\address[CPHT]{CPHT, CNRS, Ecole Polytechnique, Institut Polytechnique de Paris, Route de Saclay, 91128 Palaiseau, France}

\author[SMU]{F.~Lyonnet}
\address[SMU]{Southern Methodist University, Dallas, TX 75275, USA}

\author[BNL]{Y.~Makdisi}
\address[BNL]{Brookhaven National Laboratory, Collider Accelerator Department}

\author[LPC]{S.~Porteboeuf Houssais}
\address[LPC]{Universit\'e Clermont Auvergne, CNRS/IN2P3, LPC, F-63000 Clermont-Ferrand, France.}

\author[LIP]{C.~Quintans}

\author[DPhN]{A.~Rakotozafindrabe} 
\address[DPhN]{IRFU/DPhN, CEA Saclay, 91191 Gif-sur-Yvette Cedex, France}

\author[IJCLab]{P.~Robbe}

\author[CERN]{W.~Scandale} 
\address[CERN]{CERN, European Organization for Nuclear Research, 1211 Geneva 23, Switzerland}

\author[INR]{N.~Topilskaya}

\author[IPNL]{A.~Uras} 
\address[IPNL]{IPNL, Universit\'e Claude Bernard Lyon-I and CNRS-IN2P3, Villeurbanne, France}

\author[NCBJ]{J.~Wagner}
\address[NCBJ]{National Centre for Nuclear Research (NCBJ), Pasteura 7, 02-093 Warsaw, Poland}

\author[IJCLab,Kennesaw,Umass,Kyoto]{N.~Yamanaka}
\address[Umass]{Amherst Center for Fundamental Interactions, Department of Physics, University of Massachusetts Amherst, MA 01003, USA}
\address[Kyoto]{Nishina Center for Accelerator-Based Science, RIKEN, Wako 351-0198, Japan}
\address[Kennesaw]{Department of Physics, Kennesaw State University, Kennesaw, GA 30144, USA}

\author[CHEP]{Z.~Yang}
\address[CHEP]{Center for High Energy Physics, Department of Engineering Physics, Tsinghua University, Beijing, China}

\author[BNL]{A.~Zelenski}

\vspace*{-2.5cm}

%% file: introduction.tex
\section{Introduction}

The objective of this review is to highlight the physics opportunities of
 using the most energetic proton and ion beams ever in the fixed-target mode and to review the 
feasibility of a rich physics program for heavy-ion, hadron, spin and 
astroparticle physics with existing (LHCb or ALICE) or new set-ups  allowing one to perform such studies
 parasitically for the LHC collider program.

Let us first recall that the fixed-target mode offers several unique assets~\cite{Brodsky:2012vg}
compared to the collider mode which are particularly relevant 
with the LHC beams in the context of high-energy physics :
\begin{itemize}
\item A high luminosity thanks to the high density of the target at no cost for the LHC collider-mode experiments. 
Both an internal gas target or a bent-crystal-extracted beam from the beam halo allow for
yearly luminosities well above those of similar machines, in particular RHIC, in the ballpark of the LHC and Tevatron collider luminosities;
\item The accessibility with standard detectors, thanks to the boost between the colliding-nucleon 
centre-of-mass system (\cms) and the laboratory system,
 to the far backward \cms\ region which remains completely uncharted with hard reactions until now. 
A pseudo-rapidity acceptance of $1 \leq \eta_{\rm (lab.)} \leq 5$, combined with high luminosities, essentially allows one to measure
any probe down to the very end of the backward phase space;
\item An extended number of species for the target, including deuteron and $^3$He allowing for unique neutron studies, with the possibility to change them in a reduced amount of time for short runs;
\item The \cms\ energy per nucleon-nucleon collision (\sqrtsNN) is identical for all 7 TeV proton and  2.76 TeV lead induced collisions, namely 115 GeV for $pp$, $pd$, $pA$ systems and 72 GeV for Pb$p$, Pb$d$, Pb$A$ systems\footnote{\sqrtsNN for lighter ion beams remains on the order of 70 GeV.}. This  allows for nuclear-modification-factor measurements with drastically reduced systematic uncertainties in an energy domain between the SPS and RHIC experiments in an unexplored rapidity domain;
\item The target polarisation --whereas the LHC beams are unpolarised. This offers uncountable opportunities for single spin asymmetry (SSA) measurements 
--at large momentum fractions-- which have been the object of a growing attention in the recent years at RHIC, at CERN and at Fermilab. 
\end{itemize}

Owing to these advantages, we have identified three main topics for a strong physics case motivating a complete
fixed-target program at the LHC (referred to \AFTER in what follows) with one or more detectors.
These cover studies of 
\begin{itemize}
\item the high momentum fraction ($x$) frontiers in nucleons and nuclei with a specific emphasis on the gluon and heavy-quarks and the implication for astroparticle physics; 
\item the spin content of the nucleons with a focus on azimuthal asymmetries 
generated by the spin of the partons and Single Transverse-Spin Asymmetries (STSAs) generated by the correlation between the nucleon spin and the momentum of partons;
\item the hot medium created in ultra-relativistic heavy-ion collisions with novel quarkonium and heavy-quark observables in a new energy domain and with identified light hadrons down to the target-rapidity region.  
\end{itemize}

The structure of this review is as follows. In the section~\ref{sec:motivations}, we quickly review the context in which such a \AFTER program would take place and highlight the motivations for the three main aforementioned research axes. In the section~\ref{sec:implementation}, we provide a
state-of-the-art overview of the different available technologies to initiate collisions of the LHC beams with fixed targets. In the section~\ref{section:detector}, we elaborate on the detector aspects both for an ideal detector and for existing detectors, \ie\ those of the LHCb and ALICE collaborations\footnote{In what follows, \AFTER will refer to a generic experimental fixed-target set-up using the LHC beams, \AFTERLHCb, \AFTERALICE, \AFTERALICEMU and \AFTERALICECB to specific implementations using the LHCb or ALICE detectors.}. In the section~\ref{sec:physics}, we extensively review the projected performances for flagship studies and the studies proposed within the community for each of the 3 main topics. The section \ref{sec:conclusion} gathers our conclusions.

%% file: motivation/motivation.tex
\section{Motivations} \label{sec:motivations}

\input{motivation/motivation-high-x.tex}

\input{motivation/motivation-spin.tex}

\input{motivation/motivation-heavy-ion.tex}

%% file: motivation/motivation-high-x.tex
\subsection{The high-$x$ frontier}

Whereas the need for precise measurements of the partonic structure of nucleons and nuclei at small momentum fractions $x$ is 
usually highlighted as a strong motivation for new large-scale experimental facilities, such as the Electron-Ion Collider~\cite{Accardi2016} (EIC) 
or Large Hadron-electron Collider~\cite{AbelleiraFernandez:2012cc}
(LHeC) projects, the structure of nucleons and nuclei at {\it high} $x$
is as poorly known at both low and high scales (see \cf{fig:high_x_gluon_PDF}). Let us mention the long-standing puzzles such as the origin of the nuclear EMC\footnote{Named after its observation in 1983 by
the European Muon Collaboration~\cite{Aubert:1983xm}.} effect in nuclei or a possible non-perturbative source of charm or beauty quarks 
in the proton which would carry a relevant fraction of its momentum. With an extensive coverage of the backward region
corresponding to  high $x$ in the target, \AFTER\ is probably the best programme for such physics with hadron beams.

\begin{figure}[!hbt]
\centering
\subfigure[~]{\includegraphics[width=0.38\textwidth]{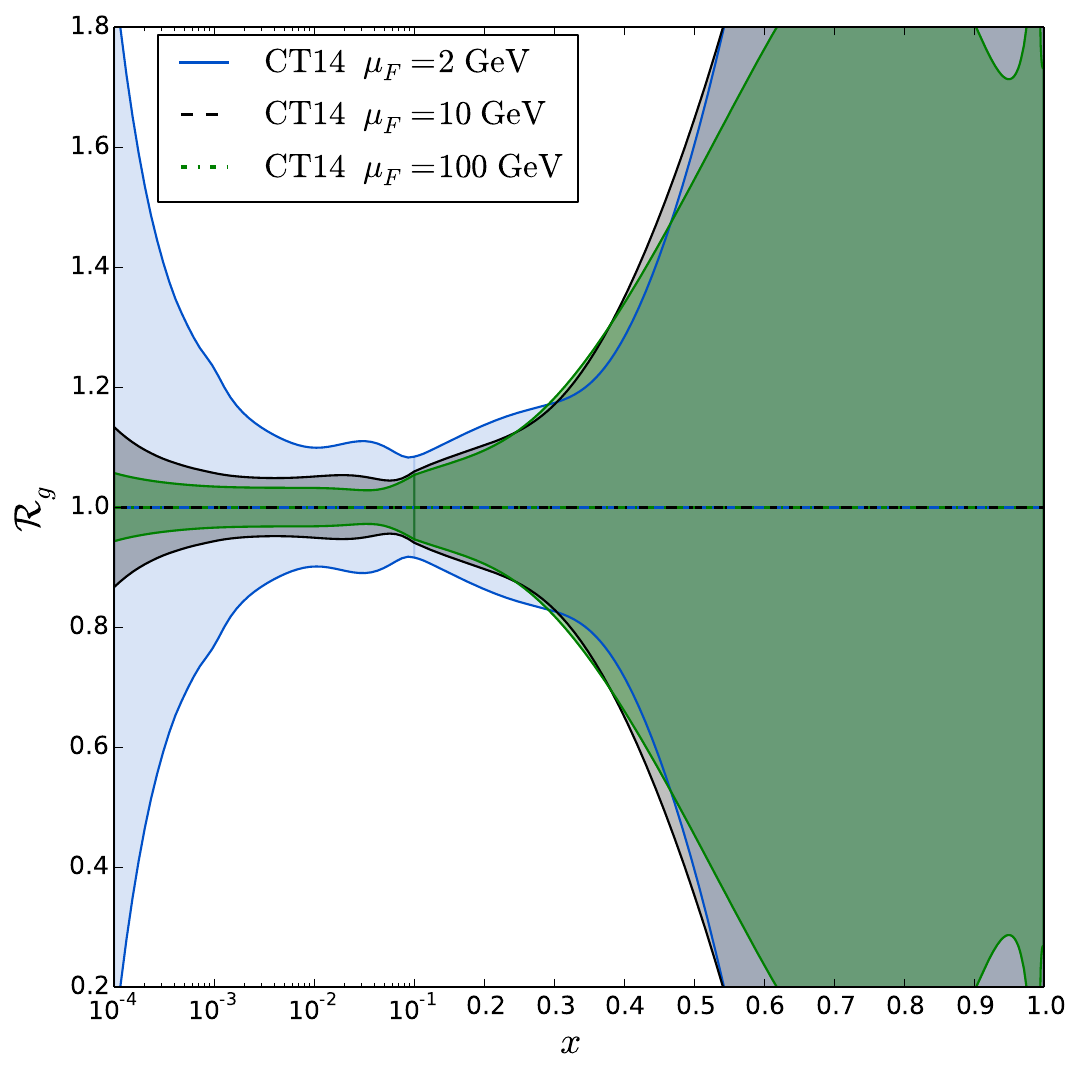}
\label{fig:high_x_gluon_PDF}}
\subfigure[~]{\includegraphics[width=0.6\textwidth]{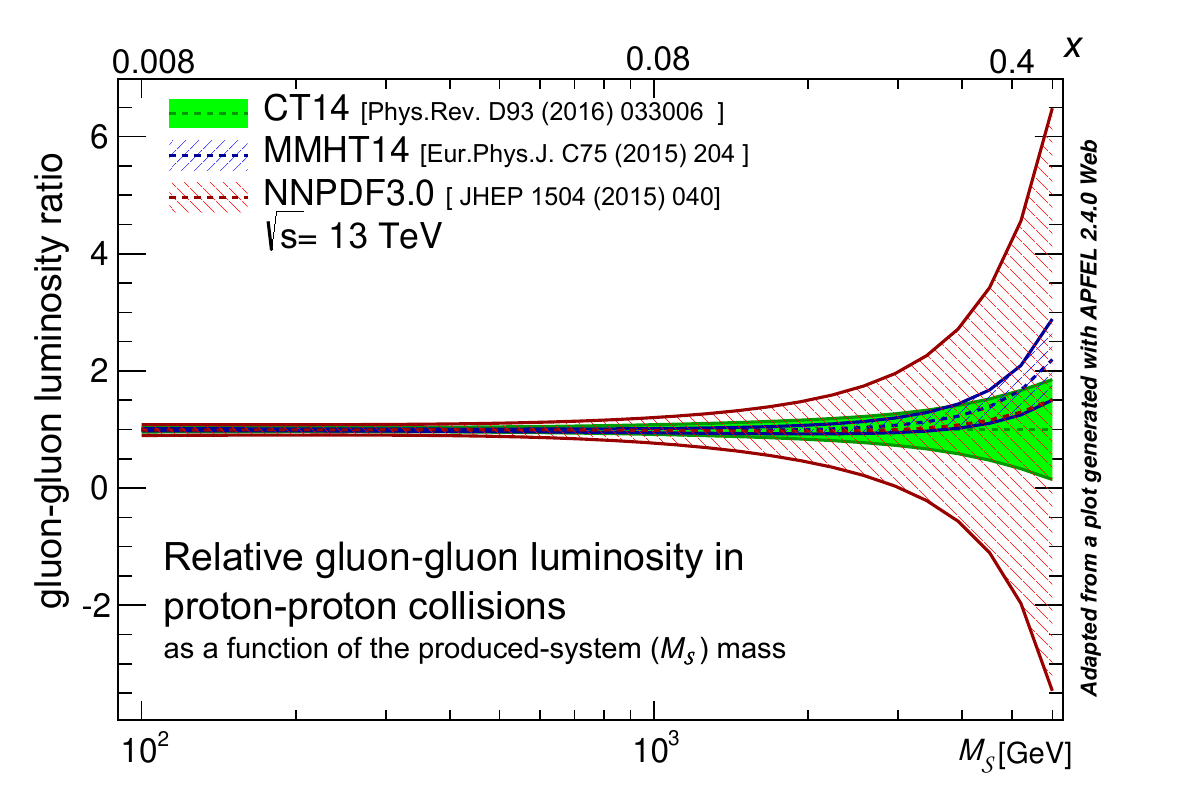}
\label{fig:gg-lumi}}
\caption{(a) CT14nlo gluon PDF relative uncertainties~\cite{Dulat:2015mca} in a proton as a function of the gluon
momentum fraction $x$ at three values of the factorisation scale, $\mu_F$, 
(b) Gluon-gluon-luminosity uncertainty computed for three sets of proton PDFs as a function of the invariant mass ($M_{\cal S}$)  of a to-be produced system at $\sqrt{s}=13$ TeV. For $y\sim 0$,
$x \simeq M_{\cal S}/\sqrt{s}$ at the LHC (indicated on the upper $x$ axis). The kinematics of the \AFTER\ programme is mainly that of high $x$ where the uncertainties blow up. Plot done thanks to the APFEL programme~\cite{Bertone:2013vaa}.
}
\label{fig:high_x_1}
\end{figure}

Studying the so-called high-$x$ physics also provides us with novel decisive means to advance our experimental
knowledge of the still poorly understood confinement properties of the strong interaction, which is one of the last
 open questions about the Standard Model. Indeed, studying 
high-$x$  fluctuations of a nucleon, where a single gluon carries the majority of the confined-system momentum,
certainly tests QCD in a new limit never explored before. On the quark side, an improved experimental determination
of the $d/u$ PDF ratio  for $x\to 1$ is also crucial to tell which picture is valid between an SU(6) symmetric one where $d/u \to 1/2$, 
the dominance of a quark-scalar diquark where $d/u \to 0$, quark-hadron duality where $d/u \to 0.42$ or a simple perturbative QCD one where $d/u \to 1/5$.
Beside such fundamental issues touching upon our understanding of confinement, charting  the high-$x$ structure of nucleons and nuclei has very practical 
implications, for instance to improve our knowledge of parton luminosities at existing and future hadron colliders (LHC, 
RHIC, Tevatron, FCC, ...) (see \cf{fig:gg-lumi}) but also of Ultra-High-Energy-Cosmic Rays (UHECR), in particular the neutrino, in the PeV range. 

\begin{figure}[!hbt]
\centering
\includegraphics[width=0.65\textwidth]{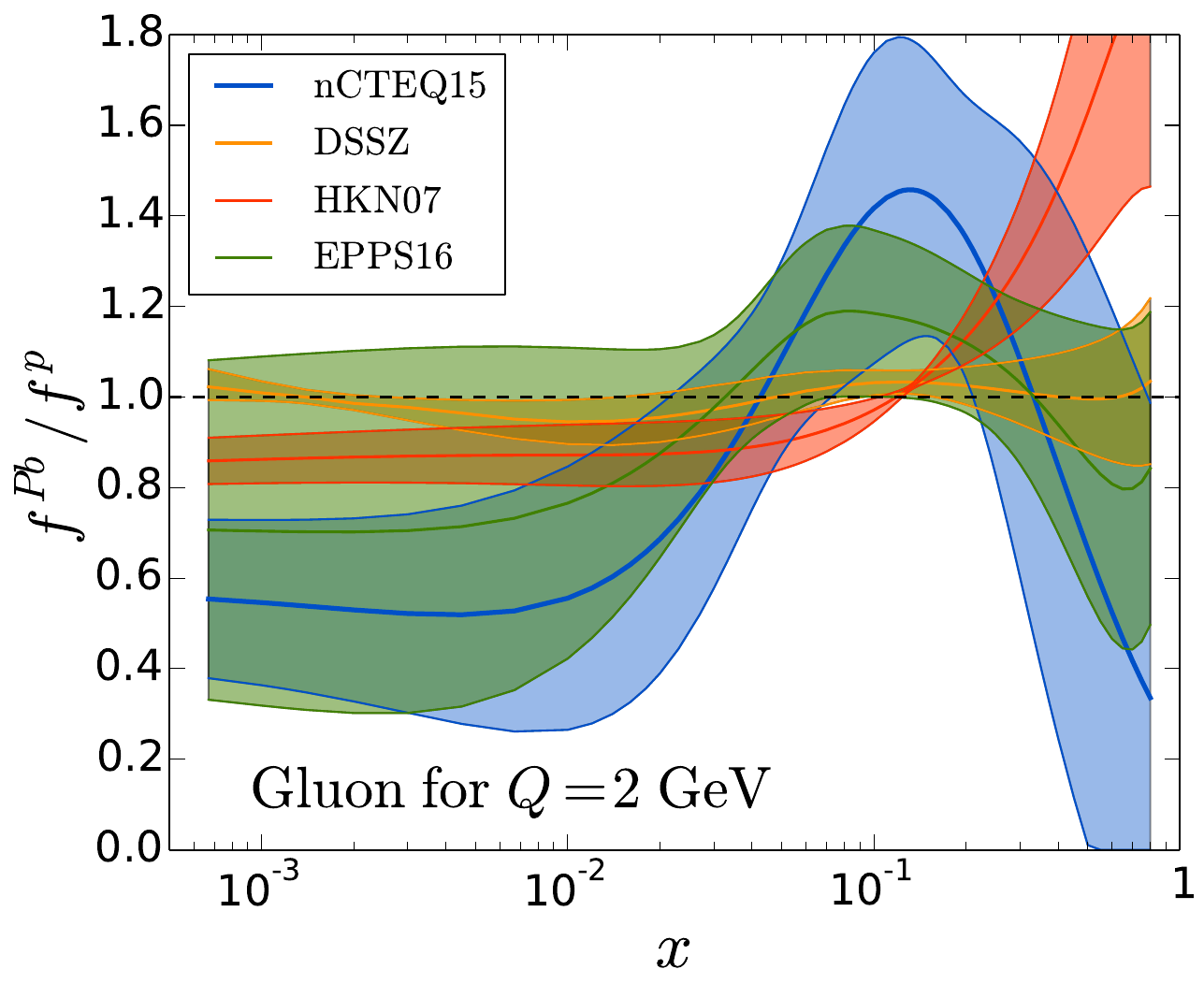}
\caption{Compilation of the gluon nuclear PDF relative uncertainties~\cite{Eskola:2016oht,Kovarik:2015cma,deFlorian:2011fp,Hirai:2007sx} in a lead nucleus at a factorisation scale (here denoted $Q$) of 2 GeV. 
}
\label{fig:high_x_gluon_nPDF}
\end{figure}

Beyond unbound nucleons, our understanding of the gluon and quark content of the nuclei is also very limited at high $x$. Since the first observation via DIS measurements of a nuclear suppression of the quark momentum distribution --the aforementioned EMC effect--, DIS data got more precise and confirmed the suppression. Yet, we still do not understand its physical origin. Recently, it was argued that the $x > 1$ scaling plateaux of some nucleus structure functions, attributed to short-range nucleon-nucleon correlations related to high local densities in nuclei, could be related to the EMC effect~\cite{Weinstein:2010rt}. In this context, the complete lack of data constraining the gluon density in this region (see \cf{fig:high_x_gluon_nPDF}) is probably very detrimental. Only indirect constraints from the scaling violation of the quark distribution exist, which obviously do not give additional experimental information. We are also lacking precise nuclear Drell-Yan (DY) data which provide a unique window on the sea quarks. A precise measurement of the {\it gluon} EMC and of its nuclear number ($A$) dependence, combined with precise DY data at high $x$, would provide decisive insights into the origin of the EMC effects which goes along with  understanding how quark and gluons behave in the nuclear medium.

\begin{figure}[!t]
\centering
\subfigure[LHCb-like]{\includegraphics[width=0.45\textwidth]{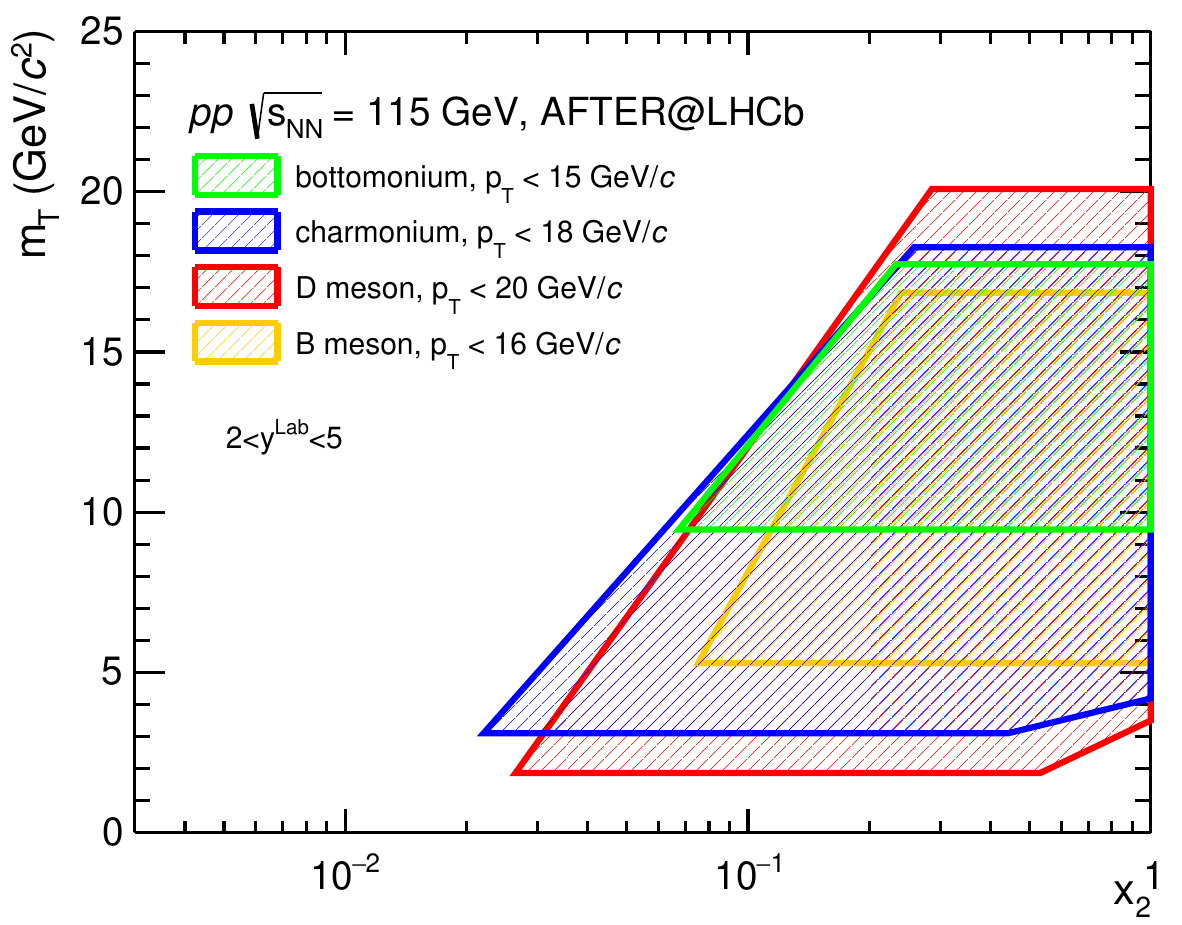}
\label{fig:high_x_kinematical_range_LHCb}}
\subfigure[ALICE-like]{ \includegraphics[width=.45\textwidth]{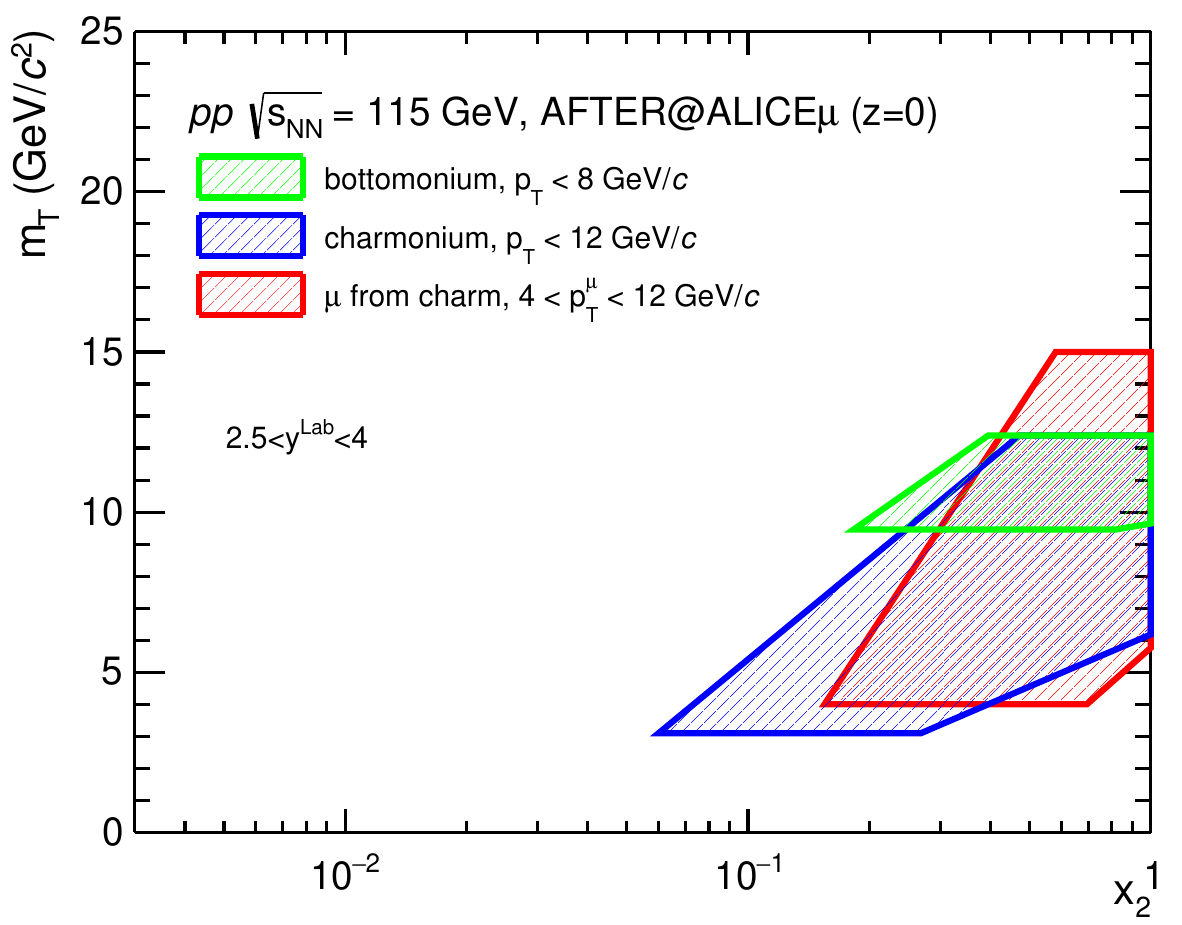}
\label{fig:high_x_kinematical_range_ALICE}}
\caption{Typical kinematical reach with heavy-hadron production in $x_2$ (our proxy to the momentum fraction of the parton in the target) and the scale (chosen to be $m_T=\sqrt{M_{\rm hadron}^2+P_{T,\rm hadron}^2}$) of the fixed-target mode with a detector acceptance like (a) LHCb and (b) ALICE.}
\label{fig:high_x_2}
\end{figure}

At the interface between the proton and nuclear cases, the deuteron and $^3$He
have a particular place. On the one hand, they can provide us with quasi-free neutron targets allowing us to test assumptions
such as $u^p = d^n$, $\bar u^p = \bar d^n$ based on the isospin symmetry, or whether gluons behave differently in protons than in neutrons.
On the other hand, it is an appealing playground to test our understanding of 
the dynamics of simple nucleon-bound systems and hidden colour configurations 
of six quarks (see \eg~\cite{Brodsky:2018zdh}).

For all these reasons, the unique opportunities offered by \AFTER\ to probe with high precision and with reliable as well as novel perturbative probes the high-$x$ domain are very interesting. \cf{fig:high_x_2} schematically illustrates, with a small set of selected (gluon-sensitive) probes, how the fixed-target mode with multi TeV beams allows one to methodically  probe the high-$x$ region in the target, namely at high $x_2$. This obviously applies equally for quark, heavy-quark, gluon and antiquark sensitive probes in protons, deuterons and light or heavy nuclei. In addition, when heavy ions are used as beam particles, one can probe their contents down to $x$ value as small as 0.005. The possibility to study the large-$x$-charm content in the (nucleon and nucleus) target will also allow one to severely reduce the uncertainty of the prompt neutrino fluxes. 

Finally, let us highlight the opportunities connected to antiproton ($\bar p$) measurements in new kinematical 
ranges and for new systems in order to further constrain the modelling of the conventional  $\bar p$ astrophysical sources. 
One of the possible very original measurements is that of $\bar p$ nearly at rest from fixed He, C, N or O targets bearing on the Particle IDentification (PID) capabilities of the ALICE central barrel. This would constrain highly-energetic $\bar p$ from (He,C,N,O)$+p\to \bar p +X$.

%% file: motivation/motivation-spin.tex
\subsection{Unraveling the nucleon spin}

Despite decades of efforts, the internal structure of the nucleons, in particular their constituent 
distribution and dynamics, is still largely unknown. One of the most significant issues is our limited understanding of the spin structure of the nucleon, 
specifically how its elementary constituents 
(quarks and gluons) bind into a spin-$\frac{1}{2}$ object. Essentially, 
there are two types of contributions to the nucleon spin from quarks and gluons: their spin and their Orbital Angular Momentum (OAM). 
For a longitudinally polarised nucleon, \ie\ with helicity $+\frac{1}{2}$, one has
\begin{equation}
\frac{1}{2} = \frac{1}{2}\Delta \Sigma + \Delta G + {\cal L}_{g} + {\cal L}_{q}
\,,
\label{e:spindec}
\end{equation}
where $\frac{1}{2}\Delta\Sigma$ denotes the combined spin contribution of quarks and antiquarks, 
$\Delta G$ the gluon spin, and ${\cal L}_{q,g}$ the quark and gluon OAM contributions 
(see \eg\ \cite{Jaffe:1989jz,Ji:1996ek,Leader:2013jra,Wakamatsu:2014zza,Wakamatsu:2016pqo}). Eq.~(\ref{e:spindec}) 
is in principle valid at any energy scale at which the nucleon is probed and this calls for the study 
of the evolution of the individual contributions at different scales. 
These questions naturally generalise to any spin-$J$ hadron.

\begin{figure}[!hbt]
\centering
\subfigure[~]{\includegraphics[width=0.49\textwidth]{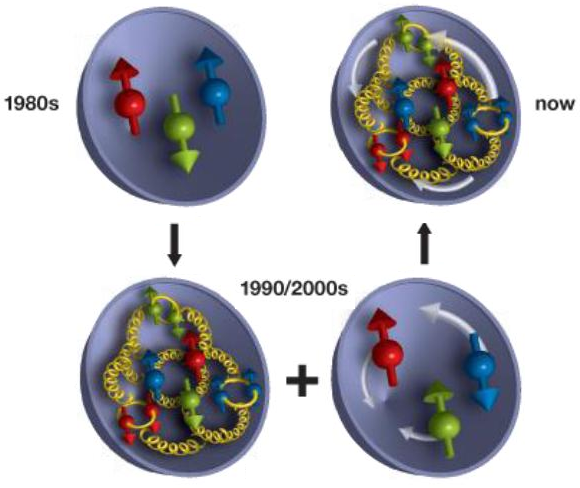}
\label{fig:spin_crisis}}
\subfigure[~]{\includegraphics[width=0.49\textwidth]{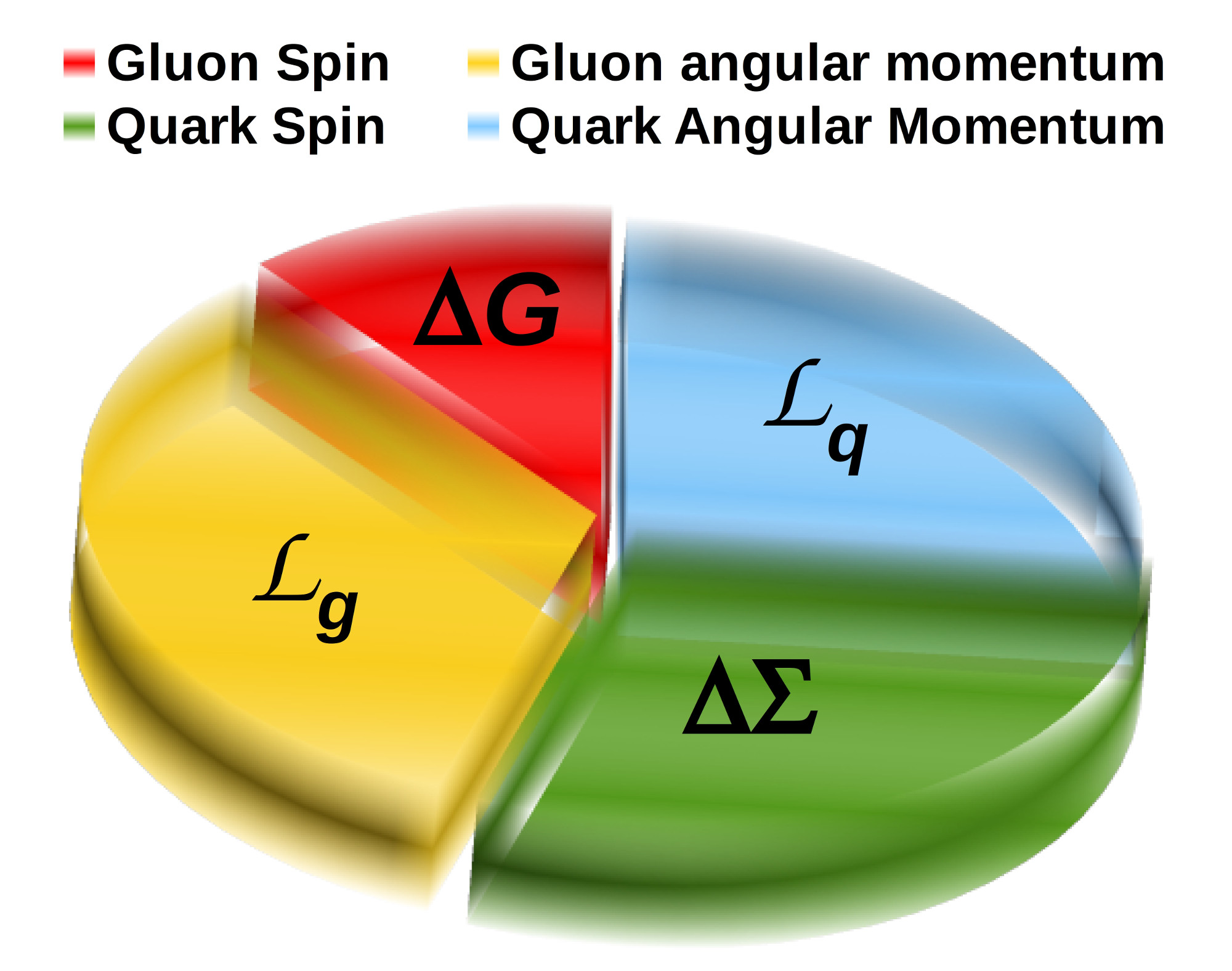}
\label{fig:spin_puzzle}}
\caption{\footnotesize{
(a) Evolution of our understanding of the spin content of the nucleon [adapted from~\cite{Burkardt:2008jw}].
(b) Decomposition of the nucleon spin relevant for high-energy processes.
}}
\end{figure}

We have come a long way since the so-called \emph{spin crisis} (Fig.~\ref{fig:spin_crisis}), faced in the 
eighties when the measurements performed by the European Muon Collaboration (EMC) revealed that only a small 
fraction of the nucleon spin was generated by the spin of quarks~\cite{Ashman:1987hv}. The crisis
has evolved into a puzzle, which consists now in determining how large the different quark and gluon contributions
to the nucleon spin are, in disentangling them and in explaining them from first principles in QCD.

Recent experimental data have shown that the quark and antiquark spins account for only about 30\% of proton total 
longitudinal spin~\cite{Airapetian:2004zf,Airapetian:2007mh,deFlorian:2008mr,Adolph:2015saz} and the gluon contribution could be as large as 40\%~\footnote{The latter has 
only been measured in the region $x > 0.05$~\cite{Adamczyk:2014ozi,deFlorian:2014yva}. Additional studies at smaller $x$ 
are thus required.}. It is thus expected that a significant part of the proton spin arises from the 
transverse dynamics of quarks and gluons (\ie\ via ${\cal L}_{q}$ and ${\cal L}_{g}$), which has however not yet been measured.
This emphasises how crucial our understanding of the transverse motion of quarks and gluons inside the proton is,
in order to validate our current picture of the nucleon structure. 

In order to measure the parton OAM, one should in principle access observables which are sensitive to the parton position and momentum. 
This is the realm of the Generalised Parton Distributions (GPDs), relevant for exclusive processes. 
Yet, one can also indirectly access information on the orbital motion of the partons bound inside hadrons via Single Spin Asymmetries (SSAs) in different hard-scattering processes, in particular with a transversely polarised hadron (see~\cite{D'Alesio:2007jt,Barone:2010zz} for reviews). 
In these Single Transverse-Spin Asymmetries (STSAs), one can access left-right asymmetries in the parton distributions with respect to the plane formed by the proton momentum and spin directions. 
These asymmetries are naturally connected to the transverse motion of the partons inside the polarised nucleons. 

Historically large STSAs (also denoted $A_N$ or $A_{UT}$) have been observed in single 
forward $\pi$~\cite{Adams:1991cs,Arsene:2008aa,Abelev:2008af} and $K$ 
production~\cite{Arsene:2008aa} in high-energy $p^\uparrow p$ collisions at 
Fermilab and Brookhaven National Laboratory (BNL), towards the valence region. 
They have also been observed in Semi-Inclusive DIS (SIDIS) by the HERMES~\cite{Airapetian:2004tw} and COMPASS~\cite{Alexakhin:2005iw} collaborations. 
Studies to look for STSAs in $J/\psi$ production in $p^\uparrow p $ collisions at RHIC~\cite{Adare:2010bd} and $J/\psi$ leptoproduction on proton target by the COMPASS experiment~\cite{Matousek:2017yhq} have also been carried out. 
Intense theoretical works have resulted in a widely accepted picture according 
to which these (large) asymmetries, which were expected to vanish at high energies,
are due to re-scatterings of the quarks and gluons with the remnants of the 
hadron undergoing the  interaction~\cite{Brodsky:2002rv,Collins:2002kn,Brodsky:2002cx}. 
It is also accepted that they vanish for partons which do not carry any transverse momentum.

As for now, these STSAs can be treated via two dual approaches~\cite{Koike:2007dg}. 
The first is an extension of the collinear parton model of Feynman and Bjorken with the introduction of three-parton (Efremov-Teryaev-Qiu-Sterman) correlations~\cite{Efremov:1981sh,Efremov:1984ip,Qiu:1991pp}, namely the Collinear Twist-3 (CT3) formalism~\footnote{Recently, it has been found that STSA for pion production in proton-proton collisions may be dominantly driven by three-parton fragmentation functions~\cite{Kanazawa:2014dca,Gamberg:2017gle}.}. 
The second, called the Transverse-Momentum Dependent (TMD) factorisation (see e.g.~\cite{Collins:1981uk,Collins:1984kg,Collins:1989gx,Ji:2004wu,Collins:2011zzd,GarciaEchevarria:2011rb,Echevarria:2012js,Echevarria:2014rua,Bacchetta:2000jk,Mulders:2000sh,Boer:2016xqr,Angeles-Martinez:2015sea}), relies on a complete three-dimensional mapping of the parton momentum and encodes all the possible spin-spin and spin-orbit correlations between the hadron and its constituents.
In particular, the effect of the aforementioned re-scatterings, responsible for the STSAs, is encoded into the gauge links in the definition of the TMD PDFs.
A particular case is the so-called Sivers function~\cite{Sivers:1989cc,Sivers:1990fh}. 
Within the CT3 formalism, the re-scattering effects are explicitly considered in the hard-scattering coefficients.

Both formalisms have their proper range of applicability.
While the CT3 approach holds when one single hard scale is present in the process (such as the transverse momentum $p_T$ of a pion produced in hadronic collisions with $\lqcd < p_T$), the TMD formalism applies when there are two separate scales (such as the transverse momentum $q_T$ and the invariant mass $M$ of the lepton pair in DY production, with $\lqcd\leq q_T<M$).
Thus the CT3 approach is better suited to describe $A_N$ for inclusive hadron production~\cite{Qiu:1998ia} but can also be applied to DY pair~\cite{Hammon:1996pw,Boer:1997bw,Boer:1999si} 
or isolated photon~\cite{Qiu:1991wg} production. 
With regards to the \AFTER physics case, this mainly concerns 
heavy-flavour and quarkonium production, whose STSAs are extremely poorly known, if not unknown at all. 
TMD factorisation is usually used for processes where the transverse momentum of the 
initial partons is accessible owing to the absence of hard final-state radiations. In the case of \AFTER, 
this includes pseudo-scalar quarkonium, quarkonium-pair and other associated production of colourless particles, and of course the DY process.
With its high luminosity, a highly polarised target and an access towards the large momentum fraction, $x^\uparrow$,
in the target, \AFTER is probably the best set-up to carry out an inclusive set of $A_N$ measurements
both to improve existing analyses and to perform studies which would simply be impossible otherwise.
Let us recall that nearly nothing is known from the experimental side about the gluon Sivers function (see e.g. the review~\cite{Boer:2015vso} and recent measurements~\cite{Adare:2010bd,Adare:2013ekj,Adolph:2017pgv,Aidala:2018gmp}).
The polarisation of not only hydrogen but also deuterium and helium targets allows for an even more ambitious spin program bearing on the neutron and spin-1 bound states (see e.g.~\cite{Bacchetta:2000jk,Boer:2016xqr} and references therein).

As explained, the TMD approach allows us to investigate the structure of hadrons in a three-dimensional momentum space (see \cite{Angeles-Martinez:2015sea} and references therein) in a rigorous and systematic 
way. 
It is important to note that it is not restricted to the study of $A_N$. 
It can help us probe in a direct manner the transverse dynamics of the partons as well as their own polarisation in {\it unpolarised} nucleons. 
The latter in particular generates observable azimuthal asymmetries in the final state. 
These are related to another TMD PDF of great interest, the Boer-Mulders function~\cite{Boer:1997nt}: it 
describes the correlation between the quark transverse spin and its transverse momentum in an unpolarised hadron,
and it might help explain the well-known violation of the Lam-Tung relation~\cite{Lam:1980uc} in unpolarised DY reaction~\cite{Lambertsen:2016wgj}.
Its counterpart for the gluon content is the distribution of linearly polarised
gluons in unpolarised protons which affects, for instance, the $H^0$ transverse-momentum distribution at the LHC.
At \AFTER, such a distribution can be probed and extracted, for instance, via pseudo-scalar quarkonium production~\cite{Boer:2012bt,Signori:2016jwo} and associated quarkonium 
production~\cite{Dunnen:2014eta,Lansberg:2014myg}, whereas the quark Boer-Mulders functions can be accessed
in DY-pair production. 

It is important to note that all these measurements ($A_N$, azimuthal 
asymmetries or transverse-momentum spectra) have to be measured in $pp$ collisions as mandatory complementary pieces of information to analogous studies in lepton-induced reactions.
Indeed, contrary to usual PDFs, the TMD (PDFs) are not universal. 
For instance, some, as the quark Sivers function, are predicted to change sign when generating the STSA in SIDIS and in DY reactions. 
In other words, the time-reversal odd TMD distributions are process-dependent, but the process dependence is calculable (from the symmetry properties of the theory).
This feature, sometimes referred to as \emph{a generalised universality}, is driven by the nature of the re-scatterings generating the STSA, which can be from the initial or final states. 
More technically, they are connected to the gauge link in the definition of the 
TMD (PDFs)~(see \cite{Buffing:2012sz,Buffing:2013kca,Buffing:2015tja} for recent references). 
This ``sign change'' 
represents one of the most important predictions of the TMD factorisation and 
dedicated experiments have been proposed to check 
it~\cite{Klein:zoa,Quintans:2011zz,Adamczyk:2015gyk,Aschenauer:2016our}. Such a process dependence 
is also explicit in the CT3 formalism.  
Recently the first measurement in DY was performed by the COMPASS collaboration~\cite{Aghasyan:2017jop} and the first one on $W$ bosons by the STAR collaboration~\cite{Adamczyk:2015gyk}, hinting at the sign change.
With \AFTER, one could go further than the 
current proposals and perform quantitative tests 
of this generalised universality, deeply connected to the symmetries of QCD. 
\AFTER will be a unique place to probe such 
aspects of the parton transverse dynamics in the gluon sector, which requires 
even more experimental inputs both from lepton- and hadron-induced reactions than the quark sector. 
\AFTER will also provide a unique playground to explore in details the connection 
between the TMD and CT3 
approaches~\cite{Boer:2003cm,Ji:2006ub,Ji:2006vf,Ji:2006br,Koike:2007dg}, 
in particular in the gluon sector. This will open the way for a full 
three-dimensional mapping of the parton momentum and, in turn, for more insights 
on the orbital angular momentum of the quarks 
and gluons.

%% file: motivation/motivation-heavy-ion.tex
\subsection{The nuclear matter in new rapidity and energy domains}

One of the prime objectives of Heavy-ion (HI) physics at high-energy facilities is the search and 
characterisation of a novel state of matter where the quark and the gluons are deconfined. 
In this state, which should be the prevailing one of the Universe a few microseconds after the Big Bang, quarks and 
gluons roam nearly freely over distances of a few femtometers, \ie\ distances much larger than 
the hadron sizes in which they are normally confined.

The existence of such a state of matter is a natural consequence of the asymptotic freedom property of  QCD, 
whereby the strong interaction becomes weak at small distances and high momentum transfers. 
It is expected to be reached when the surrounding hadronic matter is extremely 
compressed or heated -- resulting in high momentum transfers in the system. These conditions 
can be achieved in ultra-relativistic collisions of nuclei and the resulting new phase can be observed 
using specific probes. Such probes are essentially of three kinds, namely radiated particles from the plasma itself 
(\ie\ photons), the destruction of heavy-quark bound states and the momentum-spectrum modification of various
particles.

\begin{figure}[!hbt]
\centering
\subfigure[~]{\includegraphics[width=0.4\textwidth]{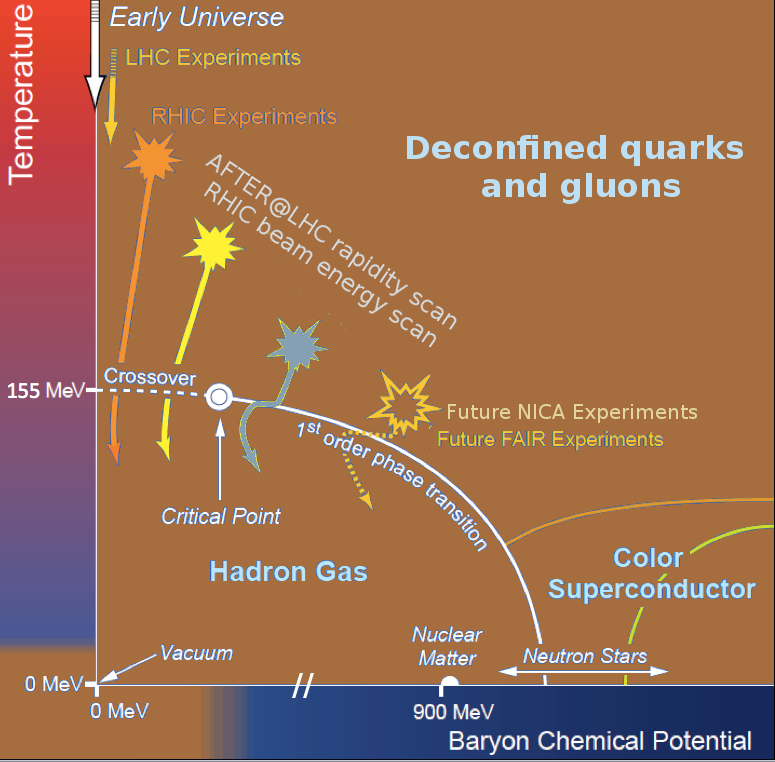}
\label{fig:phase_diagram}}
\subfigure[~]{\includegraphics[width=0.58\textwidth]{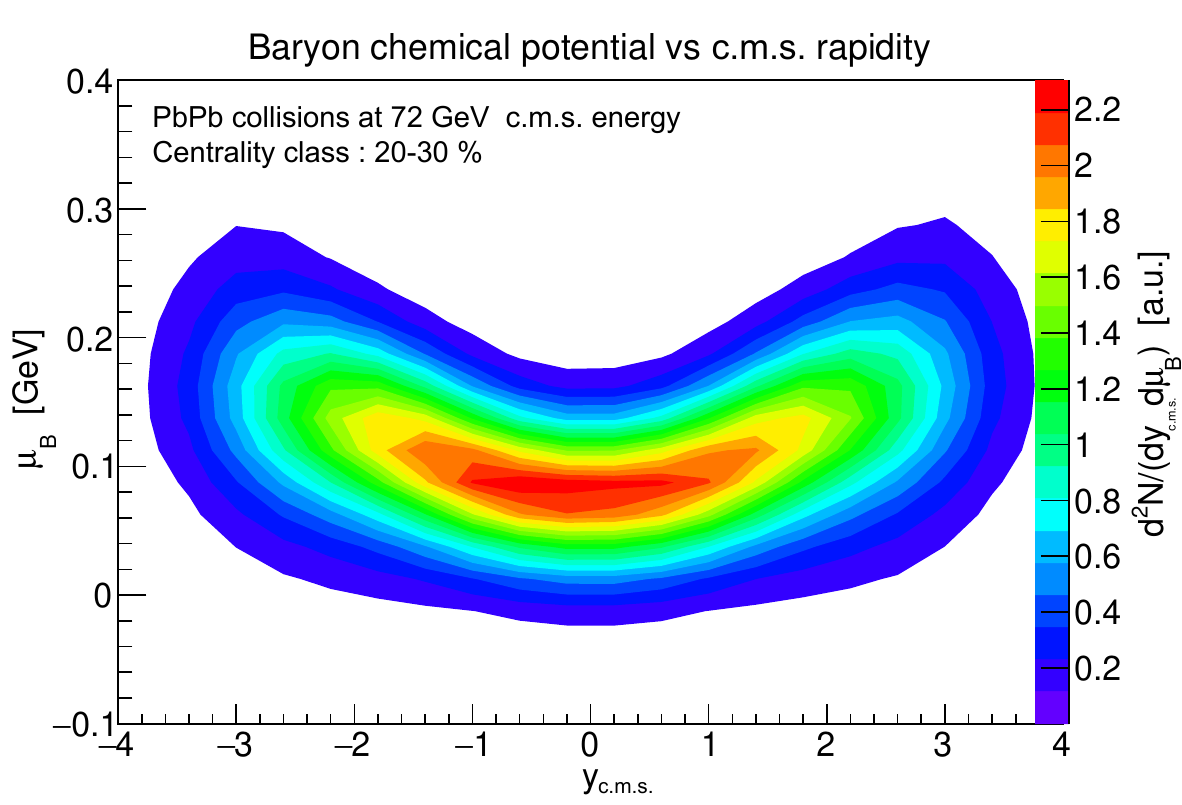}
\label{fig:muB_vs_y}}
\caption{
(a) Phase diagram of the strongly interacting matter and the reach of \AFTER\ HI programme. 
(b) The baryonic chemical potential $\mu_B$ as a function of the rapidity \ycms in mid-central \PbPb\ \sqrtsNN = 72 GeV collisions predicted by the viscous hydro+cascade model vHLLE+UrQMD [Adapted from~\cite{Karpenko:2018xam}]. The colour represents the differential density of produced particles as a function of \ycms and $\mu_B$.}
\end{figure}

\cf{fig:phase_diagram} shows the most up-to-date phase diagram of the 
strongly-interacting matter, gathering all our knowledge that was progressively acquired since
the very first relativistic heavy-ion collisions  -- nearly thirty years ago. With the LHC
multi-TeV heavy-ion beams (for now, lead and xenon ions with an energy per nucleon of 2.76 TeV), \AFTER\ 
with a \cms\ energy of 72 GeV provides a complementary coverage to the RHIC- 
and SPS-based experiments in the region of high temperatures and low baryonic chemical 
potentials where a Quark-Gluon Plasma (QGP) is expected to be produced. Moreover, model calculations indicate that the baryonic chemical potential $\mu_B$ and the temperature $T$ depend on the rapidity~\cite{Becattini:2007qr, Begun:2018efg,Karpenko:2018xam}. Figure~\ref{fig:muB_vs_y} shows an example of such a $\mu_B$ vs. \ycms relation for mid-central \PbPb\ \sqrtsNN = 72 GeV collisions by the vHLLE+UrQMD model~\cite{Karpenko:2018xam}. Measurements conducted as a function of \ycms will give access to different   $\mu_B$ and $T$ values. As such, the \AFTER\ HI programme can bear on a ``rapidity scan'' to study both the deconfined regime and the expected phase transition to the hadronic gas. It would be a new approach to investigate the QCD phase diagram, complementary to the RHIC Beam Energy Scan (BES) programme~\cite{Aggarwal:2010cw, Adamczyk:2017iwn}.

Let us also recall here a key advantage of the fixed-target mode, namely the possibility
to study different colliding systems with short transition periods while keeping high 
collected luminosities.\footnote{At the collider LHC, only 4 colliding systems have been studied 
during Run 1 and 2: \pp, $p$Pb, XeXe and \PbPb. Even though the RHIC complex
is more flexible in this regard, only \pp, \dAu, CuCu and AuAu collisions 
could be studied in the first 10 years of running with enough statistics to study 
heavy-flavour production. Upgrades were for instance needed to look at CuAu, UU, $p$Al, $p$Au or $^3$HeAu collisions. In comparison, the LHCb SMOG system (see section~\ref{subsubsec:Direct_gas}) --which could only so far 
take data in small periods-- already collected data 
for 5 systems ($p$He, $p$Ar, $p$Ne, PbAr and PbNe).} Another key advantage is the 
obvious capacity to instrument the nucleus-target region, namely  \ylab\ or \etalab close to 0:  
this would be the reach of most of the collider detectors used in the fixed-target mode.
Based on these assets, the physics case for HI physics with \AFTER\ can be outlined as follows.

Given the \cms-energy range, HI measurements at \AFTER\ have the potential to provide us with crucial 
information about the QGP properties and the nature of the phase transition to the hadronic gas regime. To do so, three experimental degrees of freedom
are at our disposal: (i) scanning the longitudinal extension of this hot medium, (ii) colliding systems of different sizes
and (iii) analysing the centrality dependence of these collisions. To our knowledge, no other experimental programme 
could fully rely on these three variables. Together, they should give us a unique 
lever arm to probe the hot medium at different enough temperatures and energy densities.
They will allow one, for instance, to probe the temperature dependence of the shear viscosity to entropy density ratio ($\eta/s$) of the created matter by measuring the rapidity dependence of the anisotropic flow, as it was recently shown in~\cite{Denicol:2015nhu} (see~\cf{fig:vn_rapidity}).

\begin{figure}[!hbt]
	\centering
    \subfigure[~]{ \includegraphics[width=0.48\textwidth]{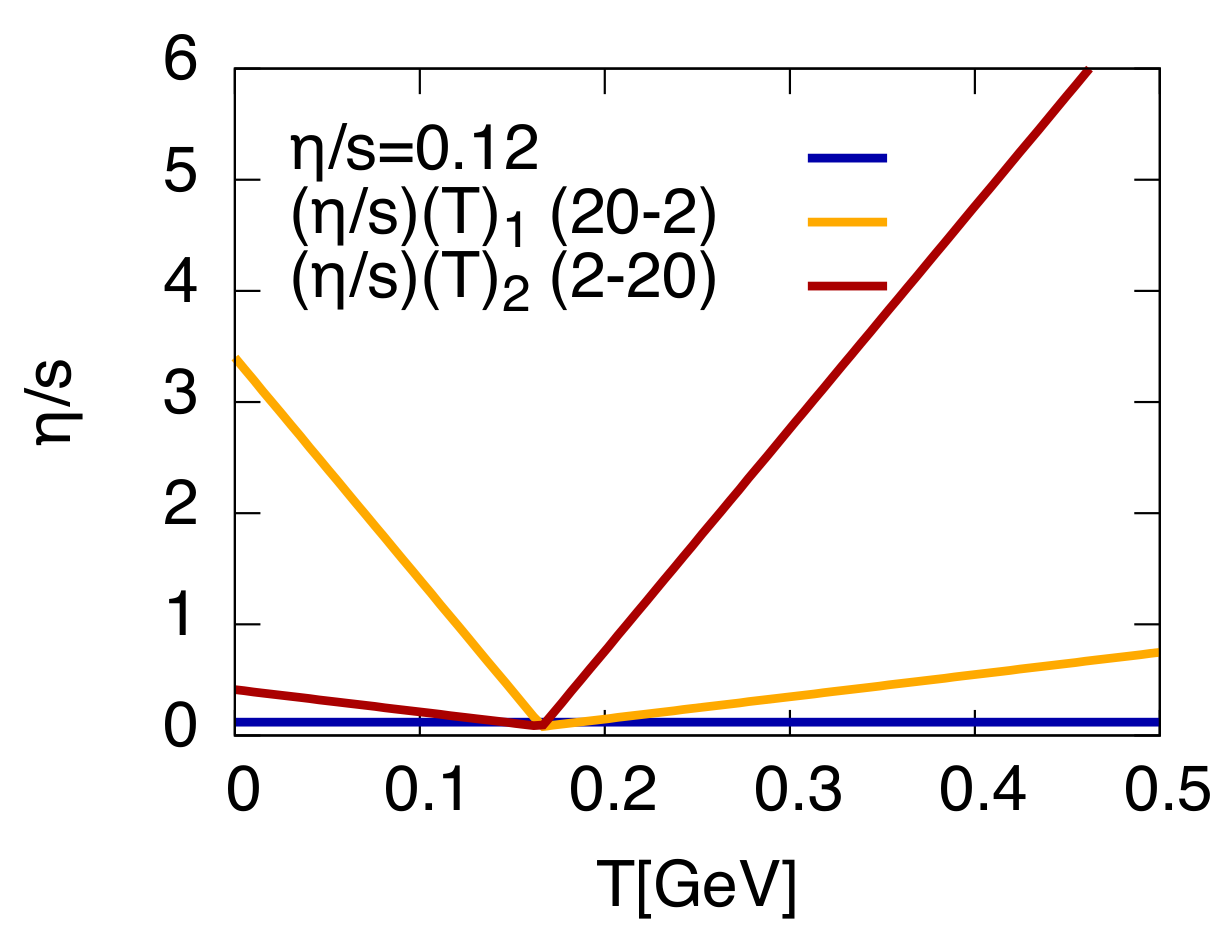} }
	\subfigure[~]{ \includegraphics[width=0.48\textwidth]{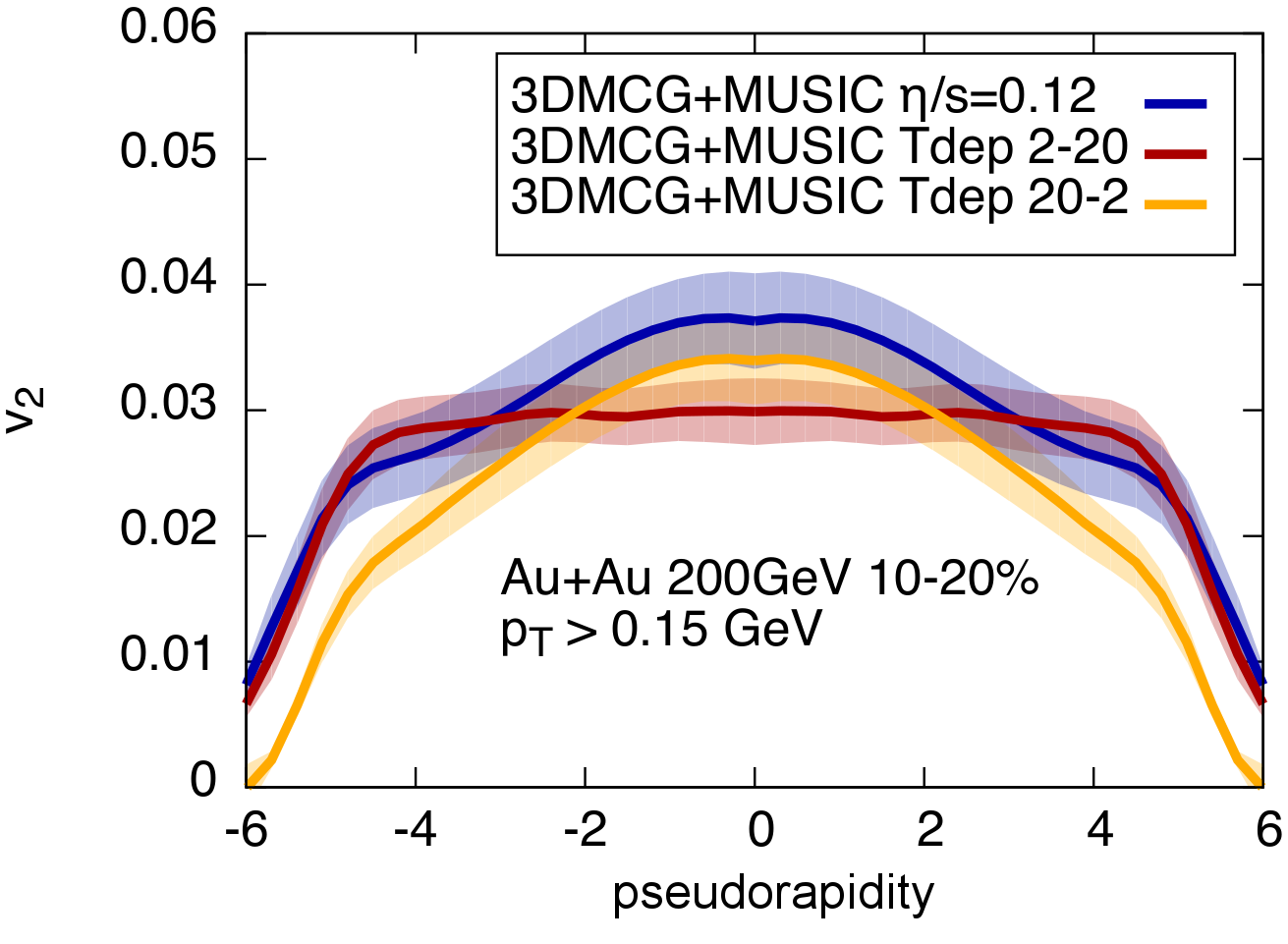} }
	\caption{\label{fig:vn_rapidity} (a) Different scenarios of the temperature dependence of the ratio of the medium shear viscosity to the entropy density ($\eta/s$) at $\mu_{B}=0$; (b) Sensitivity of the second Fourier coefficient ($v_2$) of the azimuthal asymmetry measured as a function of the pseudorapidity with ($\eta/s$) and the temperature given by 3D+1 viscous hydrodynamic calculations. [Adapted from~\cite{Schenke:QM2015}].}
\end{figure}

More generally, one can perform a tri-dimensional tomography of the hot medium produced 
in  ultra-relativistic HI collisions at $\sqrtsNN=72$ GeV. Such a tomography will rely on a set of 
precise measurements in a kinematical region ($y_\cms$ and $\sqrtsNN)$ which could so 
far only be studied with low accuracy and with an extremely limited set of probes~\cite{Lourenco:2006vw,Rapp:2008tf}. They will give us new information on the QGP properties in the longitudinal dimension and help us settle
long-standing debates about probes such as heavy-quark(onium) production in the range from SPS to RHIC energies~\cite{Andronic:2015wma,Kluberg:2005yh,Adcox:2004mh}. These
issues admittedly hinder their potential as golden probes of the QGP~\cite{Matsui:1986dk}.

Measurements at $y_\cms \simeq 0$ are also at reach with detector coverages about $\eta_\lab \simeq 4.0$. 
Studying an extensive set of hard probes in this region, where the yields are the highest,
is meant to provide us with a very reliable calibration of the properties of the system (temperature, viscosity)
in order to initiate the scan in the longitudinal direction via the rapidity dependence.

%% file: implementation-fixed-target/implementation-fixed-target.tex
\section{How to make fixed-target collisions at the LHC?}
\label{sec:implementation}
\subsection{Overview}

\input{implementation-fixed-target/Overview.tex}

\subsection{Internal gas-target solutions}
\label{subsec:Internal_gas}

\subsubsection{SMOG: a feasibility demonstrator}
\label{subsubsec:Direct_gas}
\input{implementation-fixed-target/contributions-Direct-Injection.tex}

\subsubsection{Gas-jet target}
\label{subsubsec:gas_jet}
\input{implementation-fixed-target/contributions-Gas-Jet.tex}

\subsubsection{Storage-cell gas target} 
\label{subsubsec:storage_cell}
 
\input{implementation-fixed-target/contributions-Storage-Cell.tex}
 
\subsection{Internal solid target intercepting the beam halo}
\label{subsec:internal_solid}

\input{implementation-fixed-target/contributions-Solid-Target.tex}

\subsection{External/internal target solution with a slow beam extraction using a bent crystal}
\label{subsec:external_target}

\subsubsection{Crystal-assisted extraction of the LHC beams}
\label{crystal}

\input{implementation-fixed-target/contributions-Bent-Crystal.tex}

\subsubsection{Unpolarised targets}
\label{sec:unpol-targets}

\input{implementation-fixed-target/contributions-Unpolarised_Targets.tex}

\subsubsection{Polarised targets}
\label{polarisation}

\input{implementation-fixed-target/contributions-Combined_Andi_Norihiro.tex}

\subsection{Comparison of technologies}
\label{comparison}

\input{implementation-fixed-target/contributions-comparison.tex}

%% file: implementation-fixed-target/Overview.tex
Several technological options are currently under investigation to perform dedicated fixed-target experiments at the LHC. One can indeed initiate collisions of the LHC beam particles with nucleons or nuclei at rest~:
\begin{itemize}
\item{by letting the full LHC beam go through a gas target in the LHC beam pipe,}
\item{by extracting halo particles by means of a bent-crystal deflector onto a target positioned outside the beam pipe with a dedicated beam line or inside the beam pipe,} 
\item{or by placing a wire target intercepting the faint beam halo in the beam pipe.}
\end{itemize}

The aim of this section is to summarise the advantages, the performances as well as the challenges of each solution. We first discuss them 
and then compare their performances and limitations.

The LHCb collaboration with its SMOG~\cite{FerroLuzzi:2005em} and VELO~\cite{LHCbVELOGroup:2014uea} systems has demonstrated~\cite{Maurice:2017iom} that gas injection within a certain range is tolerable and does not lead to vacuum instabilities thus paving the way toward a genuine gas target with the polarisation of light nuclei with higher densities, as well as for heavy noble gases. It allows one to use an existing LHC detector, resulting in limited costs and a relatively short time-scale installation. As we will discuss in the following sections, the acceptances of the ALICE and LHCb detectors are, in general, well suited for data taking in the fixed-target mode. 

A long narrow tube --commonly referred to as a storage cell~\cite{0034-4885-66-11-R02}-- placed on-axis of the LHC beam and fed by a polarised or unpolarised gas clearly presents an interesting opportunity to reach high luminosities for the gas-target option (see section \ref{subsubsec:storage_cell}). Such a configuration allows for different beam-target combinations including highly polarised ones -- pending the cell-coating compatibility with the LHC vacuum constraints. The possibility for a density calibration of the storage-cell targets has still to be studied. If successful this would, together with the known LHC beam current, allow one to measure absolute luminosities and cross sections.
For a given (areal) target density, the storage cell requires the lowest -- but still sizeable -- gas flow into the LHC vacuum system. 

Alternatively, higher gas fluxes can directly be injected orthogonally to the beam (see section \ref{subsubsec:gas_jet}). This is how the RHIC H-jet polarimeter~\cite{Zelenski:2005mz}, with (highly) polarised injected hydrogen, operates. In general, this leads to lower luminosities than the storage-cell option. However, if used in conjunction with the ALICE detectors, the luminosities obtainable with a gas-jet system would already be reaching the limit of the detector data-taking capabilities. In the case of polarised $^3$He and other unpolarised gases, more intense sources can in principle be used to compensate for the smaller target-areal density. The advantage of this option is the very high reachable polarisation of the target and the limited need of R\&D for an installation in the LHC complex.

On the other hand, bent crystals are being developed as part of the collimators protecting the machine~\cite{Scandale:2016krl}. They may also be placed near the beam in order to deflect halo particles and guide them onto an external target (see section \ref{crystal}). Beside thick unpolarised targets, cryogenic polarised targets could be employed. This approach involves a considerable amount of civil engineering, including a new cavern and a new detector. Another possibility based on such a bent crystal is to directly use the deviated particles of the halo on an internal target system, inside the beam pipe of an existing LHC experiment. The feasibility of such a solution is currently explored by the UA9 collaboration at CERN~\cite{Montesano:2018}. Open problems in this case are how to dump the deflected beam, and how to get a suitable polarised target.

Finally, a wire or foil target may be placed in the halo of the LHC beam~\cite{Kurepin:2011zz} in order to provide collisions at a near-axis position (see section \ref{subsec:internal_solid}). This method has been employed at HERA-B~\cite{Ehret:2000df} and STAR~\cite{Meehan:2016qon} and is particularly useful for heavy-nucleus targets. However, it may affect the main LHC beam and this solution will require dedicated simulations. In addition, it is probably not compatible with light-nucleus targets --certainly not hydrogen ones-- and basically prevents one to perform any direct luminosity measurement. 

In the next sections, the aforementioned options are detailed. In section \ref{comparison}, the Figure-of-Merit (FOM) for collisions with unpolarised and polarised target is defined and numerical values are given, allowing for a comparison of performances. A qualitative summary table of the performances of the various technological solutions to initiate fixed-target collisions at LHC is also discussed. Finally, expected integrated luminosities for each solutions are compared.

\subsection{Relevant LHC parameters and definitions}
\label{lhc_param}

Let us first recall (in \ct{lhc_param_tab}) some nominal LHC parameters~\cite{Bruning:2004ej,Benedikt:2004wm} 
as well as other generic quantities  which have been used in the comparison of the various 
technological solutions.

\begin{table}[!htb]
\center\renewcommand{\arraystretch}{1.2}
\begin{tabular}{c | c | c || c} 
      &  Proton beam & Lead beam  & Upgraded lead beam \\ \hline \hline
  Number of bunches in the LHC ($N_{b}$) & 2808 & 592 & 1232\\ \hline
  Number of particles per bunch ($N_{p}$)    & 1.2 $\times$ 10$^{11}$ & 7.0 $\times$ 10$^{7}$ & 1.8 $\times$ 10$^{8}$ \\ \hline
  LHC Revolution frequency ($\nu$) [Hz] & \multicolumn{3}{c}{11245}   \\ \hline
  Particle flux in the LHC ($\phi_{\text{beam}}$) [s$^{-1}$] & $3.6 \times 10^{18}$ & $4.7 \times 10^{14}$ & $2.5 \times 10^{15}$  \\ \hline
  LHC yearly running time ($\Delta$t)~[s] & 10$^{7}$  & 10$^{6}$ & 10$^{6}$\\ \hline
  Nominal energy of the beam (E$_{\text{beam}}$) [TeV]  & 7 & 2.76 & 2.76 \\ \hline
  Fill duration considered ($\Delta\tau$)~[h] & 10 & 5 & 5 \\ \hline 
  Usable particle flux in the halo ($\phi_{\text{ usable halo}}$)~[s$^{-1}$]& 5 $\times$  10$^{8}$ & 10$^{5}$ & $5\times 10^{5}$ \\ \hline
\end{tabular}
\caption{LHC-related quantities used in the calculations in the following sections along the expected parameters 
for an upgraded Pb beam mode. 
The usable particle flux in the halo is assumed to be half of the estimated beam losses in the proton and lead beams. The LHC yearly running times quoted are maxima.}
\label{lhc_param_tab}
\end{table}

Beside these nominal running conditions, the possibility of $p$ runs at the Pb 
energy for reference measurements is also possible as already done during the 
Run 1 and 2. These runs typically last one week, \ie\ ${\cal O}(10^5)$ seconds.
When relevant, these run durations will be specified. We also note that other 
beam species (Kr, Ca, O) can be considered for an injection in the LHC (in particular those used in the SPS, \ie\ Xe and Ar), 
as it was the case with the short XeXe run which took place by the end of 2017. The injection of other species, like He, would require
another ion source than the current one. In most cases, the instantaneous luminosities should be equal or
larger than those for Pb beams (see later).

Let us now recall useful quantities to compare the various technological implementations. First, 
we define the instantaneous luminosity, $\cal{L}$, in terms of the particle flux, $\phi_{\text{projectile}}$, impinging the target of
areal density $\theta_{\text{target}}$ of nuclei:
\begin{equation}
\cal{L} = \phi_{\text{projectile}} \times \theta_{\text{target}}.
\end{equation}
If $\theta_{\text{target}}$ is expressed in atoms $\times$ cm$^{-2}$, $\cal{L}$ is naturally expressed in cm$^{-2}$s$^{-1}$.
For a gas target, the flux is $\phi_{\text{beam}}$, the particle flux in the LHC. For a solid wire target or a solid target
put into an extracted beam, it is $\phi_{\text{usable halo}}$, the usable particle flux in the halo, which is on the order
of the beam losses (or less) to allow for a parasitic-working mode.

To compare the performances of the target for spin-related measurements, in particular Single-Spin Asymmetries (SSAs) discussed
in section \ref{section:Spin-Physics}, 
we define $\cal{F}$, the spin figure of merit of the target (and beam)
\begin{equation}
{\cal F} ={\cal P}^{2}_{\text{eff}} \times {\cal V}.
\end{equation} 
where  ${\cal P}_{\text{eff}}$, which we call the effective polarisation of the target,
contains the information about the polarisability
of the material and ${\cal V}$ contains the information about the rates (up to the cross section of the considered reaction).
It is important to understand that our comparisons will involve different target systems and materials, with different beam fluxes. A figure of merit only accounting for the target properties is therefore not sufficient.

Depending on the target material type, ${\cal P}_{\text{eff}}$ can be expressed as:
\begin{equation}
{\cal P}_{\text{eff}} = P_{T} \times f \textrm{     or     } {\cal P}_{\text{eff}} = P_{T} \times \alpha,
\end{equation} 
where $P_{T}$ is the polarisation of the nucleons in the target, $f$ a dilution factor  and $\alpha$ a depolarisation factor.
The dilution factor $f$ is in principle defined as the fractions of {\it events} scattered off the polarisable nucleons in the target. As such, it can be approximated to the fraction of the polarised nucleons in the molecules constituting the target. However, whereas this approximation is sufficient to compare similar systems, it may not encompass nuclear or isospin effects which depend on the kinematics or the probe to be studied. Following the former definition, one has~\cite{Adams:1999qy}
\begin{equation}
f=\frac{n_N \sigma_N}{n_N \sigma_N+\sum_i n_{A_i} \sigma_{A_i}},   
\end{equation} 
where $n_{N,A}$ are respectively the number of polarised nucleons and of nuclei $A$ in the target. 
 $\sigma_{N,A}$ are the corresponding cross sections for the considered observables. Because of nuclear effects, we know that $\sigma_{A} = A \times \sigma_{N}$ does not hold in general and approximating $f$ to the fraction of polarisable nucleons is not strictly correct. In the case of DIS at large $x$, the variation of such corrections arises from the EMC effect which results in visible variations of $f$~\cite{Adams:1999qy} as a function of $x_B$. As a matter of fact, the experimental derivation of $f$ amounts to measuring the cross sections off the different constituents of the target. When this is not possible, such ratios can be evaluated based on theoretical computations. In the following, we will only refer to an average value, denoted $\langle f \rangle$, computed from the fraction of polarisable nucleons.

Since the precision of a given measurement does depend on the counting rates, it is important to realise
that the luminosity ${\cal L}$ does not account for the number of the {\it nucleons} per atom or molecule in the target, but instead for that of the {\it nuclei}. For instance, at the LHC in the collider mode, similar yields are obtained in $pp$ and $p$Pb collisions when the luminosity for the $pp$ case is about 200 times larger than that of the $p$Pb case --thus not at all at similar luminosities.
This should in principle be reflected in a figure of merit. However, the luminosity is not a good one for the counting rates when different colliding systems are considered.

 In the case of "rare" or "hard" processes (such as quarkonium, heavy-quark or DY pair production), each nucleon approximatively contributes additively ($\sigma_{A} \sim A \times \sigma_{p}$) and ${\cal V}$ can be expressed as :
\begin{equation}
{\cal V} = {\cal L} \times \sum_i A_i
\end{equation} 
where $\sum_i A_i$ is the sum of the atomic masses of the target constituents (for instance 1 for H, 3 for $^3$He and 17 for NH$_3$)\footnote{Note that $\sum_i A_i$ does not reduce to the sum of the atomic masses of the target constituents when $\theta_{\text{target}}$ is expressed in {\it atoms} per cm$^2$ for {\it molecular} gases, such as H$_2$.}. 
In the case of SSAs of light hadrons, where the yields do not follow the binary-collision scaling in a single nucleus ($\sigma_A$ strongly deviates from $A\times \sigma_p$), or for collisions induced by a nuclear beam, there is no simple equivalent for ${\cal V}$ and the rates should be computed in order to perform meaningful comparisons between polarised targets made of different species. It is of course also the case if the beam energy happens to be significantly different such as to induce very different production cross sections.

Only for hard probes and when nuclear effects can be neglected, it follows that ${\cal F}$ defined as above is inversely proportional to the time needed to reach a fixed precision on SSAs. In other words $\cal{F}$ is inversely proportional to the variance
of the SSAs. 

Since different set-up cannot be used during the same amount of time over
a LHC year, it also obviously useful to make comparisons at the level of  $\int dt\, {\cal L}$ 
and $\int dt\, {\cal F}$. The yearly run duration are taken to be $10^7$ s for 
the proton beam and $10^6$ s for the lead beam. These are maxima.

%% file: implementation-fixed-target/contributions-Direct-Injection.tex
The direct injection in the LHC beampipe (\ie\ in the VELO vessel) of noble gases at a pressure on the order of 10$^{-7}$ mbar is already being used by the LHCb collaboration with the SMOG (System for Measuring the Overlap with Gas) device. Initially developed inside LHCb to allow for a precise determination of the luminosity with an uncertainty below 4\%, SMOG is a system whereby a gas can be injected inside the beam vacuum at the interaction point. The luminosity for the collider mode is then determined thanks to a Beam Gas Imaging (BGI) method, which relies on the interaction vertices between the circulating beam and the gas present at the interaction point \cite{FerroLuzzi:2005em,Barschel:2014iua,Aaij:2014ida}.

The system and its technology have been extensively tested. A pilot run of $p$ beam (Pb beam) on a Neon gas target was successfully performed in 2012 (2013) at a c.m.s energy of $\sqrt{s_{NN}}$ = 87 GeV (54 GeV). This first SMOG campaign was followed by several successful data taking periods in 2015-2017, for which the injection of other gases than Neon was explored. The system is currently limited to noble gases, namely He, Ne, Ar and, possibly, to Kr and Xe which have not yet been tested. Their impact on the machine is currently under discussion with the LHC vacuum experts. The limitation to noble gas is to avoid altering the Non-evaporable-getter (NEG) coating properties of the beam pipe and this is why SMOG is equipped with a NEG cartridge to ensure the purity of injected gas. These noble gases can thus travel from the injection point (IR8) to the ion pump stations at $\pm 20$ m and some can reach the warm-to-cold transitions of the Q1 magnets, where they can accumulate during extended periods of injection.  The beam-induced effects due to gas cryosorbed on those surfaces is still being investigated along the successive tests.  However, in 2015, Ar was injected in LHCb for about one week in a row, during the heavy-ion run which took place before the Year End Technical Stop (YETS). No decrease of the LHC performances was observed. It therefore opens good perspectives for data taking periods of up to a month per year without additional pumping systems. It would preferably take place at the end of the year, while the beam intensity is low, and before the YETS which could permit to get rid of the accumulated gas if needed.  
With the current SMOG setup, the gas pressure is about 1.5 $\times$ 10$^{-7}$ mbar, \ie\ two orders of magnitude higher than the LHC vacuum pressure ($\sim 10^{-9}$ mbar). The pressure might be increased up to 10$^{-6}$ mbar without severe hardware changes along with the support of the LHC vacuum experts. An estimate of the instantaneous luminosity obtained with the SMOG device in p-gas and Pb-gas collisions is given in \ct{tableSMOG}. 

\begin{table}[!hbt]
\renewcommand{\arraystretch}{1.2}
\begin{tabularx}{\textwidth}{ c | C | c | c | c | c || C | c}
Beam & Target & $P$ &  $\ell$ & $\theta_{\rm target}$ & $\mathcal{L}$  &  $\sigma^{\rm had}_{\rm beam-target}$ & \it{b}  \\
           &                   &    [mbar]   &  [cm]  & [atoms $\times$ cm$^{-2}$]  & [cm$^{-2}  \times$ s$^{-1}$]  & [barn] &  [\%]  \\           
                &                   &     &  &  &  &  &    \\        
                             &                   &     &  &  &  &  &   \\  
\hline\hline
$p$  & { \{He, Ne, Ar, Kr, Xe\}} & 1.5 $\times 10^{-7}$ & 40  & {$1.6\times 10^{11}$} & {$5.8\times 10^{29}$}    & {\{0.2, 0.6, 1.0, 1.6, 2.2\}}  & $  \ll  $ 0.1 \\
\hline
Pb   & { \{He, Ne, Ar, Kr, Xe\}}   & 1.5 $\times 10^{-7}$ & 40  & {$1.6\times 10^{11}$} & {$7.4\times 10^{25}$}   & {\{4.5, 7.0, 8.4, 10, 12\}}  & $ \ll$  0.1  \\


\hline
\end{tabularx}
\caption{ Beam type, target type, gas pressure ($P$), usable gas length ($\ell$), target areal density ($\theta_{\rm target}$), instantaneous luminosity ($\mathcal{L} = \phi_{\text{beam}}\theta_{\text{target}}$), hadronic beam-target cross section ($\sigma^{\rm had}_{\rm beam-target}$) and beam fraction lost over a fill ($\it{b}$) for a SMOG-like device. The areal density can be expressed for perfect gases as $\theta_{\rm target} ({\rm cm}^{-2}) =  P  \ell  \mathcal{N_{A}} / 22697$ where $P$ is in bar, $\ell$ is in cm and $\mathcal{N_{A}}$ is the Avogadro number in mol$^{-1}$. The usable gas zone assumed is $\ell$~=~40~cm. We considered the nominal parameter from \ct{lhc_param_tab} for the beam energy, the particle flux in the LHC and the fill duration. The obtained instantaneous luminosities should therefore be considered as maxima. The hadronic cross-sections used in this section to compute the beam fraction lost over a fill have been obtained using the EPOS MC generator~\cite{Pierog:2013ria}. The values are compatible with those from Fluka generator~\cite{Ferrari:2005zk} within 6$\%$.}
\label{tableSMOG}
\end{table}

The beam fraction consumed over a fill is negligible. 
There are prospects to replace SMOG with a multigas system allowing to change the type of injected gas without human intervention onsite. As for now, the gas pressure is not well known with a good precision. The installation of a calibrated Vacuum Gauge Ion (VGI) 6 m from the VELO has been performed during the YETS of 2016-2017. 
The luminosity is estimated by the parameters of the beam (number of bunches and bunch intensity), as well as the gas-target pressure measured by the pressure gauge. The uncertainty is dominated by the measured pressure, which varies as a function of the $z$ position along the beam. The luminosity is also determined from the yield of electrons scattering off the target atoms, with a precision of about 6\%~\cite{Aaij:2018svt}. 
In addition to the luminosity determination challenge, it is also worth noting that one has to cope with colliding-bunch events if the detector is also used in the collider mode. Fixed-target heavy-flavour analyses are currently limited to the analyses of non-colliding bunch events, with a vertex position within -20 and 20 cm\footnote{The vertex requirement is extended to $-70 < z < 10$ cm in ~\cite{Aaij:2018svt}.}, \ie\ in a region where the detector efficiencies are mostly constant with respect to the $z$-vertex position~\cite{LHCb:2017qap}.
Such requirements affect the effective recorded luminosities. One has to keep in mind that besides the limitation of the SMOG system itself in terms of the delivered gas pressure, the contamination of the collider events by fixed-target events also has to be considered as a constraint on the maximum pressure which can be delivered to LHCb. This consideration applies only in parasitic operation mode (during \pp runs), in order not to affect the current main LHCb B physics programme.

%% file: implementation-fixed-target/contributions-Gas-Jet.tex
Out of all the proposed solutions, the internal gas-target cell is a genuine solution to make fixed target at the LHC, at limited cost by re-using an existing LHC experiment, and with a variety of polarised and unpolarised gas targets.
From the experience gained with SMOG, and further tests of gas injection in Long Straight Section 4 (LSS4) and near ALICE, ATLAS and CMS, we consider the option of direct-injection of gas in the LHC beampipe as a viable solution to be run parallel to the collider mode. As highlighted above, the SMOG system is however not optimised and faces some limitations: the gas pressure --thus the target density-- is limited, only noble gases can be injected, the heavier noble gases may only be used before long YETS and the injection periods are currently significantly limited. Last but not least, a SMOG-like system does not allow one to inject polarised gases.

Some of these limitations can be lifted by the installation of specific pumping systems, which would however reduce the portability of the system\footnote{A SMOG-like system essentially reduces to a gas bottle, some valves and pressure gauges, a NEG filter and a small capillary to the beam pipe.}. In addition, a specific injection system would allow one to inject heavier noble gases as well as a jet of polarised gases such as H, D and $^3$He. One illustrative example of such an option is the H-jet system~\cite{Zelenski:2005mz} used on the BNL-RHIC collider as a proton-beam polarimeter (see Appendix \ref{appendix:Gas_jet}). In short, it offers much higher target densities than the current SMOG system, opens the possibilities for highly polarised target and can be coupled to a collider.

Let us briefly describe its main characteristics: it consists of a free atomic beam vertically crossing the collider beam at a speed of approximately 1560 $\rm{m\cdot s^{-1}}$. With the current Atomic Beam Source (ABS)~\cite{Zelenski:2005mz}, operated at 70~K, the polarised H inlet flux was measured to be $(1.24 \pm 0.02) \times 10^{17}$ H$^\uparrow$ s$^{-1}$. With a redesigned system, it may be doubled~\cite{Zelenski:private}. Similar numbers should be reachable for a polarised deuterium.  Higher fluxes can easily be obtained with polarised $^3$He but would require a dedicated system~\cite{Zelenski:private}. Using a deuterium target would require proper RF cavities which may prevent optimising both hydrogen and deuterium target performances with a single system. 

At the interaction point, the H-jet target profile is nearly Gaussian with a full width at half maximum of 5.5 mm. This is a significant advantage compared to a SMOG-like system where the gas diffuses along the beampipe and results in beam-gas collisions over distances of a few meters. A vacuum in the RHIC ring of $2 \times 10^{-8}$ mbar 
was  achieved with the H-jet system in operation at its nominal $10^{17}$ atoms s$^{-1}$ inlet flux thanks to  turbomolecular pumps (TMP) with a 1000~l/s pumping speed. The current H-jet system size is 375 cm high (225 cm for ABS above and 150 cm for Breit-Rabi polarimeter below the beampipe), 110 cm wide and 70 cm long. 
A redesigned system could be made more compact down to 200 (75) cm high above (below) the beam pipe. The current free ABS corresponds to a target areal density of $(1.2 \pm 0.2) \times 10^{12}$ H$^\uparrow$.cm$^{-2}$.

The H-jet system is designed to be moveable. It can be uninstalled and reinstalled in 2-3 days. It is coupled to a Breit-Rabi polarimeter monitor to measure with a high accuracy the atomic hydrogen 
polarisation which is as high as 0.96. The gas jet is however contaminated by 
hydrogen atoms bound into unpolarised proton molecules which slightly 
dilutes the average jet-proton polarisation down to $P^{\rm H-jet} = 93\%$~\cite{Okada:2005gu}.

   It was recently demonstrated~\cite{Poblaguev:2017PSTP} that the molecular hydrogen atomic mass fraction is about 0.4\% in the jet center. However since the molecular-hydrogen distribution is a factor 30-40 wider than the jet one, an integral dilution of the jet polarisation may be as large as 0.85. Detecting recoil protons from elastic $pp$ scatterings, it was possible to reconstruct the $z$-coordinate of the vertex which allowed the {\it in-situ} normalisation of the molecular-hydrogen contribution and its proper subtraction. For the actually used  event-selection cuts, the effective jet-target polarisation was found to be $95\pm 0.5 \%$. To employ this method of the jet-polarisation control, the vertex $z$-coordinate has to be measured with an accuracy of $\sigma_z\lesssim\ 1$ mm. We further note~\cite{Zelenski:2005mz} that the polarisation could reach 96-98\% if coupled with a higher holding magnetic field than the present one of 0.14 T. For the H-jet system, the field strength was also chosen to minimise the bending of the scattered recoil protons for the use as a polarimeter.

The gas-jet parameters, the instantaneous luminosities using typical LHC beam currents, and the beam fraction consumed over a fill are shown in \ct{tablegasjet}, for each type of beam-gas collisions. 

\begin{table}[!hbt]\center\renewcommand{\arraystretch}{1.2}
\begin{tabularx}{\textwidth} { c | c | c | c | c || c | c }
Beam & Target &  $\varphi^{\rm inlet}_{\rm target}$ & $\theta_{\rm target}$ & $\mathcal{L}$  &  $\sigma^{\rm had}_{\rm beam-target}$& $\it{b}$ \\ 
 & &  [atoms $\times$ s$^{-1}$]& [atoms $\times$ cm$^{-2}$]&  [cm$^{-2} \times$ s$^{-1}$] & [barn] & [$\%$]\\
\hline\hline
$p$  & H$^\uparrow$ &  $1.3\times 10^{17}$ & $1.2\times 10^{12}$ &  $4.3 \times 10^{30}$ & $48 \times 10^{-3}$ & $ \ll  $ 0.1 \\ 
$p$  & D$^\uparrow$ &  $1.3\times 10^{17}$ & $1.2\times 10^{12}$ & $4.3\times 10^{30}$ & $90 \times 10^{-3}$ & $ \ll  $ 0.1  \\
$p$  & $^3$He$^\uparrow$ &  $1.0\times 10^{19}$ & $1.0\times 10^{14}$ & $3.6\times 10^{32}$  & $12 \times 10^{-2}$ & 0.5 \\
\hline
$p$  & H$_2$ &  $1.1\times 10^{20\div 21}$ & $1.0 \times 10^{15\div16}$ &  $3.6 \times 10^{33\div34}$ & 48 $\times 10^{-3}$ & $1.9 \div 19$ \\ 
$p$   & Xe         & {$(1.0\div 5.0)\times 10^{18}$}  &{$(1.0\div5.0)\times 10^{13}$} &   {$(3.6\div18) \times 10^{31}$} &  2.2  &  $0.9\div4.4$ \\
\hline
Pb   & H$^\uparrow$ &  $1.3\times 10^{17}$  & $1.2\times 10^{12}$ & $5.6\times 10^{26}$   & 3.0   & $ \sim  $ 0.1  \\
Pb   & D$^\uparrow$ &  $1.3\times 10^{17}$  & $1.2\times 10^{12}$ & $5.6\times 10^{26}$   & 4.0   & $ \sim  $ 0.1  \\
Pb   & $^3$He$^\uparrow$ &  $1.0\times 10^{19}$  & $1.0\times 10^{14}$ & $4.7\times 10^{28}$   & 4.3   & 8.7  \\
\hline
Pb   &  H$_2$ &  $6.5\times 10^{16}$  & $2.5\times 10^{14}$ & $1.2\times 10^{29}$   & 3.0  & 15  \\
Pb   & Xe  & {$(1.0\div 5.0)\times 10^{18}$}  &{$(1.0\div5.0)\times 10^{13}$} &   {$(0.5\div2.3) \times 10^{28}$} &  12  &  $2.6\div12$ \\
\hline
\end{tabularx}
\caption{
Beam type, target type, inlet flux ($\varphi^{\rm inlet}_{\rm target}$), target areal density ($\theta_{\rm target}$), instantaneous luminosity ($\mathcal{L} =  \phi_{\rm beam} \theta_{\rm target}$ ), hadronic beam-target cross section ($\sigma^{\rm had}_{\rm beam-target}$) and beam fraction lost over a fill ($\it{b}$) for a gas-jet target inspired from the RHIC proton-beam polarimeter. We considered the nominal parameter from \ct{lhc_param_tab} for the beam energy, the particle flux in the LHC and the fill duration. The hadronic cross-sections are obtained as described in \ct{tableSMOG}. The obtained instantaneous luminosities should be considered as maxima. We have however levelled the gas inlet flux such that the lifetime of the beam is not shorten by an unrealistically large amount (we have considered 15$\%$).
}
\label{tablegasjet}
\end{table}

In particular, an estimation of the target areal density is given for polarised $^3$He and unpolarised jet targets. In the former case, a higher areal density is achievable since $^3$He, if cooled down, can be injected with lower velocity than H.  For what concerns unpolarised heavier ions, a target areal density of $1\div5 \times 10^{13}$ cm$^{-2}$ can be envisioned~\cite{Zelenski:2005mz}.

%% file: implementation-fixed-target/contributions-Storage-Cell.tex
Besides a gas-jet solution, using an internal gas-target cell is another genuine solution to make fixed-target experiments at the LHC, at a limited cost by re-using existing LHC detectors, and with a variety of polarised and unpolarised gas target.

In such a case, the target \cite{Barschel:2015mka} consists of a polarised or unpolarised gas source in combination with an open storage cell (see \cf{targetcell}) which increases the target density by more than two orders of magnitude compared to a free atomic beam jet. 

\begin{figure}[!htb]
\centering
\subfigure[~]{\includegraphics[scale=0.35]{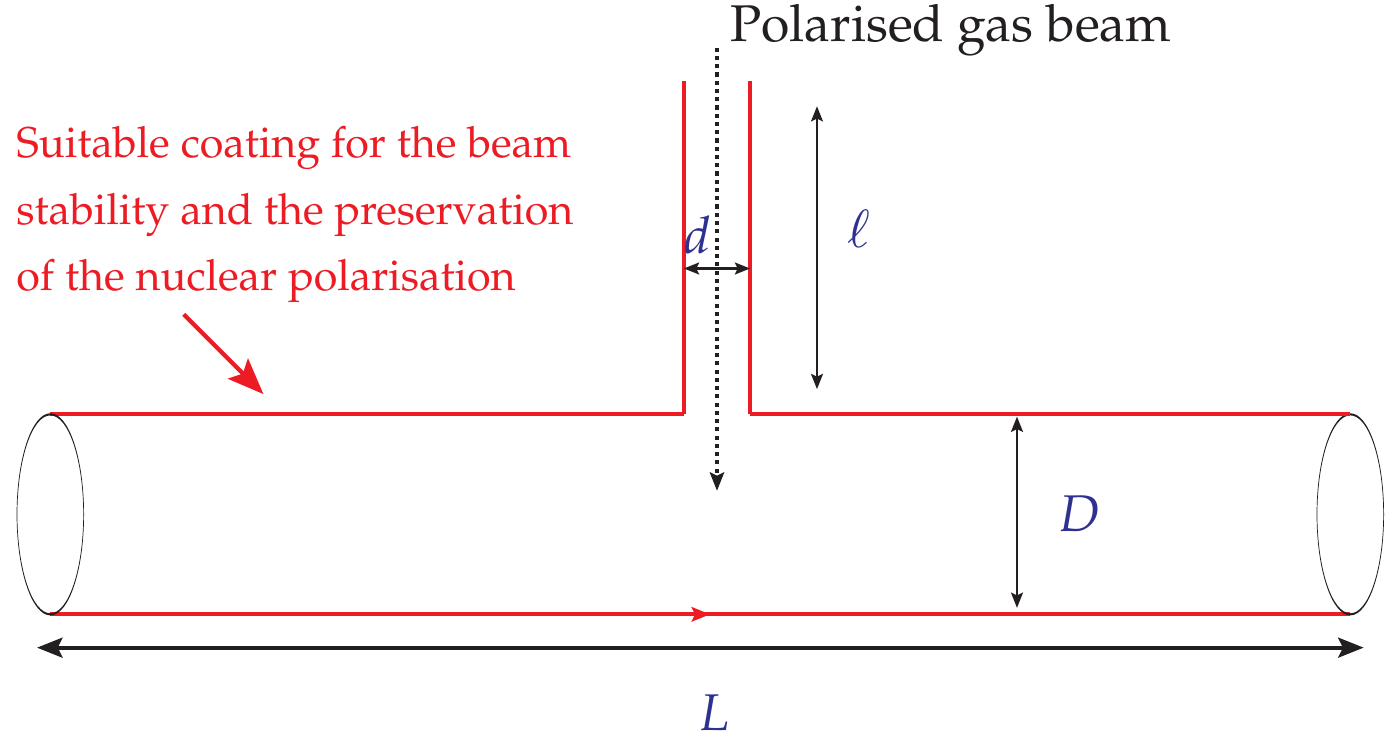}}
\subfigure[~]{\includegraphics[scale=0.35]{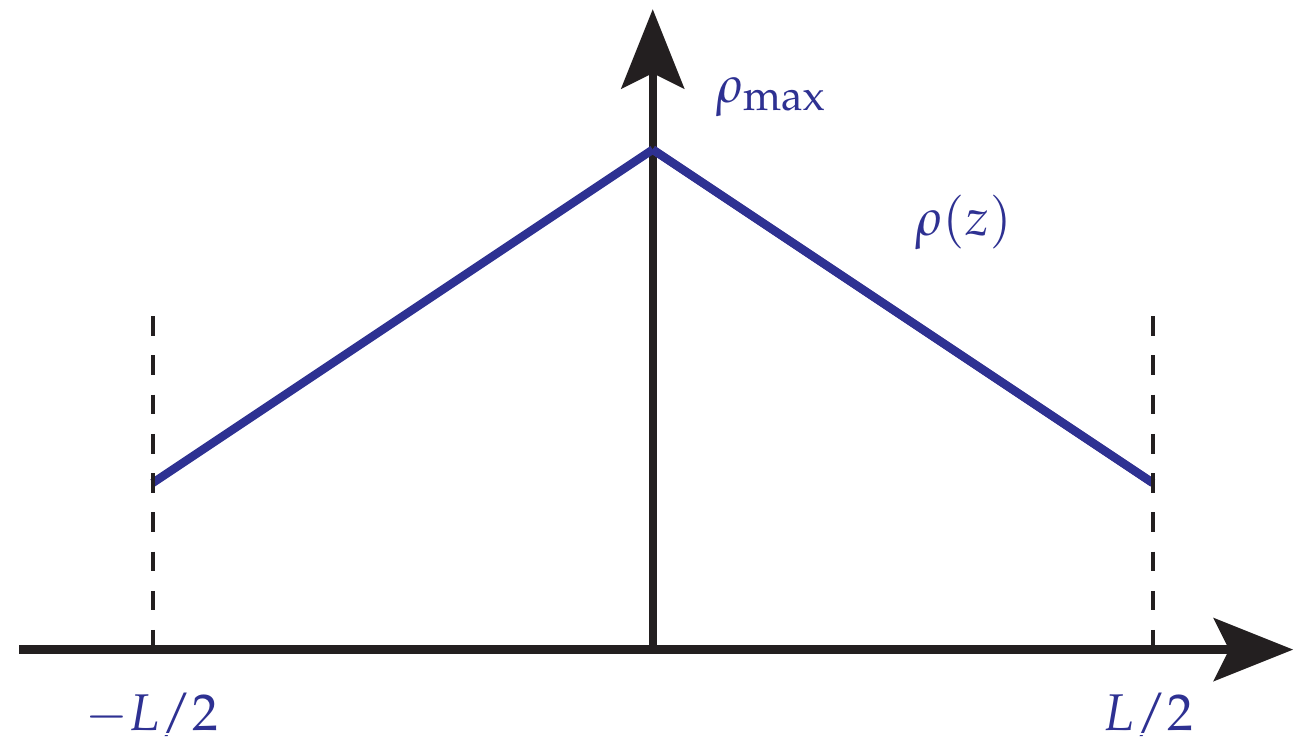}}
\caption{(a) Schematic view of the target cell. Polarised gas can be injected
  ballistically - unpolarised gas via a capillary - into the cell center; (b) Gas density profile, $\rho(z)$ along the beam $z$ axis.
}
\label{targetcell}
\end{figure}

The storage cell consists of a narrow
straight tube with thin walls located in the machine vacuum along the beam axis, into which the gas is injected at the center, in two modes:
\begin{itemize}
\item polarised atomic beam (H, D or $^3$He) into a feed tube of low gas
  conductance;
\item unpolarised gas (\eg\ H$_2$, He, Ne, Ar, Kr, Xe) via a capillary from a
  gas handling system.
\end{itemize}
The gas diffuses through the cell openings into the machine vacuum system,
which usually requires a powerful differential pumping system. The cell
consists of two movable halfs that can be opened in order to provide space 
for beam injection or manipulation. Polarised gas targets for
storage rings are reviewed in \cite{0034-4885-66-11-R02}. Targets for proton beams
at intermediate energies have been applied at the cooler ring COSY (FZ
J\"ulich) \cite{Mikirtychyants:2012nh,Grigoryev:2009zza,Ciullo:2016nhv} as well as at 
the IUCF Cooler Ring \cite{Dezarn:1995im}; and for electron beams at HERA (DESY
Hamburg) \cite{Airapetian:2004yf}.

Such a polarised H or D target (the HERMES target)\cite{Airapetian:2004yf} 
consists of three main components: the polarised
{\it Atomic Beam Source} (ABS), the {\it Storage Cell} in a
longitudinal or vertical holding field $B$ and a {\it Diagnostic System} for
analyzing a small sample of gas from the target cell, consisting of the
polarimeter (BRP) measuring the sub-state population of the atoms, and the
target gas analyzer (TGA) detecting the molecular fraction and thus the degree
of recombination within the cell. From these parameters, the target
polarisation $P$ as seen by the beam is deduced (typical values $P \approx
0.85$, corrected for magnetic guide field and degree of dissociation). It was 
however recently noted that the cell coating used then does not comply with the LHC requirements. R\&D
is thus needed to find a proper coating which would not significantly degrade the polarisation
performances.

The ABS injects polarised H into the feed tube (here 100~mm long, 10~mm inner
diameter) of a storage cell of 1000~mm length with a proposed inner diameter
of 14 mm (see \cf{targetcell}). A smaller size of 300~mm by 10~mm however seems to be more adapted 
to match the LHC requirements~\cite{DiNezza:PBC}.
The maximum areal density $\theta_{\rm target}$ is limited by the flux of the ABS and the geometry of the storage cell which needs to permit the transmission of the LHC beam and the injection of the
polarised atomic beam. The target described here is also able to deliver
polarised D with similar densities. Injecting optically-pumped polarised
$^3$He into the storage cell could be another possibility. In such a case, 
the holding field must be very homogeneous in order to prevent motional depolarisation.

Unpolarised target gas could also be injected into the cell via a capillary
for the study of HIC. The gas flow from a gas feed system and
thus the gas density in a target cell can be very high. Limits are imposed \eg\ by the maximum gas
flow that a possible LHC target station can take, and by the capability of the
detector system. Another requirement is that the gas target must not shorten the
lifetime of a Pb beam fill by an unrealistically large amount (we have considered $\sim$ 15$\%$), when running parasitically. Indeed, higher luminosities could be reached for dedicated runs. On top of hadronic interactions, photo-nuclear interactions including Bound-Free Pair Production, in 
which an $e^+e^-$ pair is created with the electron bound to one of the colliding nuclei, and Electromagnetic 
Dissociation~\cite{Jowett:2004EPAC} can affect the Pb beam lifetime. The photo-nuclear 
interactions are enhanced with the target gas atomic number and can be up to twenty times the 
hadronic cross section in case of Xe gas with Pb beam~\cite{Jowett:private}. 
The storage cell parameters, the instantaneous luminosities obtained 
using typical LHC beam currents, and the beam fraction consumed over a fill, considering hadronic interactions, 
are shown in \ct{tabellewerte}, for each type of beam-gas collisions \cite{Barschel:2015mka}.

\begin{table}[!htb]\renewcommand{\arraystretch}{1.2}
\begin{tabularx}{\textwidth} { c | C | c | c | c || C | C }
Beam & Target & $\varphi^{\rm inlet}_{\rm target}$ & $\theta_{\rm target}$ & $\mathcal{L}$ &  $\sigma^{\rm had}_{\rm beam-target}$ & \it{b} \\
 & &  [atoms $\times$ s$^{-1}$] &  [atoms $\times$ cm$^{-2}$] & [cm$^{-2} \times$ s$^{-1}$] & [barn]  & [$\%$] \\
\hline\hline
$p$  & H$^\uparrow$ &  $6.5\times 10^{16}$ & $2.5\times 10^{14}$ &  $9.2 \times 10^{32}$ & $48 \times 10^{-3}$ & 0.4  \\ 
$p$  & D$^\uparrow$ &  $5.2\times 10^{16}$ & $2.9\times 10^{14}$ & $1.1\times 10^{33}$ & $90 \times 10^{-3}$ &  1.1 \\
$p$  & $^3$He$^\uparrow$ &  $1.5\times 10^{17}$ & $1.0 \times 10^{15}$ & $3.7\times 10^{33}$  & $12 \times 10^{-2}$ & 4.4 \\
\hline
$p$  & H$_2$ &  $2.8\times 10^{17}$ & $1.6\times 10^{15}$ &  $5.8 \times 10^{33}$ &
$48 \times 10^{-3}$ & 3.1 \\ 
$p$  & { \{Ne, Ar, Kr, Xe\}}         & {$1.3\times 10^{15}$}  & {$6.4\times 10^{13}$}    & {$2.3 \times10^{32}$}  & {\{0.6, 1.0, 1.6, 2.2\}}  & {\{1.6, 2.6, 4.2, 5.7\}}  \\
\hline
Pb   & H$^\uparrow$ &  $6.5\times 10^{16}$  & $2.5\times 10^{14}$ & $1.2\times 10^{29}$   & 3.0   & 15 \\
Pb   & D$^\uparrow$ &  $3.4\times 10^{16}$  & $2.2\times 10^{14}$ & $8.8\times 10^{28}$   & 4.0   & 15 \\
Pb   & $^3$He$^\uparrow$ &  $2.7\times 10^{16}$  & $1.8\times 10^{14}$ & $8.3\times 10^{28}$   & 4.3   & 15 \\
\hline
Pb   & H$_2$   & {$6.5\times 10^{16}$}  &{$2.5\times 10^{14}$} &   {$1.2 \times10^{29}$} & 3  & 15 \\
Pb   & Xe         & {$1.3\times 10^{15}$}  &{$6.4\times 10^{13}$} &   {$3.0 \times10^{28}$} &  12  & 15 \\

\hline
\end{tabularx}
\caption{
Beam type, target type, inlet flux ($\varphi^{\rm inlet}_{\rm target}$), target areal density ($\theta_{\rm target}$), instantaneous luminosity ($\mathcal{L} =  \phi_{\rm beam} \theta_{\rm target}$ ), hadronic beam-target cross section ($\sigma^{\rm had}_{\rm beam-target}$) and beam fraction lost over a fill ($\it{b}$) for a storage cell target with a cell length of 1~m and temperature of 300~K. We considered the nominal parameter from \ct{lhc_param_tab} for the beam energy, the particle flux in the LHC and the fill duration. The hadronic cross-sections are obtained as described in \ct{tableSMOG}. The obtained instantaneous luminosities should be considered as maxima. We have however levelled the gas inlet flux such that the lifetime of the beam is not shorten by an unrealistically large amount (we have considered 15$\%$).}
\label{tabellewerte}
\end{table}

%% file: implementation-fixed-target/contributions-Solid-Target.tex
Another possible internal target solution for the LHC is to use a wire, a ribbon or a foil positioned in the halo of the beam as first proposed in~\cite{Kurepin:2011zz}. 
Such an approach was adopted by the HERA-B~\cite{Ehret:2000df} experiment with the prerequisite of not affecting experiments functioning on the 920 GeV proton beam. The HERA-B system consisted of 2 stations of 4 wires each, made of Ti and W, positioned in a square shape around the beam. For C, flat ribbons were used\footnote{The diameter of the wires was 50 $\mu$m whereas the ribbons were 100 $\mu$m wide in the direction perpendicular to the beam and 500 $\mu$m along the beam (which can thus be considered as the target thickness).}. These could be positioned independently and adjusted with respect to the beam condition in order to scrap the beam halo. Other materials were also considered such as Al, Fe and Cu. One of the main limitations for HERA-B was the widening of the beam due to multiple Coulomb scattering in the target, which is reduced with low $Z$ materials. Multiple Coulomb scattering is not a limitation anymore when the beam halo is used, due to the lower halo intensity. For the LHC, simulations and tests (as done for the bent crystals discussed in the next section) are therefore needed to completely assess the feasibility and then the performance of such a system.

In principle, such a target system can be placed in the vicinity (or even inside) ALICE or LHCb at a moderate cost. If the system was found to be incompatible with high intensity runs, it could be used only for heavy-ion runs or during special runs; this would in turn reduce the collected luminosity in $p$A collisions. Another limiting factor is the impossibility to carry out $pp$ collisions which are extremely important to quantify the nuclear effects. In the absence of such reference, it is mandatory to have at one's disposal measurements made with species with sufficiently different $A$. However, there are strong constraints from the LHC on the species which can be placed inside the LHC beam pipe. Therefore the mechanics for the positioning of the target should allow one to shift the target during the tuning of the beam. We have considered so far\footnote{Note that the usage of a solid Pb target is a priori excluded by LHC experts due to the low melting temperature of the Pb.} C, Ti, and W as possible species but further studies might be required (also for other species). Finally, let us note that such a system does not allow one to use a polarised target. The solid target parameters and the instantaneous luminosities obtained using typical LHC beam halo fluxes are shown in~\ct{tablewire}, for each type of beams and targets.  
      
\begin{table}[!hbt]\renewcommand{\arraystretch}{1.2}
\begin{center}
\begin{tabular}{ c | c | c | c | c | c | c | c }
Beam & Target & $\rho$ & $M$ &  $\ell$ & $\theta_{\rm target}$ & $\phi_{\text{ usable halo}}$ & $\mathcal{L}$  \\
 &  & [g $\times$ cm$^{-3}$]   &  [g $\times$ mol$^{-1}$]  & [$\mu$m] & [cm$^{-2}$] &  [s$^{-1}$]  & [cm$^{-2} \times$ s$^{-1}$] \\
\hline\hline
$p$ & C & 2.25 & 12 & 500  & $5.6\times 10^{21}$ & $5.0\times 10^8$ & $2.8\times 10^{30}$ \\
$p$ & Ti & 4.43 & 48 & 500  & $2.8\times 10^{21}$  &  $5.0\times 10^8$ & $1.4\times 10^{30}$  \\
$p$ & W & 19.3 & 184 & 500  & $3.1\times 10^{21}$  & $5.0\times 10^8$ & $1.6\times 10^{30}$  \\
\hline
Pb   & C & 2.25 & 12  & 500   & $5.6\times 10^{21}$ & $1.0\times 10^5$   & $5.6 \times 10^{26}$  \\
Pb   & Ti & 4.43 & 48  & 500   & $2.8\times 10^{21}$  & $1.0\times 10^5$    & $2.8 \times 10^{26}$  \\
Pb   & W & 19.3 & 184 & 500   & $3.1\times 10^{21}$ & $1.0\times 10^5$     & $3.1 \times 10^{26}$  \\
\hline
\end{tabular}
\end{center}
\caption{
A selection of beam type, target type, target density ($\rho$), target molar mass ($M$), target thickness ($\ell$), target areal density ($\theta_{\rm target}$), usable particle flux in the halo $\phi_{\text{ usable halo}}$ and instantaneous luminosity ($\mathcal{L} = \phi_{\text{ usable halo}}\theta_{\rm target}$) for internal wire targets positioned in the halo of the LHC beams. The target areal density $\theta_{\rm target}$ is equal to $\mathcal{N_{A}} \ell \rho / M$ with $\mathcal{N_{A}}$ the Avogadro number in mol$^{-1}$.
}
\label{tablewire}
\end{table}

%% file: implementation-fixed-target/contributions-Bent-Crystal.tex
The idea of a controlled non-resonant slow extraction of the CERN LHC beams --on the order of $10^{8}$ protons per second--  to be used for 
fixed target physics is not new. Already, in the early 90's, the LHB collaboration submitted a letter of intent to the LHCC to get an experiment based on bent-crystal extraction approved. At the time, it was not 
clear whether such a technique could be used in the LHC conditions. 

Since then, significant progresses were achieved with successful tests for protons at the SPS~\cite{Arduini:1997kh}, 
Fermilab~\cite{Asseev:1997yi}, Protvino~\cite{Afonin:2012zz} and for Pb ions at the SPS~\cite{Scandale:2011za}.
These were made possible by numerous experimental advances, like the improvement of the crystal quality with a production technique allowing to reach a channeling efficiency close to the theoretical one, or like the development of goniometers matching the critical angle requirements for a 6.5 TeV beam channeling. Thanks to this legacy, the UA9 collaboration 
proposed this technique as a smart alternative for the upgrade of the LHC collimation 
system~\cite{Scandale:2010zzb,Scandale:2012wy} following the concept developed in the frame of the INTAS programme 03-51-6155, see for instance Ref.~\cite{Biryukov:2002vn} issued in 2002. 
Tests were recently successfully carried out at the betatron cleaning insertion (IR7) both at injection and top energy (6.5 TeV) in 2015 \cite{Scandale:2016krl}. Tests have also been performed with lead beam at injection and top energies (2.5 TeV) at the end of 2016.
They clearly demonstrated the feasibility of crystal-assisted collimation and, in turn, gave a new momentum 
in the plans for crystal-assisted extraction (see also \cite{Scandale:2018cdf}). Let us also stress that the crystal degradation due to radiation, 
once thought to be an issue, is negligible as demonstrated  by 
tests~\cite{Baurichter:2000wk} with the HiRadMat facility of the SPS.

The generic requirements for such an extraction system were already outlined in 1990 at
the Aachen LHC workshop where it was identified that~\cite{Scandale:1990rr}: 
\begin{itemize}\setlength{\itemsep}{-4pt}\setlength{\itemindent}{-0.5cm}
\item an extraction outside the ring is preferred to avoid interferences with the main tunnel and the experimental cavern;
\item an extraction in the horizontal plane would probably be more favourable since an enhanced deflection\footnote{The enhanced deflection would be beyond the one resulting from the crystal pass.} could be achieved by an appropriate re-design of the separation recombination dipoles D1 and D2;
\item the extraction would be at one of the odd points of the LHC;
\item the crystal location would be between the quadrupoles Q3 and Q4 and it would provide a deflection of about 1 mrad; 
\item a further deflection up to 20 mrad could be achieved at 250 m;
\item increasing the size of the halo would result in a higher extraction efficiency with more particles crossing the
crystal.
\end{itemize}

To be more quantitative about the required modifications of the beam pipe, 
since the halo is located at approximately 3~mm from the beam pipe axis, a deflection by an angle of 1 mrad would result in these particles exiting the beam pipe at 30 m downstream, considering a LHC beam pipe radius of 3-4 cm in the LHC tunnel.  Another proposal was made in 2005~\cite{Uggerhoj:2005xz} 
consisting in "replacing" the kicker-modules in LHC section IR6 (the beam dump) by a bent crystal that 
would provide the particles in the beam halo with a sufficient kick to overcome the septum 
blade and to be extracted. It is however not clear to which extent the beam-dump area, even with obvious modifications to move the beam-dump facility,  can be used for experiments. 
As for now, beyond the generic requirements above, the possible locations of a possible extraction zone have not been listed. 
Currently, there is an active project, CRYSBEAM \cite{CRYSBEAM,Mazzolari:2018hsu}, whose objective is to demonstrate the feasibilty of crystal-assisted extraction on the LHC and which
should upon its completion give us better insight on the technical realisation of this solution. Operating the crystal in real parasitic mode still requires further studies. 

Another possibility, which would avoid further manipulating the beam downstream of the crystal, is to directly use the extracted --and highly collimated-- particles of the halo (which we refer to as the split beam) on an (semi) internal target system which would however not interact with the main LHC beam. Such a solution would probably allow one to limit the civil-engineering work to a minimum and also to use an existing experiment. Two caveats have however to be adressed. First, only a fraction of the split beam would interact with the target and the remaining $10^8$ protons per second should be absorbed or deviated from the experiment. Second, this would probably induce a non-negligible azimuthal asymmetry in the experimental system. For spin-related analyses, this may be a serious limitation in the cancellation of some systematical uncertainties. A first setup (see Appendix \ref{appendix:UA9}), compatible with the LHCb detector, has been proposed to measure the electric and magnetic dipole moments of charmed charged baryons at LHC top energies \cite{Bezshyyko:2017var, Bagli:2017foe, Botella:2016ksl}. A first bent crystal, located at 5$\sigma$ from the center beam line, deflects the halo particles by about 150 $\mu$rad at about 100 meter upstream of the LHCb interaction region, in order to separate them from the circulating beam. A target, 1cm long and 5~mm thick along the beam direction, inserted in the pipe intercepts the halo, producing short-lived baryons. A second crystal\footnote{Note that this second crystal is needed in the case of electric or magnetic dipole moment measurements but not for the physics cases developed in that paper.} channels part of the baryons and deflects them by 7 mrad\footnote{Let us remind that the deflection angle is related to the critical radius of the bent crystal, which increases linearly with the energy. At high energy, to reach larger angles, the length of the crystal has to be increased.} into the LHCb detector, where the spin orientation is measured. An additional absorber intercepts the halo particles non-interacting with the target, thereby allowing the possibility of fixed-target operation in parasitic mode. An initial test of the beam-splitting concept and of the double crystal use in an accelerator was made in the SPS by the UA9 collaboration in 2017 \cite{Montesano:2018}.
However, there is still a long way before achieving a fully effective scenario, compatible with the LHC-collimation system and with the LHCb detector.

%% file: implementation-fixed-target/contributions-Unpolarised_Targets.tex

As aforementioned, the LHC beams can be extracted by means of a bent crystal with typical 
fluxes on the order of 5 x $10^8$ s$^{-1}$ for the proton beam\footnote{Recent LHC collimation studies show that, for parasitic operation, the layout proposed for LHCb could deliver at least $10^6$ protons per second~\cite{Barschel:2653780}. Another scenario with controlled excitation of beam losses
for selected proton bunches would provide higher flux but would need further studies.}  and $10^5$ s$^{-1}$ for the lead beam. We will consider 5mm-thick targets, which are in principle compatible with both
the split beam option and a dedicated beam line. \ct{tab:lumi:solid-target} displays
the corresponding instantaneous luminosities with the same species as for the internal wire
target as well as for solid hydrogen to illustrate the case of light elements. 


\begin{table}[!hbt]
\center\renewcommand{\arraystretch}{1.2}
\begin{tabular}{ c | c | c | c | c | c | c | c }
Beam & Target &  $\rho$  & $M$  & $\ell$ & $\theta_{\rm target}$ & $\phi_{\text{ usable halo}}$ & $\mathcal{L}$  \\
 & & [g $\times$ cm$^{-3}$]  & [g $\times$ mol$^{-1}$] & [mm]  & [cm$^{-2}$]& [s$^{-1}$] &  [cm$^{-2}$s$^{-1}$]\\
\hline\hline
$p$ & solid H & 8.80 $\times 10^{-2}$ & 1 &  5   & $2.6\times 10^{22}$  & $5.0\times 10^8$ & $1.3\times 10^{31}$   \\
$p$ & C & 2.25 & 12 & 5   & $5.6\times 10^{22}$  & $5.0\times 10^8$ & $2.8\times 10^{31}$   \\
$p$ & Ti & 4.43 & 48 &  5   & $2.8\times 10^{22}$  & $5.0\times 10^8$ & $1.4\times 10^{31}$   \\
$p$ & W & 19.3 & 184 & 5   & $3.1\times 10^{22}$  & $5.0\times 10^8$ & $1.6\times 10^{31}$   \\
\hline
Pb   & solid H & 8.80 $\times 10^{-2}$ & 1 &  5  & $2.6\times 10^{22}$ & $1.0 \times 10^5$ & $2.6\times 10^{27}$   \\
Pb   & C & 2.25 & 12 &  5  & $5.6\times 10^{22}$ & $1.0 \times 10^5$  & $5.6\times 10^{27}$   \\
Pb   & Ti  & 4.43 & 48 & 5   & $2.8\times 10^{22}$ & $1.0 \times 10^5$ & $2.8\times 10^{27}$   \\
Pb   & W & 19.3 & 184 & 5   & $3.1\times 10^{22}$ & $1.0 \times 10^5$ & $3.1\times 10^{27}$   \\
\hline
\end{tabular}
\caption{A selection of beam type, target type, target density ($\rho$), target molar mass ($M$), target thickness ($\ell$), target areal density ($\theta_{\rm target}$), usable particle flux in the halo $\phi_{\text{ usable halo}}$ and instantaneous luminosity ($\mathcal{L} = \phi_{\text{ usable halo}} \theta_{\rm target}$) for an extracted beam of protons or of lead ions by means of a bent crystal and impinging a solid target. The target areal
density is computed as in \ct{tablewire}. With a beam splitting option, there is currently no clear solution to allow for the usage of a light target (solid/liquid H, D).}
\label{tab:lumi:solid-target}
\end{table}

We stress that for an experiment with a specific target system away from the beampipe, targets
as thick as a meter can be used for light species like H or D. Thicker 
targets can also be used for heavier species provided that the effect of multiple scatterings in the target can be mitigated. 
In general, the targets being thicker than in the internal-wire-target case (5 vs. 0.5~mm), instantaneous luminosities are larger.

%% file: implementation-fixed-target/contributions-Combined_Andi_Norihiro.tex
One of the main thrusts of the proposed physics with AFTER@LHC will be the measurement of the Sivers asymmetry which will require a transversely polarised target. In the following, we present two possible target systems for \AFTER, one inspired from the E1039 project at Fermilab~\cite{Klein:zoa} and the other inspired by the polarised target of the COMPASS experiment at CERN~\cite{Pesek:2014uua,Doshita:2004ee,Ball:2003vb}. In both cases, the polarisation of the target relies on the Dynamic Nuclear Polarisation (DNP) method, whose general principle is described in \cite{Goertz:2002vv}. Both targets could be envisioned in the case of crystal-assisted extraction of the LHC beams into a new cavern, however, if redesigned, the E1039 target might also be usable with the crystal beam splitting solution. 

\paragraph{The E1039 target}
This  target  consists of a split coil superconducting magnet, operating at 5~T. The coils are arranged such that the B field is either parallel or antiparallel to the vertical direction, resulting in a transverse polarisation. Inside the magnet there is a refrigerator, which provides the necessary cooling power to keep the target at 1K. In the centre of the whole system resides the target stick, which contains the target cells, the microwave horn and the Nuclear-Magnetic-Resonance (NMR) coils to measure the actual polarisation. The target insert has 2 active cells filled with frozen NH$_{3}$ beads, one empty and one with a carbon target. The cells  have an elliptical cross section, with one half axis being 1.9~cm diameter and the other one 2.1~cm while the length is 8 cm. The target material is positioned in a liquid He bath, cooled to 1~K, by lowering the vapour pressure of liquid He to 0.117 Torr. This is achieved with a dedicated system which have a capacity of pumping 15 000~m$^{3}$/h He gas. The microwave horn is sitting above the two top cells, which contain NH$_{3}$ and  ND$_{3}$, thus allowing to measure polarised $p$ and $n$ under identical run and target conditions. This greatly reduces the systematic uncertainties by comparing $p$ and $n$. While both cells see the radiation from the microwave, only the one in the centre region of the coils will be polarised. Since the material has to be uniformly cooled, the ammonia is in the form of small frozen beads, which reduces the maximum density by a packing fraction of about 0.6. From the planned beam intensities, we estimate that the material has to be changed every 140 days, due to radiation damage.
In \cf{E1039_target} of Appendix~\ref{appendix:E1039_tg}, a schematic view of the E1039 target is shown.  

\paragraph{The COMPASS target}
The target consists of two identical, 55 cm long cylindrical cells with a diameter of 4 cm. Each cell has 5 NMR coils to measure the polarisations ~\cite{Kondo:2004dz} (see \cf{target_COMPASS} of Appendix~\ref{appendix:compass_tg} for a schematic view of the COMPASS target). The cells can be polarised in opposite direction and there is a 20 cm long gap between the cells, in order to cleanly separate interactions from the respective target cells. The orientations are reversed by changing microwave frequency at 2.5 T at regular intervals in order to reduce the systematic error. The  polarisation is obtained by the DNP method with a high cooling power dilution 
refrigerator with a 13500 m$^{3}$h$^{-1}$ pumping speed of {8 Pfeiffer Roots blowers} in series, a 2.5 T solenoid magnet and two microwave systems of about 70 GHz~\cite{Ball:2003vb}. The spin can be oriented perpendicular to the beam direction by using a 0.6 T dipole magnet. Under this magnetic condition the polarisation can not be enhanced by the DNP method, but can be maintained at a lattice temperature below 100 mK. The proton polarisation achieved in 2015 with NH$_{3}$ was 80 \% in 1 day and about 90 \% after 2 days\footnote{The proton polarisation for the NH$_{3}$ target and for the 2015 analyzed data sample was in average 73\%~\cite{Aghasyan:2017jop}.}.
In a beam intensity of $8 \times 10^{7}$ pions s$^{-1}$ the beam intensity for each NH$_{3}$ bead of 2--3 mm is about $10^{6}$ s$^{-1}$ which will not lead to a significant depolarisation in the frozen spin mode. No significant radiation damage could be observed in more than half a year of data taking in 2015 at COMPASS. If needed for the case of 5 $\times 10^{8}$ p/s beam intensity with a more focused beam, the target material may be considered to be annealed twice or thrice a year\footnote{
The process of the annealing consists of removing the $^3$He gas, warming up to 70 K, cooling down and re-filling with $^3$He. In total it takes one week for the COMPASS system.} to recover the optimal target performances.
A spin relaxation time of about 1000 hours was measured at 0.6 T and 50 mK for the proton in NH$_{3}$.

\paragraph{Comparison}
\ct{comp_combined} displays the parameters of the E1039 target and of the COMPASS target, which could be used for \AFTER. The COMPASS target permits to reach higher luminosities than the E1039 target, however the E1039 target offers several advantages. The latter is smaller and could therefore be used in beam splitting mode if a significant modification of the beam line and target is performed. Both also permit the usage of polarised deuterium target (ND$_{3}$ or $^6$LiD), complementary to the hydrogen one (\eg\ NH$_{3}$).

\begin{table}[!htb]
\small
\renewcommand{\arraystretch}{1.4}
\begin{tabularx}{\textwidth}{ p{1.6cm} | p{1cm} | c | p{0.7cm} | p{1.4cm} | C | c | C | C }
   & \centering Target  & $\rho$ & \centering $\ell$  & \centering $M$ & $\theta^{\rm (eff.)}_{\rm target}$ &   $\phi_{\text{ usable halo}}$  &  $\mathcal{L}$  & $y \times z$ \\
   & & [g  cm$^{-3} $]   &  \centering [cm]  & \centering [g  mol$^{-1}$]        & [cm$^{-2}$]         &  [s$^{-1}$]     &  [cm$^{-2}$ s$^{-1}$]  &  [mm $\times$ mm] \\  \hline \hline

\multirow{2}{*}{\parbox{1.9cm}{\centering E1039 \linebreak target}}    & \centering NH$_{3}^\uparrow$  & 0.86 & \centering 8 & \centering 17 &  $1.4  \times 10^{23}$ & $5.0 \times 10^{8}$ &  $7.2 \times 10^{31}$ & {$1853 \times 975$} \\ \cline{2-8}
                                            & \centering ND$_{3}^\uparrow$  & 1.01 & \centering 8 & \centering 20 &  $1.4 \times 10^{23}$ & $5.0 \times 10^{8}$  & $7.2 \times 10^{31}$   \\ \hline
 \multirow{3}{*}{\parbox{1.9cm}{\centering COMPASS target}} & \centering NH$_{3}^\uparrow$ &  0.86 & \centering 110 & \centering 17 &  $2.0 \times 10^{24}$ & $5.0 \times 10^{8}$  & $1.0 \times 10^{33}$ & $2820 \times 3120$ \\  \cline{2-8}
 												  & \centering $^{6}$LiD$^\uparrow$  &  0.84 & \centering 110 & \centering 8 &  $3.9 \times 10^{24}$ &  $5.0 \times 10^{8}$  &  $1.9 \times 10^{33}$ \\ \cline{2-8}
& \centering butanol$^\uparrow$  &  0.99 & \centering 110 & \centering 74 &  $5.3 \times 10^{23}$ &  $5.0 \times 10^{8}$  &  $2.7 \times 10^{32}$ \\ \hline

\end{tabularx}
\normalsize
\caption{
Target type, target density ($\rho$), target thickness ($\ell$), target molar mass ($M$), effective target areal density ($\theta^{\rm (eff.)}_{target}$), usable particle flux in the halo $\phi_{\text{ usable halo}}$, instantaneous luminosity ($\mathcal{L} = \phi_{\text{ usable halo}} \theta_{\rm target}$) and target dimension in the $y$ and $z$ directions for the E1039 and COMPASS targets. As what regards the target geometry, we note that the E1039 target is cylindrical. As in \ct{tablewire}, the target areal density is defined as $\theta_{target}$ = ($\mathcal{N_{A}}  \ell  \rho$) / $M$. However, in order to account for the fact that the target material does not occupy all the volume of the cell due to technicalities during filling of the cell and the presence of small amounts of coolant (He), additional factors need to be considered. The cell density is expressed as $\rho_{cell} = \rho_{NH_{3}}  PF + \rho_{He}(1-PF)$, where $PF$ is the packing factor (we considered $PF$ = $0.6$ for the NH$_{3}$, ND$_{3}$ and butanol targets, and 0.55 for $^{6}$LiD). The second term, $\rho_{He}(1-PF)$, however is neglected in our calculation. It is called the rest space factor and amounts to about 10$\%$. We thus introduced the effective target areal density such that $\theta^{\rm (eff.)}_{target}=\theta_{target}\hspace{0.1 true cm}PF$.}
\label{comp_combined}
\end{table}

%% file: implementation-fixed-target/contributions-comparison.tex
In this section, we will discuss a qualitative comparison of the various technological solutions which have been developed. More quantitative comparisons of the instantaneous luminosities which could be achieved and performances for STSA measurements for the various solutions will also be presented. 

\subsubsection{Qualitative comparison of the various technological solutions} 

\begin{table}[htb]\renewcommand{\arraystretch}{1.2}
\begin{tabular}{p{2.7cm}|c|c|c|c|c|c}
  & \multicolumn{3}{c|}{Internal-gas target} & Internal-solid target & Beam    & Beam  \\  \cline{2-4} 
   \centering Characteristics               & SMOG & Gas Jet & Storage Cell  &  with beam halo  & splitting & extraction\\ \hline \hline
\centering Run duration\savefootnote{$\star \star \star$: no limitation;  $\star \star$: possible limitation;$\star$: data taking for special runs only} 
& $\star$ & $\star \star$ & $\star \star$ & $\star$ & $\star \star$ & $\star \star \star$  \\ \hline
\centering Parasiticity\savefootnote{$\star \star \star$: no impact on the other LHC experiments;  $\star \star$: no impact pending constraints; $\star$: significant impact} 
& $\star \star$ & $\star \star$  & $\star \star$ & $\star$ & $\star \star$ & $\star \star \star$ \\ \hline
\centering Integrated luminosity\savefootnote{$\star \star \star$: highest;  $\star \star$: high; $\star$: moderate [corresponding to 1 LHC year]}  
& $\star$ & $\star \star \star$  & $\star \star \star$  & $\star$  & $\star \star$  & $\star \star \star$ \\ \hline
\centering Absolute luminosity determination\savefootnote{$\star \star \star$: direct with small uncertainty;  $\star \star$: direct with moderate uncertainty;$\star$: indirect}  
& $\star$ & $\star \star$ & $\star \star$ & $\star$ & $\star \star$ & $\star \star \star$ \\ \hline
\centering Target versatility\savefootnote{$\star \star \star$: no limitation;  $\star \star$:  some target types are not possible; $\star$: only a few target types are possible} 
&  $\star$ & $\star \star $ & $\star \star$ & $\star$  & $\star \star$ & $\star \star \star$\\ \hline
\centering (Effective) target polarisation\savefootnote{$\star \star \star$: highest; $\star \star $: high; $\star$: moderate}  
& - & $ \star \star \star$ & $\star \star$ & - & - / $\star$\savefootnote{with a redesigned E1096 target} &   $\star$ \\ \hline
\centering Use of existing experiment\savefootnote{$\star \star\star$: without any experiment/beam-pipe modifications;  $\star\star$:  with slight experiment/beam-pipe modifications;  $\star$:  with slight experiment/beam-pipe modifications and potential non-optimal acceptances} 
& $\star \star \star$ &$\star \star $  &$\star $  & $\star \star$ &$\star\star$ & - \\ \hline
\centering Civil engineering or R\&D \savefootnote{$\star \star \star\star$: none;  $\star \star\star$: some R\&D;$\star\star$: some R\&D with possible show stoppers; $\star$ significant civil engineering} 
& $\star \star \star \star$ & $\star \star \star$  & $\star \star$  & $\star\star$  & $\star \star$ & $\star$ \\ \hline
\centering Cost & $\star \star \star$ & $\star \star$ & $\star \star$ & $\star \star\star$&  $\star \star$  & $\star$  \\ \hline
\centering Implementation time & $\star \star \star$ & $\star \star$ & $\star \star$ & $\star \star \star$&  $\star \star$  & $\star$  \\ 
\hline \hline
\centering High $x$
& $\star$  & $\star \star \star$ & $\star \star \star \star$ & $\star$       &  $\star \star$  & $\star \star \star \star$ \\ \hline
\centering Spin Physics
& -        &   $\star \star \star$      & $\star \star \star$ & -             &  - / $\star \star $       & $\star \star \star$ \\ \hline
\centering Heavy-Ion
& $\star$  &  $\star \star \star$  & $\star \star \star$ & $\star \star $&  $\star \star$  & $\star \star \star \star$  \\ \hline
\end{tabular}
\caption{Qualitative comparison of the various technological solutions.}
\label{comp_qual}
\end{table}

\ct{comp_qual} gathers our qualitative judgement of the different solutions with regards to a number of decisional criteria and to the reach in the three physics cases developed in the sections~\ref{section:High-x},~\ref{section:Spin-Physics} and~\ref{section:heavy Ion Physics}.

In particular, we stress that it is assumed here that one uses a detector without specific data-taking-rate limitations. The physics reach when using the ALICE and LHCb detectors will be discussed in the section~\ref{section:detector}. 


For the internal-gas-target solutions, the current SMOG system in LHCb has the advantage to be mostly parasitic to other LHC experiments due to its low gas density. Also, various noble gases up to argon have already been used. However, its duration time is limited as well as the possible yearly integrated luminosity and it can not run with polarised gas. Furthermore, the luminosity can barely be directly estimated. 

To achieve the physics reach proposed in this report with gas-target solutions, it is important to increase the gas density with respect to SMOG and to opt for a more flexible gas system with polarised gases for spin physics. Moreover, running with  hydrogen gas allows one to obtain a reference measurement with protons as target for high-$x$ and heavy-ion physics. For that purpose, the gas-jet and storage-cell solutions are probably the most promising -- with the highest luminosity for the storage-cell solution. Note, however, that the cell coating has to be compatible with the LHC vacuum constraints, which seems not to be the case of the original HERMES target. In the target chamber of the gas-jet system, one can also inject nuclear-target gases, however this was not tested at RHIC and the possible gas density in that case was estimated. 

The internal-solid-target solution directly on the beam halo has the advantage to be compatible with various target species. This solution suffers however from a low luminosity and will likely impact the LHC-beam stability. The beam-splitting solution, by using a slow beam extraction with a bent crystal, will probably have less impact on the LHC beam and will allow one to run for a longer period and with thicker targets. If coupled with a redesigned E1039 target, it would allow for spin physics. 

Finally the beam-extraction solution is more suitable for the physics reach. However the necessary civil engineering would largely increase the cost by more than one order of magnitude and the implementation time with respect to the other solutions that are at reach with limited technical developments.

\subsubsection{Comparison of the luminosities achieved for AFTER@LHC with the various technological solutions}


\begin{table}[!htp]
\center
\begin{adjustbox}{angle=90}
{\renewcommand{\arraystretch}{1.2}
\begin{tabular}{p{1.8cm}|c|c|c|c|c|c|c|c}
\multicolumn{3}{c}{ } & \multicolumn{6}{|c}{Beam} \\
\cline{4-9}
\multicolumn{3}{c}{ } & \multicolumn{3}{|c|}{$p$} & \multicolumn{3}{c}{ Pb} \\
\cline{4-9}
\multicolumn{3}{c|}{Target} & $\cal L$ & $\Delta t$  & $\int\cal L$  & $\cal L$  & $\Delta t$ & $\int\cal L$ \\
\multicolumn{3}{l|}{ } & [cm$^{-2}$ s$^{-1}$] & [s] & [nb$^{-1}$]  & [cm$^{-2}$ s$^{-1}$] & [s] & [nb$^{-1}$]  \\
\hline
\hline
\multirow{10}{*}{\parbox{1.8cm}{\centering Internal gas target}} & \multirow{1}{*}{SMOG} & He, Ne, Ar & 5.8 $\times 10^{29}$   & 2.5 $\times 10^{5}$  & 1.5 $\times 10^{2}$ & 7.4 $\times 10^{25}$  & 1.0 $\times 10^{6}$  & 7.4 $\times 10^{-2}$   \\
									\cline{2-9}
								    &\multirow{4}{*}{Gas-Jet} & H$^\uparrow$  & 4.3 $\times 10^{30}$   &  $ 1.0 \times 10^{7}$  & $4.3 \times 10^{4}$  & $5.6 \times 10^{26}$   & $1.0 \times 10^{6}$  & 5.6 $\times 10^{-1}$ \\
								    	&							   & H$_{2}$ & $3.6 \times  10^{33\div34}$  & 1.0 $\times 10^{7}$   & $3.6 \times 10^{7\div 8}$  & $1.2 \times 10^{29}$  & $1.0 \times 10^{6}$  & 1.2 $\times 10^{2}$    \\
									&							   & D$^\uparrow$ & $4.3 \times 10^{30}$   &  1.0 $\times 10^{7}$  & 4.3 $\times 10^{4}$  & 5.6 $\times 10^{26}$   & 1.0 $\times 10^{6}$  & 5.6 $\times 10^{-1}$ \\
									&							   & $^3$He$^\uparrow$ & $3.6 \times 10^{32}$ & 1.0 $\times 10^{7}$  & 3.6 $\times 10^{6}$  & 4.7 $\times 10^{28}$ & 1.0 $\times 10^{6}$ & 47 \\
	&							   & Xe &$(3.6 \div 18)  \times 10^{31}$  & 1.0 $\times 10^{7}$  & $(3.6\div18) \times 10^{5}$ & $(0.5\div 2.3) \times 10^{28}$  & 1.0 $\times 10^{6}$ & $5.0 \div 23$  \\
							     	\cline{2-9}
									&\multirow{5}{*}{Storage Cell} & H$^\uparrow$ & $9.2 \times 10^{32}$  & 1.0 $\times 10^{7}$  & 9.2 $\times 10^{6}$   & 1.2 $\times 10^{29}$   & 1.0 $\times 10^{6}$  & 1.2 $\times 10^{2}$ \\
									&							   & H$_{2}$ & $5.8 \times 10^{33}$  & 1.0 $\times 10^{7}$  & 5.8 $\times 10^{7}$   & 1.2 $\times 10^{29}$ & 1.0 $\times 10^{6}$  & 1.2 $\times 10^{2}$   \\
									&							   & D$^\uparrow$ & $1.1 \times 10^{33}$   & 1.0 $\times 10^{7}$  & 1.1 $\times 10^{7}$  & 8.8 $\times 10^{28}$ & 1.0 $\times 10^{6}$ & 88    \\
									&							   & $^3$He$^\uparrow$ & $3.7 \times 10^{33}$  & 1.0 $\times 10^{7}$  & 3.7 $\times 10^{7}$ & 8.3 $\times 10^{28}$ & 1.0 $\times 10^{6}$ & 83 \\
									&							   & Xe &$2.3 \times 10^{32}$  & 1.0 $\times 10^{7}$  & 2.3 $\times 10^{6}$ & 3.0 $\times 10^{28}$  & 1.0 $\times 10^{6}$ & 30  \\ \hline
\centering Target &  Wire  & C & 2.8 $\times 10^{30}$  & 1.0 $\times 10^{7}$  & 2.8 $\times 10^{4}$  & 5.6 $\times 10^{26}$  & 1.0 $\times 10^{6}$ & 5.6 $\times 10^{-1}$\\
\centering on the 									&		Target					                                                & Ti & 1.4 $\times 10^{30}$  & 1.0 $\times 10^{7}$  & 1.4 $\times 10^{4}$ & 2.8 $\times 10^{26}$ & 1.0 $\times 10^{6}$ & 2.8 $\times 10^{-1}$ \\		\centering beam halo	
									&						(0.5 mm) 	                                                & W & 1.6 $\times 10^{30}$ & 1.0 $\times 10^{7}$ & 1.6 $\times 10^{4}$  & 3.1 $\times 10^{26}$ &  1.0 $\times 10^{6}$ & 3.1 $\times 10^{-1}$ \\		\hline
\multirow{5}{*}{\parbox{1.8cm}{\centering Beam splitting}}& \multirow{2}{*}{E1039} & NH$_3^\uparrow$ & 7.2 $\times 10^{31}$ & 1.0 $\times 10^{7}$  & 7.2 $\times 10^{5}$ & 1.4 $\times 10^{28}$  & 1.0 $\times 10^{6}$ & 14\\
							   &  &  ND$_3^\uparrow$ & 7.2 $\times 10^{31}$ & 1.0 $\times 10^{7}$  & 7.2 $\times 10^{5}$ & 1.4 $\times 10^{28}$  & 1.0 $\times 10^{6}$ & 14\\
						     	\cline{2-9}
						     	             &  Unpolarised  & C & 2.8 $\times 10^{31}$  & 1.0 $\times 10^{7}$  & 2.8 $\times 10^{5}$  & 5.6 $\times 10^{27}$  & 1.0 $\times 10^{6}$ & 5.6\\
									&							   solid                             & Ti & 1.4 $\times 10^{31}$  & 1.0 $\times 10^{7}$  & 1.4 $\times 10^{5}$ & 2.8 $\times 10^{27}$ & 1.0 $\times 10^{6}$ & 2.8   \\			
									&							          target (5 mm)                                       & W & 1.6 $\times 10^{31}$ & 1.0 $\times 10^{7}$ & 1.6 $\times 10^{5}$  & 3.1 $\times 10^{27}$ &  1.0 $\times 10^{6}$ & 3.1   \\		\hline
\multirow{4}{*}{\parbox{1.8cm}{\centering Beam extraction}}& \multirow{2}{*}{E1039} & NH$_3^\uparrow$ & 7.2 $\times 10^{31}$ & 1.0 $\times 10^{7}$  & 7.2 $\times 10^{5}$ & 1.4 $\times 10^{28}$  & 1.0 $\times 10^{6}$ & 14\\
							   &  &  ND$_3^\uparrow$ & 7.2 $\times 10^{31}$ & 1.0 $\times 10^{7}$  & 7.2 $\times 10^{5}$ & 1.4 $\times 10^{28}$  & 1.0 $\times 10^{6}$ & 14\\
							                     \cline{2-9}
							                    &  \multirow{2}{*}{COMPASS} & NH$_3^\uparrow$ & 1.0 $\times 10^{33}$ & 1.0 $\times 10^{7}$ & 1.0 $\times 10^{7}$  & 2.0 $\times 10^{29}$ & 1.0 $\times 10^{6}$ & 2.0 $\times 10^{2}$ \\
							                     &  & $^{6}$LiD$^\uparrow$ & 1.9 $\times 10^{33}$  & 1.0 $\times 10^{7}$ & 1.9 $\times 10^{7}$ & 3.9 $\times 10^{29}$ & 1.0 $\times 10^{6}$ & 3.9 $\times 10^{2}$  \\ 	
								                &  & butanol $^\uparrow$ & 2.7 $\times 10^{32}$  & 1.0 $\times 10^{7}$ & 2.7 $\times 10^{6}$ & 5.3 $\times 10^{28}$ & 1.0 $\times 10^{6}$ & 53  \\ \hline		
\end{tabular} 
}
\end{adjustbox}	
\vspace{-0.6 true cm}
\caption{Summary table of the achievable integrated luminosities for the various technical solutions described in this section.}
\label{tab_lumi_comp}
\end{table}%
\vspace{+0.6 true cm}

\ct{tab_lumi_comp} compares the instantaneous luminosities, the expected running time with the proton or lead beam and the integrated luminosities achievable in one LHC year of data taking, for the various technical solutions described in this section. These numbers should be interpreted as maxima, and can be decreased according to data-taking-detector capabilities (see section \ref{section:detector}). For the internal-solid target with beam-halo, beam-splitting and beam-extraction solutions, the fluxes of proton and lead on target are assumed to be 5 $\times 10^{8}$~s$^{-1}$ and 10$^{5}$~s$^{-1}$, respectively. 
As can be seen from \cf{tab_lumi_comp}, the highest luminosity which can be achieved in pH$^{\uparrow}$ collisions is about 10 fb$^{-1}$ with a storage-cell gas target. Integrated luminosities of about the same order of magnitude could be reached with the E1039 target with a beam-splitting or beam-extraction option. The gas-jet solution gives luminosity about two orders of magnitude smaller than the storage cell for polarised hydrogen. For unpolarised $p$H$_{2}$ collisions, the performances of the storage-cell and gas-jet targets are similar, of the order of 40-50 fb$^{-1} $. In proton-nucleus collisions for large nuclei, the storage-cell and gas-jet targets also give the best integrated luminosities (on the order of the fb$^{-1}$). In order to obtain similar luminosities in $p$W collisions with an internal solid target would require the width of the target to be of about 50 mm (or the usage of serial targets). In lead-nucleus collisions, the performances of the storage-cell and gas-jet targets are also the best, allowing for the collection of approximately 1 fb$^{-1}$ in PbH$_{2}$ and 30 nb$^{-1}$ in PbXe collisions. 

\flushfootnote

\subsubsection{Comparison of the polarised-target performances for STSA measurements}

\ct{TabSTSA} shows the comparison of the figure of merit for STSA measurements, for the various polarised targets described in this section. The nominal LHC-proton-beam flux is considered for the gaseous targets while the expected proton-beam flux extracted by means of a bent crystal is considered for the E1039 and COMPASS targets.  While the absolute systematic error on STSA measurements is governed by the precision on the luminosity measurement for the two polarisation states (strictly speaking, the determination of their {\it relative} luminosity), the relative error will mainly come from the knowledge of the polarisation (usually determined with a precision of about 3-4$\%$, see \eg\ \cite{Airapetian:2004yf,Adams:1999qy}).

\begin{table}[h]\center
\renewcommand{\arraystretch}{1.3}
{
\begin{tabular}{p{3.5cm}|c|c|c|c|c|c|c} 
\centering Target & $P_{T}$   & $\langle f \rangle$  or $\alpha$  & $\sum_i A_i$ &  $\theta_{\rm target}$ &  $\mathcal{L}$ & ${\cal P}^{2}_{\text{eff}}$ & $ {\cal F}$   \\	
						 &           &  & & [cm$^{-2}$] & [cm$^{-2}$s$^{-1}$] &  & [cm$^{-2}$ s$^{-1}$] \\
\hline \hline
\centering NH$_3$  E1039 & 0.85 &  0.17 & 17 & 1.4 $\times 10^{23}$ & 7.2 $\times 10^{31}$ & 0.021 &  2.6 $\times 10^{31}$ \\ \hline
\centering ND$_3$  E1039 & 0.32 & 0.30  & 20 & 1.4 $\times 10^{23}$ & 7.2 $\times 10^{31}$ & 0.009 &  1.3 $\times 10^{31}$ \\ \hline
\centering NH$_3$  COMPASS &0.90 & 0.18 & 17 & 2.0 $\times 10^{24}$ & 1.0 $\times 10^{33}$ & 0.025 &  4.3 $\times 10^{32}$ \\ \hline
\centering Butanol COMPASS & 0.90 & 0.14 & 74 & 5.3 $\times 10^{23}$& 2.7 $\times 10^{32}$  & 0.015 &  3.0 $\times 10^{32}$ \\ \hline
\multirow{2}{*}{\hspace{+0.3 true cm} $^6$LiD COMPASS} & 0.46\tablefootnote{For D.} & \multirow{2}{*}{0.250} & \multirow{2}{*}{8} & \multirow{2}{*}{3.9 $\times 10^{24}$}& \multirow{2}{*}{1.9 $\times 10^{33}$}  & \multirow{2}{*}{0.050} &  \multirow{2}{*}{7.6 $\times 10^{32}$} \\ 
& 0.43\tablefootnote{For $^6$Li.} & &  & &  & &   \\ \hline
\centering H~HERMES like  storage cell \tablefootnote{$T=300$~K, $\ell=100$ cm.} &  0.85 & 0.95 & 1 & 2.5 $\times 10^{14}$ & 9.2 $\times 10^{32}$  & 0.650 & 6.0 $\times 10^{32}$ \\ \hline
\centering $^3$He~HERMES like storage cell\tablefootnote{$T=300$~K, $\ell=100$ cm.} & 0.70 & 0.33  & 3 &  1.0 $\times 10^{15}$ & 3.7 $\times 10^{33}$  & 0.053 & 5.9 $\times 10^{32}$ \\ \hline
\centering D~HERMES like storage cell\tablefootnote{$T=300$~K, $\ell=100$ cm.} & 0.85 & 0.92  & 2 &  2.9 $\times 10^{14}$ & 1.1 $\times 10^{33}$  & 0.610 & 1.3 $\times 10^{33}$ \\ \hline
\centering H RHIC-like  gas jet &  0.96 & 0.97 & 1 & 1.2 $\times 10^{12}$ & 4.3 $\times 10^{30}$  & 0.860 & 3.7 $\times 10^{30}$ \\ \hline
\centering $^3$He RHIC-like gas jet & 0.70  & 0.33 & 3 & 1.0 $\times 10^{14}$ & 3.6 $\times 10^{32}$  & 0.053 &   5.8 $\times 10^{31}$ \\ \hline
\centering D RHIC-like gas jet & 0.85  & {\cal O}(1) & 2 & 1.2 $\times 10^{12}$ & 4.3 $\times 10^{30}$  & 0.720 &   6.2 $\times 10^{30}$ \\ \hline
\end{tabular}
}
\caption{Comparison of the target performances for STSA measurements. From left to right: target, target polarisation ($P_{\rm T}$), average dilution factor ($\langle f \rangle $) or depolarisation factor ($\alpha$), total number of nucleons in the target ($\sum_i A_i$), target areal density ($\theta_{\rm target}$), instantaneous luminosity ($\mathcal{L}$), effective polarisation (${\cal P}_{\text{eff}}$) and spin figure of merit of the target and beam ($ {\cal F}$).}
\label{TabSTSA}
\end{table}

The 'HERMES' H-gas target has the best figure of merit, on the order of 6 $\times 10^{32}$ cm$^{-2} s^{-1}$. Using a cooled storage cell at $T=100$~K, the gas density would increase by a factor of $\sqrt{3}$, leading to an increase in the instantaneous luminosity to 1.59 $\times 10^{33}$ cm$^{-2}$s$^{-1}$, which is about 16$\%$ of the pp collider luminosity. The figure of merit $ {\cal F}$  would increase up to 1.04 $\times 10^{33}$ cm$^{-2}$s$^{-1}$. The 'RHIC' H-jet-gas-target figure of merit is smaller by two orders of magnitudes with respect to the 'HERMES' H-gas target, because of the smaller achievable gas density. Similar performances as the 'RHIC' H-jet gas target can be reached with the solid E1039 and COMPASS targets. Due to the larger length of the COMPASS target, the figure of merit of the COMPASS target is better with respect to the E1039 one, at the cost of less portability and the likely impossibility to couple it to the beam-split option. 

\flushfootnote


%% file: detector/detector.tex
\section{Detector requirements and expected performances}
\label{section:detector}

The ambitious physics case outlined in this document imposes significant requirements on the detector needed for such an experiment.
The particle production is shifted towards larger angles,
and the rapidity shifts are $\Delta y = $ 4.2 and 4.8 for a beam energy per nucleon 
of $2.76$ and $7$ TeV, respectively. \cf{fig:acceptanceComparison-LHC-RHIC} shows 
the rapidity acceptances of ALICE~\cite{Aamodt:2008zz} and LHCb~\cite{Alves:2008zz} detectors in the collider and fixed-target modes for a given target position (see caption), in comparison 
with the STAR~\cite{Anderson:2003ur,Beddo:2002zx,Allgower:2002zy,Jacobs_2009} and PHENIX~\cite{Adcox:2003zm} detectors at RHIC. 

\begin{figure}[hbt!]
\centering
\includegraphics[width=0.8\textwidth]{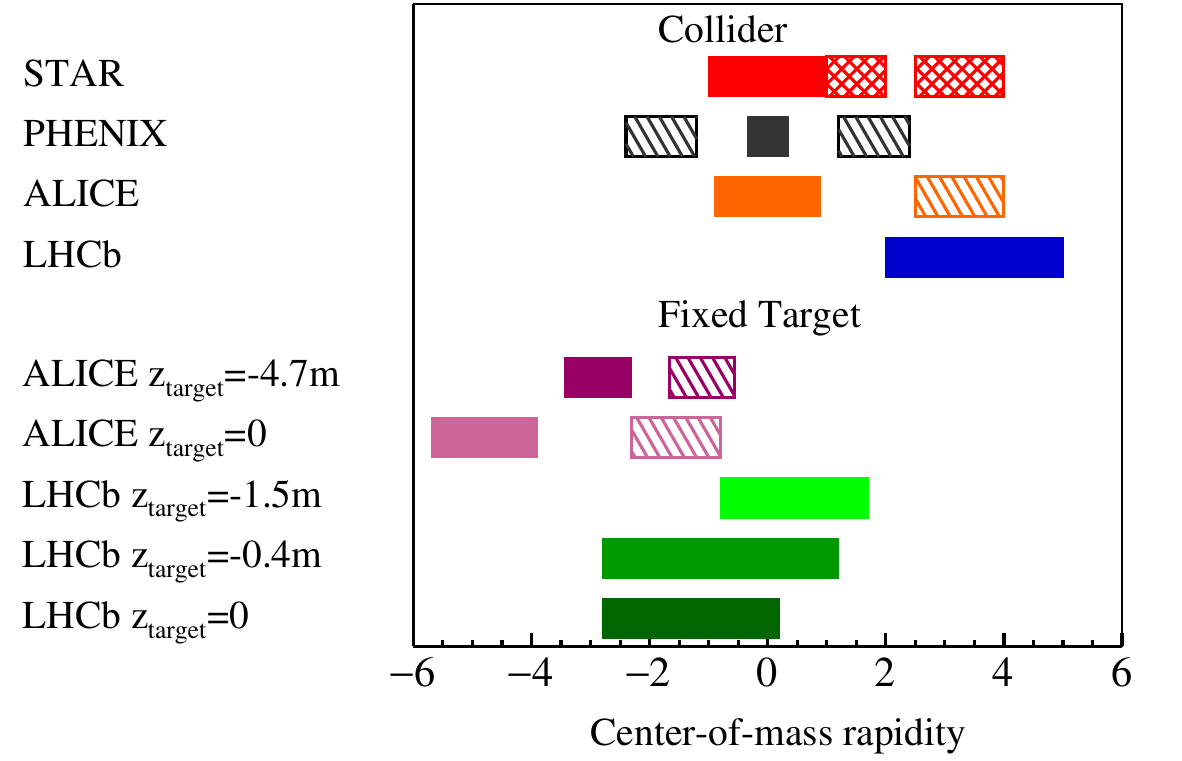} 
\caption{Comparison of the kinematical coverages of the ALICE and LHCb detectors at the LHC and the STAR and PHENIX detectors at RHIC. 
	For ALICE and LHCb, the acceptance is shown in the collider and the fixed-target modes for a 7 TeV proton beam. For LHCb, the target position is at the nominal Interaction Point (IP), \ie~$z_{\rm target} = 0$, and the acceptance are also shown for two other positions corresponding to $z_{\rm target} = -0.4$~m and $-1.5$~m on the opposite side of the spectrometer (see text for more details). For ALICE, the acceptances are shown for a target located at the IP as well as at  $z_{\rm target} = -4.7$~m on the opposite side of the Muon spectrometer. The fully filled rectangles refer to detectors with particle identification capabilities, the double-hatched rectangles to electromagnetic calorimeters and the hatched rectangles to muon detectors.} 
\label{fig:acceptanceComparison-LHC-RHIC}
\end{figure} 

As outlined by this comparison of LHC detectors used in a fixed-target mode, 
the major advantage of a fixed-target experiment is that 
particle production can be easily measured at very large values of negative-\ycms with standard detector technologies. On the other hand, the full
forward hemisphere is compressed into a very small solid angle area. 
The instantaneous luminosities with a fixed-target experiment by using the LHC beams are expected to 
be high, as described in the previous section, leading to large inelastic rates and allowing one to probe the 
full rapidity range with high statistics for many processes.  
In this section, we will first describe the general detector requirements in order to achieve the rich physics programme proposed in that paper for 
a fixed-target experiment at the LHC, and we will then discuss more specifically possible implementations with 
the existing detectors of the ALICE and LHCb experiments. The two implementations will be compared in terms of rapidity coverage, integrated luminosities 
and physics reach. 

\subsection{Detector requirements}

The rapidity range in the laboratory frame of a fixed-target experiment should be as broad as possible covering the regions of 
backward and mid-rapidity in the \cms, i.e. from $\ylab=0$ to $\ylab=4.2$ and $4.8$ 
with a beam energy per nucleon of $2.76$ and $7$ TeV, respectively. 
A multi-purpose experiment with detectors able to identify particles such as electrons, 
hadrons, photons as well as muons down to low $p_T$ would fit better the rich physics 
programme proposed here. 
A high-resolution vertex detector would allow one to measure precisely the primary and secondary vertices
associated to the production of heavy-flavour hadrons.
A polarised target requires space, e.g. for pumping system and diagnosis apparatus in the case of a gas target, 
and it is challenging, but possible, to couple it with a large angle detector.

\begin{figure}[ht!]
\centering
{\includegraphics[width=\textwidth]{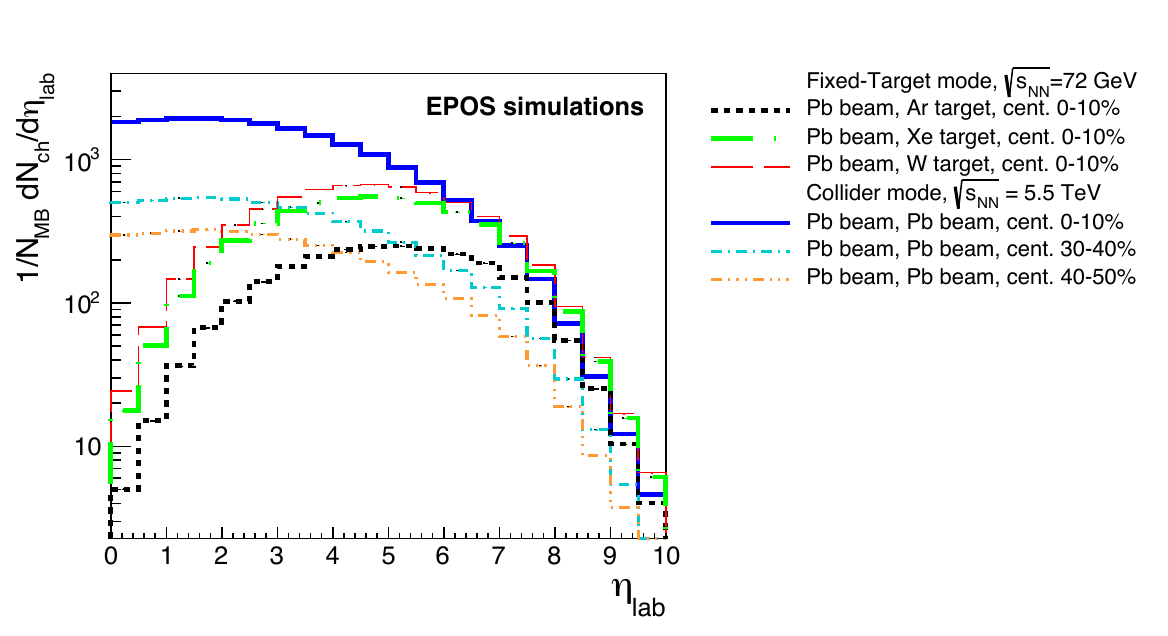}}
\caption{The averaged charged-particle multiplicity as a function of the pseudorapidity in the 
laboratory frame for the most central collisions in \PbAr, \PbXe and \PbW collisions at \sqrtsNN~= 72~GeV in a fixed-target mode and for various event centrality intervals in \PbPb collisions at \sqrtsNN~=5.5~TeV in a collider mode. 
}
\label{fig:multiplicity}
\end{figure}

The physics case comprises lead-nucleus collisions with instantaneous
luminosity that can reach up to $3\times10^{28}$ cm$^{-2}$s$^{-1}$ as well as proton-proton and proton-nucleus collisions with 
instantaneous luminosity up to 10$^{33}$ cm$^{-2}$s$^{-1}$. These numbers correspond to the maximum luminosities in each colliding system quoted in Table~\ref{tab_lumi_comp}.
The detectors must be able to cope with the occupancies and fluences for
both of these configurations. 
In the case of the heaviest nuclear collisions foreseen,
\PbXe and \PbW collisions at \sqrtsNN $=$ 72 GeV, 
the average number of charged particles is maximal at $\etalab \sim 4.2$ and amounts to $dN_{ch}/d\eta \sim 600-700$ 
for the most 10\% central 
collisions according to EPOS~\cite{Pierog:2013ria,Werner:2005jf}.
The charged-particle multiplicity is shown in \cf{fig:multiplicity} 
for various heavy-ion systems as a function of the pseudorapidity in the laboratory frame and is compared to the one 
obtained at the LHC in a collider mode. In a fixed-target mode, the multiplicity does not exceed the 
one obtained in \PbPb collisions at \sqrtsNN = 5.5 TeV in a collider mode. 
If one considers the maximum instantaneous luminosities quoted in Table~\ref{tab_lumi_comp} and the inelastic cross-sections from EPOS \footnote{In order to compute the inelastic rate, we use the inelastic cross sections from EPOS, $\sigma_{\rm inel.} = 39$ mb in \pp collisions at $\sqrts = 115$~GeV, $\sigma_{\rm inel.} = 1.3$ b in \pXe collisions at \sqrtsNN = 115 GeV and $\sigma_{\rm inel.} = 6.2$ b in \PbXe collisions at \sqrtsNN = 72 GeV, and the instantaneous luminosities of 10$^{33}$~cm$^{-2}$s$^{-1}$, $2 \times 10^{32}$~cm$^{-2}$s$^{-1}$ and $3\times 10^{28}$~cm$^{-2}$s$^{-1}$ in \pp, \pXe and \PbXe collisions, respectively.}
one ends up with inelastic rates corresponding to 36 MHz, 300 MHz and 190 kHz in \pp, \pXe and \PbXe collisions, respectively. These 
numbers are, for the \pp\ and \AA\ cases, of the same order of magnitude than the maximum rates planned for LHC in a collider mode in Run 3 and Run 4. 

\subsection{Possible implementations with existing apparatus}

The proposed physics programme is rich and it is clear that building a completely new experiment would allow 
one to cope with the various requirements briefly detailed above. 
However one could already use an existing detector 
at the LHC in order to cover a large part of the physics programme. In this respect, 
the right panel of \cf{fig:acceptance} shows the evolution of the 
rapidity coverage in the \cms\ frame with \sqrtsNN. The rapidity phase-space decreases 
while lowering the energy. 
While in a collider mode a forward-angle detector with 
$2 < \eta < 5$ covers approximately a forward-rapidity region of $2< \ycms < 5$, in a fixed-target mode the same 
detector covers the mid-rapidity region as well as half of the backward-rapidity region. 
In the case of a proton beam of 7 TeV on a fixed target, 
the rapidity coverage for the mentioned pseudorapidity range is $-2.8 <\ycms < 0.2 $. 

\begin{figure}[ht!]
\centering
{\includegraphics[width=0.6\textwidth]{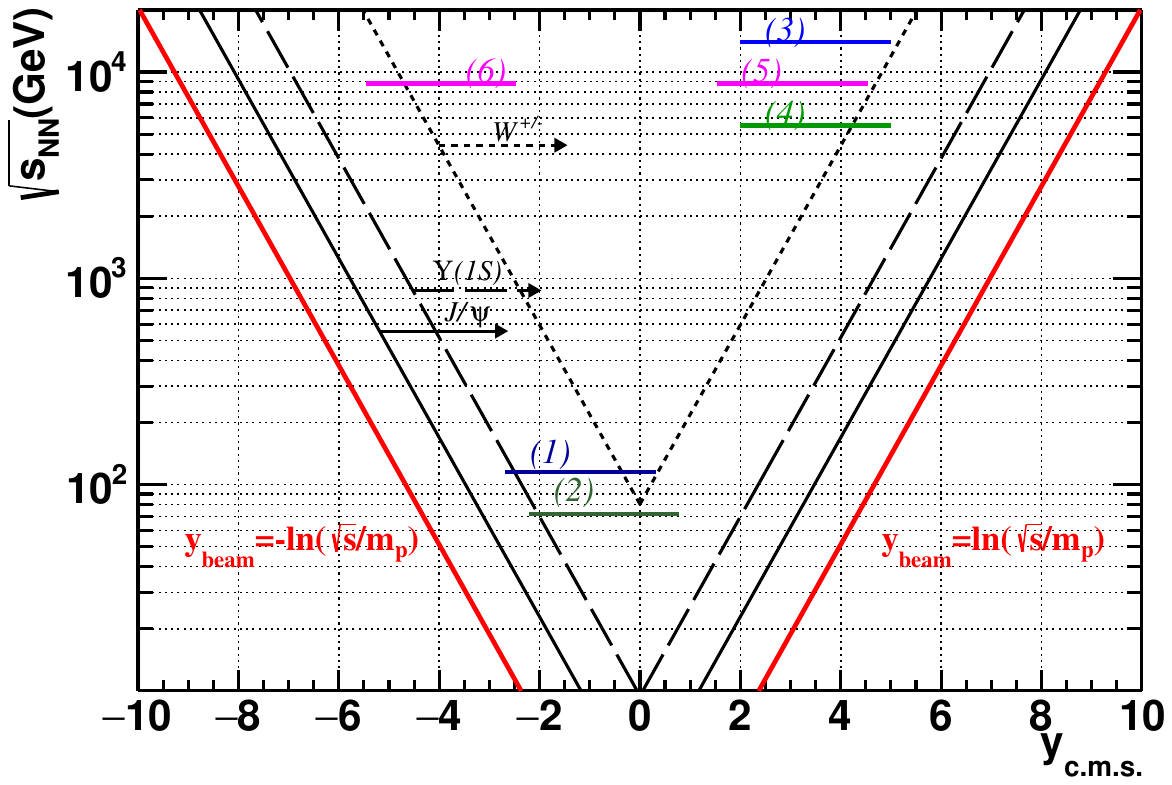}}
\caption{The \ycms\ coverage 
as a function of the colliding energies per nucleon pair (\sqrtsNN). The red solid lines represent the beam rapidity. 
The solid, dashed and dotted black lines respectively show the \ycms coverage for the \jpsi, $\Upsilon(1S)$ 
and $W^{+/-}$ production. The horizontal lines show the \ycms acceptance of a detector with a pseudorapidity coverage in the 
laboratory frame of $2<\eta<5$ for different colliding systems and modes using 
the 7 TeV proton and 2.76 A.TeV Pb LHC beams: (1) \pp and \pA collisions in the fixed-target mode at $\sqrtsNN = 115$ 
GeV, (2) \PbA collisions in the fixed-target mode at $\sqrtsNN = 72$ GeV, (3) \pp collisions in the collider mode at 
$\sqrts = 14$ TeV, (4) \PbPb collisions in the collider mode at $\sqrtsNN = 5.5$ TeV, (5) \pPb collisions in the collider mode at $\sqrtsNN = 8.8$ TeV, (6) \Pbp collisions in the collider mode at $\sqrtsNN = 8.8$ TeV.
}
\label{fig:acceptance}
\end{figure}

In the following sections, we will discuss possible implementations of the fixed-target programme at the LHC 
with two existing experiments: ALICE and LHCb. 
The detectors will be briefly presented in both cases as well as their upgrades planned for LHC Run 3 and 4 and
 we will discuss their ability to cover the physics programme described in this document. In the case of LHCb, 
a fixed-target programme has recently started with a reduced luminosity and some aspects of the fixed-target 
mode will be described. For both experiments, the rapidity acceptance, the achievable luminosities as well as the physics reach will be discussed 
for various fixed-target systems and based on experimental constraints.

\input{./detector/alice.tex}

\input{./detector/lhcb.tex}

\input{./detector/overview.tex}

%% file: detector/alice.tex
\subsubsection{ALICE as a fixed-target experiment}
\label{section:detector:alice}

The detectors of ALICE~\cite{Aamodt:2008zz,Abelev:2014ffa} are optimised for studying the QCD matter 
created in high-energy collisions of lead nuclei. 
They are able to cope with high-multiplicity events and to track charged particles down to $p_T\sim0.15$ GeV$/$c 
at mid-rapidity. 

The Central Barrel (CB) detectors are embedded into the L3 solenoid magnet that provides a field of 
0.5~T parallel to the beam line. The inner most detector, the Inner Tracking System (ITS), tracks charged particles 
within $|\eta|<0.9$ and allows one to reconstruct primary and secondary vertices. The two innermost layers of the ITS 
cover $|\eta|<2$ and $|\eta|<1.4$ for the first and second layer, respectively. 
The resolution on the longitudinal position of the primary vertex ranges from 10 to 150 $\mu m$ decreasing 
with the charged-particle multiplicity. The Time Projection Chamber (TPC) provides track reconstruction as well as 
particle identification (PID) via the measurement of the specific ionisation energy loss ${\rm d}E/{\rm d}x$ in the 
gas volume. The phase space covered by the TPC in pseudorapidity is $|\eta|<0.9$ with full radial track length. 
The TPC acceptance can be extended by considering only 1/3 of the full radial track length (also denoted as 
``TPC reduced track length'' in the following) at 
the cost of worsening the momentum resolution. In that case, the pseudorapidity acceptance is $|\eta|<1.5$. 
The Time Of Flight (TOF) detector extends the PID via the measurement of the flight time of the charged 
particles from the Interaction Point (IP). Its pseudorapidity coverage is $|\eta|<0.9$. 
For that purpose the T0 detector located along the beam line 
measures the event collision time. 
The CB includes also High Momentum Particles Identification Detector (HMPID), calorimeters 
(Electromagnetic Calorimeter: EMCal and Photon Spectrometer: PHOS) and Transition Radiation Detector (TRD) for particle identification purpose. 
The transverse momentum relative resolution measured with both ITS and TPC ranges from 0.8 to 2\% for $p_T=1$ 
to 10 GeV.

At forward rapidity, the Muon Spectrometer (MS) covers the pseudorapidity range $2.5 < \eta < 4$ in the laboratory
frame. It includes a dipole magnet with an integrated field of 3~Tm, five tracking
stations
and two trigger stations.
A system of absorbers located in front of the tracking and trigger stations and around the beam pipe 
is used for filtering out the hadrons and to protect the chambers from secondary particles produced 
during interactions of large-$\eta$ primary particles with the beam pipe.  
The combined effect of the front absorber and of the iron wall implies the detection of tracks matching the 
trigger chambers with $p > 4$ GeV. The relative muon momentum resolution is $\delta p/p \approx 1\%$.

The ALICE upgrade~\cite{Abelev_et_al_2014} is scheduled for the second LHC Long Shutdown (LS2) that will take place in 2019 and 2020, and will exploit the LHC Run 3 and Run 4. In order to allow for a continuous readout at 
an interaction rate of 50 kHz in \PbPb collisions at $\sqrtsNN=5.5$~TeV, many detectors or their electronics will be upgraded. In \pp and \pA collisions, the detector upgrade will allow one to record data with a rate of 200 kHz. \cf{fig:Alice:Det:Run3} presents a schematic view of the ALICE detectors for Run 3.   
A new detector, 
the Muon Forward Tracker (MFT), a 
Si-tracking detector, is designed to add vertexing capabilities to the MS by measuring charged tracks 
with a high spatial resolution. It is positionned along the beam axis between the ITS inner barrel 
and the MS front absorber. The MFT will cover the pseudorapidity acceptance $2.5 < \eta < 3.6$. 
The MFT capability to identify tracks coming from secondary vertices is measured by experimental resolution 
on the track offset to the primary vertex, the latter being measured by the ITS. 
Resolutions below 100 and 1000 $\mu m$ are found 
for $p_T > 1 $ GeV in the transverse and longitudinal direction, respectively. 
These resolutions decrease with increasing $p_T$ down to 25 and 180 $\mu m$ at large $p_T$ 
in the transverse and longitudinal direction, respectively. 
It is worth noting that the mass resolution will be greatly improved for the low-mass dimuon 
($M_{\mu\mu} < 1.5$ GeV) by adding the MFT to the MS. The mass resolution has been evaluated 
to be lower than 20 MeV for $\eta$, $\phi$ and $\omega$ mesons.

\begin{figure}[ht!]
\centering
\subfigure[~]{\includegraphics[width=0.49\textwidth]{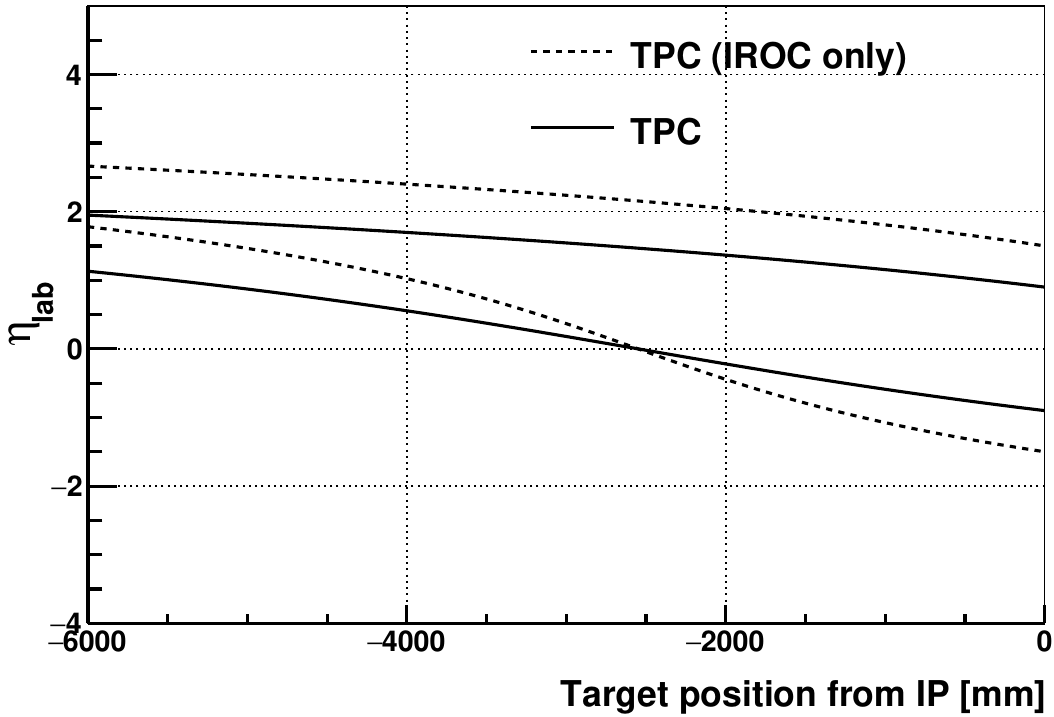}}
\subfigure[~]{\includegraphics[width=0.49\textwidth]{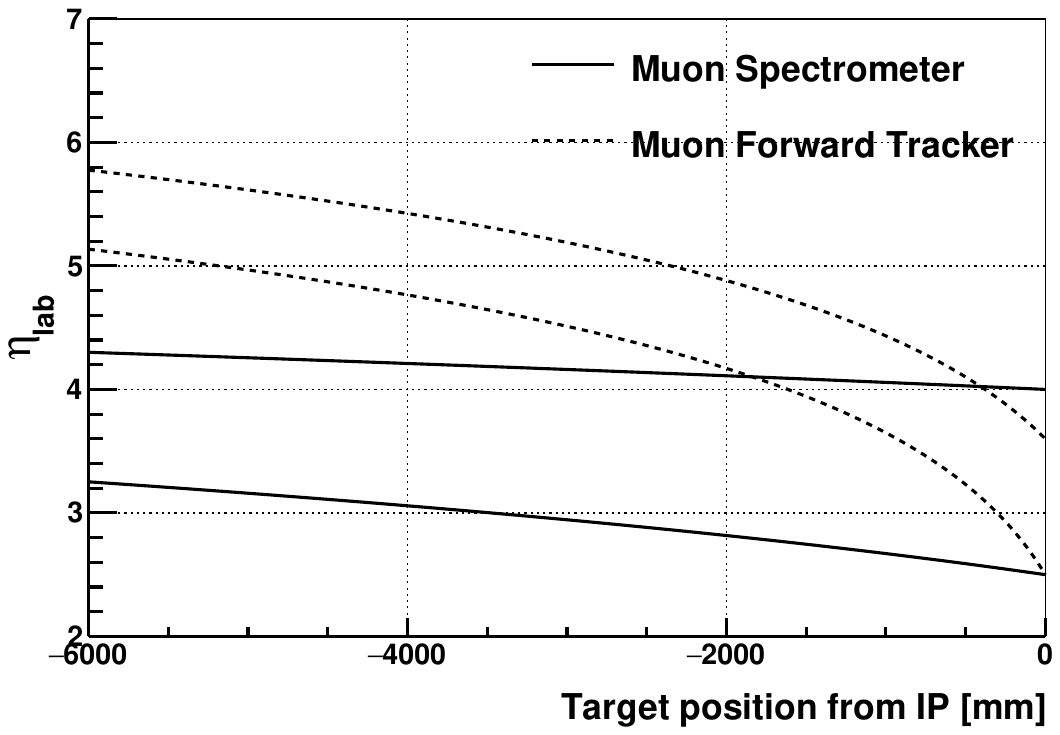}}
\caption{(a) $\eta_{\rm lab.}$ acceptance (between 2 curves) of the ALICE TPC as a function of the target position ($z_{\rm target}$) upstream from the nominal IP. Full (dashed) lines refer to reduced (full) radial track length in the TPC; (b) $\eta_{\rm lab.}$ acceptance of the ALICE MS (full line) and MFT (dashed line) as a function of $z_{\rm target}$.
}
\label{fig:acceptanceALICE}
\end{figure}

In a fixed-target mode and with a target positioned at the IP, the acceptance of the 
MFT and MS allow for measurements in the rapidity regions of $-2.3 < \ycms < -1.2$ 
with a $7$ TeV proton beam and of $-1.7 < \ycms < -0.6 $ with a $2.76$ A.TeV Pb beam. The CB covers 
the very rear region with a center-of-mass rapidity of $\ycms<-3.9$ and $\ycms<-3.3$, respectively, i.e. it 
allows one to access the very high-$x$ region. 
The target can also be displaced upstream of the nominal IP (on the A-side of ALICE, opposite to the MS). \cf{fig:acceptanceALICE} 
shows the pseudorapidity acceptance of some ALICE detectors as a function of the target position, where the acceptance 
is computed considering the geometry of the active detectors. The acceptances are shifted towards the forward region when the target is displaced upstream 
of the nominal IP, in the opposite direction of the MS. 
If the target is displaced by a large amount and if one wants to 
measure the primary and secondary vertices precisely, a new vertex detector close to the target is then required.
In case of a polarised target, its polarisation may be modified with the magnetic field of the L3 magnet. 
This brings additional constraints on the target position. 

The occupancy of the \AA~collision systems is not an issue in ALICE since the detectors were designed to 
measure \PbPb collisions at \sqrtsNN = 5.5 TeV and the average charged-particle multiplicity in a fixed-target mode 
does not exceed the one in a collider mode as shown in \cf{fig:acceptance}.

As specified above, the detectors in ALICE will be upgraded in Run 3 in order to cope with an inelastic rate of $50$ kHz in \PbPb collisions and $200$ kHz in \pp and \pA collisions, in a collider mode. The rate is essentially limited by the detector occupancy and, since similar occupancy is expected in $50$ kHz \PbPb and 4.5 MHz \pp collisions, it might be possible to run up to 1 MHz in \pp and \pA collisions in a collider mode. This is under discussion in ALICE. %
In addition, in a fixed-target mode, a lower detector occupancy is expected and the rate will be limited by the occupancy in the MS acceptance, where the multiplicity is the largest. By scaling the average charged-particle multiplicities of the fixed-target to the collider mode in the MS acceptance, an increase in the ALICE readout 
rate by factors of about two in \PbXe and ten in $p$H collisions, respectively, can be projected. Further studies are needed to demonstrate if such a higher rate is sustainable. 
 
The MS and the MFT will cover the physics programme described above with the detection of single
muon from heavy-flavours, muon pairs (such as DY) and quarkonia down to low \pt. Further works are needed 
to estimate the level of background for the critical analyses such as DY in \AA~collisions and 
in case the target is displaced upstream from the ALICE IP. 
The CB can detect and identify neutral 
and charged particles in the very backward region. Further studies will determine if the achievable luminosities 
allow one to complete some of the physics cases in this rear region.

There are ongoing feasibility studies on the installation of an internal solid target in the ALICE experiment~\cite{Barschel:2653780}. 
The beam splitting option is currently investigated, where the beam halo is deflected by a crystal placed $\sim70$~m upstream from the nominal IP, and the deflected particles hit the target located inside the L3 magnet. The target holder is envisioned as an adjustable device, which facilitates moving the target from the parking position (outside of the beam pipe) to the working point, 13 mm from the beam axis~\cite{Galluccio:2671944}. The mechanical design of the system is under study. 
The target system could be integrated during LS3 at approximately 5~m from the interaction point (opposite side of the MS), in front of an existing valve that will be located at 4.8~m from the interaction point in Run 3. The crystal and target devices could then be used in Run 4.

%% file: detector/lhcb.tex
\subsubsection{LHCb as a fixed-target experiment}
\label{section:detector_lhcb}

The LHCb detector~\cite{Alves:2008zz,Aaij:2014jba} is a single-arm forward spectrometer,
designed for studies of hadrons containing $b$ and/or $c$ quarks. 
Its pseudorapidity coverage in the laboratory frame is $2<\eta<5$. Such a geometrical coverage offers great opportunities when it is used in the fixed-target mode. 
It comprises a high precision tracking system, two ring-imaging Cherenkov detectors for the identification of different types of charged hadrons, a calorimeter system to identify photons, electrons and hadrons, and a muon system for the muon identification.
The tracking system includes a silicon-strip vertex locator (VELO) and four stations with a dipole magnet between the first and the other three stations.
It can achieve a relative momentum uncertainty of charged particles varying from $0.5\%$ to $0.8\%$ for the momentum between a few \gev to $100\gev$.
The calorimeter system is composed of a scintillating pad, a preshower detector, an electromagnetic calorimeter, and a hadron calorimeter.
The muon system consists of five muon stations with alternating layers of
iron and multiwire proportional chambers. As described below, some detectors will be upgraded during the LS2 and the LS3, in order to exploit the LHC Run 3 and Run 4 periods. A schematic view of the upgraded detectors is shown in Fig.~\ref{fig:LHCb:det}. The different subdetectors are as follows:

\begin{itemize}
\item VELO. 
The current VELO of LHCb is composed of 84 single-sided silicon strip sensors, operated in a secondary vacuum inside the LHC beam pipe~\cite{Alves:2008zz}.
The VELO length is about $1\m$ along the beam.
The pitch of the $R$ sensors varies from $40$ to $102\mum$, and that of the $\phi$ sensors varies from $38$ to $97\mum$. 
The length of the shortest (longest) strip is $3.8\mm$ ($33.8\mm$) for R sensors;
The length of the shortest (longest) strip is $5.9\mm$ ($24.9\mm$) for $\phi$ sensors.
The resolution of the reconstructed primary vertex is $13\mum$ in the $x-y$ plane and $71\mum$ in the $z$ direction, assuming that the number of tracks of the primary vertex is $25$.
When the number of tracks reduces, the resolution becomes slightly worse.
The resolution of the impact parameter is about $15-50\mum$~\cite{LHCbVELOGroup:2014uea,Aaij:2014jba}.
The excellent vertex reconstruction ability allows one to well separate the primary vertex and the secondary vertex of B or charmed hadron decays.
For the LHCb upgrade during the LS2, the current VELO detector will be completely replaced by a new detector based on hybrid silicon pixel sensors~\cite{Collaboration:1624070}.
The pixel pitch is $50\mum\times50\mum$.
It will have the same physics performance and can deliver a readout at $40\mhz$.
Compared to the current silicon strip VELO, the new VELO can cope with events with much higher track multiplicity.

\item Tracking. 
The first station of the current tracking system is based on silicon micro-strip. The other three stations, which are located after the LHCb dipole magnet, are composed of silicon micro-strip inner trackers and straw drift tube outer trackers. The relative momentum resolution is about $0.5-1\%$. 
The mass resolution of \KS mesons is $3.5~(7)\mev$ if they decay inside (outside) the VELO.
For the LHCb upgrade, the first station will be replaced by high granularity silicon micro-strip planes, and the other stations will be replaced by scintillating fibre trackers~\cite{Collaboration:1647400}.
The momentum resolution will be about $10-20\%$ better than the current resolution. 

\item Calorimeter. 
The electromagnetic calorimeter is composed of a sampling scintillator-lead structure. The hadron calorimeter is a sampling scintillator-iron structure. 
The mass resolution of low transverse momentum \piz mesons, reconstructed with well-separated photons, is $8\mev$. For \piz mesons with transverse momentum greater than $2\gev$, the mass resolution is around $20~(30)\mev$ for those reconstructed with well separated (merged) photons.
If one uses converted photons, the resolution of the mass difference between $M(\mumu\gamma)$ and $M(\mumu)$ is around $5\mev$ for \chic states.
LHCb is now discussing an upgrade of the calorimeter that would occur in LS3 and LS4~\cite{Aaij:2244311}, replacing the current electromagnetic calorimeter by a silicon-tungsten sampling calorimeter. 

\item Muon system. 
The muon system includes five rectangular shaped stations. The first station is made of triple Gas Electron Multiplier detectors, while the other four are composed of multiwire proportional chambers.
The dimuon invariant-mass resolution is about $14~(43)\mev$ at the \jpsi (\OneS) mass. 
The muon identification efficiency is above $95\%$ for the tracks with transverse momentum above $1.7\gev$.
\item Readout. 
The current LHCb detector reduces the event rate from $40\mhz$ to $1\mhz$ at the first level hardware trigger. After the LHC LS2, the upgraded hardware trigger will have the capability to read the full event information at a rate up to $40\mhz$.

\item \herschel detector~\cite{CarvalhoAkiba:2018paq}. 

LHCb installed a \herschel (High Rapidity Shower Counters for LHCb) subdetector for Run 2 of the LHC. 
It is a system of forward shower counters consisting of five scintillator planes with PMTs.
These five stations are installed perpendicular to the beam, their $z$ coordinates are $-114\m$, $-19.7\m$, $-7.5\m$, $+20\m$, and $+114\m$, respectively. 
The $z$ direction of the coordinate system is from the VELO to the muon system along the beam, and the origin is the interaction point inside the VELO. 
Combining with other LHCb subdetectors, \herschel greatly extends the sensitivity to detect charged particles at high pseudorapidities: $-10<\eta<-5$, $-3.5<\eta<-1.5$, $2<\eta<5$, and $5<\eta<10$.
\end{itemize}

The capability of LHCb to cope with high-multiplicity events and the limit on the event charged track multiplicity will be defined with the ongoing data reconstruction of \PbPb collisions at $5\tev$ and \PbAr collisions at $69\gev$. 
The occupancy of the VELO is essential to determine the track reconstruction, since the VELO provides the best position precision among all subdetectors of the tracking system. 
As shown in Fig. 19 of Ref.~\cite{LHCbVELOGroup:2014uea}, the cluster occupancy in the current VELO varies from $0.4\%$ to $0.6\%$ depending on the positions of the silicon strips.
This result is obtained with a data sample passing a random trigger on beam crossing, and the average number of visible interactions per beam crossing is $\mu=1.7$.
The occupancy of the upgraded VELO will be significantly reduced owing to the replacement of silicon strips by hybrid silicon pixels.
As shown in Fig.~20 of Ref.~\cite{Collaboration:1624070}, the cluster occupancy is $0.08\%$ for the pixels closest to the IP ($0.5\mm$). It drops rapidly below $0.01\%$ as the radius increases. This result is obtained using simulated minimum-bias events at $\mu=5.2$.
When the difference of the data samples is taken into account, the occupancy of the upgraded VELO is expected to be reduced by a factor of approximately $20$. 

As discussed in Section~\ref{subsubsec:Direct_gas}, since the pilot runs of $p$Ne and PbNe in 2012 and 2013, the direct injection system SMOG~\cite{Barschel:2014iua} is used in the fixed-target mode by injecting different gases inside the VELO vessel. 
For the $6.5\tev$ proton beams, only special runs, like van der Meer scans or the period corresponding to the ramp up of the beam energy, were used. The time duration of $p$He, $p$Ne and $p$Ar collisions was typically 10--20 hours for each year, respectively.
For $2.5\tev$ proton beams, the proton-gas data were taken in parallel with \pp collisions; the time duration of $p$He collisions was around 100 hours, and that of $p$Ne was around $200$ hours.
Only Neon and Argon gases were used with lead beams. During the \PbPb LHC run in 2015, \PbAr data were taken with a time duration of approximately 50 hours.

Even though the data taking time was limited, SMOG delivered physics results~\cite{Aaij:2018svt,Aaij:2018ogq}. The LHCb collaboration reported the anti-proton differential-cross-section measurement in $p$He collisions and the $p$Ar run demonstrated the LHCb capabilities for charmed meson and charmonium reconstruction in the fixed-target mode. Both the $\jpsi \rightarrow \mu^+\mu^-$ and $D^0 \rightarrow K^{\mp}\pi^{\pm}$ were measured with an excellent mass resolution and an adequate efficiency~\cite{LHCb:2017qap}. \cf{fig:SMOG:pAr:data} shows the ratio of \jpsi and $D^0$ yields evaluated in $p$Ar collisions at \sqrtsNN = 110 GeV. 

When the SMOG data were taken in parallel with \pp or \PbPb collisions, the trigger of the SMOG related events was designed to fully utilise the data acquisition (DAQ) potential. When the DAQ busy time was too large, tighter cuts were used for the SMOG trigger. 

In a fixed-target mode and with a target positioned at the IP, the LHCb detectors probe a rapidity region of $-2.7 < \ycms < 0.3$ with a 7 TeV proton beam and of $-2.2 < \ycms < 0.8$ with a 2.76 A.TeV Pb beam. If the target is shifted by 0.4 (1.5) m on the opposite side of the spectrometer, one obtains the following acceptance by simply considering the active parts of the detectors including the VELO: $ -2.7< \ycms < 1.3$ ($-0.7 < \ycms < 1.8$) with the proton beam and of $ -2.2 < \ycms < 1.8$ ($-0.2 < \ycms < 2.3$) with the Pb beam. 

\begin{figure}[!hbt]
\centering
    \subfigure[~]{\includegraphics[width=0.48\textwidth]{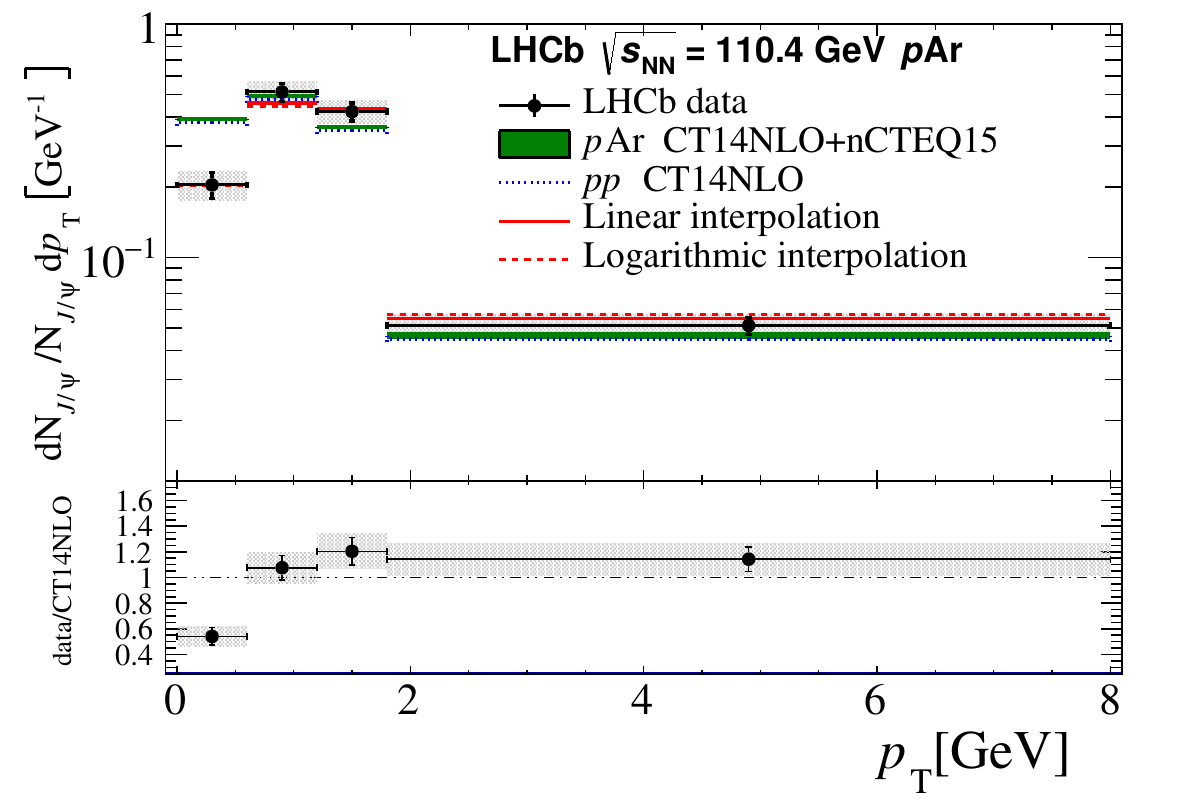}} 
    \subfigure[~]{\includegraphics[width=0.48\textwidth]{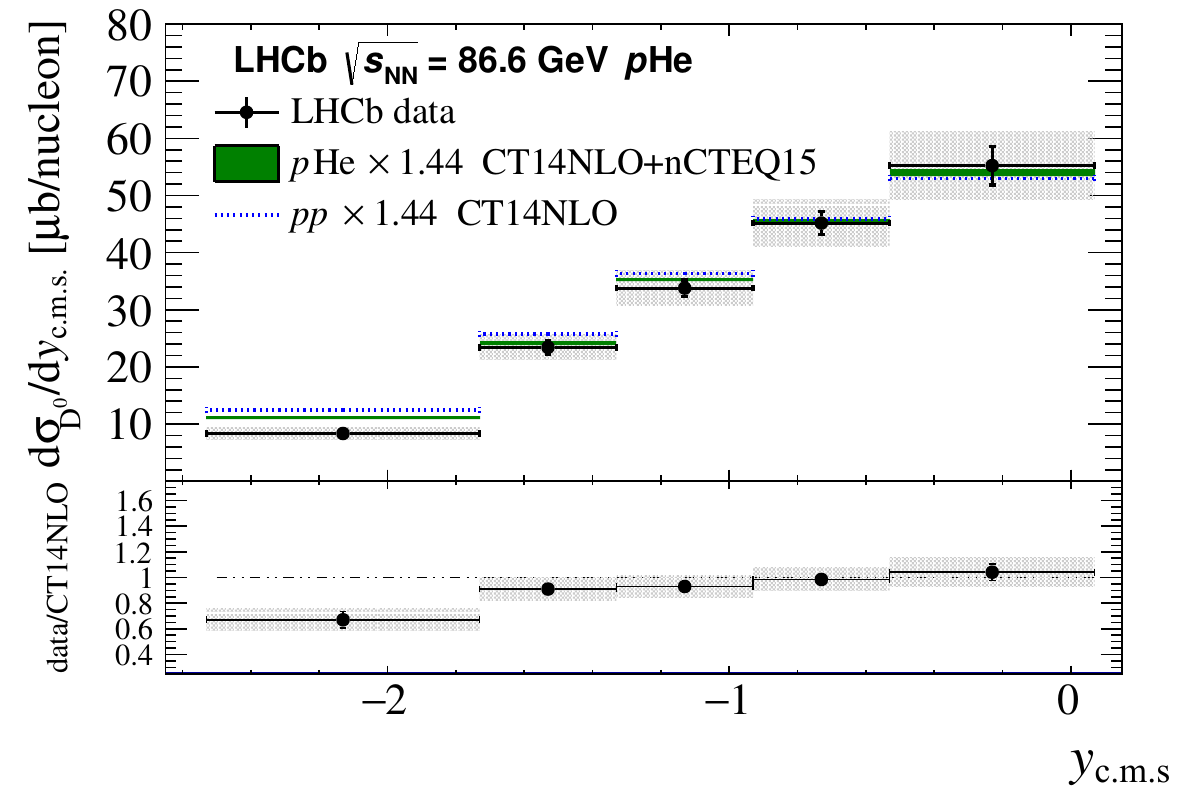}}
    \caption{(a) \jpsi yield as a function of transverse momentum in $p$Ar collisions. (b) $D$ meson cross-section as a function of center-of-mass rapidity in $p$He collisions. In both cases, the data are collected by the LHCb detector in the fixed-target mode. [Adapted from~\cite{Aaij:2018ogq}].}
    \label{fig:SMOG:pAr:data}
\end{figure}

Studies were carried out for the installation of polarised and unpolarised high-density gaseous target in the LHCb experiment~\cite{Barschel:2653780}.
It was proposed to inject the gases in a storage cell (as described in Section~\ref{subsubsec:storage_cell}) attached to the end of the VELO RF shields and located at 0.4~m from the nominal IP. This project, denoted SMOG2~\cite{DiNezza:2651269, LHCbCollaboration:2673690}, would significantly increase the luminosity, by up to a factor of 100. A baseline scenario was discussed in~\cite{Graziani:2019SeptemberPBC} for Run 3 with the following integrated luminosities: ${\cal L}_{pH}=150~$pb$^{-1}$, ${\cal L}_{pD}=9~$ pb$^{-1}$, ${\cal L}_{pAr}=45~$ pb$^{-1}$, ${\cal L}_{pKr}=30~$ pb$^{-1}$ and ${\cal L}_{pXe}=22~$ pb$^{-1}$at \sqrtsNN$=115$~GeV as well as ${\cal L}_{PbAr}=50~$nb$^{-1}$,  ${\cal L}_{PbH}=10~$nb$^{-1}$and ${\cal L}_{pAr}=5~$pb$^{-1}$ at \sqrtsNN$=72$~GeV. SMOG2 will be a first step towards the fixed-target programme described in this paper. 
The storage cell was also proposed to be used with a polarised setup~\cite{Aidala:2019pit} with an atomic-beam source, a gas diagnosis system and a magnet that would provide a 0.3~T field, transversally to the beam axis. In this case, the target position has to be defined according to the space availability in front of the VELO vessel. The target cannot be inside the VELO vessel because of the high gas flow that requires differential pumping system in a separate target chamber. Studies are ongoing with a target displaced by $1-2$~m from the IP in front of a sector valve which isolates the target chamber from the VELO~\cite{Pappallardo:2019FTEatLHC}. In these studies, a vertex detector is proposed behind the target. An installation of such a polarised target system is not foreseen before LS3.

%% file: detector/overview.tex
\subsubsection{Comparison of possible implementations}

\begin{figure}[hbt!]
\centering
\includegraphics[width=0.6\textwidth]{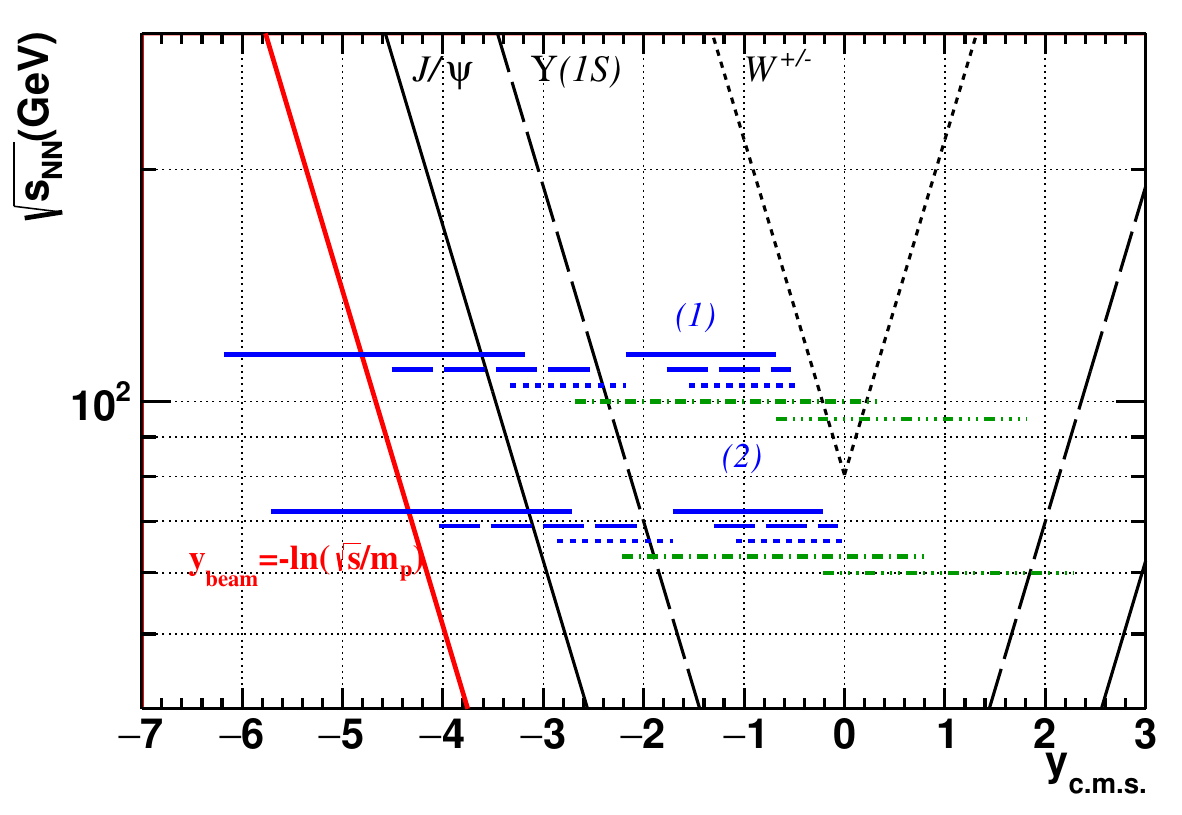}
\caption{Centre-of-mass-rapidity (\ycms) coverage as a function of the colliding energies per nucleon pair 
(\sqrtsNN) as in \cf{fig:acceptance}. The blue lines represent the acceptance of the TPC and MS 
of ALICE. The full, long-dashed and short-dashed 
lines correspond to targets located at the IP, upstream of the IP, at $z_{\rm target}$ = -2.75 and -4.7 m, 
respectively. The dash-dotted green lines represent the acceptance of the LHCb detector with a target 
at the IP and the dash-triple-dotted green lines with a target upstream of the IP by $z_{\rm target}=1.5$~m. 
The long-dashed and short-dashed blue lines as well as the dash-dotted and dash-triple-dotted green lines are shifted in energy 
for a better visibility. 
}
\label{fig:acceptanceComparison-FTLHC}
\end{figure}

\cf{fig:acceptanceComparison-FTLHC} shows the \ycms acceptances of the 
ALICE and LHCb detectors for two fixed-target energies, namely $\sqrtsNN = 72$ GeV and 115 GeV. The rapidity coverages with a 
target position at the nominal IP are shown as full line for ALICE and dash-dotted line for LHCb. 
The latter corresponds to the case of LHCb used as a fixed-target detector
with the SMOG system. The acceptances are also shown with two (one) other target positions for ALICE (LHCb), where the acceptance is determined considering the geometry of the active part of the detector as detailed in Section~\ref{section:detector:alice} and ~\ref{section:detector_lhcb}. For a target located at the IP, the 
ALICE MS and the LHCb detectors cover the central \ycms region as well as 
half of the backward rapidity acceptance ($\ycms<0$), while the ALICE CB has the particularity to probe the target-rapidity region and 
the end of the phase space ($x_2 \to 1$ and $x_F \to -1$). When the target is shifted on the opposite side of the detector, the rapidity interval probed is shifted forward. In the ALICE and LHCb cases, the wide \ycms range makes these detectors suitable 
to study the rapidity dependence of various probes. 

\begin{table}[!hbt]
\center
\begin{adjustbox}{angle=90}
\renewcommand{\arraystretch}{1.2}
\begin{tabularx}{0.87\textheight}{p{1.5cm}|c|c|c|c|c|c|c|c|c|c}
\multicolumn{3}{c}{ } & \multicolumn{8}{|c}{ALICE}  \\
\multicolumn{3}{c}{ } & \multicolumn{4}{|c|}{Proton beam (\sqrtsNN = 115 GeV)} & \multicolumn{4}{c}{Pb beam (\sqrtsNN = 72 GeV)} \\
\cline{4-11}
\multicolumn{3}{c|}{Target} & $\cal L$ & $\sigma_{inel.}$  & $\Gamma_{inel.}$ &   $\int\cal L$ &  $\cal L$   & $\sigma_{inel.}$  & $\Gamma_{inel.}$  & $\int\cal L$ \\
\multicolumn{3}{c|}{} & [cm$^{-2}$ s$^{-1}$] & [mb]  &  [kHz] & [pb$^{-1}$]  & [cm$^{-2}$ s$^{-1}$] & [b] &   [kHz] & [nb$^{-1}$]  \\

\hline
\hline
\multirow{9}{*}{\parbox{1.5cm}{\centering Internal gas target}} %
								    &\multirow{4}{*}{\centering Gas-Jet} & H$^\uparrow$  &  4.3 $\times 10^{30}$  & 39  & 168  & 43   &  5.6 $\times 10^{26}$  & 1.8 & 1.0  & 5.6 $\times 10^{-1}$  \\
								    	&							   & H$_{2}$ & 2.6 $\times 10^{31}$   &  39 & 1000 & 2.6 $\times 10^{2}$ & 2.8 $\times 10^{28}$   & 1.8 & 50  & 28  \\
									&							   & D$^\uparrow$ & 4.3 $\times 10^{30}$   & 72 & 309  & 43 & 5.6 $\times 10^{26}$  & 2.2 & 1.2 & 5.6 $\times 10^{-1}$  \\
									&							   & $^3$He$^\uparrow$ & 8.5 $\times 10^{30}$   & 117  & 1000  & 85 & 2.0 $\times 10^{28}$  & 2.5 & 50 & 20 \\
&							  									   & Xe &  7.7 $\times 10^{29}$  & 1300 & 1000  &  7.7   &  8.1 $\times 10^{27}$ & 6.2  & 50  &   8.1   \\
							     	\cline{2-11}
									&\multirow{5}{*}{\parbox{1.5cm}{\centering Storage Cell}} & H$^\uparrow$  & 2.6 $\times 10^{31}$   &  39 & 1000 & 2.6 $\times 10^{2}$ & 2.8 $\times 10^{28}$   & 1.8 & 50  & 28  \\
									&							   & H$_{2}$  & 2.6 $\times 10^{31}$   &  39 & 1000 & 2.6 $\times 10^{2}$ & 2.8 $\times 10^{28}$   & 1.8 & 50  & 28  \\  
									&							   & D$^\uparrow$ & 1.4 $\times 10^{31}$  &  72 & 1000   & 1.4 $\times 10^{2}$  & 2.2 $\times 10^{28}$  & 2.2  &  50 & 22  \\
									&							   & $^3$He$^\uparrow$ & 8.5 $\times 10^{30}$   & 117  & 1000  & 85 & 2.0 $\times 10^{28}$  & 2.5 & 50 & 20 \\
									&							   & Xe &  7.7 $\times 10^{29}$  & 1300  & 1000  & 7.7   & 8.1 $\times 10^{27}$ & 6.2  & 50  &  8.1   \\ \hline
\multirow{3}{*}{\parbox{1.5cm}{\centering Internal solid target on beam halo}}& \multirow{4}{*}{\parbox{1.2cm}{\centering Wire Target}} & C (500 $\mu$m) & 2.8 $\times 10^{30}$ & 271   & 760  & 28  & 5.6 $\times 10^{26}$ & 3.3 & 1.8 & 5.6 $\times 10^{-1}$ \\
								&		      & Ti (500 $\mu$m) &  1.4 $\times 10^{30}$   &  694 & 971 & 14  & 2.8 $\times 10^{26}$  & 4.7 & 1.3  &  2.8 $\times 10^{-1}$  \\			
									                                                 &	          & W (184 $\mu$m) &   5.9  $\times 10^{29}$ & 1700  & 1000  & 5.9  & --& -- & --& --  \\	
                            &	          & W (500 $\mu$m) & --   & --  & -- & -- & 3.1 $\times 10^{26}$ & 6.9  & 2.1 & 3.1 $\times 10^{-1}$ \\		\hline
\multirow{6}{*}{\parbox{1.5cm}{\centering Beam splitting}}& \multirow{2}{*}{\centering E1039} & NH$_3^\uparrow$ & 2.4 $\times 10^{30}$   &  420 & 1000 & 24  & 2.7 $\times 10^{27}$  &  19 & 50 & 2.7  \\
							   &  &  ND$_3^\uparrow$ & 1.9 $\times 10^{30}$ & 519  & 1000  &  19 & 2.2 $\times 10^{27}$  & 22 & 50 & 2.2 \\
						     	\cline{2-11}
						     	             &   & C (658 $\mu$m ) & 3.7 $\times 10^{30}$ & 271  & 1000 & 37  & --& -- & -- & -- \\
          &   & C (5 mm) & --  & -- & -- & -- & 5.6 $\times 10^{27}$ & 3.3 & 18 & 5.6  \\
									&			\multirow{3}{*}{\parbox{1.2cm}{\centering Unpol\-arised solid target}}				                                              & Ti (515 $\mu$m) &  1.4 $\times 10^{30}$  & 694  & 1000 & 14  & --  & -- & -- &  -- \\	
&							                                              & Ti (5 mm) &  --  &-- & -- & -- & 2.8 $\times 10^{27}$  & 4.7 & 13 &  2.8 \\	
									&							                                              & W(184 $\mu$m) &   5.9  $\times 10^{29}$ & 1700  & 1000  & 5.9 & -- & --  & - & --    \\ 
& & W(5 mm) &   -- & -- & --  & -- & 3.1 $\times 10^{27}$ & 6.9  & 21 & 3.1 \\ \hline
\end{tabularx} 
\end{adjustbox}
\caption{Summary table of the achievable integrated luminosities with the ALICE detector accounting for the data-taking-rate capabilities in the collider mode and by considering the luminosities of \ct{tab_lumi_comp}. As detailed in the text, a higher inelastic rate ($\Gamma_{inel.}$) depending on the collision system could be envisioned. The inelastic cross sections ($\sigma_{inel.}$) are taken from EPOS~\cite{Pierog:2013ria,Werner:2005jf}.}							
\label{tab_lumi_comp_alice}
\end{table}

\flushfootnote

\ct{tab_lumi_comp_alice} and \ct{tab_lumi_comp_lhcb} 
show the achievable luminosities using the ALICE and LHCb detectors during one LHC year, if one considers as a limitation
the aforementioned experimental data-taking rates in the collider mode and by considering the luminosities of \ct{tab_lumi_comp}. 
As specified in Section~\ref{section:detector:alice}, a higher rate could be envisioned in some cases since the charged-particle multiplicity 
is lower in the fixed-target mode.  
We have also assumed that ALICE and LHCb could run in the fixed-target mode during the full year with proton ($10^7$~s) and lead beams ($10^6$~s) with the corresponding instantaneous luminosities. 
In some cases, namely the gas-jet, the storage-cell and the solid target coupled to the beam splitting by a crystal, the resulting interaction rates are
high and close to those expected in the collider mode for the LHC Runs 3 and 4. Additional limitations will arise from
various constraints such as the disk storage, the high-level trigger, the radiation level, the simultaneous running with the collider mode, etc. The Run 3 and Run 4 programmes for ALICE and LHCb are well settled in the collider mode and any fixed-target-running scenario will have to comply with these programmes. These 
constraints are not discussed in this review as they deserve dedicated studies within the experiments. In the following, we discuss the luminosities obtained with the three technical implementations in both the ALICE and LHCb set-ups and comment on the luminosity needs for some of the physics cases that will be described in sections~\ref{section:High-x}, \ref{section:Spin-Physics} and \ref{section:heavy Ion Physics}. 

\begin{table}[!hbt]
\center
\begin{adjustbox}{angle=90}
\renewcommand{\arraystretch}{1.3}
\begin{tabularx}{0.9\textheight}{p{2.0cm}|c|c|c|c|c|c|c|c|c|c}
\multicolumn{3}{c}{ } & \multicolumn{8}{|c}{LHCb}  \\
\multicolumn{3}{c}{ } & \multicolumn{4}{|c|}{Proton beam (\sqrtsNN = 115 GeV)} & \multicolumn{4}{c}{Pb beam (\sqrtsNN = 72 GeV)} \\
\cline{4-11}
\multicolumn{3}{c|}{Target } & $\cal L$  & $\sigma_{\rm inel.}$  & $\Gamma_{inel.}$ &  $\int\cal L$ & $\cal L$  & $\sigma_{\rm inel.}$  & $\Gamma_{inel.}$  & $\int\cal L$  \\
\multicolumn{3}{c|}{} & [cm$^{-2}$ s$^{-1}$] & [mb]  &  [kHz] & [pb$^{-1}$]  & [cm$^{-2}$ s$^{-1}$] & [mb] &   [kHz] & [nb$^{-1}$]  \\
\hline
\hline
		\multirow{10}{*}{\parbox{2cm}{\centering Internal	gas target}	}						    &\multirow{5}{*}{\centering Gas-Jet}  & H$^\uparrow$  &  4.3 $\times 10^{30}$  & 39  & 168  & 43   &  5.6 $\times 10^{26}$  & 1.8 & 1  & 5.6 $\times 10^{-1}$ \\
								    	&							   & H$_{2}$ & 1.0 $\times 10^{33}$  & 39  & 40000  & 1.0 $\times 10^4$  & 1.2 $\times 10^{29}$  & 1.8 & 212  & 1.2 $\times 10^{2}$  \\
									&							   & D$^\uparrow$ &  4.3 $\times 10^{30}$  & 72 & 309  & 43  & 5.6 $\times 10^{26}$  & 2.2 & 1 & 5.6 $\times 10^{-1}$  \\
									&							   & $^3$He$^\uparrow$ & 3.4 $\times 10^{32}$  & 117  &  40000 & 3.4 $\times 10^3$  & 4.7 $\times 10^{28}$ & 2.5 & 118  & 47  \\
&							  									 & Xe &  3.1 $\times 10^{31}$  & 1300  & 40000 & 3.1 $\times 10^{2}$  & 2.3 $\times 10^{28}$  & 6.2 & 186 &  23 \\
							     	\cline{2-11}
							&\multirow{5}{*}{\parbox{1.5cm}{\centering Storage Cell}} & H$^\uparrow$ & 9.2 $\times 10^{32}$  & 39 & 35880 & 9.2 $\times 10^3$   & 1.2 $\times 10^{29}$  & 1.8 & 212 & 1.2 $\times 10^{2}$ \\
									&							   & H$_{2}$ &  1.0 $\times 10^{33}$   & 39 & 40000 & 1.0$\times 10^4$  & 1.2 $\times 10^{29}$  & 1.8 & 212 & 1.2 $\times 10^{2}$ \\
									&							   & D$^\uparrow$ & 5.6 $\times 10^{32}$   & 72  & 40000  & 5.6 $\times 10^3$   & 8.8 $\times 10^{28}$  & 2.2 & 194 & 88  \\
									&							   & $^3$He$^\uparrow$ & 1.3 $\times 10^{33}$ & 117  & 40000  & 1.3 $\times 10^4$ & 8.3 $\times 10^{28}$  & 2.5 & 206 & 83 \\
									&							   & Xe & 3.1 $\times 10^{31}$  & 1300  & 40000 & 3.1 $\times 10^{2}$ & 3.0 $\times 10^{28}$  & 6.2 & 186 &  30 \\
\hline
\multirow{3}{*}{\parbox{2cm}{\centering Internal solid target on beam halo}}  & \multirow{3}{*}{\parbox{1.2cm}{\centering Wire Target}} & C (500 $\mu$m) &  2.8 $\times 10^{30}$ & 271   & 760  & 28   & 5.6 $\times 10^{26}$ & 3.3  & 2 & 5.6 $\times 10^{-1}$\\
 	&							                   & Ti (500 $\mu$m) &  1.4 $\times 10^{30}$ & 694  & 972  & 14   & 2.8 $\times 10^{26}$  & 4.7 & 1  &  2.8 $\times 10^{-1}$  \\				
								&							                                & W (500 $\mu$m) &  1.6 $\times 10^{30}$ &  1700 & 2720  & 16    & 3.1 $\times 10^{26}$ & 6.9  & 2 & 3.1 $\times 10^{-1}$  \\		\hline
\multirow{5}{*}{\parbox{1.5cm}{\centering Beam splitting}}& \multirow{2}{*}{\centering E1039} & NH$_3^\uparrow$ & 7.2 $\times 10^{31}$ & 420 & 30240 & 7.2 $\times 10^{2}$  & 1.4 $\times 10^{28}$  &  19 & 259 & 14  \\
							   &  &  ND$_3^\uparrow$ & 7.2 $\times 10^{31}$ & 519  & 37368  & 7.2 $\times 10^{2}$  & 1.4 $\times 10^{28}$  & 22 & 314 & 14 \\
						     	\cline{2-11}
						     	             & \multirow{3}{*}{\parbox{1.2cm}{\centering Unpol\-arised solid target}}   & C (5 mm) &  2.8 $\times 10^{31}$ & 271  & 7600  & 2.8 $\times 10^{2}$ & 5.6 $\times 10^{27}$ & 3.3 & 18 & 5.6  \\
									&			 				                             				  & Ti (5 mm)  &  1.4 $\times 10^{31}$ & 694  & 9720  & 1.4 $\times 10^{2}$ & 2.8 $\times 10^{27}$  & 4.7 & 13 &  2.8 \\		
									&							                                               & W (5 mm) &  1.6 $\times 10^{31}$ &  1700 & 27200  & 1.6 $\times 10^{2}$ & 3.1 $\times 10^{27}$ & 6.9 & 21 & 3.1    \\ \hline
\end{tabularx} 
\end{adjustbox}
\caption{Same as \ct{tab_lumi_comp_alice} for the LHCb detector.}							
\label{tab_lumi_comp_lhcb}
\end{table}

With LHCb, the luminosity reach with a proton beam on an hydrogen target can be very large, up to yearly luminosity of the order of $10$~fb$^{-1}$, if one can run at 40~MHz for a full LHC year. As discussed in section~\ref{section:High-x}, this will allow one to measure hard probes such as $W$ or associated \jpsi production, to collect very large statistics for DY and probe the $D$ meson production at the most backward rapidity range. For example, DY measurements are very useful to probe the light quark and anti-quark PDFs at high-$x$ and a low scale, $\mu_F$, even with a reduced luminosity of $1$~fb$^{-1}$. By using nuclear solid or gas targets, one can reach luminosities on the order of 100~pb$^{-1}$, and even more depending on the target option, allowing one to probe the nuclear PDFs with a very high precision by measuring DY, open heavy flavour and quarkonium production. 
With proton and lead beams, the luminosity reach for ALICE is lower than for LHCb because of the lower data-taking rates. However, the rapidity coverage is complementary as well as the physics reach. Even though with ALICE the yearly luminosity can be as high as $250$~pb$^{-1}$ with a proton beam on a H-gas target, a luminosity of about $40$~pb$^{-1}$ with the ALICE detectors 
would already allow one to measure low energy $\bar p$, thanks to the very backward rapidity coverage, in $p+p \to \bar p + X$. This measurement would improve our knowledge of the cosmic $\bar p$ spectrum. The $\bar p$ spectrum could be further measured with various target types, such as He or C with the gas or solid target option, respectively.  

With a transversally polarised H-gas target, luminosities from about $40$~pb$^{-1}$ to $250$~pb$^{-1}$ can be expected with ALICE for one LHC year. As discussed in section~\ref{section:Spin-Physics}, this would allow one to access the spin asymmetry of probes such as $\Lambda$ in the CB and \jpsi in the MS. With a larger luminosity in LHCb, on the order of $10$~fb$^{-1}$, similar rare probes as those mentioned above for the unpolarised case ($W$, associated \jpsi, DY, ...) will be accessible. These studies will definitely advance our understanding of the internal spin structure of the proton and neutron.

 With a Pb beam on a heavy nuclear target, the luminosity is mainly limited by the impact on the beam lifetime for the gas target and by the usable beam flux for the beam splitting case. The luminosities do not differ by more than a factor of four between ALICE (${\cal L}=8$~nb$^{-1}$) and LHCb (${\cal L}=30$~nb$^{-1}$) for \PbXe. In the case of the beam splitting option, large luminosities are also expected (${\cal L}= 3$~nb$^{-1}$) for \PbW.  When the nuclear target is lighter,  the Pb beam lifetime is less affected and the luminosities are larger for the case of LHCb coupled to a gas target. 
A full programme of heavy-ion studies can be carried out in the fixed-target mode at the LHC in particular with precise quarkonium measurements, with studies of the heavy-quark energy-loss mechanism and a rapidity scan of the yield and elliptic flow of identified charged particles over a broad rapidity range, as discussed in section~\ref{section:heavy Ion Physics}. These studies can be performed with large statistics in PbA collisions but also in \pp and \pA collisions (see \eg\ \ct{tab:Upsilon_yields} for the $\Upsilon$ case), with the ALICE and LHCb detectors. In the latter case, it is not yet established up to which event centrality the tracks can efficiently be reconstructed, in particular for the heaviest nuclear target such as Xe. As discussed in section~\ref{section:detector_lhcb}, on-going studies will give more information on the event-centrality reach with LHCb for the Run~3 and Run~4. A heavy-ion programme is already envisioned in LHCb in Run 3 with Ar gas target that will allow one to study soft probes as well as hard probes such as open heavy flavour and quarkonium production.

%% file: physics-high-x/physics-high-x.tex
\subsection{High-$x$ frontier for particle and astroparticle physics}
\label{section:High-x}

The purpose of this section is to address the question whether a modern fixed-target
experiment with a record energy and with high luminosities can help answer
problems at the frontier of particle and astroparticle physics.
We divide this section in three parts for which the physics cases are quite distinct.
In the first part, we discuss the impact of such an experiment on our understanding of the
high-$x$ structure of nucleons.
In the second part, the physics case for the high-$x$ structure of complex nuclei is
considered.
Finally, the third part is devoted to astroparticle-physics applications.

\input{physics-high-x/physics-high-x_nucleon.tex}

\input{physics-high-x/physics-high-x_nuclear.tex}

\input{physics-high-x/physics-high-x_astro.tex}

%% file: physics-high-x/physics-high-x_nucleon.tex
\subsubsection{Nucleon structure}
\label{subsec:nucleon}
Much progress has been made in the past 30 years in our understanding of the partonic structure
of nucleons. The parton distribution functions (PDFs) are determined in global analyses 
\cite{Gao:2013xoa,Ball:2014uwa,Harland-Lang:2014zoa,Alekhin:2013nda,Owens:2012bv,Jimenez-Delgado:2014twa}
using a wealth of experimental information from fixed-target and collider experiments. The analyses are routinely
performed at next-to-leading order (NLO) and next-to-next-to-leading order (NNLO) of perturbative QCD and the uncertainties of the PDFs are carefully evaluated.
Still, at high momentum fractions $x$, the PDFs are poorly known, in particular the smaller distributions.
For example, the uncertainty of the gluon distribution gets very large at $x\gtrsim 0.4$
and the strange, charm and bottom PDFs are completely unconstrained in this kinematic range.
A better understanding of the high-$x$ structure of the nucleon is warranted for several reasons:
\begin{itemize}
\item First, while it is well-known that the gluon carries over 40\% of the
nucleon momentum, most of the gluons carry a small momentum fraction.
On the other hand, in a constituent quark picture, it is rather the gluon distribution carrying a high
momentum fraction which can be interpreted as binding the constituent quarks together.
Furthermore, light-cone models predict a relatively sizeable high-$x$ component of 
the strange, charm and bottom PDFs. This means that the higher the probed $x$-values will be the better these fundamental
aspects of QCD can be studied.
Needless to say, that progress on the high-$x$ gluon, strange and charm PDFs will lead to a refined picture
for the light-quark valence and sea distributions in this kinematic region.
\item At the same time, PDFs are a crucial input for making theoretical predictions for observables at the LHC.
In many cases, the PDF uncertainty has become the limiting factor in the accuracy of the predictions.
This is particularly true for processes involving heavy new states in 
Beyond the Standard Model (BSM) theories where the high-$x$ PDFs are probed.
Clearly, an improved understanding of the high-$x$ PDFs is crucial for BSM searches at the LHC
and any future hadron collider,
and AFTER@LHC offers the unique opportunity to study these aspects of high-$x$ hadron structure 
in detail. 
\end{itemize}

\paragraph{Kinematic coverage of lepton pair production}

As is well known, at leading order (LO) the cross section for DY lepton pair production
in a collision of two nucleons $A$ and $B$ is given by the following expression:
\begin{equation}
\frac{d^2 \sigma}{dx_1 dx_2} = \frac{4 \pi \alpha^2}{9 s x_1 x_2} \sum_i e_i^2 
\left[q^A_i(x_1) \bar q^B_i(x_2) + \bar q^A_i(x_1) q^B_i(x_2) \right]\, ,
\end{equation}
where $e_i$ is the electric charge of the quark (in units of $e$) 
and the sum runs over all active quark flavours.
Therefore, it is clear that this process provides information on the (light) quark sea.
Existing DY data which are used in global PDF analyses come from fixed-target experiments
at Fermilab (E866/NuSea, E605) and the LHC.

The cross section is usually given as a function of the invariant mass of the lepton pair
$M_{(\ell\ell)}$ (at LO $M^2 = x_1 x_2 s$) and the Feynman variable $x_F = x_1 - x_2$ 
from which the momentum fractions $x_{1,2}$ can approximately be recovered using the relation
$x_{1,2} = (\sqrt{x_F^2 + 4 \tau} \pm x_F)/2$ where $\tau = M_{\ell\ell}^2/s$. Correspondingly, one has in terms of the pair rapidity in the \cms , $Y^{\rm c.m.s.}_{\ell\ell}$, $x_{1,2} = M_{\ell\ell}/\sqrt{s} e^{\pm Y^{\rm c.m.s.}_{\ell\ell}}$. 

\cf{fig:large_x_DY_pp_simplified_acceptance} shows a simplified kinematical reach  with \AFTERLHCb\ and \AFTERALICE\ for DY measurements at $\sqrt{s}=115$ GeV considering different (pseudo)rapidity domains\footnote{In \cf{fig:large_x_DY_pp_simplified_acceptance}, we have made the approximation that $Y^{\rm lab.}_{\ell\ell}\simeq \eta^{\rm lab.}_{\ell\ell}$. 
$Y^{\rm c.m.s.}_{\ell\ell}$ is then obtained by subtracting 4.8 from $Y^{\rm lab.}_{\ell\ell}$ which corresponds to the rapidity shift between the lab. and c.m.s. frames with a 7 TeV beam.}
which follow from the acceptance of the specified detectors viewed from the different indicated target locations.
In the LHCb case, a target located at $0$ or $-0.4$m offers similar reaches whereas at $-1.5$~m the reach is clearly shifted to lower $x$. In particular, even above the bottomonium region where the statistics will be limited, one does not access $x_2=1$ any more. In the ALICE case, even with a remote target, \eg\ at $-4.7$ m, one can still reach $x_2=1$ between the charmonium and bottomonium family. The Intermediate Mass Region (IMR) below the charmonium family, which {\it de facto} offers the largest statistics, still covers the valence region in the case of the ALICE CB. Projection studies of the DY yield in this region, which is complementary to the DIS studies at JLab, would however essentially rely on the technique used to subtract the combinatorial background. As such, they would be driven rather by systematical uncertainty projections than by statistical ones. In this context, we will limit ourselves to discuss DY projections above the charmonium family, namely for $M_{\ell\ell} > 4$~GeV. However, this does not mean at all that IMR DY studies are not possible with \AFTERLHCb\ and \AFTERALICE.

\begin{figure}[!hbt]
\centering
\subfigure[~]{
\includegraphics[width=0.48\textwidth]{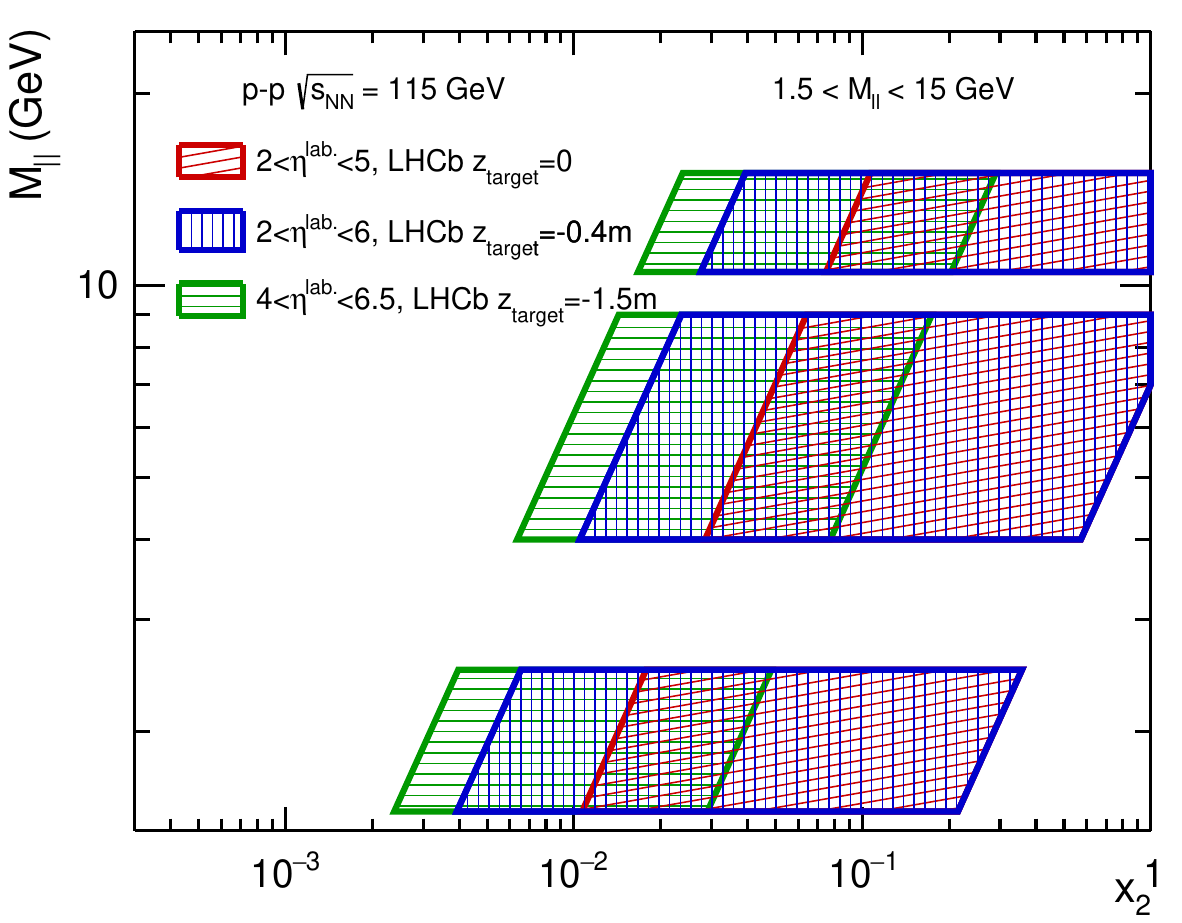}
}
\subfigure[~]{
\includegraphics[width=0.48\textwidth]{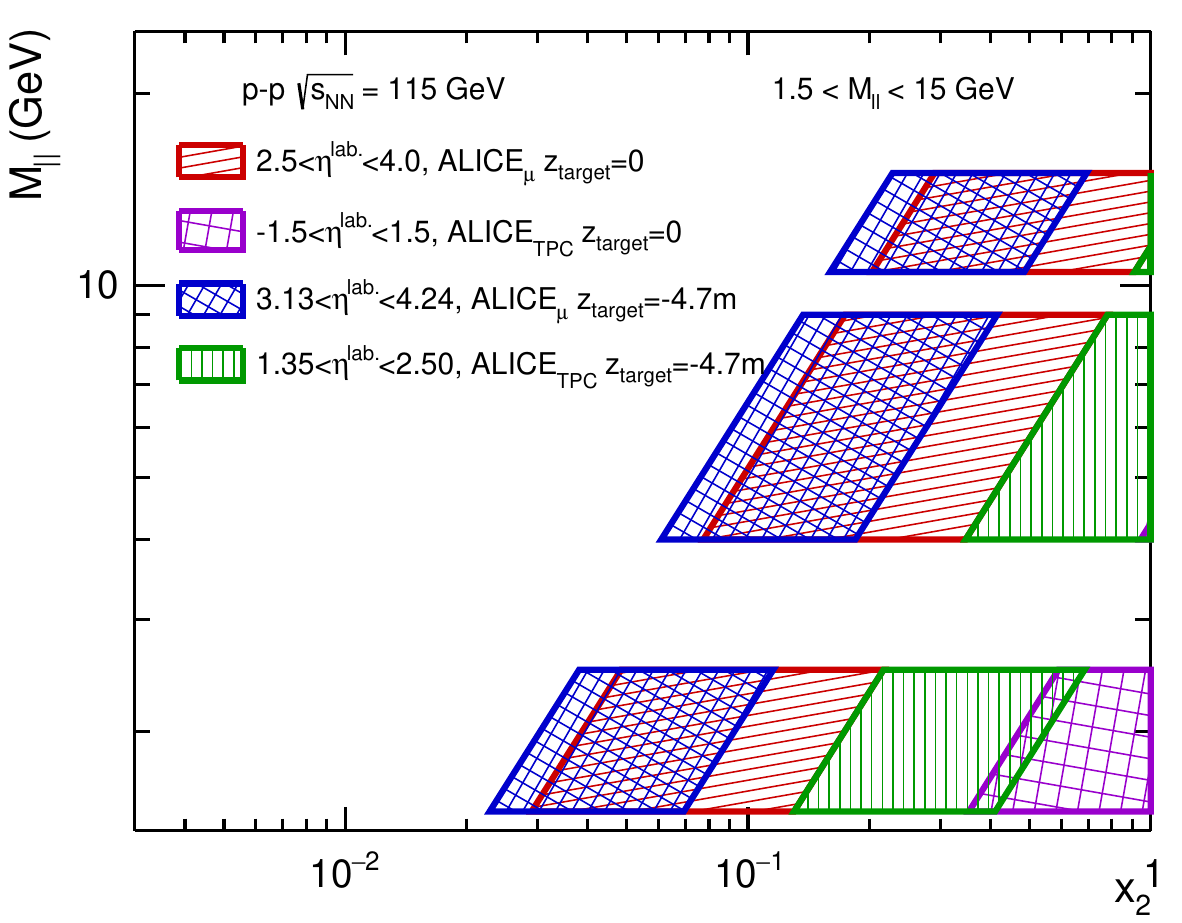}
}
\caption{
Simplified kinematical reach in $x_2$ for DY lepton-pair production with \AFTERLHCb\ and \AFTERALICE\ in $pp$ collisions at $\sqrt{s}=115$ GeV 
for different muon-pair acceptances and invariant masses. The acceptances in $\eta^{\rm lab.}_{(\ell\ell)}$ correspond to different
 target locations within (a) LHCb and (b) ALICE.
[No $p_{T,\mu}$ constraint is applied.]}
\label{fig:large_x_DY_pp_simplified_acceptance}
\end{figure}

In \cf{fig:large_x_DY_pp}, the kinematical acceptance for DY lepton-pair production is shown
assuming $pp$ collisions 
at a \cms\ energy of $\sqrt{s}=115$ GeV with an integrated luminosity of 10 fb$^{-1}$
and a single-muon acceptance of $2<\eta^{\rm lab.}_{\mu}<5$ and $p_{T,\mu}>1.2$ GeV.
It should be noted that each cell contains at least 30 events.
For comparison, the kinematic coverage of existing DY data (E605, E866/NuSea) used in global proton PDF
analyses is depicted.\footnote{We are grateful to V.\ Bertone from the NNPDF collaboration for providing us
the points.}
The NuSea data have been obtained in 800 GeV $pp$ and $pd$ collisions ($\sqrt{s}=38.8$ GeV)
covering the di-muon mass ranges from 4.2 to 8.7 GeV and 10.85 to 16.85 GeV and the Feynman-$x_F$ range
from $-0.05$ to 0.8.
As anticipated with our simplified kinematical-reach analysis, AFTER@LHC will be able to extend the coverage up to even higher $x$-values close to one.
Furthermore, while the NuSea data are dominated by statistical uncertainties reaching 100\% at the kinematic
boundaries, AFTER@LHC will considerably improve the precision due to the higher center-of-mass energy
and the higher luminosity.

\begin{figure}[!hbt]
\centering
\subfigure[~]{
\includegraphics[width=0.48\textwidth]{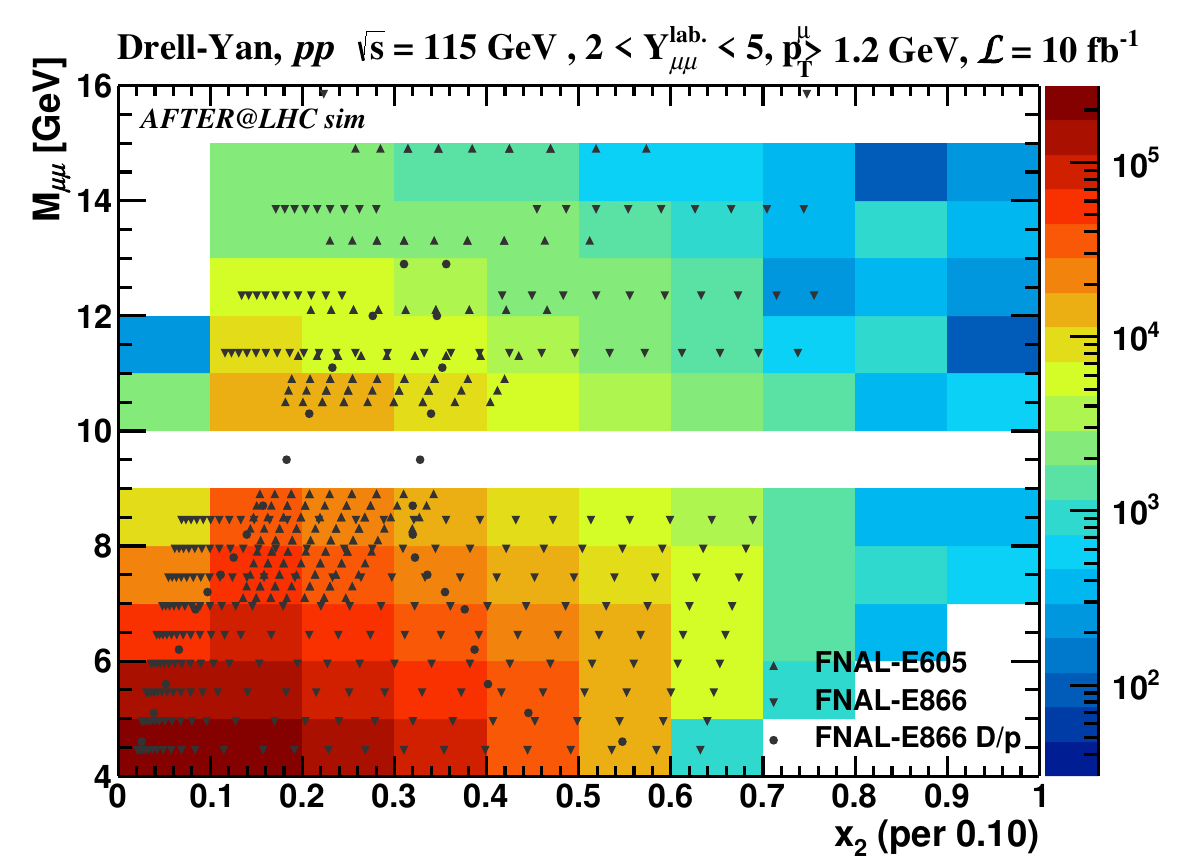}
}
\subfigure[~]{
\includegraphics[width=0.48\textwidth]{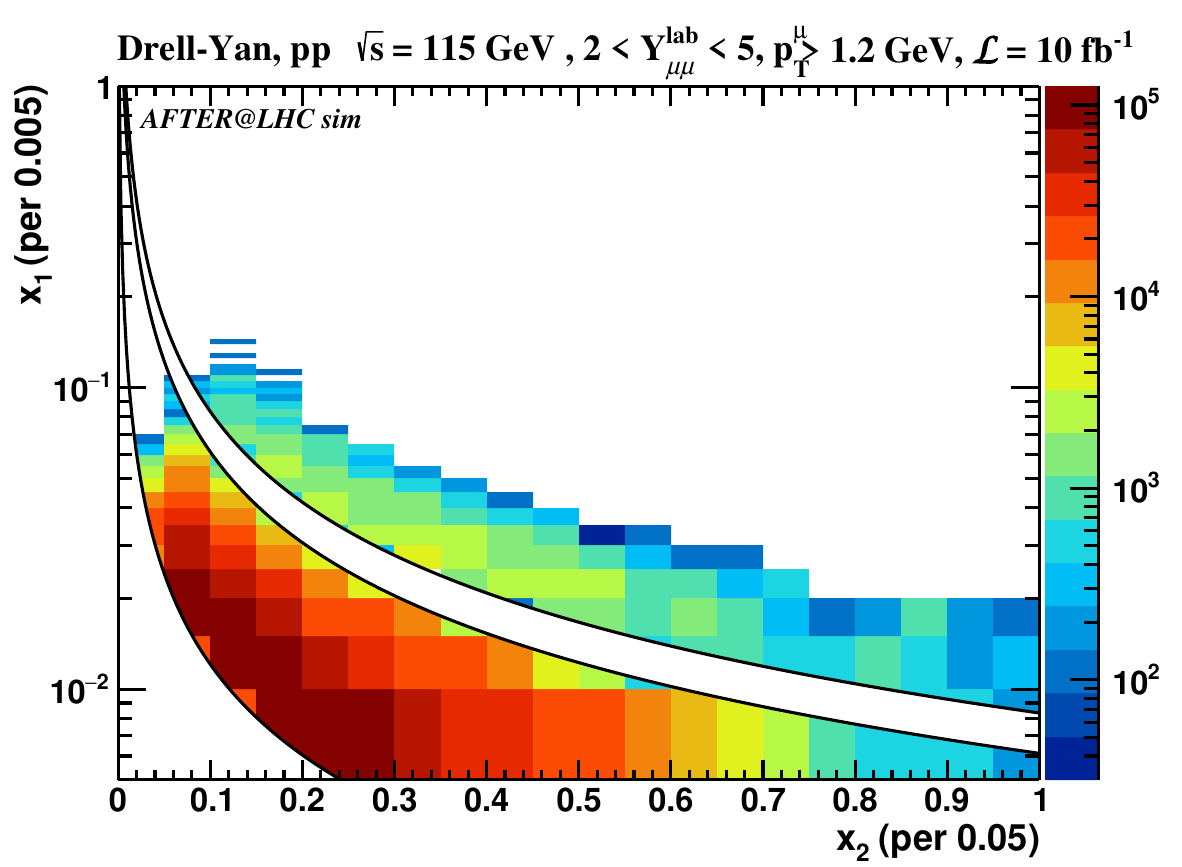}
}
\caption{
Kinematical acceptance for DY lepton-pair production with \AFTERLHCb\ in $pp$ collisions at $\sqrt{s}=115$ GeV
	with a muon-pair acceptance of $2<Y^{\rm lab.}_{\mu\mu}<5$ and single-muon requirements: $2<\eta^{\rm lab}_\mu<5$ and $p_{T}^{\mu}>1.2$ GeV. (a) Di-muon invariant mass vs. $x_{2}$ compared to the existing DY
data~\cite{Moreno:1990sf,Webb:2003ps,Webb:2003bj,Towell:2001nh}
used in current global PDF fits. (b) $x_{1}$ vs $x_{2}$ for the considered di-muon invariant-mass range of 4 GeV $< M_{\mu\mu} < $ 15 GeV excluding $\Upsilon$ mass range of 9 GeV $< M_{\mu\mu} < $ 10.5 GeV. Colours correspond to expected yields of the DY signal in each kinematical region, and each coloured cell contains at least 30 events.
We stress that the bins with a number of counts less that 30 are not shown in these
\AFTERLHCb\ projections, whereas some of the shown data points from the FNAL experiments
include bins with significantly less counts.
}
\label{fig:large_x_DY_pp}
\end{figure}

The DY measurements at AFTER@LHC provide important tests of nucleon structure.
In the limit $x_F \to -1$ and moderate or small invariant masses $M_{\ell\ell}$, we have $x_1 \simeq M_{\ell\ell}^2/(s |x_F|)$, 
$x_2 \gtrsim |x_F|$.
For example, for $x_F = -0.8$, $M_{\ell\ell}=10$ ($M_{\ell\ell}=15$) GeV we have $x_2 \simeq 0.8$ and $x_1 \simeq 0.01$ 
($x_1 \simeq 0.02$). In this kinematic region the ratio of the DY cross section in $pn$ collisions with the
one in $pp$ collisions is approximately given by (LO, $\bar u(x_2), \bar d(x_2), s(x_2), \bar s(x_2) \ll u_v(x_2), d_v(x_2)$, neglecting $Z$-exchange):
\begin{equation}
R = \frac{\sigma^{\rm DY}(pn)}{\sigma^{\rm DY}(pp)} 
\simeq \frac{4 \bar u(x_1) d(x_2) + \bar d(x_1) u(x_2)}{4 \bar u(x_1) u(x_2) + \bar d(x_1) d(x_2)}
\simeq \frac{4 d(x_2) + u(x_2)}{4 u(x_2) + d(x_2)}
= \frac{1+ 4 r_v}{4 + r_v} \, 
\end{equation}
where  $\bar d(x_1) \simeq \bar u(x_1)$ has been used to arrive at the third equality and
$r_v = d(x_2)/u(x_2) \simeq d_v(x_2)/u_v(x_2)$.
Interestingly, exactly the same LO parton model expression is found for the ratio of structure functions
$F_2^n(x,Q^2)/F_2^p(x,Q^2)$ in the limit $x \to 1$ which corresponds to elastic scattering.
As a further consequence, the ratio $R$ is bounded, $1/4 \le R \le 4$, similarly to the famous bounds 
for the ratio of Deep-Inelastic Structure (DIS) functions, $1/4 \le F_2^n/F_2^p \le 4$,
derived by Nachtmann \cite{Nachtmann:1972pc}.
The PDFs vanish for $x \to 1$ and generally, the high-$x$ behavior of the PDFs at the
initial scale $Q_0$ is parametrised as $x f_i(x,Q_0) \propto (1-x)^{b_i}$ where $b_i$ depends 
on the parton flavor ``$i$''.
Currently, only $b_{u_v}$  is relatively well constrained 
with values in the range $2.6 \lesssim b_{u_v} \lesssim 3.6$,
which is in agreement with the expectation from 
counting rules \cite{Brodsky:1973kr} ($b_{u_v}= b_{d_v}=3$),
whereas $b_{d_v}$ is known to a much lesser extent and varies 
strongly between 1.4 and 4.6 for different sets of PDFs, see Figs.\ 2 and 5 in \cite{Ball:2016spl}.\footnote{Needless to say 
that  the exponents for the gluon and the quark sea are very poorly known.}
Note also that the CJ15 analysis \cite{Accardi:2016qay} which has a particular focus on the high-$x$ region
points to a constant $d(x)/u(x) \sim 0.1$ for $x\to 1$ implying $R \to \, \sim0.34$.
However, for the time being it is reasonable to allow for the possibilities
that $r_v$ can vanish, approach a finite value $k$, or diverge in the limit $x \to 1$ (see Fig.\ 8 in \cite{Ball:2016spl}).
Consequently, we find in the limit $x_2 \to 1$
that a measurement of $R$ could constrain $r_v$ and 
provide important tests of different models of nucleon structure
\cite{Close:1973xw,Farrar:1975yb,Melnitchouk:1995fc,Close:2003wz,Holt:2010vj,Roberts:2013mja}.

Experimentally, it is the ratio of cross sections in $pd$ over $pp$ collisions which is accessible.
Neglecting any nuclear effects in deuterium the ratio can be written as
\begin{equation}
R_{d/p}(x_2) = \frac{\sigma^{\rm DY}(pd)}{\sigma^{\rm DY}(pp)}  
= 1 + \frac{\sigma^{\rm DY}(pn)}{\sigma^{\rm DY}(pp)} \simeq 5\,  \frac{1+ r_v(x_2)}{4 + r_v(x_2)} \, .
\end{equation}
Consequently, we find in the limit $x_2 \to 1$
\begin{align}
R_{d/p} & \to
\begin{cases}
2 & ; \quad  r_v = 1
\\
2.5 & ; \quad r_v =0
\\
5 & ; \quad r_v \to \infty
\end{cases}
\, ,
\end{align}
and a sufficiently precise measurement of the ratio will allow one to determine $r_v(x)=d(x)/u(x)$ at high-$x$.
In practice, a full fledged QCD analysis at NLO or NNLO of the data needs to be performed.

\paragraph{Drell-Yan lepton pair production and PDFs}
In order to estimate the possible impact of the DY lepton pair production
in $pp$ collisions at AFTER@LHC on the PDFs we have performed a profiling analysis~\cite{Paukkunen:2014zia}
using the xFitter package~\cite{Alekhin:2014irh}. 
For this purpose we have used pseudo-data constructed out of NLO QCD predictions
for the rapidity distributions in the \cms\ using the MCFM programme \cite{Boughezal:2016wmq}
and projected experimental uncertainties adding the statistical uncertainty from the DY yield
and the uncertainties from the subtraction of the combinatorial background in quadrature.
The statistical uncertainties were estimated assuming a yearly integrated luminosity
of $\L_{pp}=10$ fb$^{-1}$.
No additional systematic uncertainty has been taken into account.
The pseudo-data have been generated for several bins in the invariant mass of the muon pair
($ M_{\mu \mu} \in [4,5], [5,6], [6,7], [7,8]$ GeV and $M_{\mu \mu} > 10.5$ GeV)
and have been constructed such that the central values of 
the ``measurements'' and predictions coincide.
This is illustrated in \cf{fig:DY_pp_sig_y} for the invariant-mass bin $4 < M_{\mu \mu} < 5$ GeV.
In some cases the uncertainties are smaller than the data points and therefore not visible.\footnote{Note that the covered rapidity range 
in the \cms\ is from $-2.8 < \ycms  < 0.2$
such that there are no generated data at rapidities $\ycms >0.2$.}
As can be seen, the band showing the uncertainty of the theory prediction due to the NLO CT14 PDF uncertainties is much larger than errors of the simulated data.

\begin{figure}[htb!]
\centering
\includegraphics[width=0.4\textwidth]{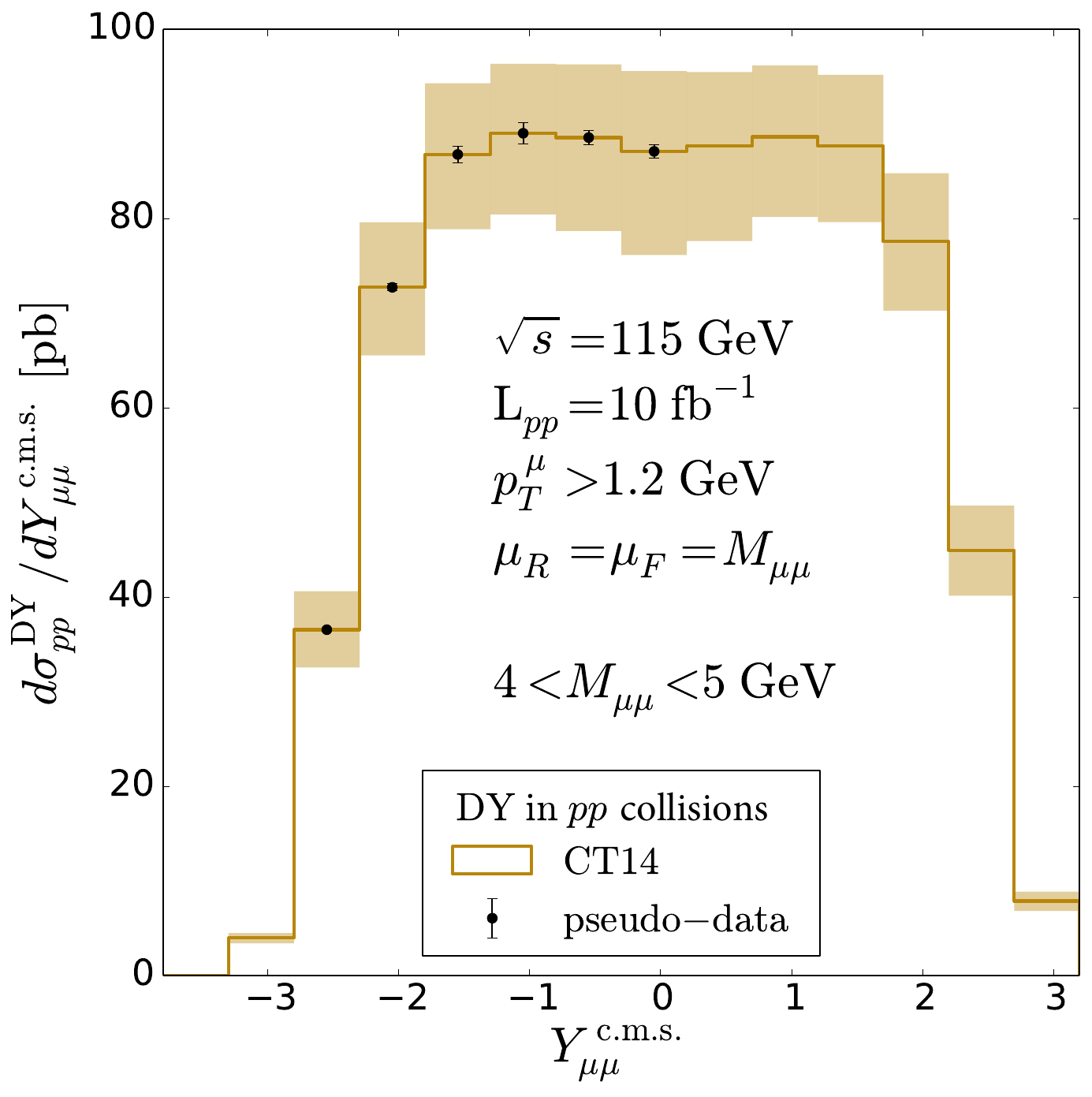}
\caption{
DY lepton-pair-production cross-section in $pp$ collisions as a function
of the muon-pair \cms\ rapidity for the invariant mass $4\, \mathrm{GeV} < M_{\mu\mu} < 5\, \mathrm{GeV}$, with $p_T^{\mu}>1.2$ GeV.
The NLO theory predictions obtained using CT14 PDFs are overlaid by pseudo-data.}
\label{fig:DY_pp_sig_y}
\end{figure}

\begin{figure}[!hbt]
\hspace*{-2mm}
 \subfigure[~$u$ PDF]{
\includegraphics[width=0.24\textwidth]{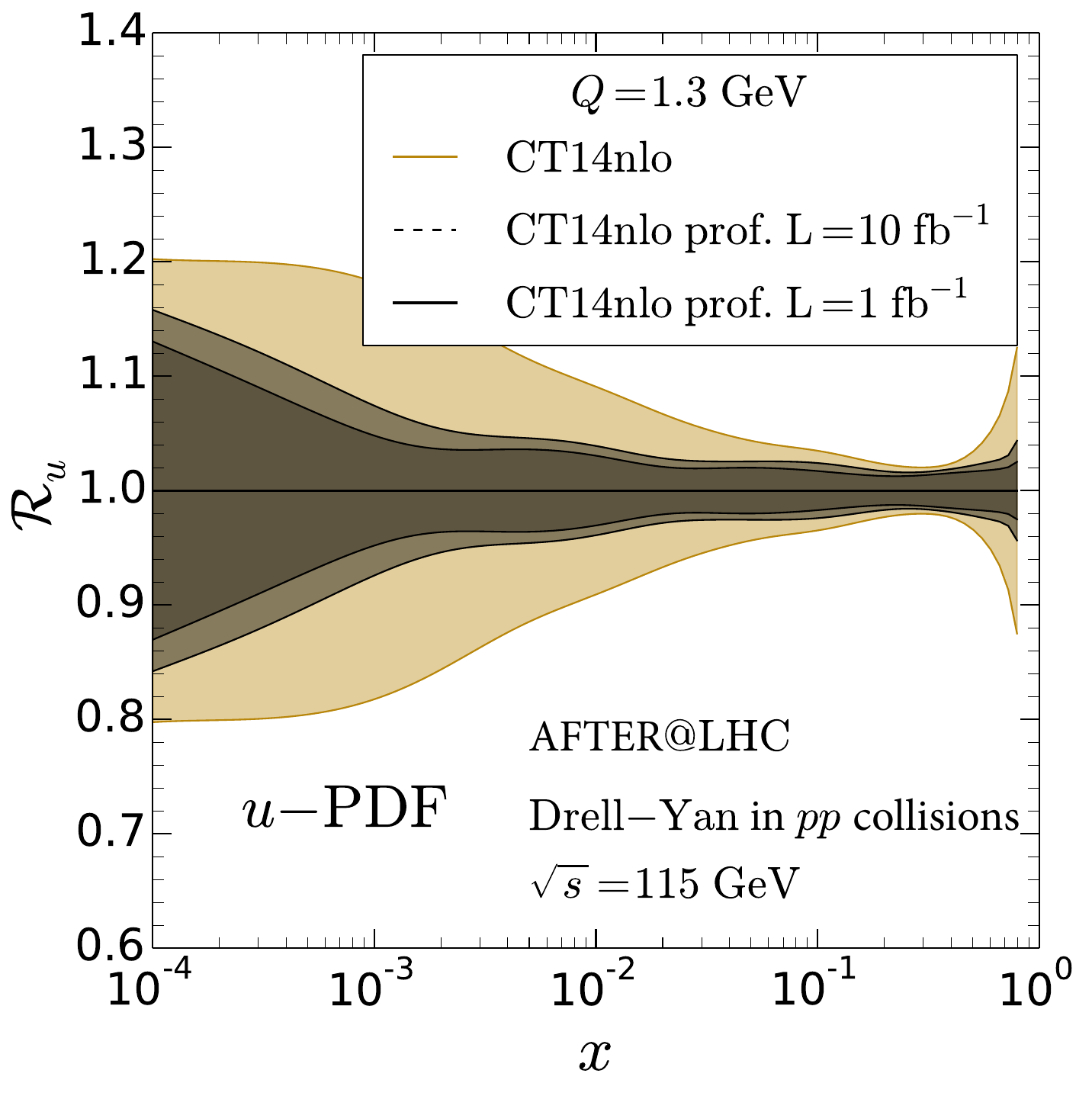}
 }\hspace*{-3mm}
 \subfigure[~$d$ PDF]{
\includegraphics[width=0.24\textwidth]{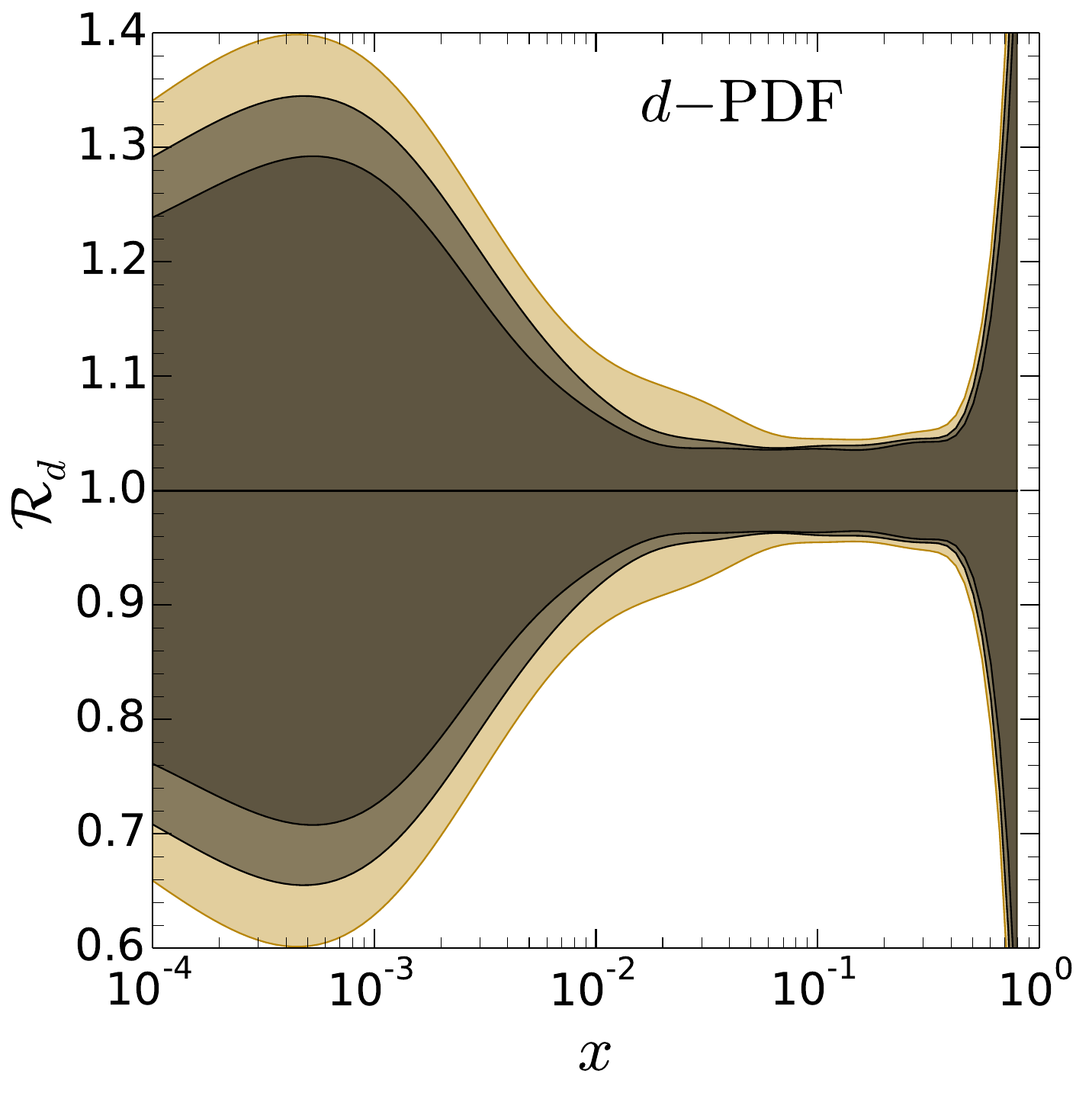}
 }\hspace*{-3mm}
 \subfigure[~$\bar{u}$ PDF]{
\includegraphics[width=0.24\textwidth]{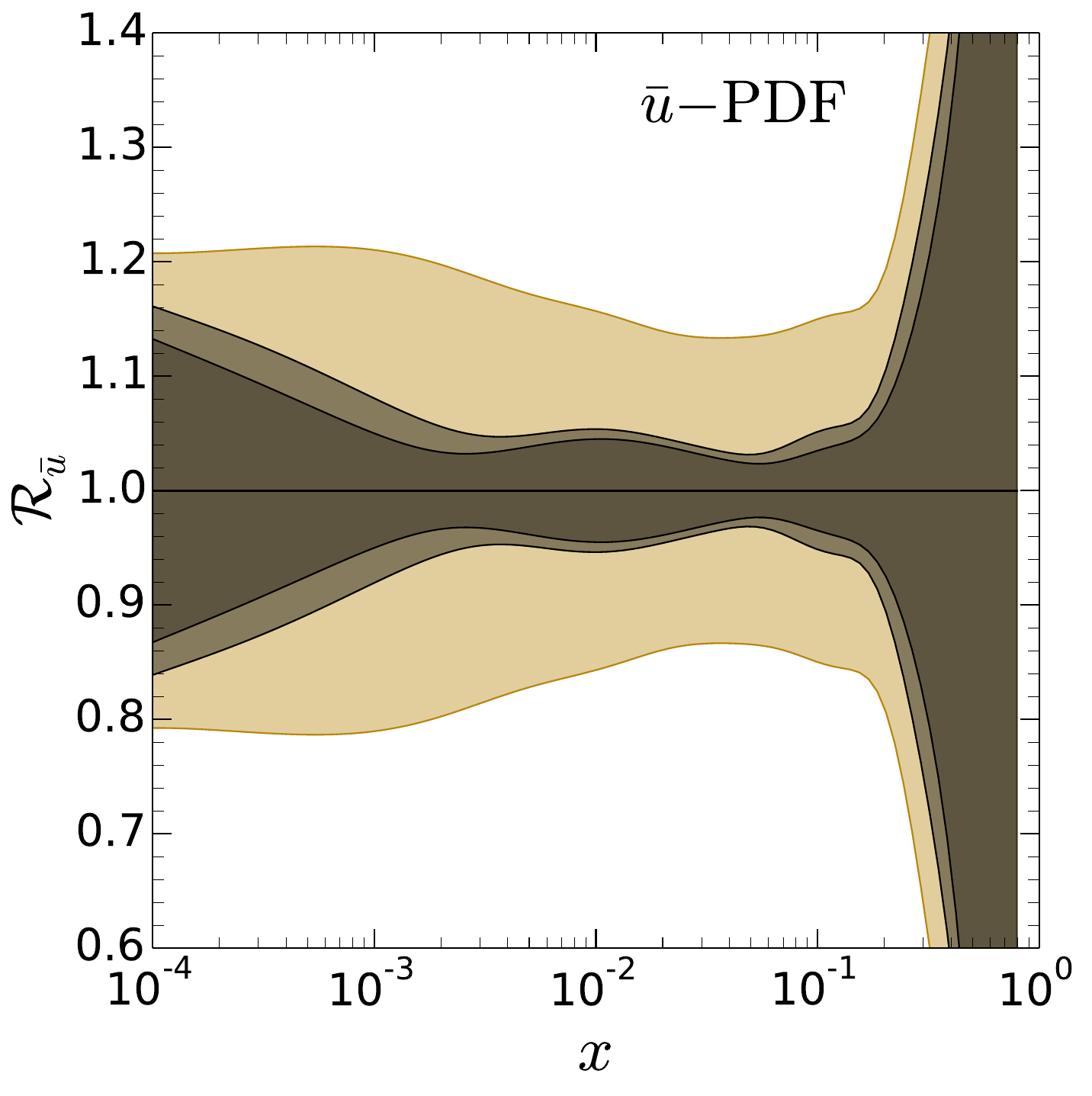}
 }\hspace*{-3mm}
 \subfigure[~$\bar{d}$ PDF]{
\includegraphics[width=0.24\textwidth]{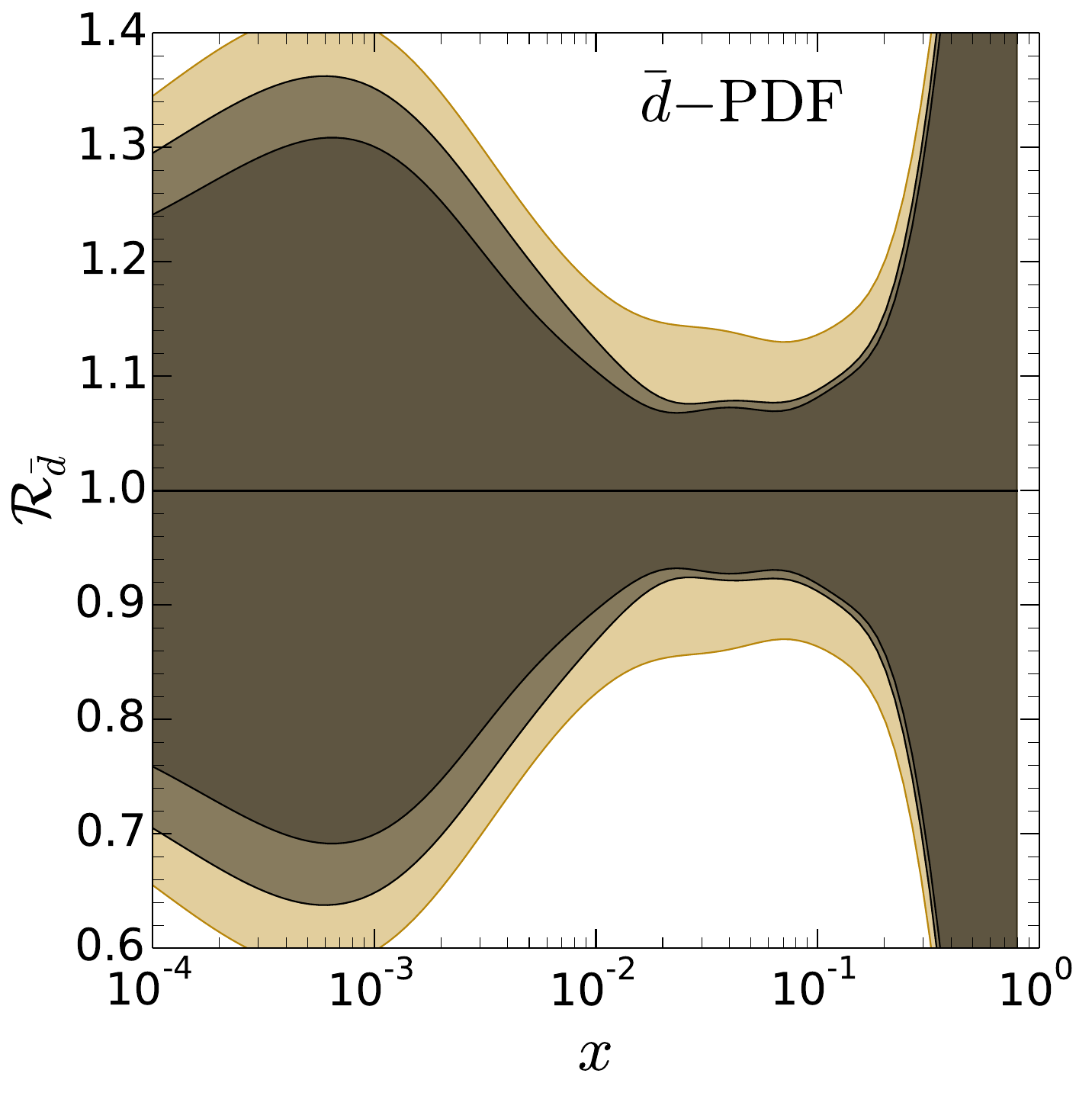}
 }\hspace*{-4.5mm}
\caption{
Impact of the DY lepton pair production in \pp\ collisions at $\sqrt{s}=115$ GeV
on the PDF uncertainties. The $u, d, \bar u$ and $\bar d$ PDFs from
CT14~\cite{Dulat:2015mca} are plotted as a function of $x$ at a scale
$Q=1.3$ GeV before and after including \AFTERLHCb pseudo-data in the global analysis
using the profiling method~\cite{Paukkunen:2014zia,Alekhin:2014irh}.
Two scenarios with different integrated luminosities were considered:
inner band: $\L_{pp}=10$ fb$^{-1}$, middle band: $\L_{pp}=1$ fb$^{-1}$ (the outer band represents
current PDF uncertainties).}
\label{fig:ct14_prof_log}
\end{figure}

\begin{figure}[!hbt]
 \hspace*{-2mm}
 \subfigure[~$u$ PDF]{
\includegraphics[width=0.25\textwidth]{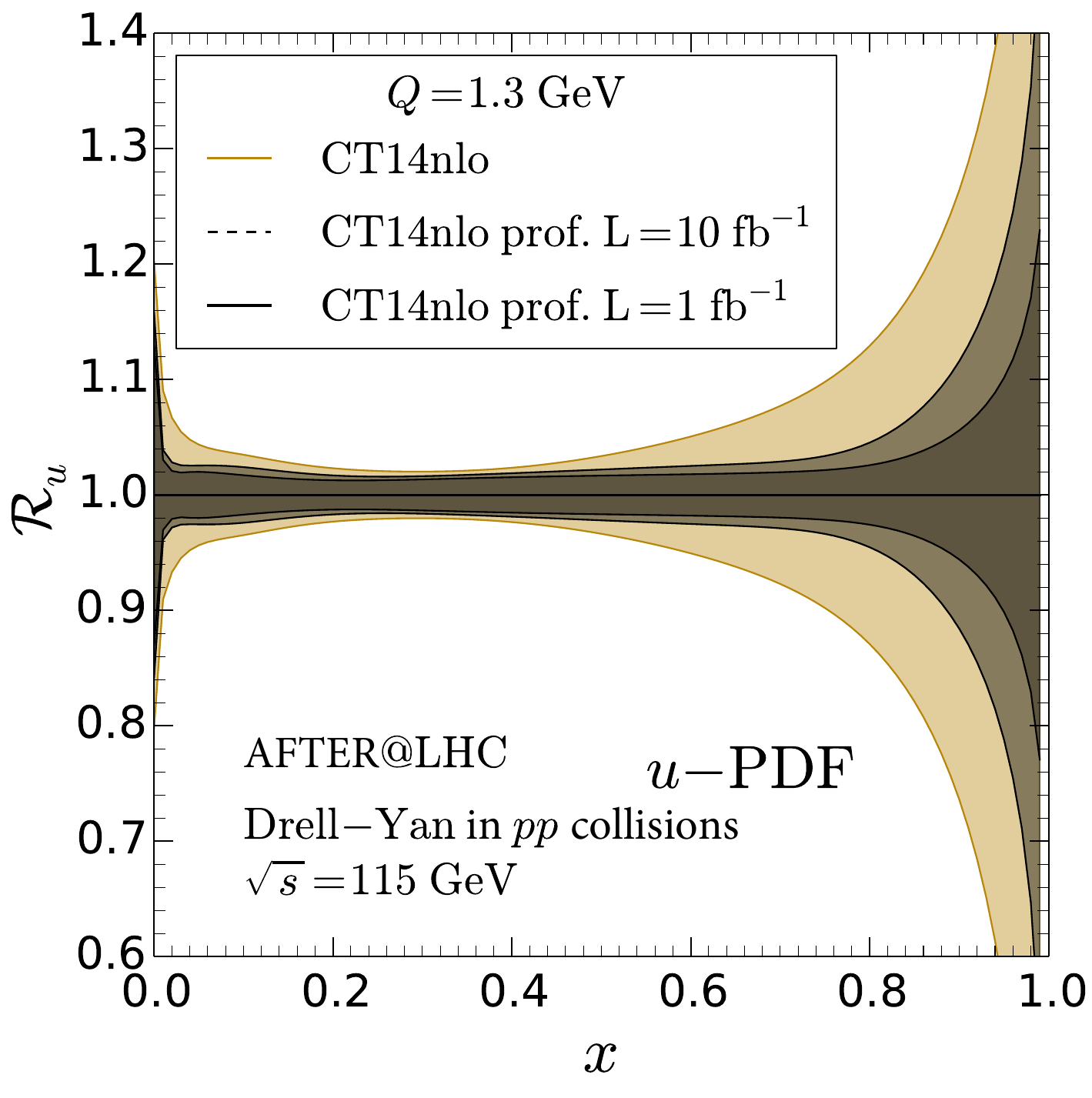}%
}\hspace*{-1.5mm}
 \subfigure[~$d$ PDF]{
\includegraphics[width=0.26\textwidth]{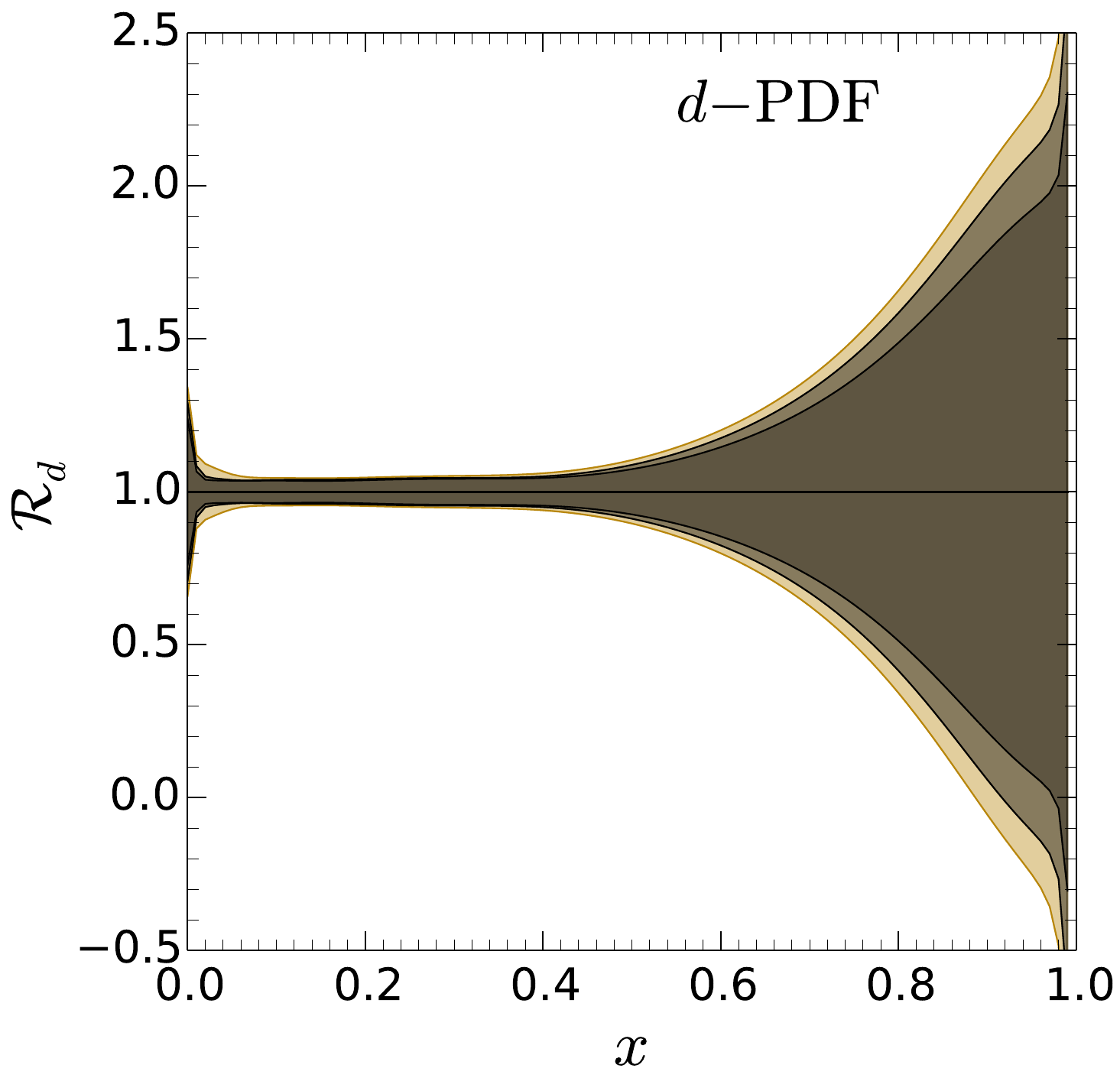}
 }\hspace*{-2.5mm}
 \subfigure[~$\bar{u}$ PDF]{
\includegraphics[width=0.25\textwidth]{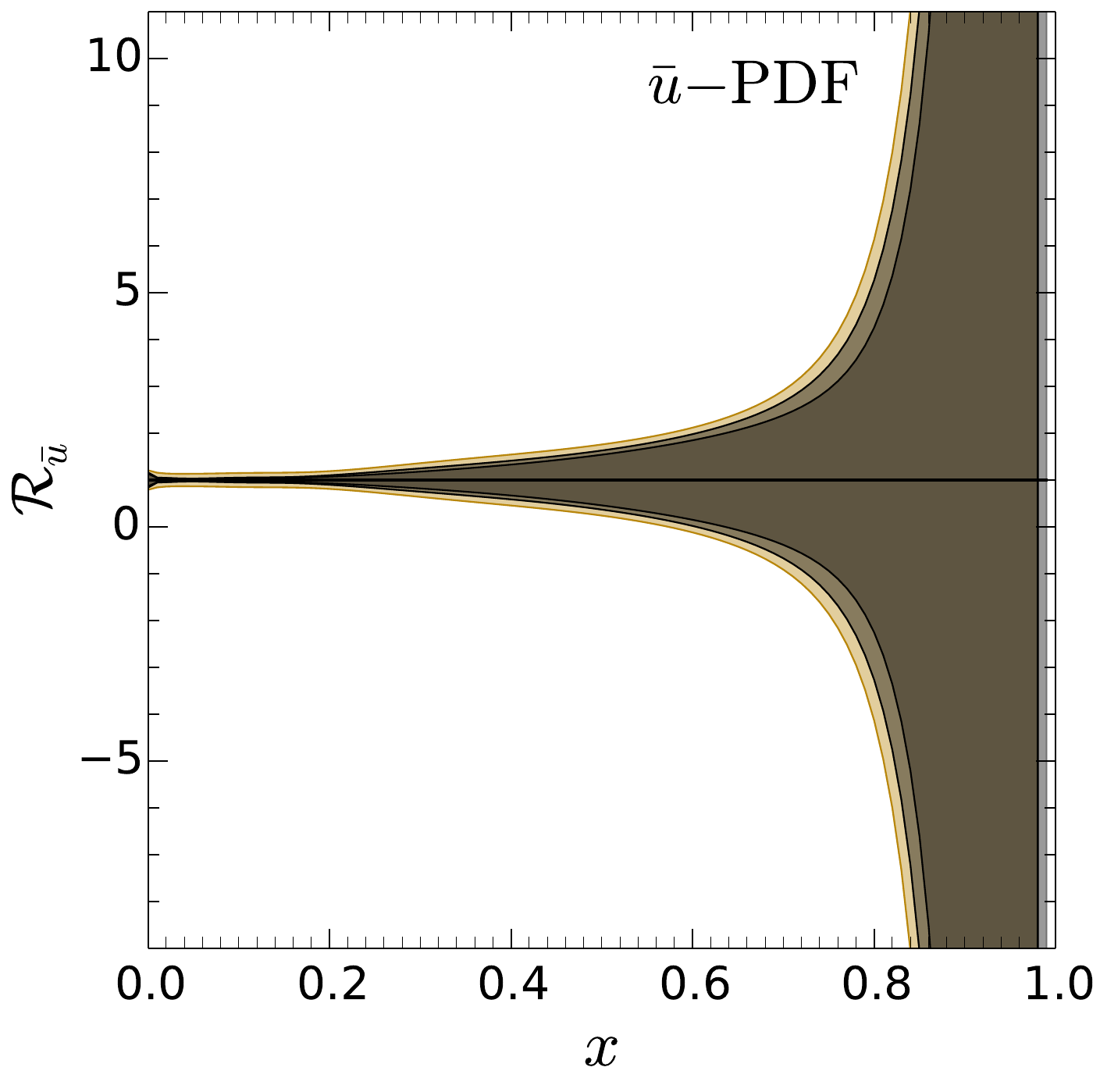}
 }\hspace*{-2.5mm}
 \subfigure[~$\bar{d}$ PDF]{
\includegraphics[width=0.25\textwidth]{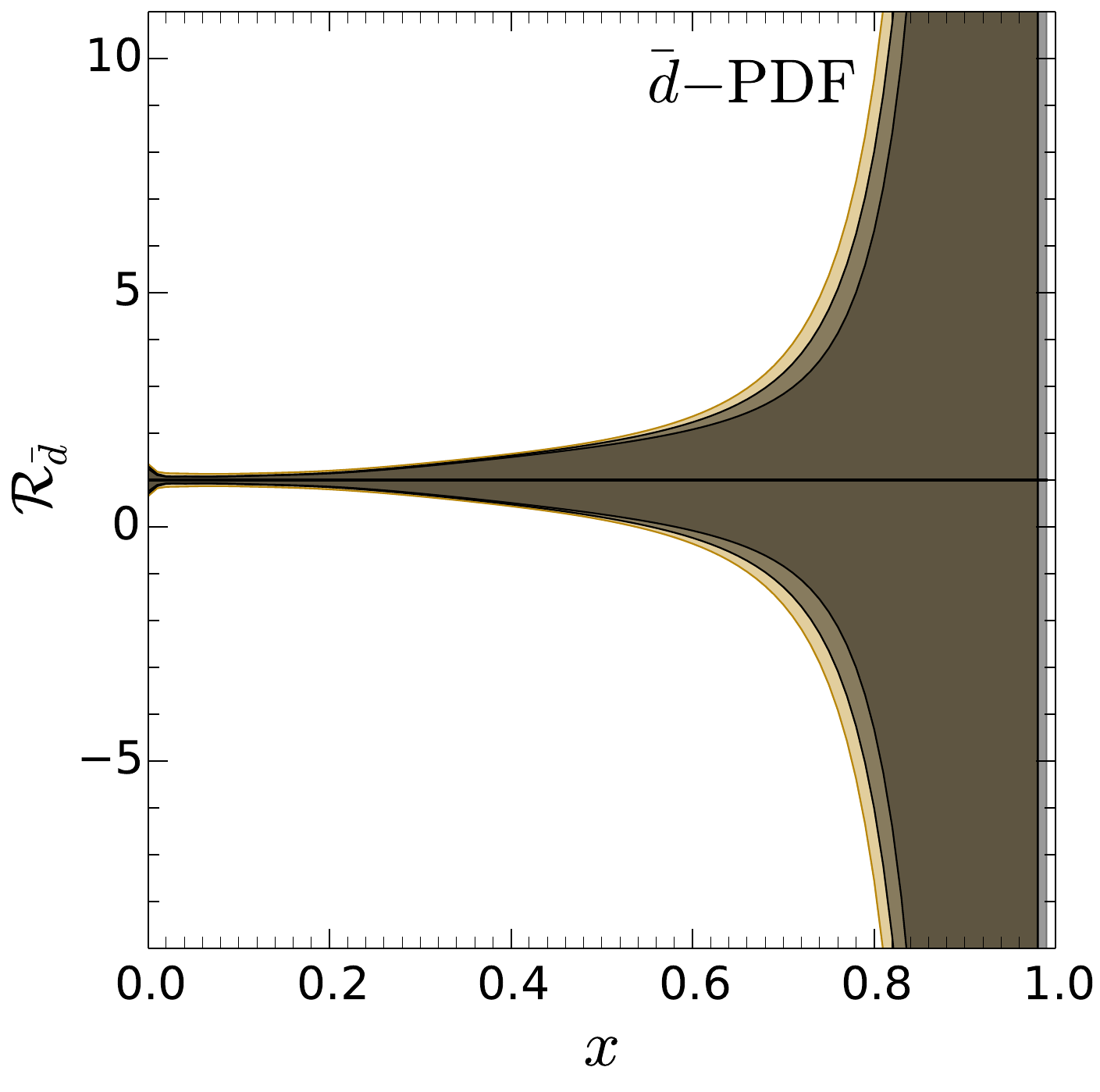}
 }\hspace*{-2.5mm}
\caption{
Same as in \cf{fig:ct14_prof_log} on a linear scale highlighting the
high-$x$ region.}
\label{fig:ct14_prof}
\end{figure}

The effect of the profiling analysis, showing the decrease of the PDF uncertainties
after including these data in a PDF global fit, is presented in 
Figs.~\ref{fig:ct14_prof_log} (logarithmic in $x$) and~\ref{fig:ct14_prof} (linear in $x$)
for the light-quark ($f=u, d, \bar u, \bar d$) distributions. 
To be precise, for each of these PDFs
the upper and lower curves delimiting the bands are defined as
\begin{equation}
R_f(x,Q) = 1 \pm  \frac{1}{2 f_0(x,Q)} \sqrt{\sum_i [f_{i+}(x,Q) - f_{i-}(x,Q)]^2}\, ,
\end{equation}
where $f_0(x,Q)$ is the central PDF and $f_{i\pm}(x,Q)$ are the 'i-th' error PDF in the plus or minus direction
and a sum over all eigenvector directions is performed.
Remarkably, \cf{fig:ct14_prof_log} shows a sizeable reduction of the PDF uncertainties
not only in the high-$x$ but also
in the intermediate and small $x$ region ($x \sim 0.1 \ldots 10^{-4}$). 
The effect is largest for the $u$ and $\bar{u}$
distributions but it is also substantial for the $d$ and $\bar{d}$ PDFs.
The main focus of this section is the high-$x$ region which is highlighted in
Fig.\ \ref{fig:ct14_prof}. Here it can be seen that our knowledge of the 
valence quark distributions can be considerably improved for $x \gtrsim 0.4$
where the effect is again more pronounced for the up quark.
However, even some information on the light quark sea at high-$x$ can be obtained.

Additionally, we have considered a scenario with luminosity reduced by a factor 10,
\ie\ $\L_{pp}=1$ fb$^{-1}$. From Figs.~\ref{fig:ct14_prof_log} and~\ref{fig:ct14_prof},
it is clear that even such a reduced statistics will allow for a substantial reduction
of the current PDF uncertainties.

\paragraph{$W$ boson production close to the threshold} 

\newcommand\Tstrut{\rule{0pt}{2.6ex}}         %
\newcommand\Bstrut{\rule[-0.9ex]{0pt}{0pt}}   %
\begin{table}  
\begin{center}
\begin{tabular}{c|cc|c|cc|c}
\multirow{2}{*}{\pp} & \multicolumn{3}{c|}{$W^+$}      & \multicolumn{3}{c}{$W^-$}  \Tstrut\\
                     &  NLO    & NNLO   & Counts/year &  NLO    & NNLO   & Counts/year \Bstrut\\
\hline
\hline
$p_T^l>10$ GeV &  $22.5^{+4.8}_{-4.3}$   & $25.9^{+4.8}_{-5.0}$    & $259 \pm 49$ & $5.5^{+1.3}_{-1.3}$       &  $6.2^{+1.1}_{-1.4}$    & $62 \pm 13$ \Tstrut\Bstrut\\
\hline
$p_T^l>20$ GeV &  $1.9^{+1.2}_{-0.7}$    & $2.3^{+1.3}_{-1.1}$     &  $23 \pm 12$           & $0.38^{+0.29}_{-0.20}$    &  $0.50^{+0.25}_{-0.25}$ & $5 \pm 2.5$  \Tstrut\Bstrut\\
\hline
$p_T^l>30$ GeV &  $0.28^{+0.91}_{-0.27}$ &  $0.27^{+0.72}_{-0.24}$ &   $2.7 \pm 4.8$          & $0.035^{+0.091}_{-0.039}$ &  $0.04^{+0.09}_{-0.04}$ & $0.4 \pm 0.7$ \Tstrut\Bstrut\\ \hline
\end{tabular}
\end{center}
\caption{Cross sections in [fb] at NLO and NNLO integrated over the rapidity range 
$2<\eta^{\rm lab.}_\mu<5$ and imposing a cut $p_{T}^{\mu}>10$ GeV. The results
have been obtained for $pp$ collisions at $\sqrt{s}=115$ GeV with
FEWZ~\cite{Gavin:2012sy} using the NLO and NNLO CT14 PDFs~\cite{Dulat:2015mca}, respectively.
The renormalisation and factorisation scales have been set to $\mu_R = \mu_F = M_W$.
The asymmetric uncertainties have been calculated using the error PDFs.
The expected number of events has been obtained with a yearly luminosity of 10 fb$^{-1}$.}
\label{tab:large_x_W-Z}
\end{table}

Due to the high \cms\ energy of 115 GeV it is possible to
study the production of $W$ bosons close to the production 
threshold.\footnote{Note that the cross section 
for $Z$ boson production is too low to be accessible at AFTER@LHC.}
Assuming a yearly integrated luminosity of 10 fb$^{-1}$
we expect roughly 250 $W^+$ and 60 $W^-$ events per year
before taking into account the experimental efficiencies.\footnote{These numbers are for one leptonic decay channel.
In a more realistic estimate it will be necessary to sum up the electron and muon channels
taking into account the different efficiencies.}
These event numbers are based on NNLO cross sections calculated 
by integrating over the rapidity range
$2<\eta^{\rm lab.}_\ell<5$ and imposing a requirement of $p_{T}^{\ell}>10$ GeV on the transverse momentum
on the $W$-decay lepton using FEWZ~\cite{Gavin:2012sy} together with NNLO CT14 PDFs ~\cite{Dulat:2015mca}. 
The factorisation and renormalisation scales have been chosen to be $\mu_R = \mu_F = M_W$.
For convenience, the cross sections at NLO and NNLO along with the event numbers and the PDF uncertainties are summarised in Table\ \ref{tab:large_x_W-Z} for a selection of  $p_T^{\ell}$ kinematic cuts.

In \cf{fig:large_x_Wp}, we show NNLO predictions for the
differential cross section for $W^+$ production in $pp$ collisions at 
AFTER@LHC as a function of the transverse mass $M_T$ 
for the case of $p_{T}^{\ell}>10$ GeV (left) and the 
transverse momentum $p_T^{\ell}$ of the produced lepton (right).
The yellow band represents the PDF uncertainty and the error bars represent the uncertainty
due to renormalisation/factorisation scale variation by a factor $2$ around the central scale choice $\mu_R=\mu_F=M_W$.
As can be seen, the PDF uncertainty dominates over the scale uncertainty for $M_T > 20$ GeV.
It is also interesting to note that the $M_T$ distributions peaks at $M_T \sim 25$ GeV
far below $M_W$.

\begin{figure}[hbt!]
\centering
\subfigure[~]{\includegraphics[width=0.49\textwidth]{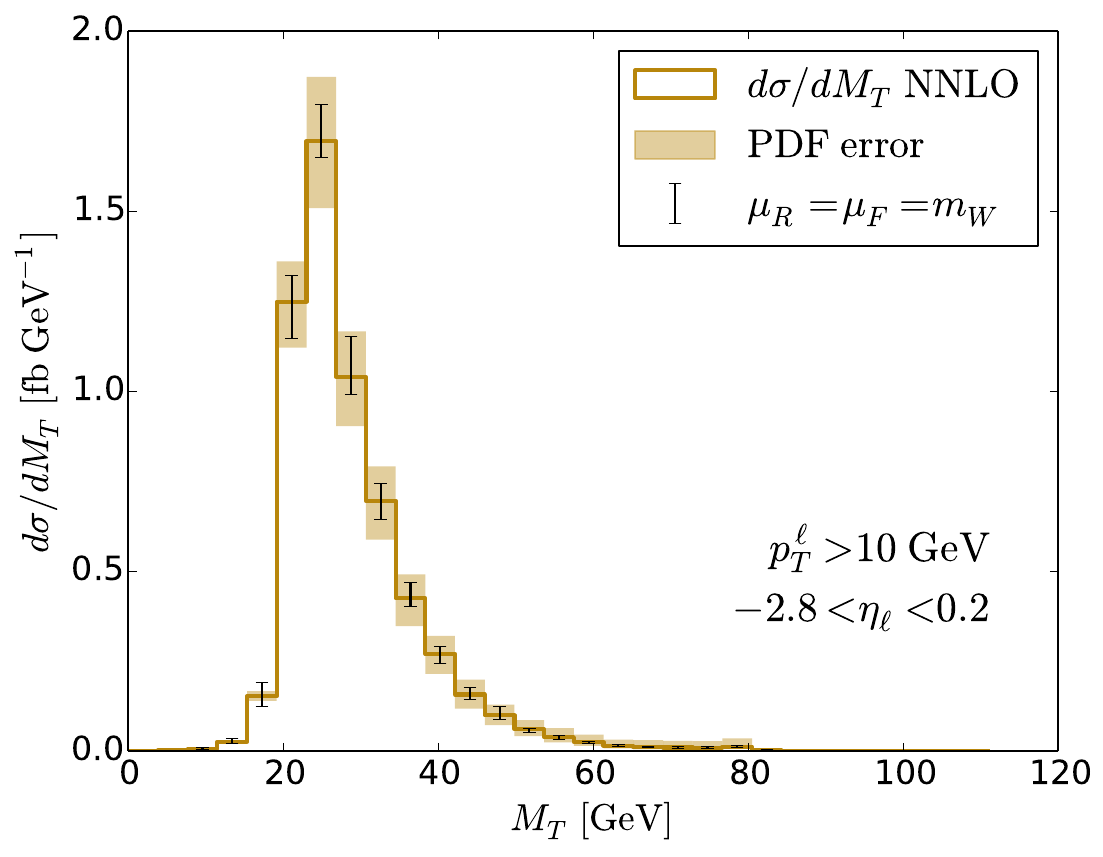}}
\subfigure[~]{\includegraphics[width=0.49\textwidth]{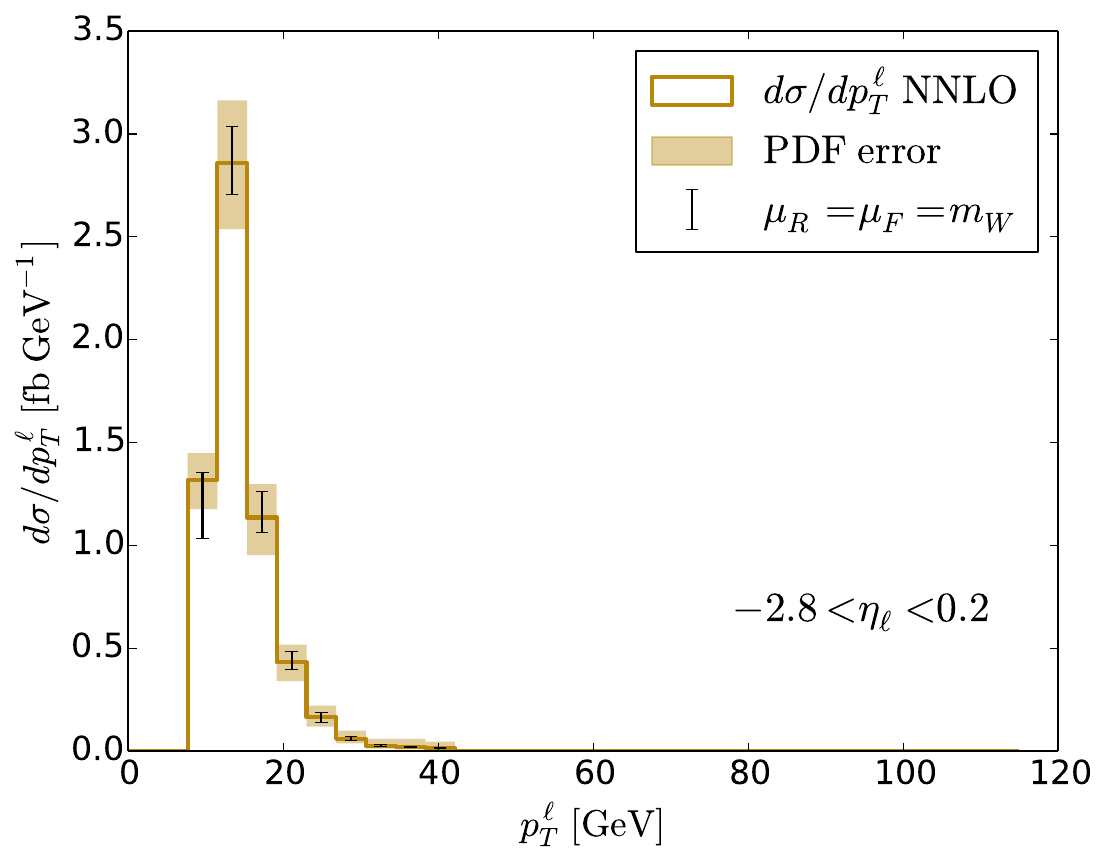}}
\caption{NNLO cross section in [fb/GeV] for $W^+$ production in $pp$
collisions at AFTER@LHC as a function of (a) the transverse mass $M_T$ and (b) 
the transverse momentum $p_T^{\ell}$ of the produced lepton.
The cross section has been obtained by integrating over the rapidity range 
$2<\eta_\ell<5$ (in the laboratory frame) using FEWZ~\cite{Gavin:2012sy}.
For the $M_T$-distribution (left) $p_{T}^{\ell}>10$ GeV has been imposed.
The yellow bands represent the PDF uncertainty and the error bars represent the uncertainty
due to renormalisation/factorisation scale variation by a factor $2$ around the central
scale choice $\mu_R=\mu_F=M_W$.} 
\label{fig:large_x_Wp}
\end{figure}

In the following, we illustrate that even a rough measurement of
the $W$ cross section at AFTER can provide interesting information
on the high-$x$ behaviour of the light sea quarks.
The LO cross section for $W$ boson production reads
\begin{equation}
\frac{d\sigma}{dy} = \frac{2 \pi}{3} \frac{G_F}{\sqrt{2}} \sum_{i,j} |V_{ij}|^2 
\left[q^A_i(x_1) \bar q^B_j(x_2) + \bar q^A_j(x_1) q^B_i(x_2) \right]\, ,
\end{equation}
where the momentum fractions $x_{1,2}$ are related to the (\cms) rapidity of the
$W$-boson in the usual way, $x_{1,2} = (M_W/\sqrt{s}) e^{\pm y}$.
Assuming a diagonal CKM matrix and neglecting the contribution from the $sc$-channel
one can easily derive the following ratio of cross sections:
\begin{align}
R^W & = \frac{\frac{d\sigma}{dy}(pn \to W^+ + W^-)-\frac{d\sigma}{dy}(pp \to W^+ + W^-)}
{\frac{d\sigma}{dy}(pn \to W^+ + W^-)+\frac{d\sigma}{dy}(pp \to W^+ + W^-)} 
= 1 - 2 \frac{\frac{d\sigma}{dy}(pp \to W^+ + W^-)}{\frac{d\sigma}{dy}(pd \to W^+ + W^-)}
\nonumber\\
& =
\frac{[u(x_1)-d(x_1)][\bar u(x_2) - \bar d(x_2)]+[\bar u(x_1)- \bar d(x_1)][u(x_2) - d(x_2)]}{[u(x_1)+d(x_1)][\bar u(x_2) + \bar d(x_2)]+[\bar u(x_1)+ \bar d(x_1)][u(x_2) + d(x_2)]} \, .
\end{align}
At central rapidity, $x_1 = x_2 = x$, 
the ratio reduces to the remarkably simple  expression\footnote {At central rapidity in the \cms, $y=0$, one has $x_1 = x_2 = M_W/\sqrt{s}$.
However, as shown in \cf{fig:large_x_Wp}, most $W$ bosons are produced off-shell.
In that case one can effectively replace $x_1 = x_2 \sim M^*/\sqrt{s}$ with $M^* \sim 35$ GeV.}
\begin{equation}
R^W(y_\cms=0) =\frac{(1-r_v)(1-r_s)}{(1+r_v)(1+r_s)},
\end{equation}
where $r_v(x) = d(x)/u(x)$ and $r_s(x) = \bar d(x)/\bar u(x)$ at $x \sim 0.3$.
Therefore, even a rough measurement of this ratio with about 30\% 
precision could provide valuable information on the barely known
ratio $r_s = \bar d/\bar u$ at high-$x$. This would already compete 
with the existing E866 measurements~\cite{Towell:2001nh} and 10\% precision
would be welcome, which could be obtained by accumulating $W$ data over a couple of years.
On the other hand, we estimate that such a measurement will not be able to compete
with the possible future SeaQuest DY measurements~\cite{Reimer:2010zza}.

Another interesting aspect is that a measurement of vector boson production close to the
threshold could serve as a proxy for searches of new heavy resonances at the LHC.
As was highlighted above the $W$ boson production at AFTER@LHC is predominantly off-shell.
Therefore, one can expect a similar behaviour for a heavy new resonance with a mass close to the \cms\ energy.
The current mass limits for such heavy resonances are typically on the order of 3 to 4 TeV depending on the model.
With increasing statistics even higher resonance masses will be probed, and we are approaching 
the region of the production threshold where the high-$x$ PDFs are probed and the PDF error becomes the dominant theoretical uncertainty in precision calculations \cite{Jezo:2014wra,Bonciani:2015hgv}.
Furthermore, soft gluon resummation effects are expected to become important \cite{Jezo:2014wra}
which could be partially tested at AFTER@LHC.

\paragraph{The charm quark PDF at high $x$}

The high-$x$ heavy-quark PDFs can be important for BSM physics in which new heavy particles
have couplings to the SM fermions which are proportional to the fermion mass or for models which
predominantly couple to the second and/or third generation \cite{Lyonnet:2015dca}.
Most global analyses of PDFs rely on the assumption that the charm and bottom PDFs are perturbatively
generated  by gluon splitting, $g \to Q \bar Q$, and do not involve any non-perturbative
degrees of freedom. It is clearly necessary to test this hypothesis with suitable QCD processes.
Conversely, a non-perturbative, intrinsic contribution to the heavy-quark PDF in the proton
comes from QCD diagrams in which the heavy-quark pair is attached by two or more gluons to the valence quarks. 
It thus depends on the non-perturbative intrinsic structure of the proton~\cite{Brodsky:1984nx,Brodsky:1980pb,Franz:2000ee}.
For a recent review, see \cite{Brodsky:2015fna}.

There are extensive indications for charm production at high $x$, which are, however, not yet fully conclusive
and new data from the LHC, a future EIC and a fixed-target experiment like AFTER@LHC
will be necessary. One example is the EMC measurement of $c(x,Q^2)$ in muon DIS~\cite{Aubert:1982tt}.
The rate observed by the EMC was found to be approximately 30 times higher at $x =0.42, Q^2$ = 75~GeV$^2$ than predicted by gluon splitting~\cite{Harris:1995jx}.
In a more recent analysis, the EMC data have been described in the context of a fitted charm quark
distribution which increase the stability of the fit with respect to variations of the charm quark
mass \cite{Ball:2016neh}. In this study, the NNPDF collaboration found a fitted high-$x$ component
peaking at $x\sim0.5$ and carrying about $1\%$ of the total proton momentum. At the same time, they
observe that the EMC data cannot be fitted with a perturbatively generated charm PDF and the additional
parameters, effectively parametrising the high-$x$ intrinsic component, are needed to describe these data.
Therefore, there is already some evidence for a high-$x$ intrinsic charm (IC) component carrying about 1\% of the total
momentum of the proton\footnote{As we mentioned in section \ref{section:detector_lhcb}, LHCb using the SMOG system has recently reported a charm production measurement~\cite{Aaij:2018ogq} which in principle should, as we advocate here, constraint the IC magnitude. Unfortunately, in the absence of an $H$ target and of a thorough data meta-analysis including theory uncertainty, such current constraints are in fact very weak.}
.
IC also predicts the observed features of the ISR data for ${d\sigma\over dx_F}(pp \to \Lambda_c X) $~\cite{Chauvat:1987kb} and more recently by SELEX~\cite{Garcia:2001xj}.  In this process, 
the comoving $c ,u$ and $d$ coalesce to produce the $ \Lambda_c$ at high $x_F$, where $x_F = x_c + x_u + x_d$.  Other observations at high $x_F$ include  $\Lambda_b, \Lambda_c$, single and double quarkonium, double-charm baryons $ccu$, $ccd$, we refer to \cite{Brodsky:2015fna} for a more complete overview.
We also note here recent works~\cite{Groote:2017szb,Koshkarev:2016ket,Koshkarev:2016acq}
providing predictions for doubly heavy baryons, $B_c$ mesons and all-charmed tetra quarks
at \AFTER based on the intrinsic heavy-quark mechanism, showing that this mechanism should
dominate at high $x_F$ allowing for its observation at \AFTER and potentially confirming the
existence of the IC.
$\Xi_{cc}$ production at \AFTER
was also studied from DGLAP-generated charm in~\cite{Chen:2014hqa}. Besides, 
the production of the hidden charm pentaquark $P_c^+$ was found to peak, for the \AFTER\ kinematics, at about $y_{\rm lab.} \simeq 1.4$ \cite{private:Siddikov}  using the same formalism as that of a  previous study for the LHC kinematics~\cite{Schmidt:2016cmd}.

\begin{figure}[!hbt]
\centering
\subfigure[~$2<y_{\rm lab.}<3$]{\includegraphics[width=0.32\textwidth]{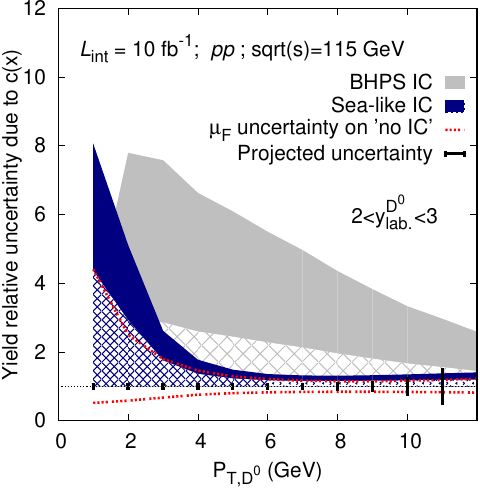}}
\subfigure[~$3<y_{\rm lab.}<4$]{\includegraphics[width=0.32\textwidth]{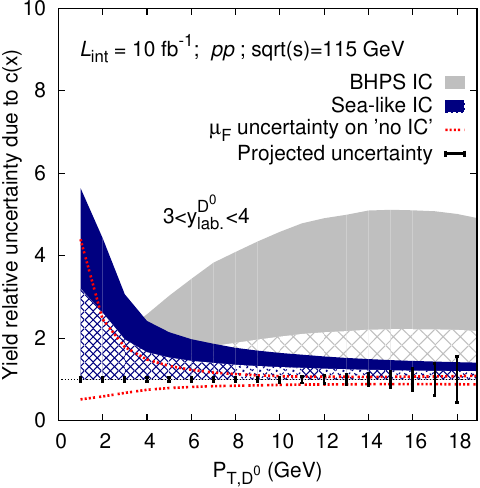}}
\subfigure[~$4<y_{\rm lab.}<5$]{\includegraphics[width=0.32\textwidth]{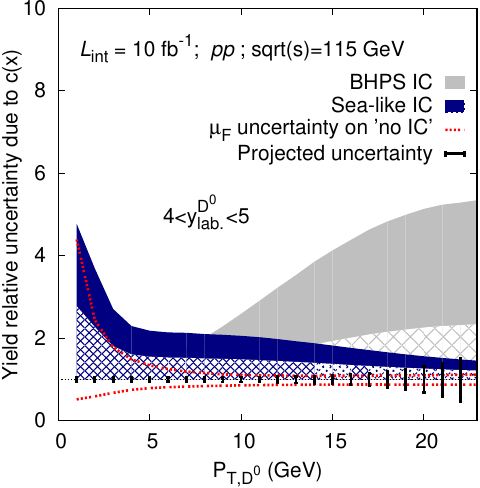}}
\caption{
Impact of the uncertainty on the charm content of the proton on the $D^0$ yield as function of $p_T$ compared to projected uncertainties from the measurement of the $D^0$ yield in \pp\ collisions at $\sqrt{s}=115$ GeV in the LHCb acceptance. The grey (resp. blue) zones correspond to yields computed with charm PDFs including  BHPS-like (resp. sea-like) IC~\cite{Pumplin:2005yf,Pumplin:2007wg}. The filled areas correspond to yields computed with up to $\langle x_{c +\bar c}\rangle= 2\%\, (\text{resp. } 2.4\%)$ and the hashed areas up to  $\langle x_{c +\bar c}\rangle= 0.57\%\, (\text{resp. } 1.1\%)$.
The dashed red curves indicate the factorisation scale ($\mu_F$) uncertainty on the 'no-IC' yield obtained by varying $\mu_F$ between $m_T$ and $2m_T$ with $m_T^2=m_c^2+p_T^2$.
Systematic uncertainties of 5\% are included and the statistical uncertainty for the background subtraction is assumed to be negligible which is reasonable assuming LHCb-like performances, see~\cite{LHCb:2017qap}.
The rates were computed by assuming an average efficiency of $\langle \varepsilon \rangle = 10\%$ and $\BR(D^0 \to K\pi)=3.93 \%$.
}
\label{fig:large_x_2}
\end{figure}

A non-perturbative IC component modifies the predictions for a number of processes at the
LHC such as inclusive $D$ meson production \cite{Kniehl:2009ar,Kniehl:2012ti}
or the associated production of a heavy quark with a photon  \cite{Bednyakov:2013zta} or
$Z+c$ production  \cite{Ball:2016neh}. However, in these examples, one has to go to relatively large transverse momenta 
or to very forward rapidities to expect a sizeable effect.

Simply owing to the large boost between the laboratory frame and the c.m.s., 
the fixed-target mode is the ideal set-up to uncover an excess of charm at high $x$.
To illustrate this statement we show in \cf{fig:large_x_2},
the relative yield uncertainty for inclusive $D^0$ meson production at AFTER
for three rapidity bins ($2<y_{\rm lab.}<3$, $3<y_{\rm lab.}<4$, $4<y_{\rm lab.}<5$) as a function of the transverse momentum ($P_{T,D^0}$)
of the $D^0$ meson. An integrated luminosity of 10 fb$^{-1}$ has been assumed to compute the expected yields -- accounting
for the expected efficiency and the branching ratio -- from the theoretical 
cross sections which have been obtained using the set-up described in \cite{Brodsky:2015fna}.
From these yields, we derived the expected uncertainties shown in the figures.
The blue and grey bands (and hatched zones) correspond to two IC models.
As can be seen, even for $P_{T,D^0}\lesssim 15$ GeV the expected precision of the measurement will clearly allow one to 
considerably constrain such IC model, by up to an order magnitude. At such large $x$, the perturbative charm is indeed suppressed.

\paragraph{The gluon PDF at high $x$}

The available data from DIS structure functions and from DY lepton-pair production provide only rather
weak constraints on the gluon PDF, particularly at high $x$. Important information on the gluon density
can be drawn from inclusive jet data and from top quark pair production data.
In the latter case, the differential distributions ($y_{t\bar t}$, $y_t$, $p_T^t$, $m_{t\bar t}$) have been shown
to considerably reduce the uncertainty of the gluon PDF at high $x$ \cite{Czakon:2016olj}.
Nevertheless, these data provide constraints at quite large factorisation scales $\mu_F \sim 100$ GeV and
the knowledge of the gluon PDF at $x \gtrsim 0.5$ remains limited.

Let us discuss here a number of possibilities to obtain information on the gluon PDF offered by
the \AFTER programme. As can be seen 
in \cf{fig:high_x_2}, open heavy-flavoured mesons and heavy quarkonia will be abundantly produced at \AFTER covering 
an important region in the $(x,m_T)$ plane. As should be clear from the discussion above on the charm quark PDF, 
the gluon and the heavy-quark distributions are inextricably linked. 
Therefore, in the context of a global analysis, the heavy-quark data have the potential to constrain both the heavy-quark and the gluon distribution at high $x$. It should also be noted that, in the case of $b$ quark production, the contribution
from an intrinsic bottom component is expected to be very small \cite{Lyonnet:2015dca}.
One should mention that normalised rapidity distributions of $D$ and $B$ mesons from LHCb
have been already used with success to study low-$x$ gluon~\cite{Zenaiev:2015rfa,Gauld:2016kpd}.

In addition to the inclusive heavy-quark observables, the associated production of a heavy quarkonium with a
photon is sensitive to the high $x$ gluon distribution.
This is illustrated in \ct{tab:psigamma} which shows that this process is largely dominated by the
gluon-gluon initiated subprocess probing $x_2$ values in the range from 0.1 to 0.6, whereas at such large $x$
single-inclusive-quarkonium production is presumably largely from (heavy) quark induced channels.

\begin{table}[!hbt]
\begin{center}\renewcommand{\arraystretch}{1.2}
\begin{tabular}{c|c|c|c|c} 
Isolated $J/\psi+\gamma$ & $\langle x_2\rangle \sim \frac{M_{\psi\gamma}}{\sqrt{s}} e^{-Y_{\psi\gamma}}$ & $\sigma_{gg} \times \BR_{\mu\mu}$ [fb] & $\sigma_{q\bar q}\times \BR_{\mu\mu}$ [fb] & Counts/year  \\
\hline \hline
$|Y^{\cms}_{\psi\gamma}|<0.5$       & 0.10 & ${\cal O}(100)$ & ${\cal O}(0.2)$ & ${\cal O}(1000)$ \\
$-1.5<|Y^{\cms}_{\psi\gamma}|<-0.5$ & 0.25 & ${\cal O}(50)$ &${\cal O}(0.2)$ & ${\cal O}(500)$ \\
$-2.5<|Y^{\cms}_{\psi\gamma}|<-1.5$ & 0.60 & ${\cal O}(10)$ &${\cal O}(0.04)$ & ${\cal O}(100)$ \\ \hline
\end{tabular} 
\caption{
Isolated $J/\psi+\gamma$ production for three bins in the \cms\ rapidity of the pair. From left to right: the average $x_2$ , the partial contributions to the cross section from the $gg$-initiated and $q \bar q$-initiated subprocesses (multiplied by the $J/\psi$ branching into di-muons) \cite{Lansberg:2014myg} respectively, and the order of magnitude of the expected number of events per year assuming a luminosity of 10 fb$^{-1}$
and a detector efficiency on the order of unity.
\label{tab:psigamma}
}
\end{center}
\end{table}

Along these lines one can also study double-$J/\psi$ production which is dominated by the $gg$-channel up to
large values of the target $x_2$ as is shown in \ct{tab:psipsi}.
It is also interesting to notice that \AFTER\ provides the unique opportunity to study double parton scatterings and double parton correlations in the nucleon at energies around $100$ GeV via di-$J/\psi$ production~\cite{Lansberg:2015lva}, where the similar studies already exist at the Tevatron and the LHC~\cite{Lansberg:2013qka,Lansberg:2014swa}.

\begin{table}[!hbt]
\begin{center}\renewcommand{\arraystretch}{1.2}
\begin{tabular}{c|c|c|c|c} 
$J/\psi+J/\psi$ & $\langle x_2\rangle \sim \frac{M_{\psi\psi}}{\sqrt{s}} e^{-Y^\cms_{\psi\psi}}$ & $\sigma_{gg}\times \BR_{\mu\mu}^2$ [fb] & $\sigma_{q\bar q}\times \BR_{\mu\mu}^2$ [fb] & Counts/year  \\
\hline \hline
$4.5<Y^{\rm lab.}_{\psi\psi}<5.0$ & 0.13 & ${\cal O}(5)$ &${\cal O}(1)$ & ${\cal O}(50)$ \\
$4.0<Y^{\rm lab.}_{\psi\psi}<4.5$ & 0.29 & ${\cal O}(50)$ &${\cal O}(10)$ & ${\cal O}(500)$ \\
$3.5<Y^{\rm lab.}_{\psi\psi}<4.0$ & 0.45 & ${\cal O}(50)$ &${\cal O}(10)$ & ${\cal O}(500)$ \\
$3.0<Y^{\rm lab.}_{\psi\psi}<3.5$ & 0.60 & ${\cal O}(10)$ &${\cal O}(10)$ & ${\cal O}(100)$ \\
$2.5<Y^{\rm lab.}_{\psi\psi}<3.0$ & 0.77 & ${\cal O}(5)$ &${\cal O}(2)$ & ${\cal O}(70)$ \\ \hline
\end{tabular} 
\caption{
$J/\psi+J/\psi$ production for five bins in the laboratory rapidity of the pair. From left to right: the average $x_2$,
the partial contributions to the cross section from the $gg$-initiated and $q \bar q$-initiated sub-processes (with the $J/\psi$ branchings into di-muons)
and the order of magnitude of the expected number of events (via di-muon pairs) per year assuming a luminosity of 10 fb$^{-1}$
and a detector efficiency on the order of unity.
\label{tab:psipsi}
}
\end{center}
\end{table}

Another interesting observable which probes the gluon distribution directly at LO
via the $qg\to \gamma q$ subprocess is inclusive prompt-photon production. 
It was shown that the theoretical pQCD predictions at NLO describe well the collider
data for \cms\ energies ranging from 200 GeV to 7 TeV \cite{dEnterria:2012kvo}.
However, a series of measurements carried out at \cms\ energies $\sqrt{s} \sim 20 \div 40$ GeV by the fixed-target
E706 experiment \cite{Alverson:1993da,Apanasevich:1997hm,Apanasevich:2004dr} 
were not well described by NLO pQCD calculations \cite{Baer:1990ra,Aurenche:1992yc,Gordon:1994ut}. 
This discrepancy was only partially cured
by the inclusion of resummed soft-gluon contributions in the theoretical predictions 
\cite{Laenen:1998qw,Catani:1998tm,Kidonakis:2000gi,Kidonakis:2003bh,deFlorian:2005wf}.
Therefore, a new measurement at \AFTER\ at $\sqrt{s}=115$~GeV would be interesting
in itself to shed additional light on the data-theory discrepancy at fixed-target energies.
Furthermore, such a measurement has the potential to improve our knowledge of the gluon distribution
at $x>0.3$ provided isolated photons can be measured with transverse momentum $p_T \gtrsim 10 \div 20$ GeV 
\cite{dEnterria:2012kvo}.

%% file: physics-high-x/physics-high-x_nuclear.tex
\subsubsection{Nuclear structure}
\label{sec:high_x_nucl_structure}
As a fixed-target experiment, \AFTER\ allows one to study $pA$ collisions with different nuclei $A$
and some fundamental open questions can be addressed in this case.
More than 30 years ago, the EMC collaboration discovered that nuclear structure functions in DIS are suppressed compared to the prediction from the naive combination of free proton and 
neutron structure functions in the high-$x$ region \cite{Aubert:1983xm}.
The physics mechanism behind this EMC effect is still not fully understood and subject of an active experimental
programme at the Jefferson Laboratory~\cite{Higinbotham:2013hta,Malace:2014uea}.

\begin{figure}[!hbt]
\centering
\subfigure[~]{
\includegraphics[width=0.48\textwidth]{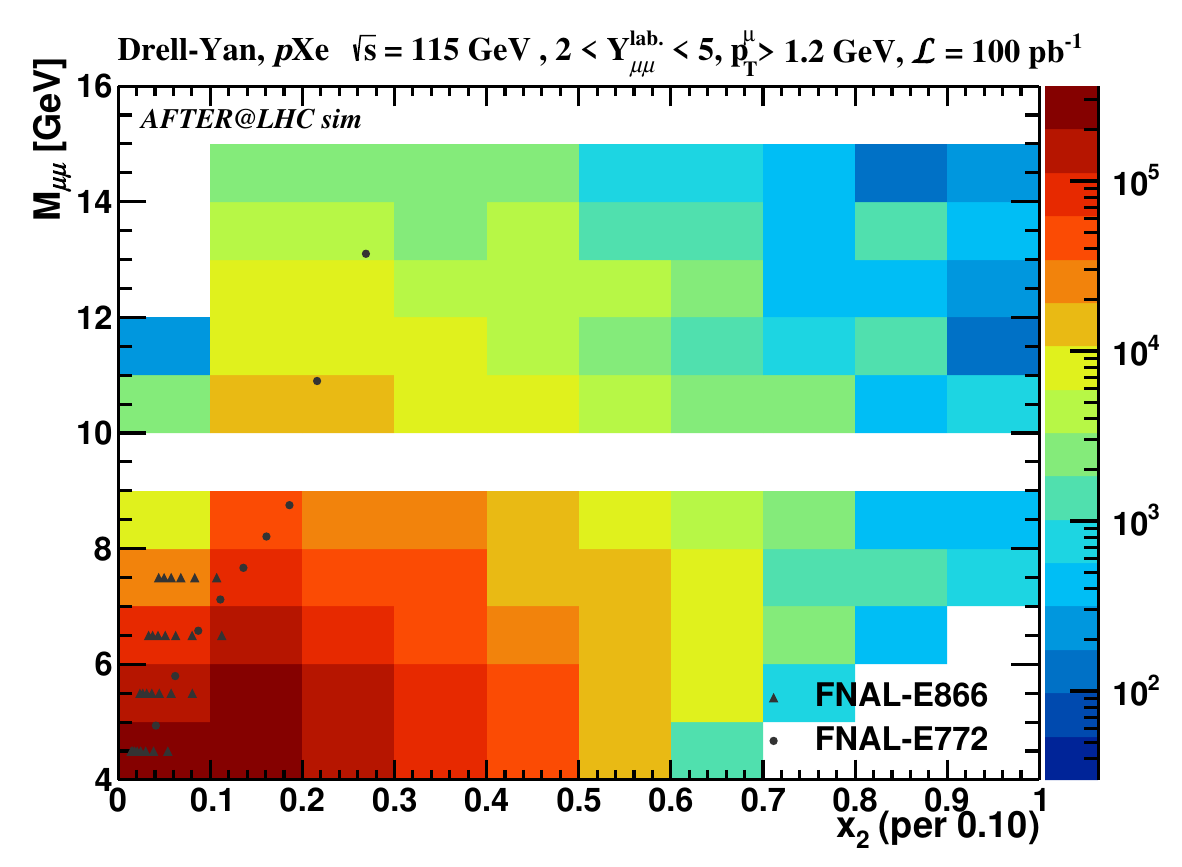}
}
\subfigure[~]{
\includegraphics[width=0.48\textwidth]{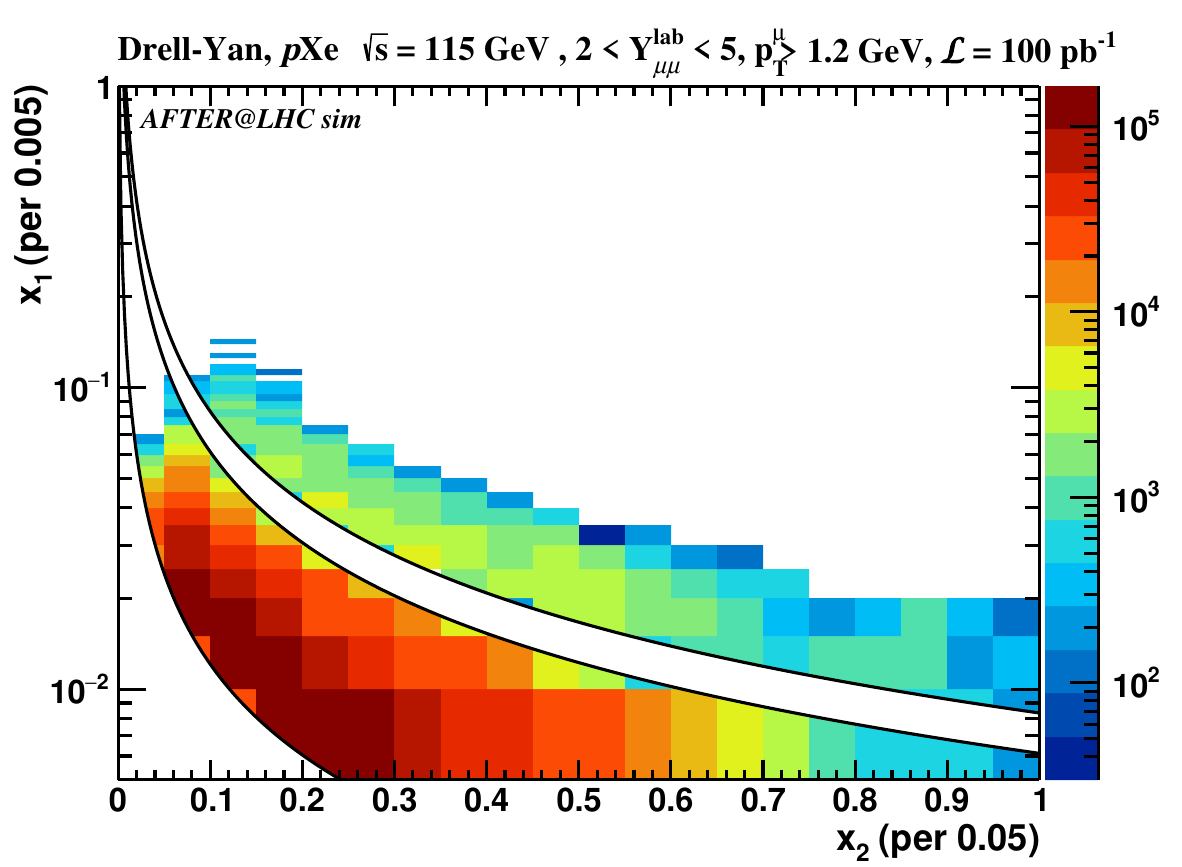}
}\vspace*{-0.5cm}
\caption{
	Kinematical acceptance for DY lepton-pair production with \AFTERLHCb\ in \pXe\ collisions at $\sqrt{s_{NN}}=115$ GeV
	with a muon-pair acceptance of $2<Y^{\rm lab.}_{\mu\mu}<5$ and single-muon requirements: $2<\eta^{\rm lab.}_\mu<5$ and $p_{T}^{\mu}>1.2$ GeV. Projections for \pXe collisions are done by applying a nuclear scaling factor ($A_{\rm Xe}$) to the cross sections obtained from the $pp$ simulations. [For ${\cal L}_{\pXe}={\cal L}_{pp}$, the $\pXe$ yields are thus $A_{\rm Xe}$ times larger than the $pp$ yields].
	(a) Di-muon invariant mass vs. $x_{2}$ compared to the existing DY
data~\cite{Moreno:1990sf,Webb:2003ps,Webb:2003bj,Towell:2001nh}
used in current global PDF fits. (b) $x_{1}$ vs $x_{2}$ for the considered di-muon invariant-mass range of 4 GeV $< M_{(\ell\ell)} < $ 15 GeV excluding $\Upsilon$ mass range of 9 GeV $< M_{(\ell\ell)} < $ 10.5 GeV. Colours correspond to expected yields of the DY signal in each kinematical region, and each coloured cell  contains at least 30 events.
}
\label{fig:large_x_DY_pA}
\end{figure}

\flushfootnote

\begin{figure}[!hbt]
\centering
\subfigure[$4\, \gev \, <M_{\mu\mu}<5$ GeV]{\includegraphics[width=0.32\textwidth]{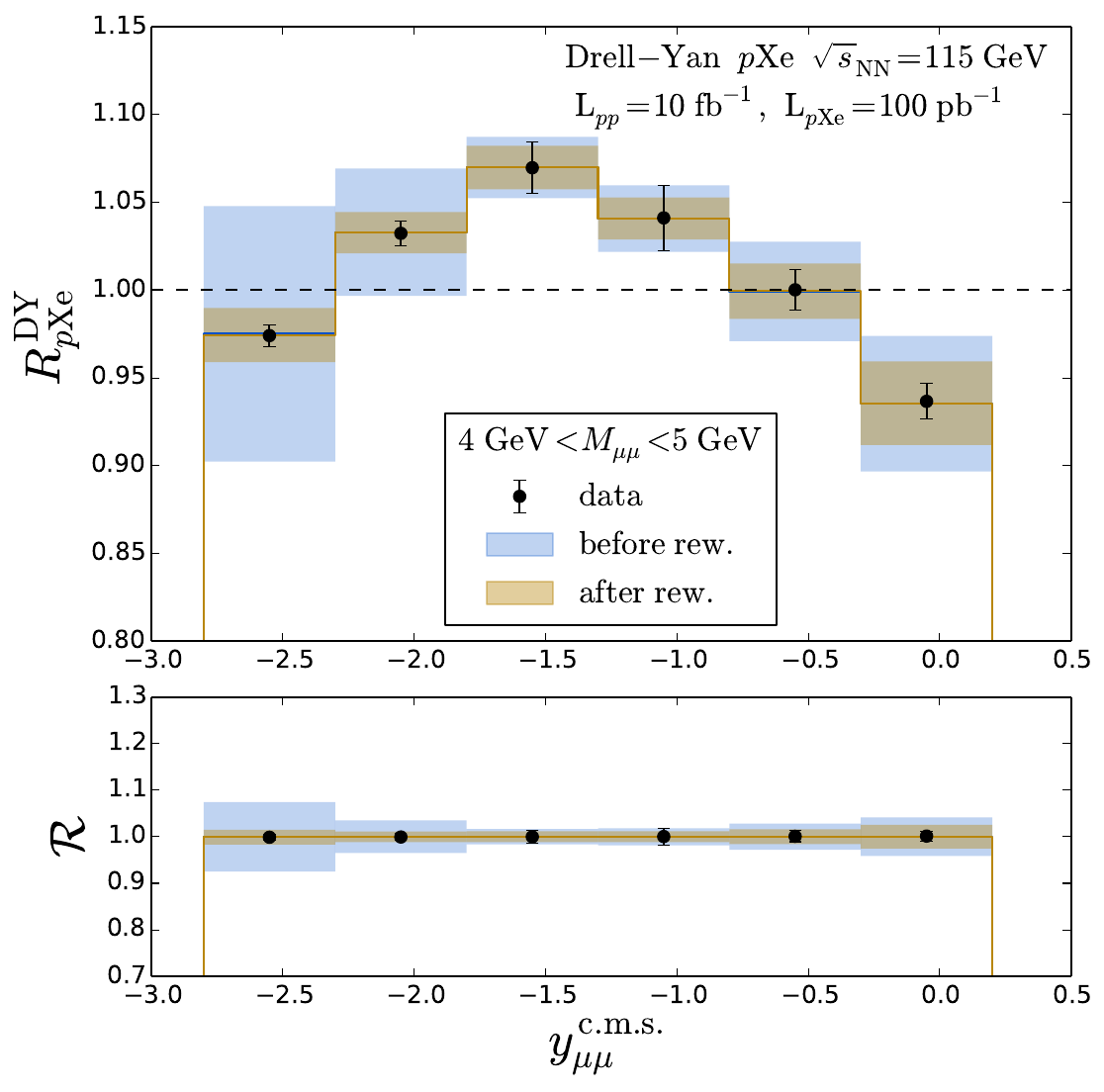}}
\subfigure[$5\, \gev \,<M_{\mu\mu}<6$ GeV]{\includegraphics[width=0.32\textwidth]{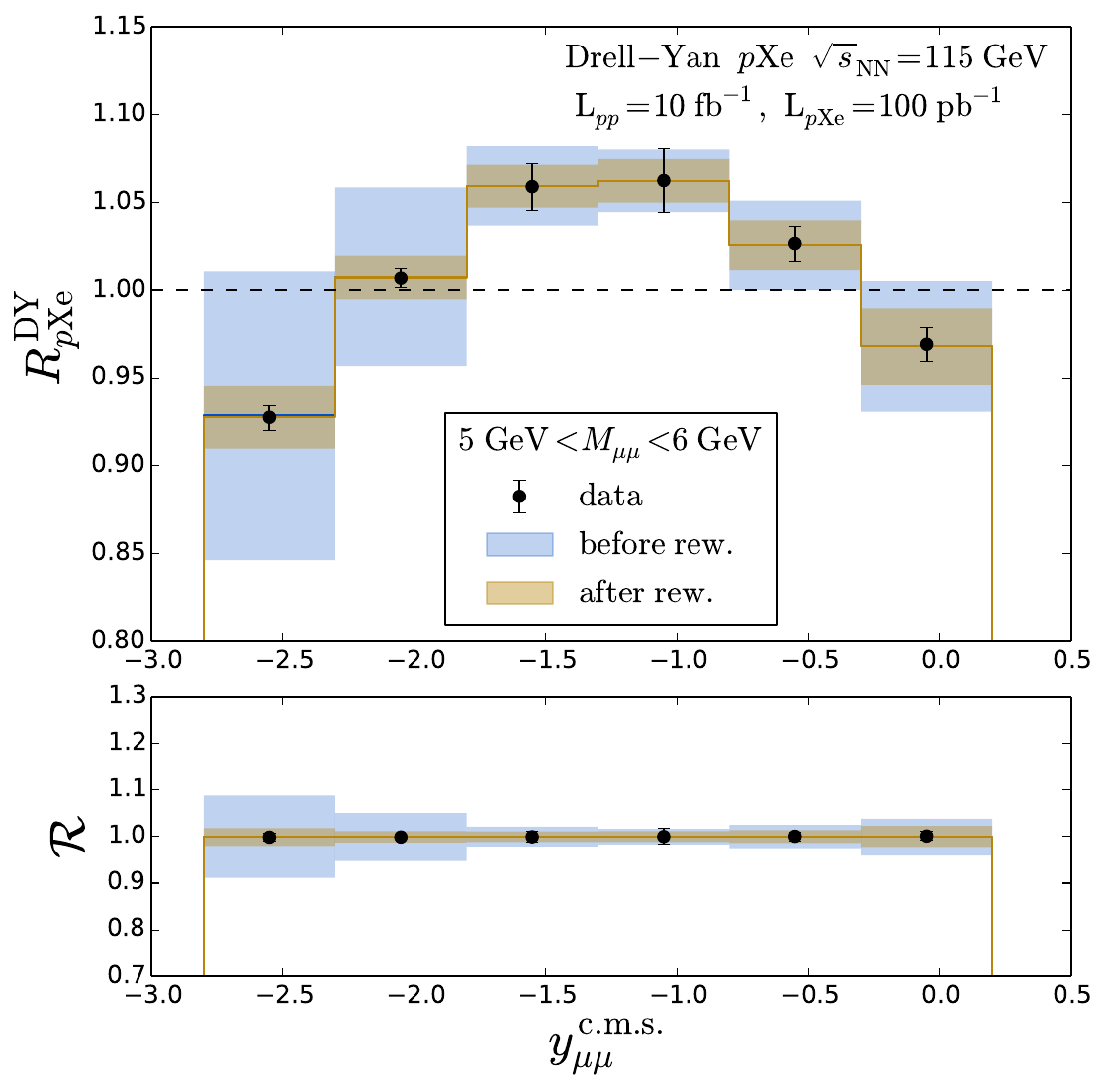}}
\subfigure[$M_{\mu\mu} > 10.5\, \gev$]{\includegraphics[width=0.32\textwidth]{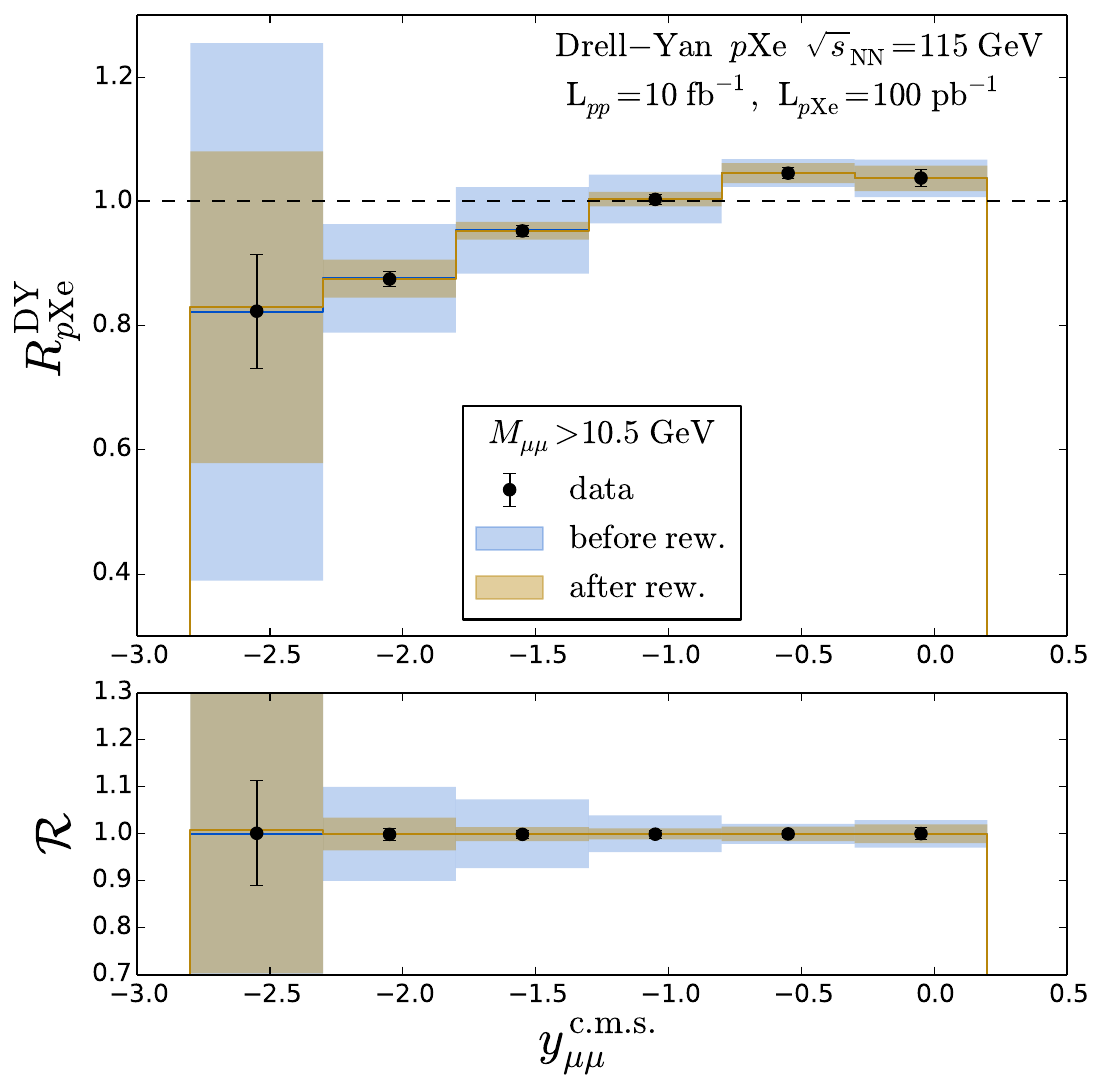}}
\caption{Projection of the statistical uncertainties on the nuclear modification factor $R_{p\text{A}}=\sigma_{\text{DY}}^{\text{Xe}}/(A\times \sigma_{\text{\rm DY}}^{pp})$ for DY lepton-pair production in \pXe\ collisions in different mass ranges compared to the uncertainties encoded in nCTEQ15 nPDFs (in blue `before rew.'), which are representative of typical nPDF uncertainties. 
The projected statistical uncertainties arise from the  subtraction of the uncorrelated, combinational background (based on the like-sign technique) and  assuming the yearly integrated luminosities of
$\mathcal{L}_{pp}=10$ fb$^{-1}$ and $\mathcal{L}_{\pXe}=100$ pb$^{-1} $. The brown band (`after rew.')  correspond to the uncertainty of the \RpA\ after a Bayesian reweighting of the nPDF using the corresponding pseudo-data.
}
\label{fig:RpA_vs_y_DY}
\end{figure}

\begin{figure}[!hbt]
\centering
\subfigure[~]{\includegraphics[width=0.24\textwidth]{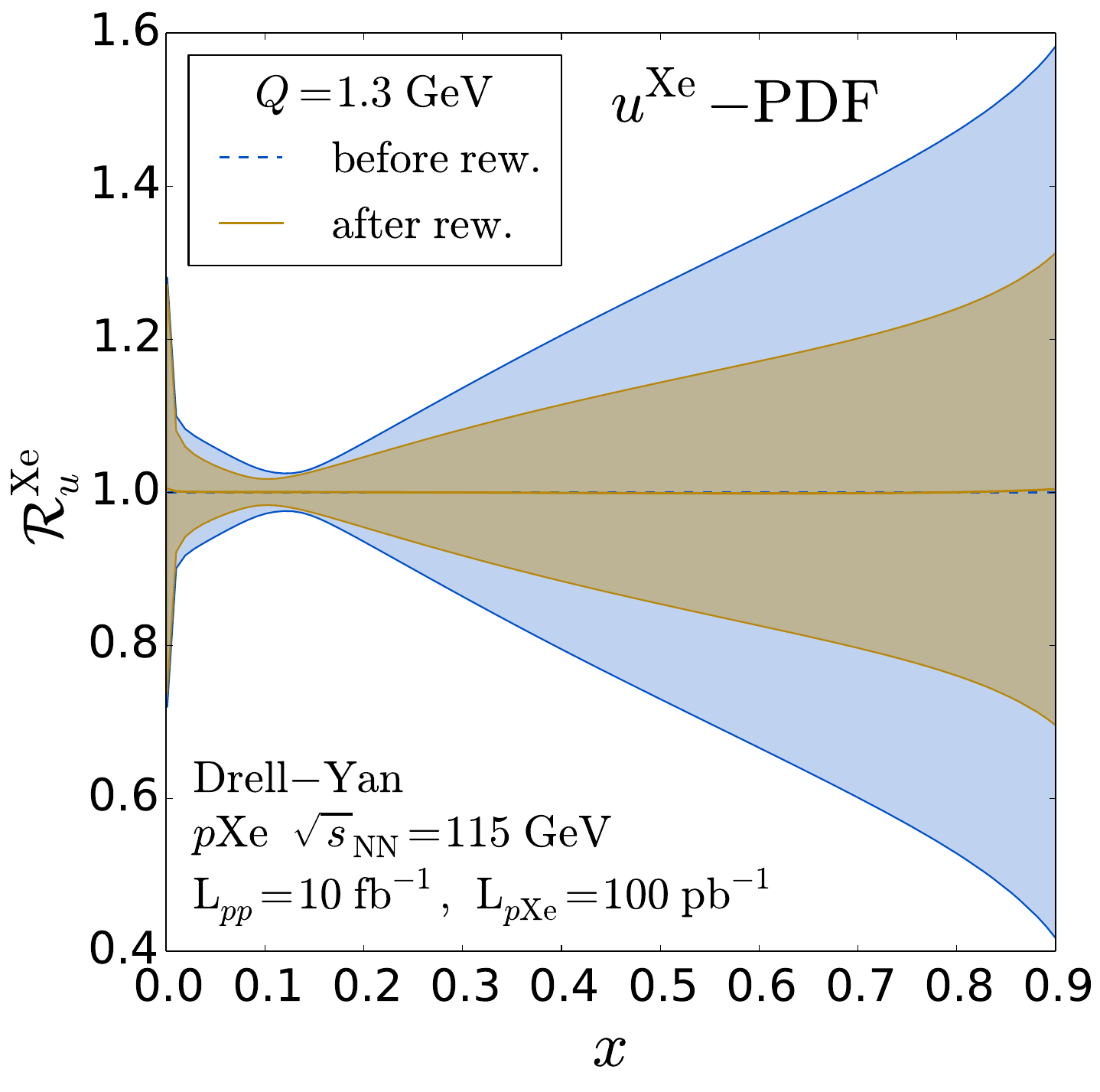}}
\subfigure[~]{\includegraphics[width=0.24\textwidth]{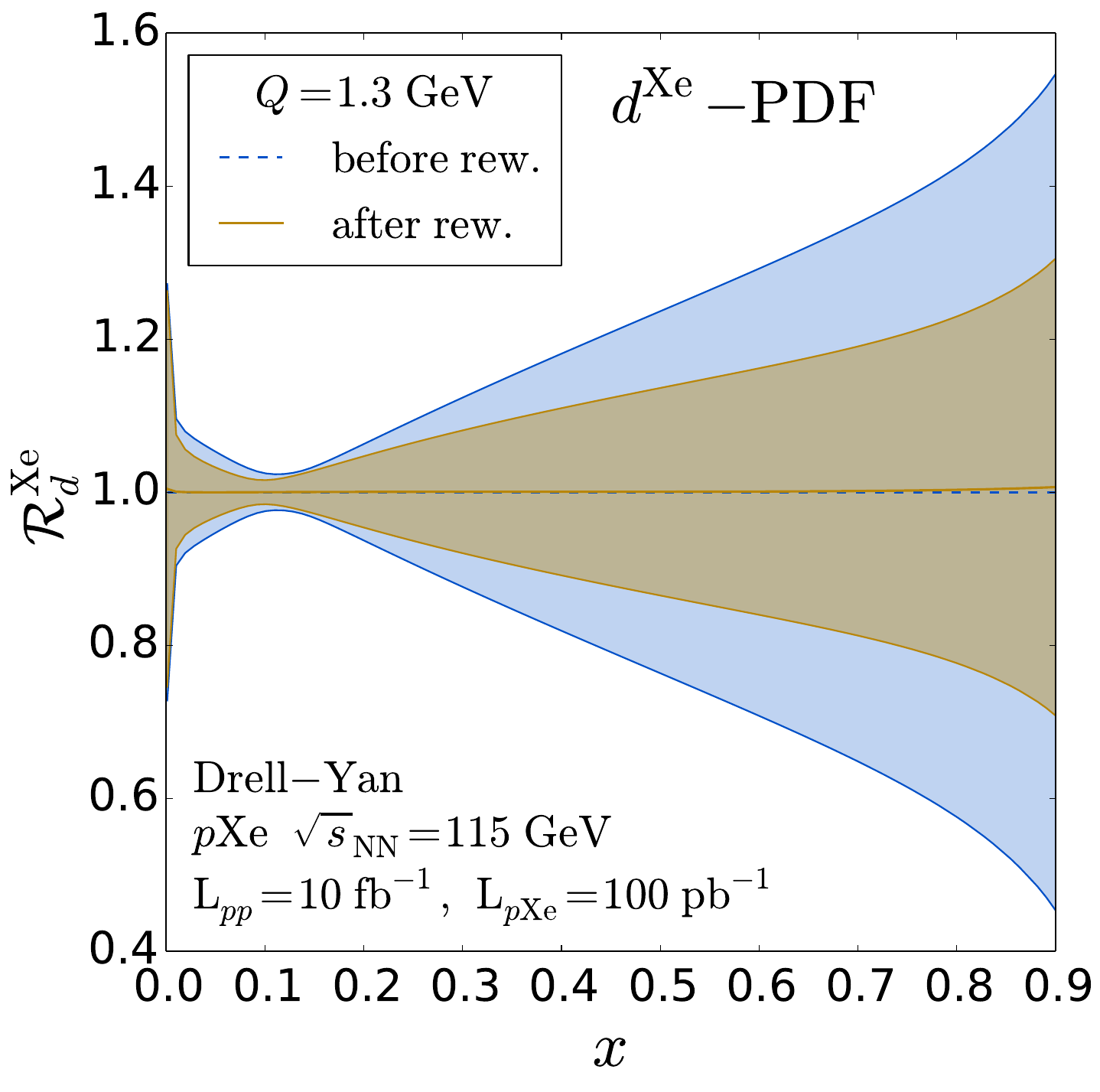}}
\subfigure[~]{\includegraphics[width=0.24\textwidth]{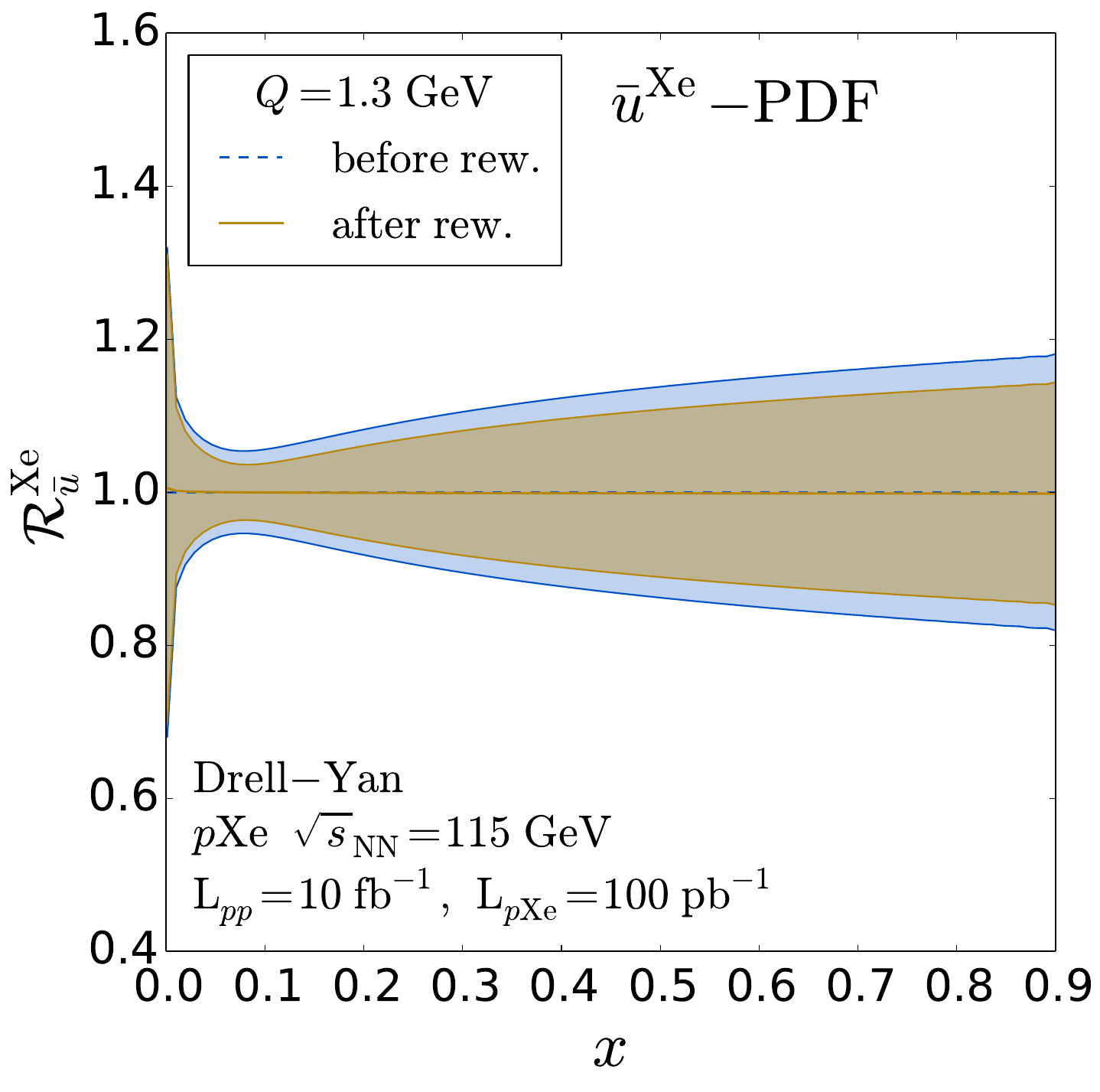}}
\subfigure[~]{\includegraphics[width=0.24\textwidth]{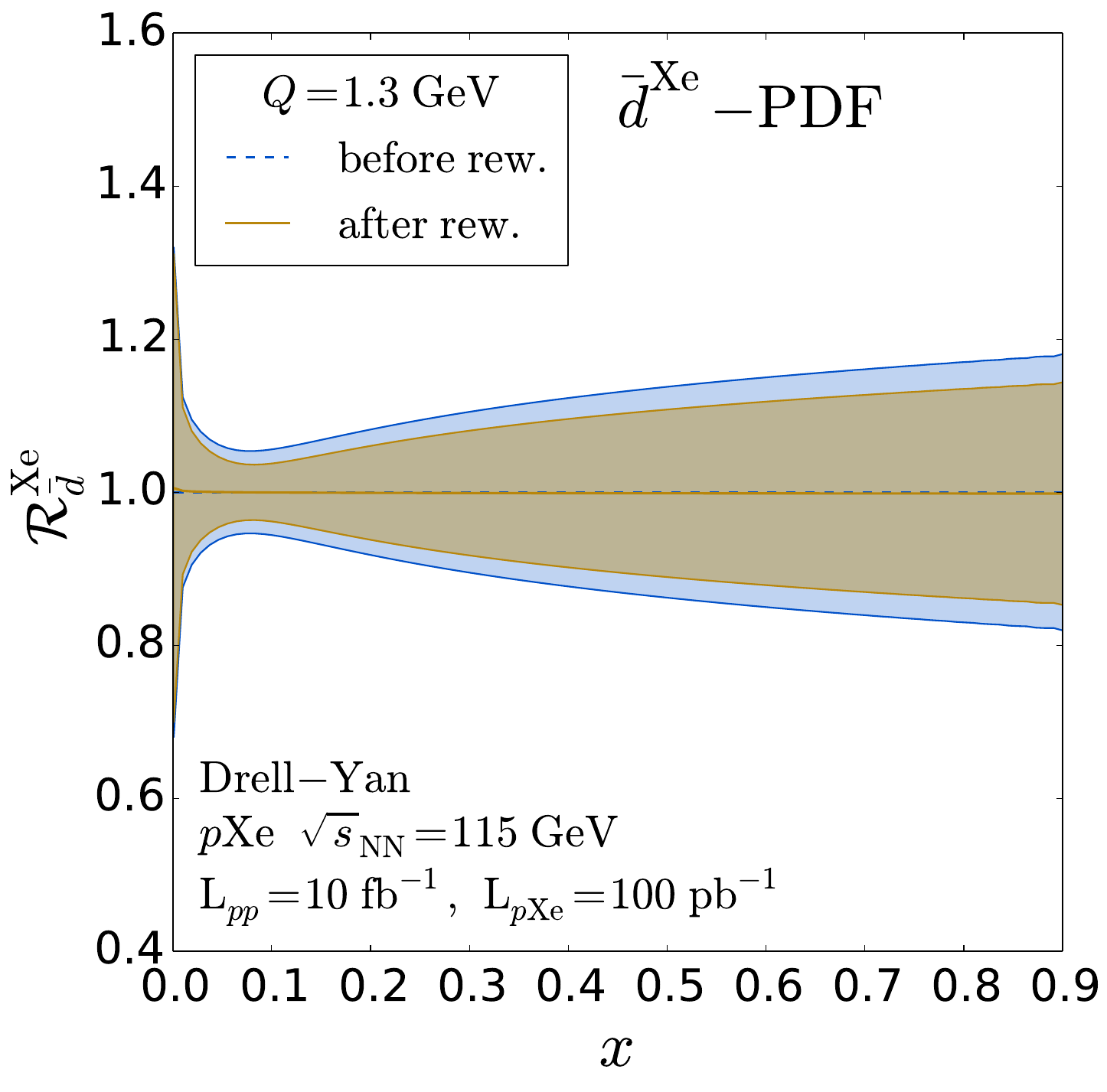}}
\caption{
Uncertainty on nCTEQ15 nPDFs before and after the reweighting using $R^{\text{DY}}_{p\text{Xe}}$
\AFTERLHCb pseudo-data in the range indicated on~\cf{fig:large_x_DY_pA}. The plots show ratio of nPDFs for tungsten ($W$) and the corresponding uncertainties
compared to the central value at the scale $Q=1.3$ GeV.}
\label{fig:npdf_rew}
\end{figure}

\begin{figure}[!hbt]
\centering
\subfigure[$4\, \gev \, <M_{\mu\mu}<5$ GeV]{\includegraphics[width=0.32\textwidth]{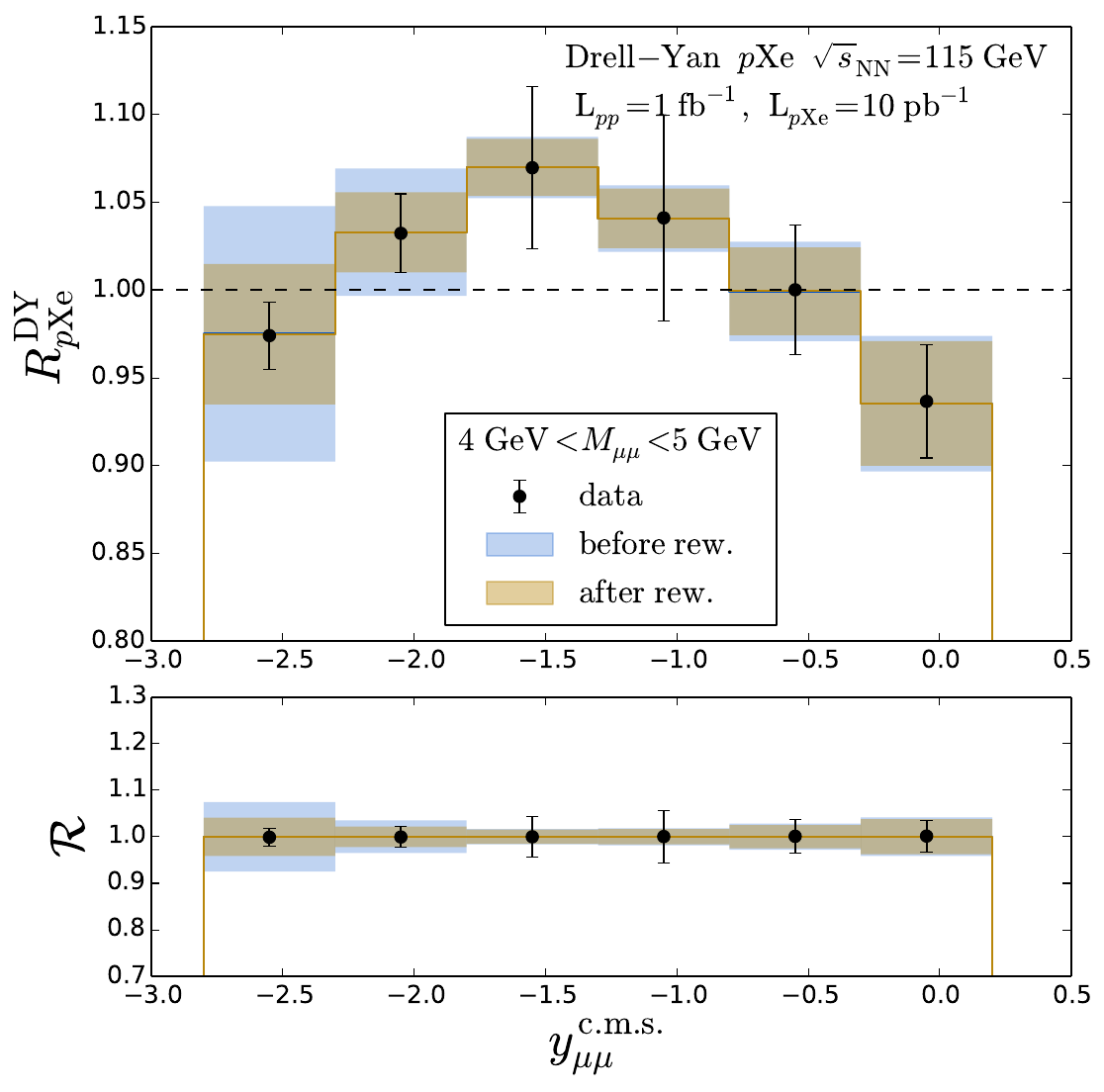}}
\subfigure[$5\, \gev \,<M_{\mu\mu}<6$ GeV]{\includegraphics[width=0.32\textwidth]{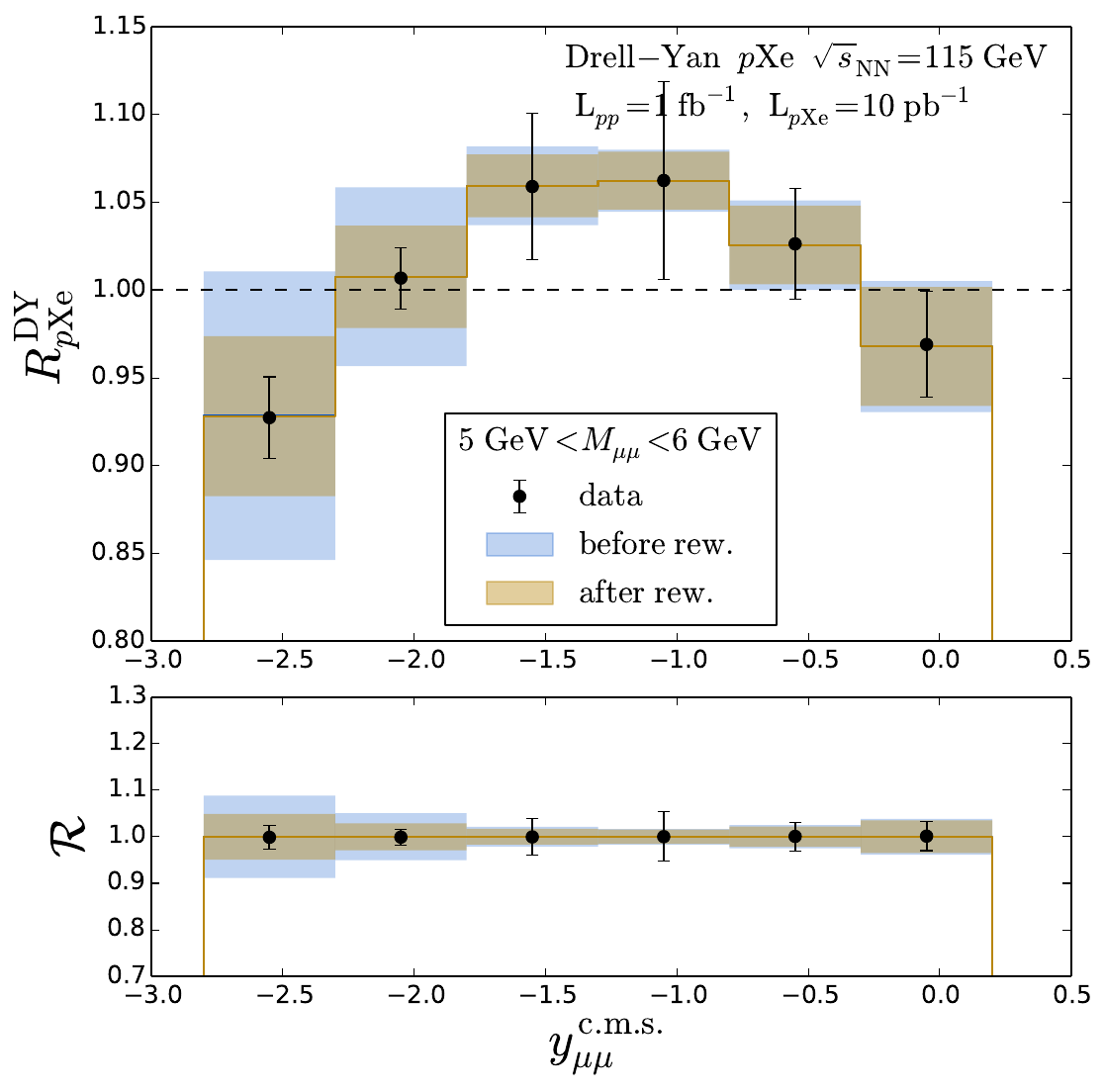}}
\subfigure[$10.5\, \gev \,<M_{\mu\mu}$ GeV]{\includegraphics[width=0.32\textwidth]{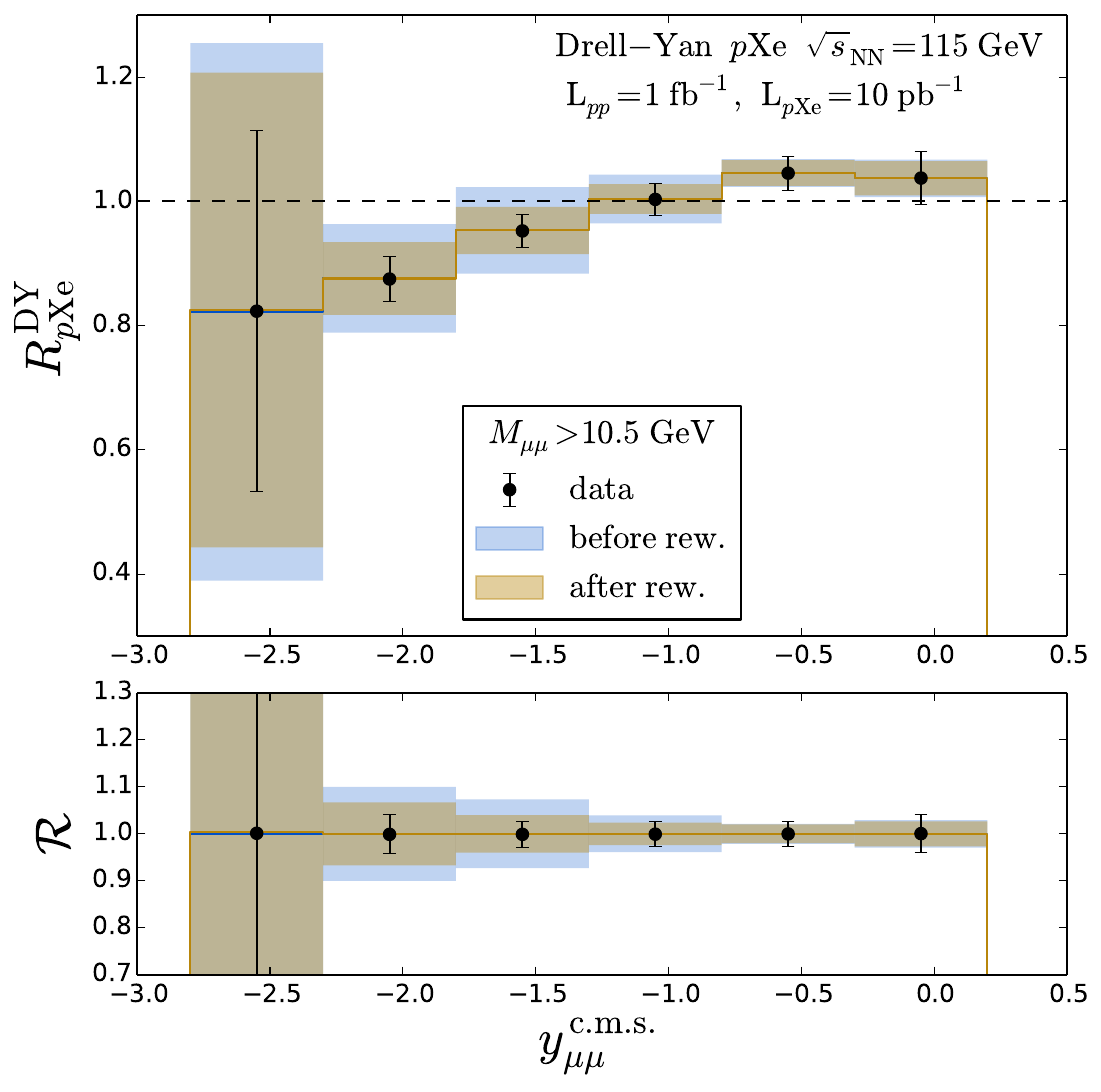}}
\\
\subfigure[~]{\includegraphics[width=0.24\textwidth]{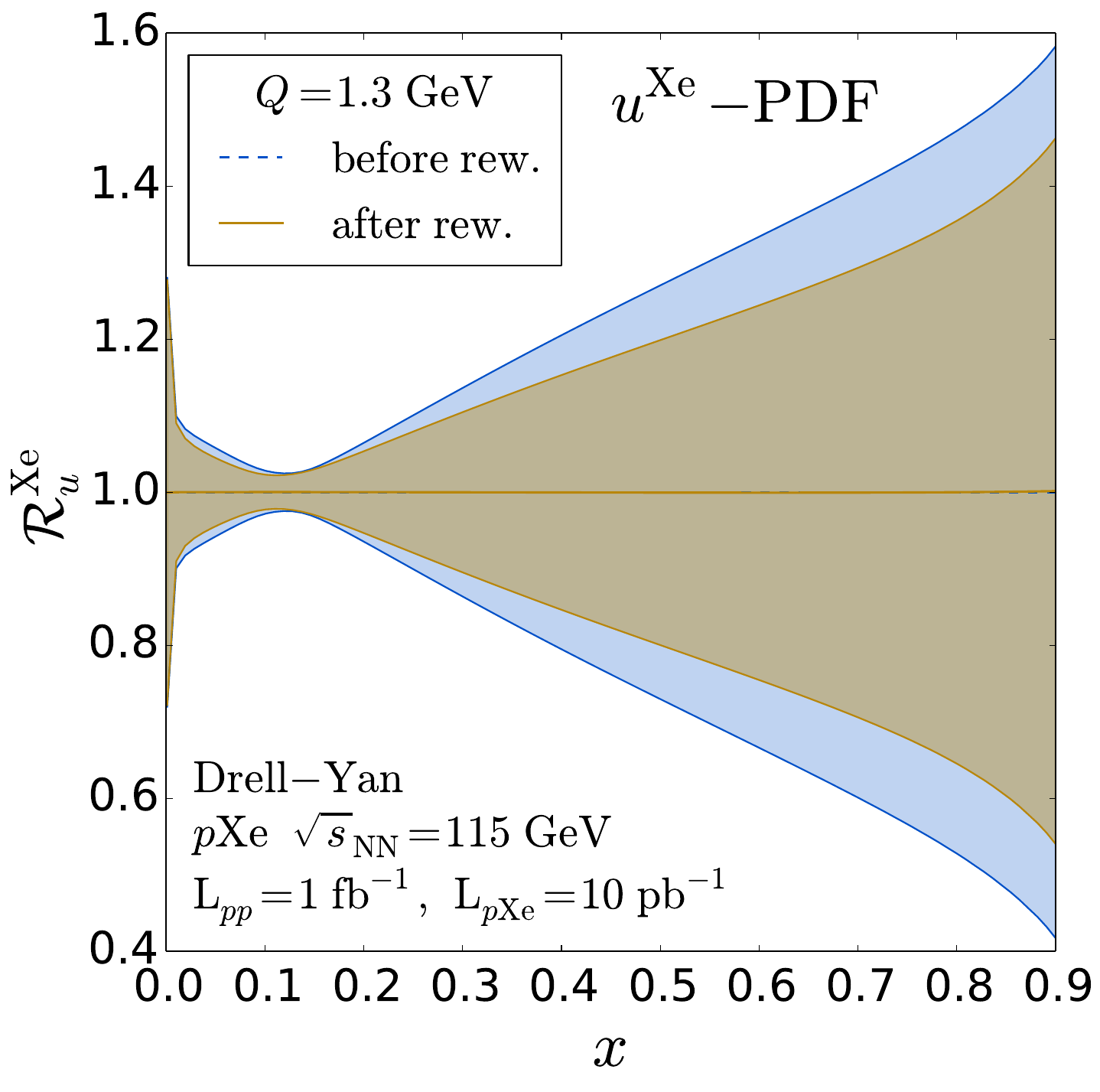}}
\subfigure[~]{\includegraphics[width=0.24\textwidth]{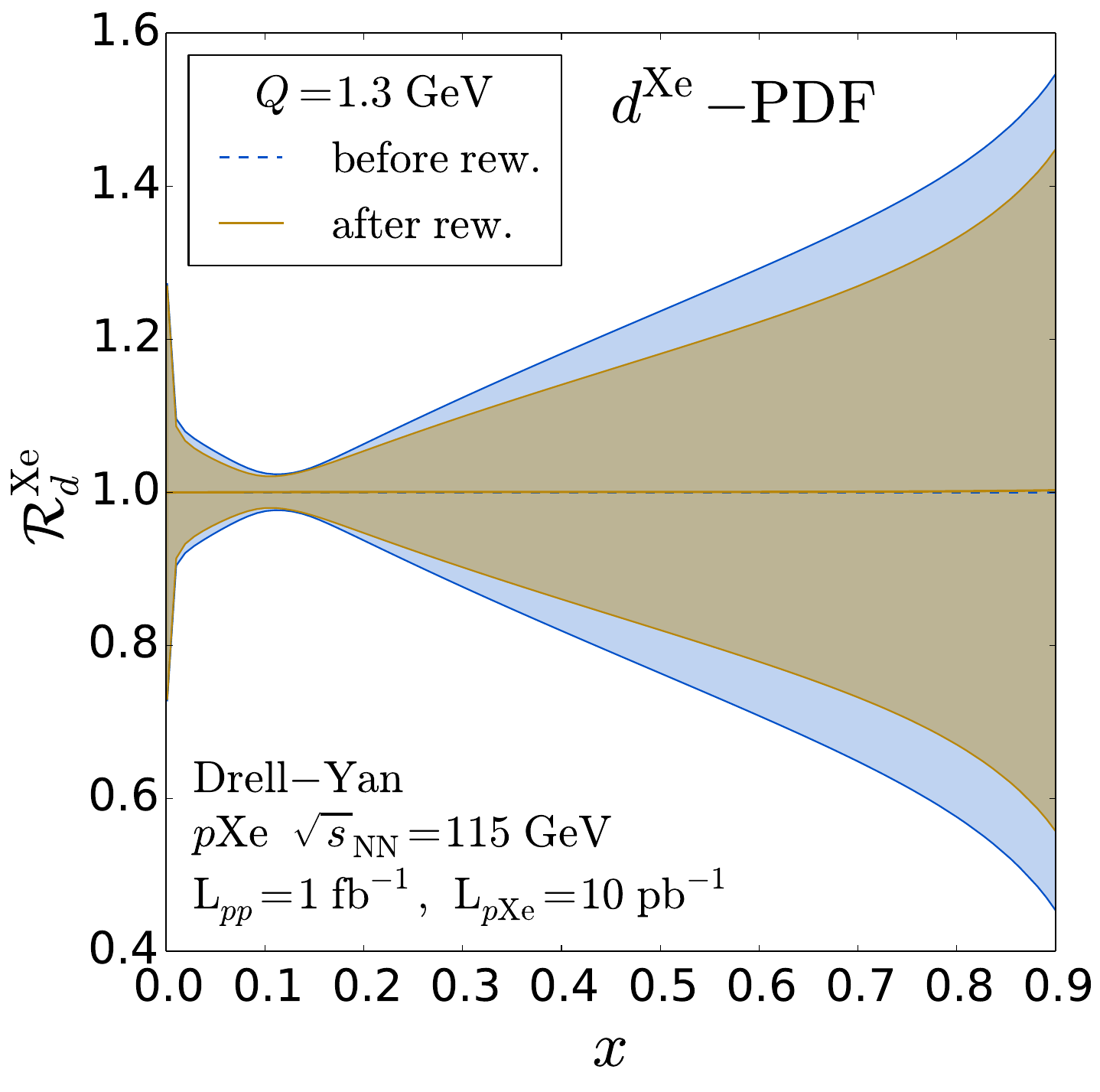}}
\subfigure[~]{\includegraphics[width=0.24\textwidth]{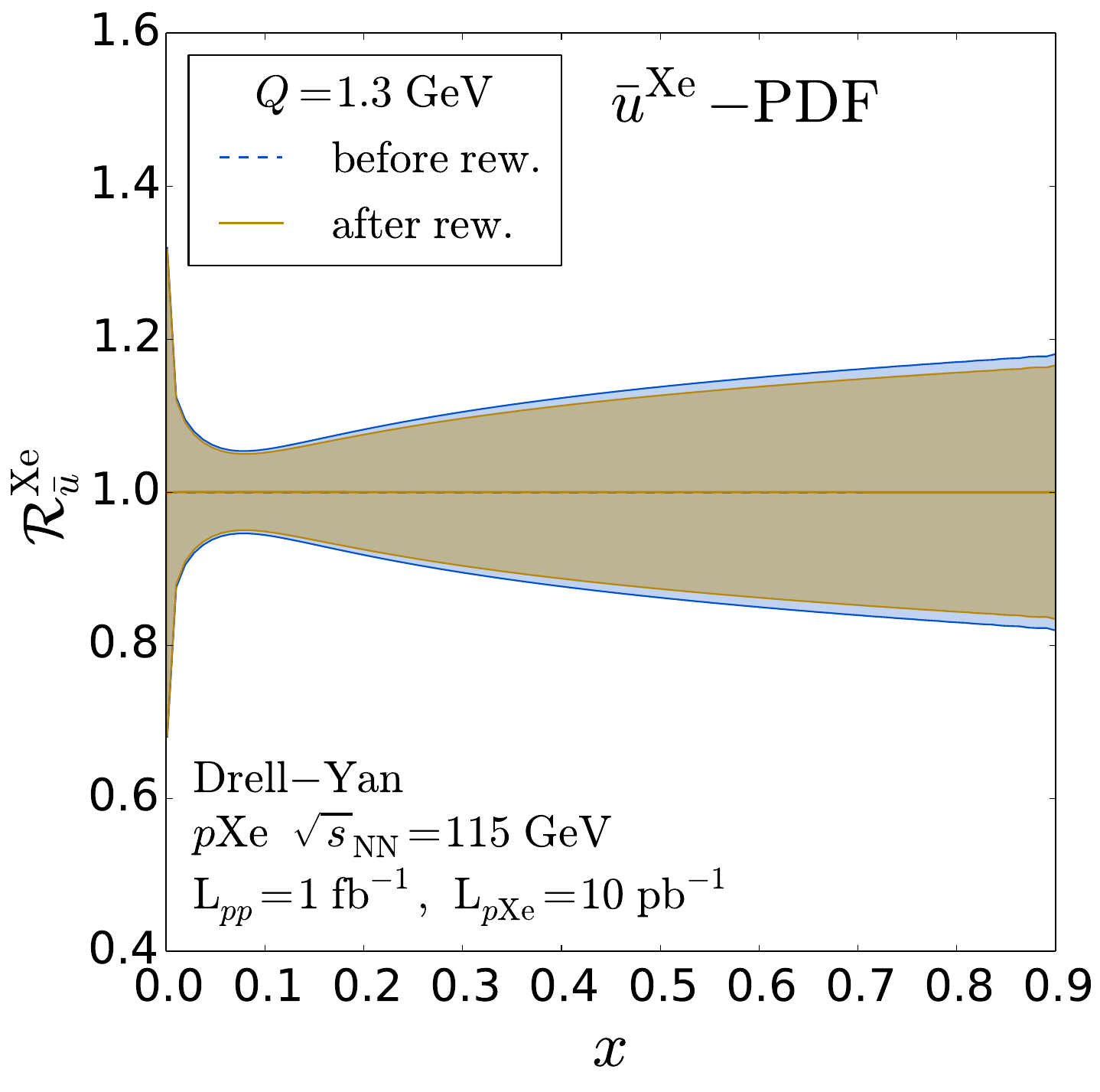}}
\subfigure[~]{\includegraphics[width=0.24\textwidth]{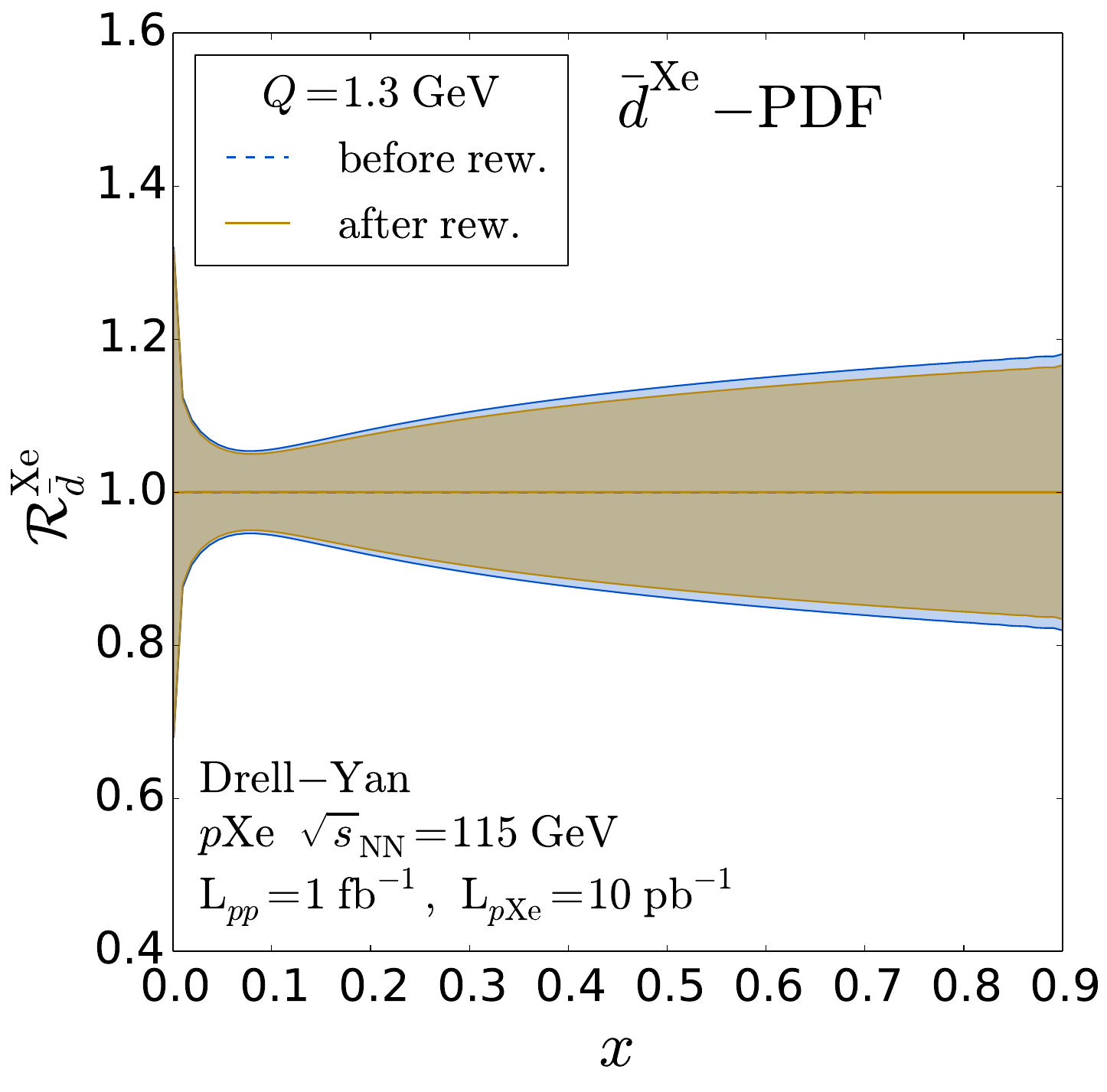}}
\caption{Same as \cf{fig:RpA_vs_y_DY} (upper row) and \cf{fig:npdf_rew}
(lower row) with luminosities reduced by a factor 10:
$\mathcal{L}_{pp}=1$ fb$^{-1}$ and $\mathcal{L}_{\pXe}=10$ pb$^{-1}$.}
\label{fig:RpA_vs_y_DY_Lred10}
\end{figure}

A modern precision measurement of the DY lepton-pair production at \AFTER\ would allow one to study the
EMC effect in this process (at high negative $x_F$ and with much higher precision) and to compare it 
to the DIS case.
In addition, as in the nucleon case, nuclear PDFs (nPDFs) are determined in global analyses of DIS and
DY data \cite{Eskola:2016oht,Kovarik:2015cma,deFlorian:2011fp,Eskola:2009uj,Schienbein:2009kk,Hirai:2007sx}
and are a crucial ingredient to predict hard processes in $pA$ and $AA$ collisions at the
LHC.
Compared to the proton PDFs, the nPDF determinations are clearly lagging behind both at
the level of sophistication but most importantly due to the much smaller number of experimental constraints.
Currently, the analyses are statistically dominated by DIS data with only about 90 data points
from the DY process entering the fits.
Incorporating data from various processes is essential for flavour separation in PDF analyses.
Therefore, access to the DY data with a wide kinematic coverage will provide a unique opportunity not
only for more precise PDF determinations but will also allow one to test their universality which
is a fundamental property of QCD and basis for all high energy hadron scattering computations.

\begin{figure}[!hbt]
\centering
\subfigure[~]{\includegraphics[width=0.42\textwidth]{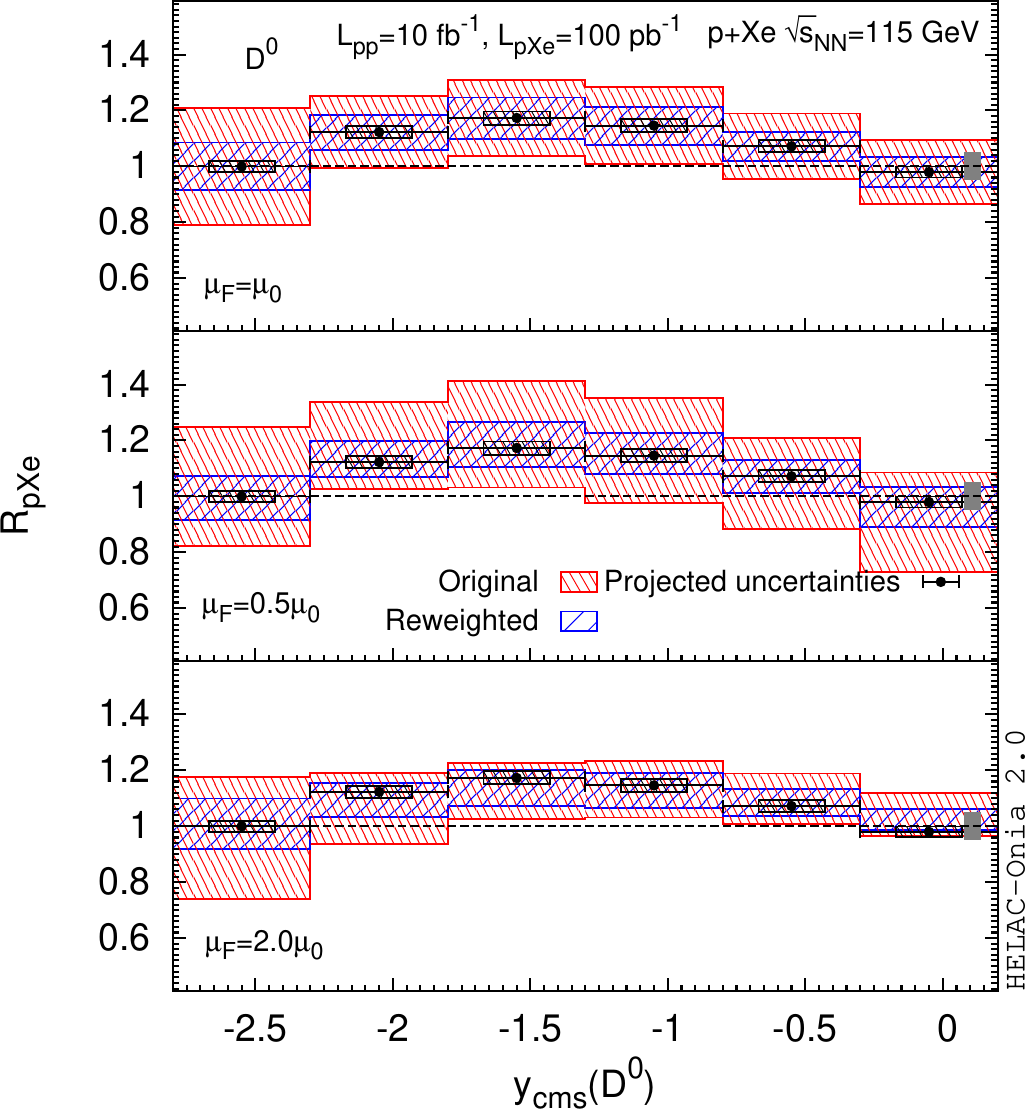}}
\subfigure[~]{\includegraphics[width=0.42\textwidth]{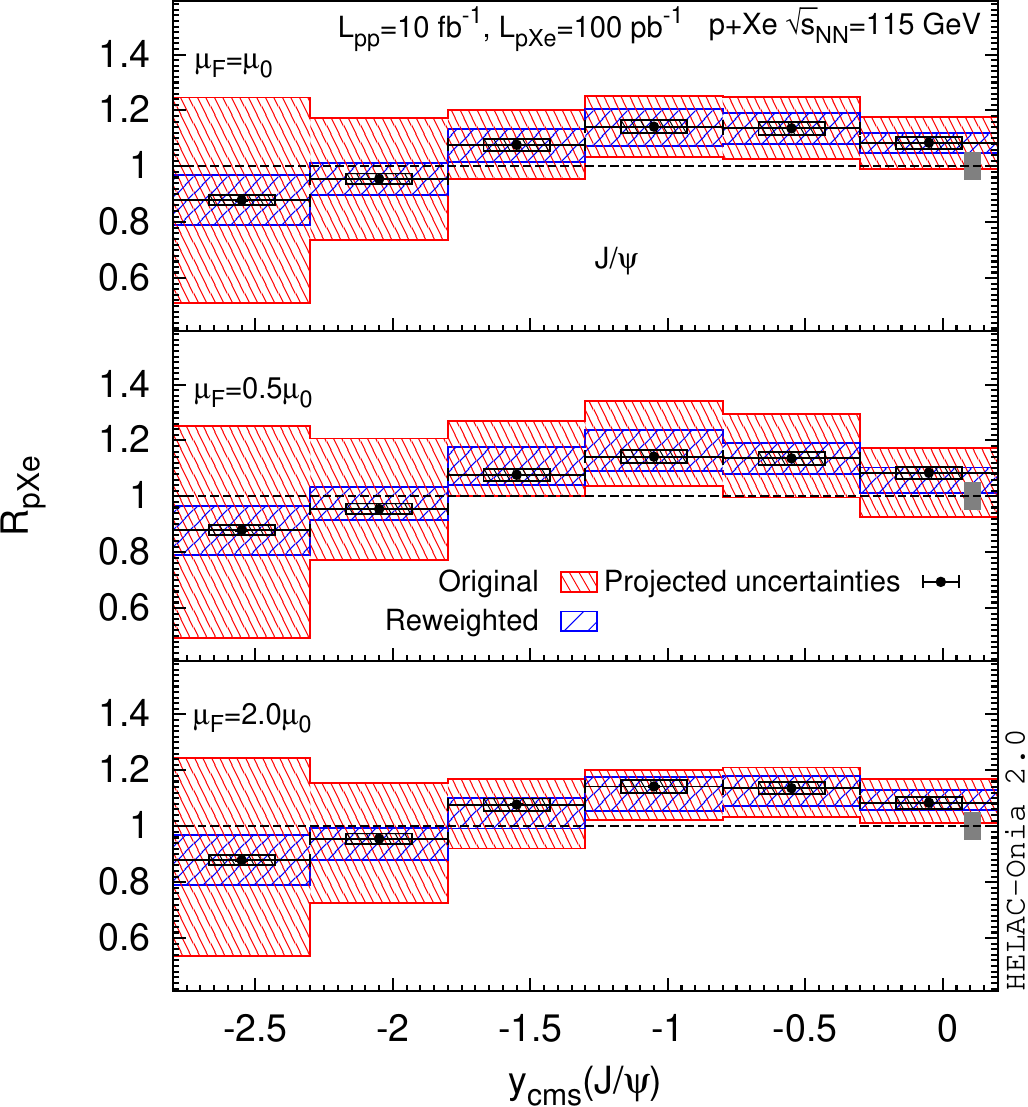}}
\subfigure[~]{\includegraphics[width=0.42\textwidth]{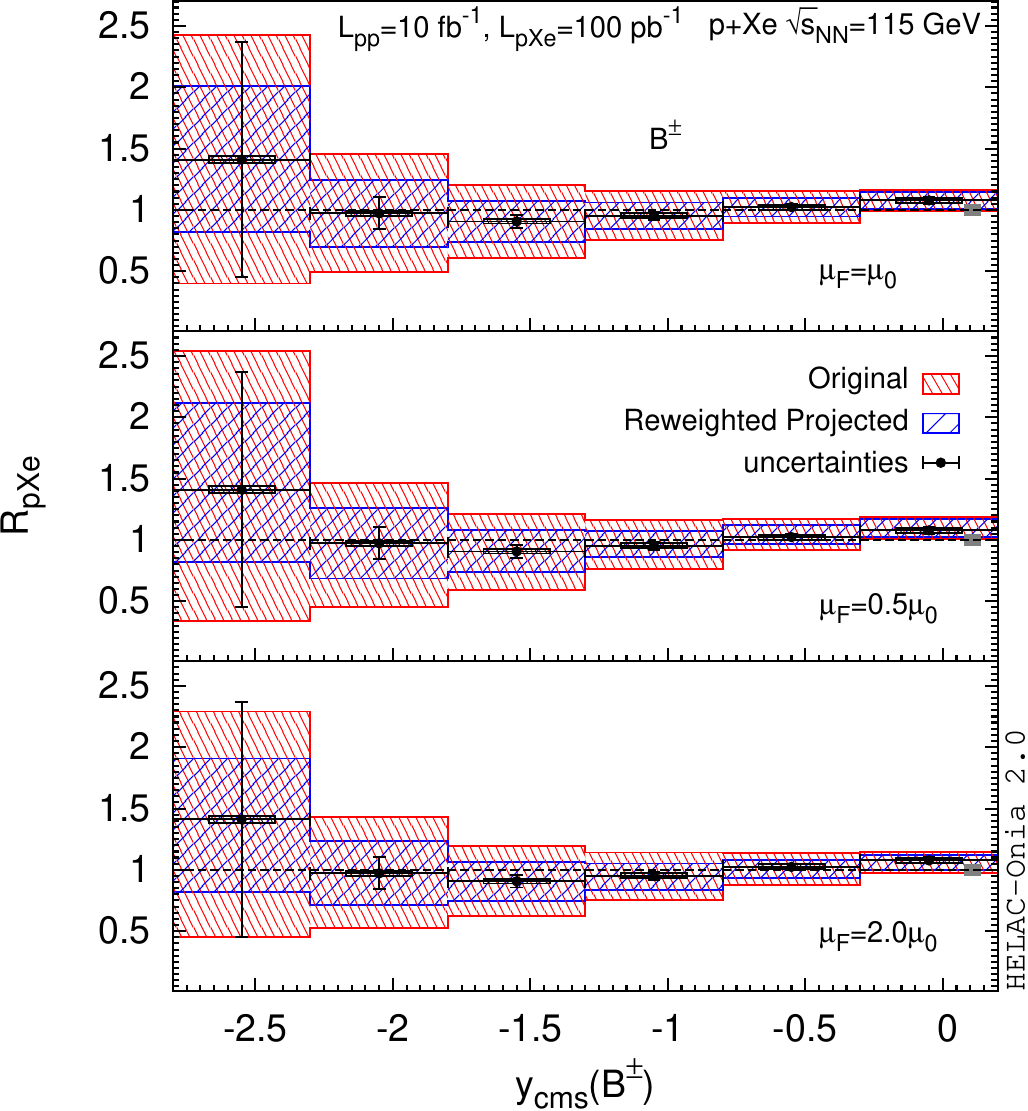}}
\subfigure[~]{\includegraphics[width=0.42\textwidth]{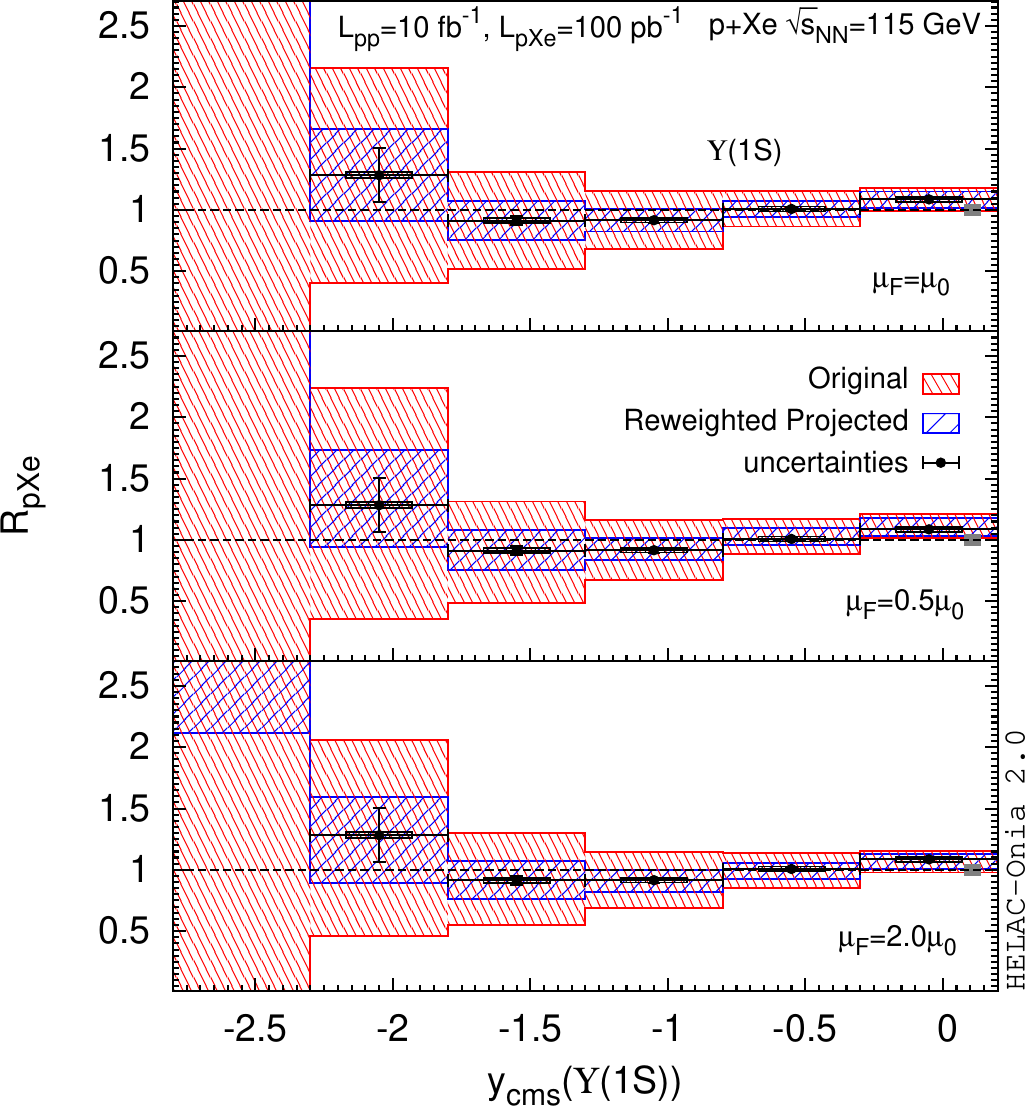}}
\caption{Projection of the statistical uncertainties on the \RpA\ for (a) $D^0$, (b) $J/\psi$, (c) $B^+$, and (d) $\Upsilon(1S)$ production in \pXe\ collisions compared to the uncertainties encoded in nCTEQ15 nPDFs, which are representative of typical nPDF uncertainties, evaluated at different typical choices of the factorisation scale, $\mu_F$, varied about $\mu_0$ like in~\cite{Kusina:2017gkz}. A 2\% uncorrelated systematic uncertainty and a 5\% global uncertainty are also shown.
The projected statistical uncertainties are estimated assuming the yearly integrated luminosities of
$\mathcal{L}_{\rm pp}=10\ {\rm fb}^{-1}$ and $\mathcal{L}_{\rm pXe}=100\ {\rm pb}^{-1}$.
}
\label{fig:RpA_vs_y_HF}
\end{figure}

The kinematic reach of \AFTER\ (\cf{fig:large_x_DY_pA}) would allow one to probe much higher
$x_2$ (target $x$) values than the currently available data (data points in \cf{fig:large_x_DY_pA})
for a variety of targets.
In particular, \AFTER\ could shed new light on the origin of the EMC effect
by verifying its presence/absence in DY lepton-pair production.
As Fig.\ \ref{fig:large_x_DY_pA} shows, 
a modern precision measurement of DY lepton-pair production at \AFTER\ covering a wide
range in invariant masses of the lepton pairs and extending to higher $x_F$ would lead to
significant improvements over the current state of the art and
would be complementary to results from a future Electron-Ion-Collider (EIC).
Clearly, it would be invaluable input for nuclear PDF determinations.

\begin{figure}[!hbt]
\centering
\subfigure[~]{\includegraphics[width=0.4\textwidth]{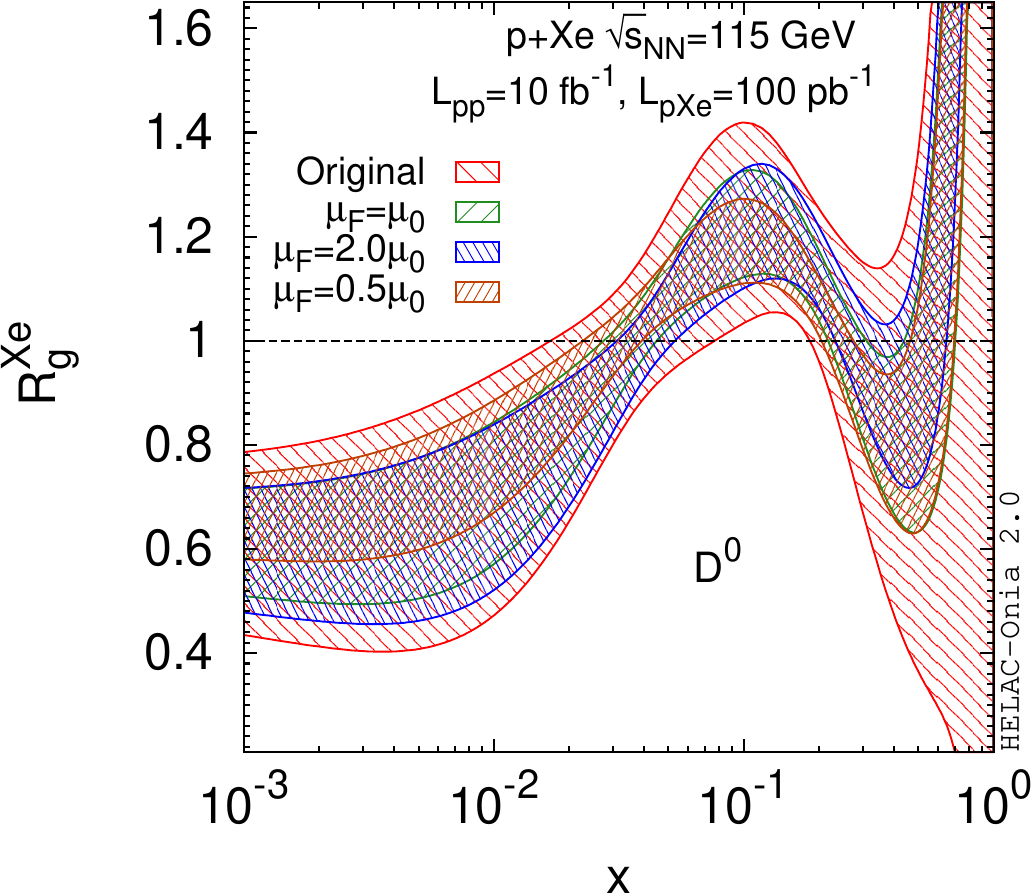}}
\subfigure[~]{\includegraphics[width=0.4\textwidth]{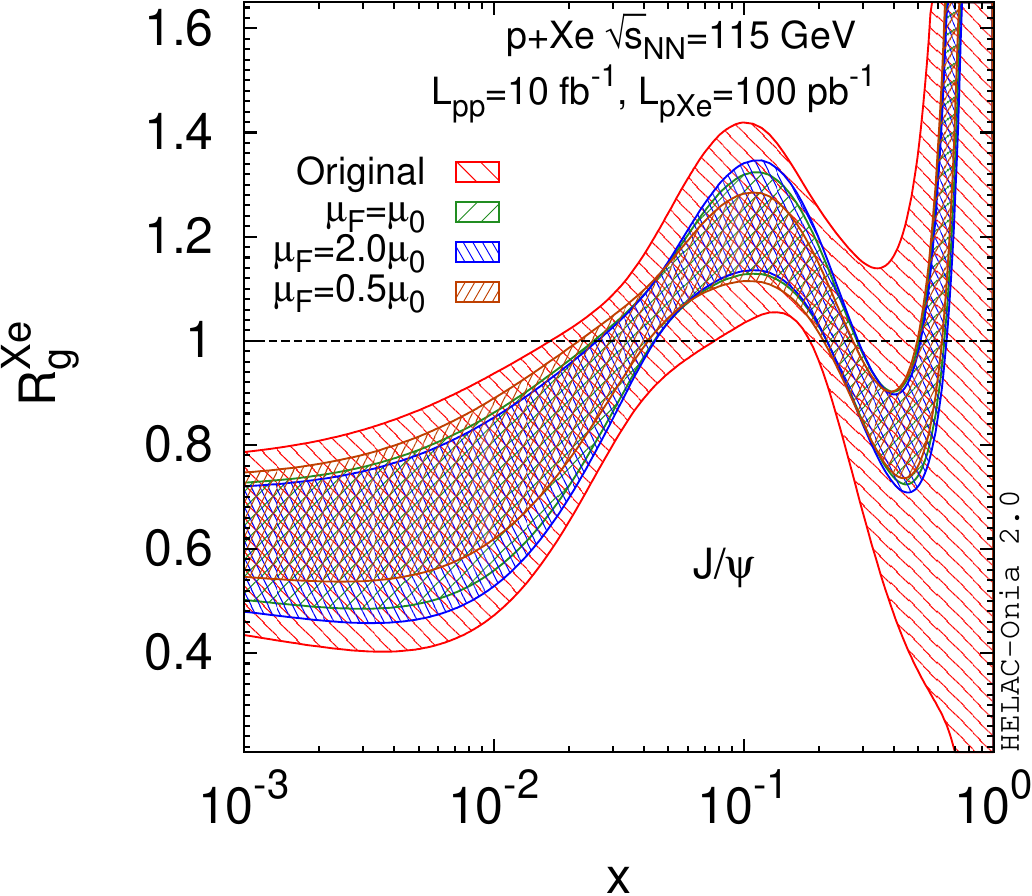}}\\
\subfigure[~]{\includegraphics[width=0.4\textwidth]{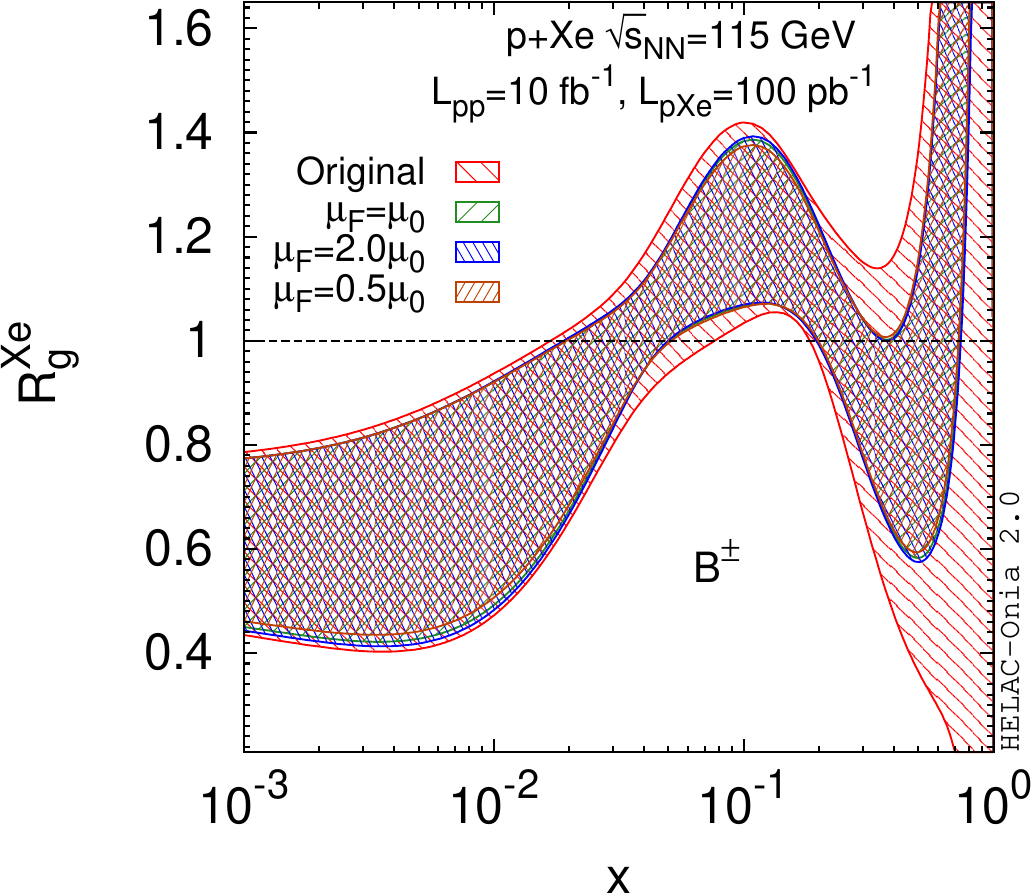}}
\subfigure[~]{\includegraphics[width=0.4\textwidth]{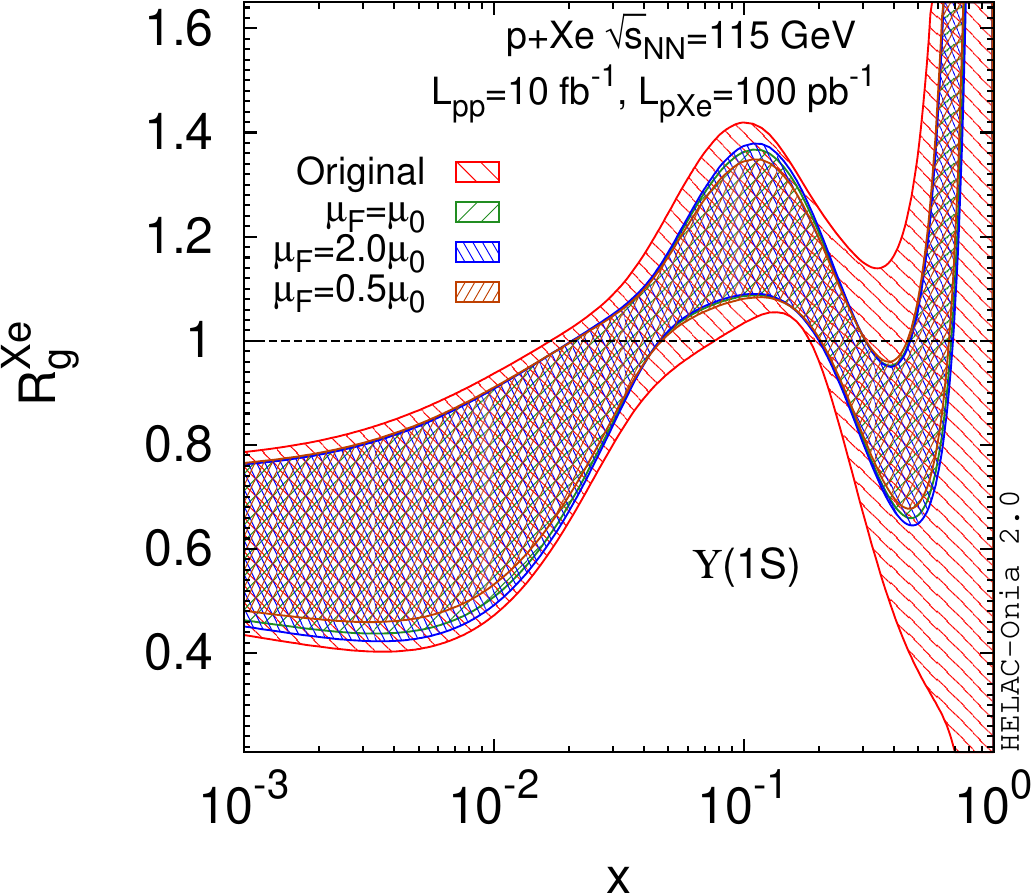}}
\caption{nCTEQ15 nPDFs before and after the reweighting using $R_{pXe}$ pseudo-data shown in \cf{fig:RpA_vs_y_HF} for (a) $D^0$,
(b) $J/\psi$, (c) $B^+$, (d) $\Upsilon(1S)$ production at \AFTERLHCb. The plots show ratios $R_g^{Xe}$
of gluon densities encoded in nCTEQ15 over that in CT14 PDFs at scale $Q=2$ GeV.}
\label{fig:gluon_from_RpA_vs_y_HF}
\end{figure}

As an example we present here a reweighting
analysis~\cite{Giele:1998gw,Sato:2013ika,Ball:2010gb,Kusina:2016fxy}
showing the potential impact of the DY lepton-pair production data from \AFTER\
in $p$Xe collisions on the nCTEQ15 nPDFs. In this analysis we use pseudo-data
for the nuclear modification factors
$R^{\rm  DY}_{p\text{Xe}} = \sigma_{\text{DY}}^{\text{Xe}}/(A\times \sigma_{\text{\rm DY}}^{pp})$
as a function of the rapidity in five bins of lepton-pair invariant mass: 
$4\, \gev \, <M_{\mu\mu}<5\, \gev \, $, $5\, \gev \, <M_{\mu\mu}<6\, \gev \, $, $6\, \gev \, <M_{\mu\mu}<7\, \gev \, $, $7\, \gev \, <M_{\mu\mu}<8\, \gev \, $ and $M_{\mu\mu}>10.5$ GeV, in order to estimate the effect these data can have on the current nPDFs. A selection of the pseudo-data together with the corresponding theory predictions is shown in \cf{fig:RpA_vs_y_DY}, where we assumed the follwoing luminosities: $\mathcal{L}_{\rm pp}=10\ {\rm fb}^{-1}$ and $\mathcal{L}_{\rm pXe}=100\ {\rm pb}^{-1}$ for the psedo-data.

\begin{figure}[!hbt]
\centering
\subfigure[~]{\includegraphics[width=0.4\textwidth]{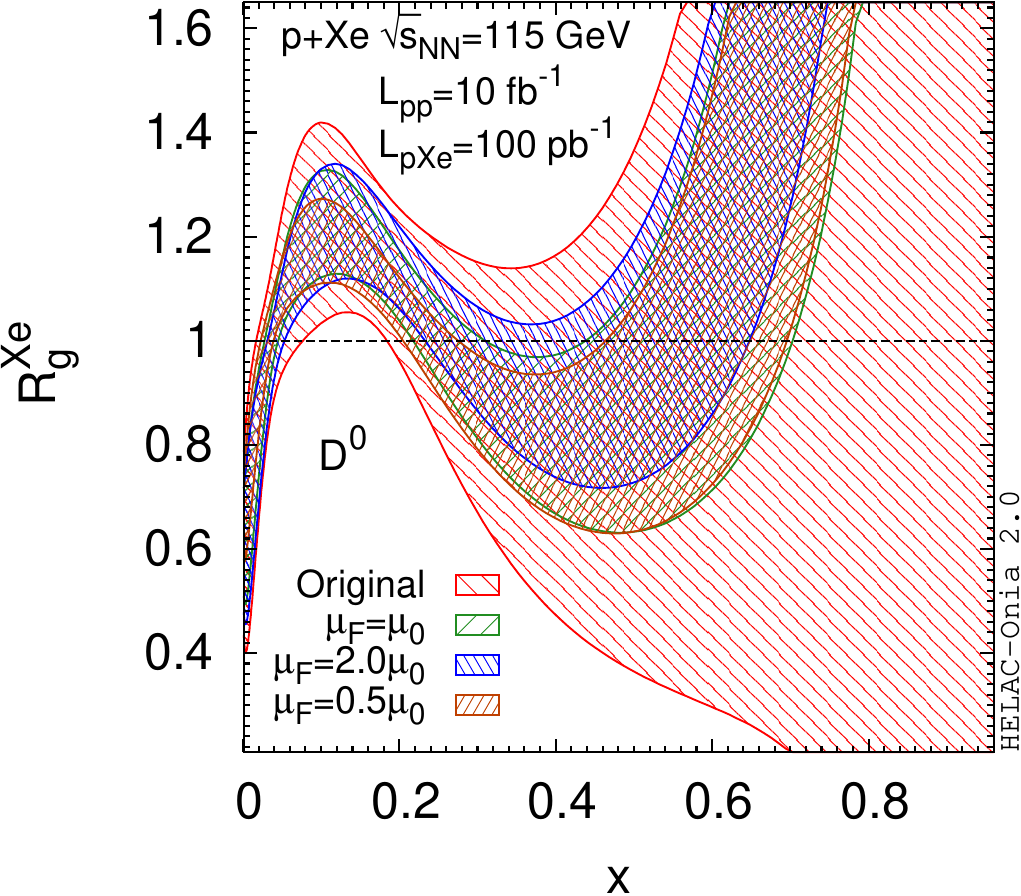}}
\subfigure[~]{\includegraphics[width=0.4\textwidth]{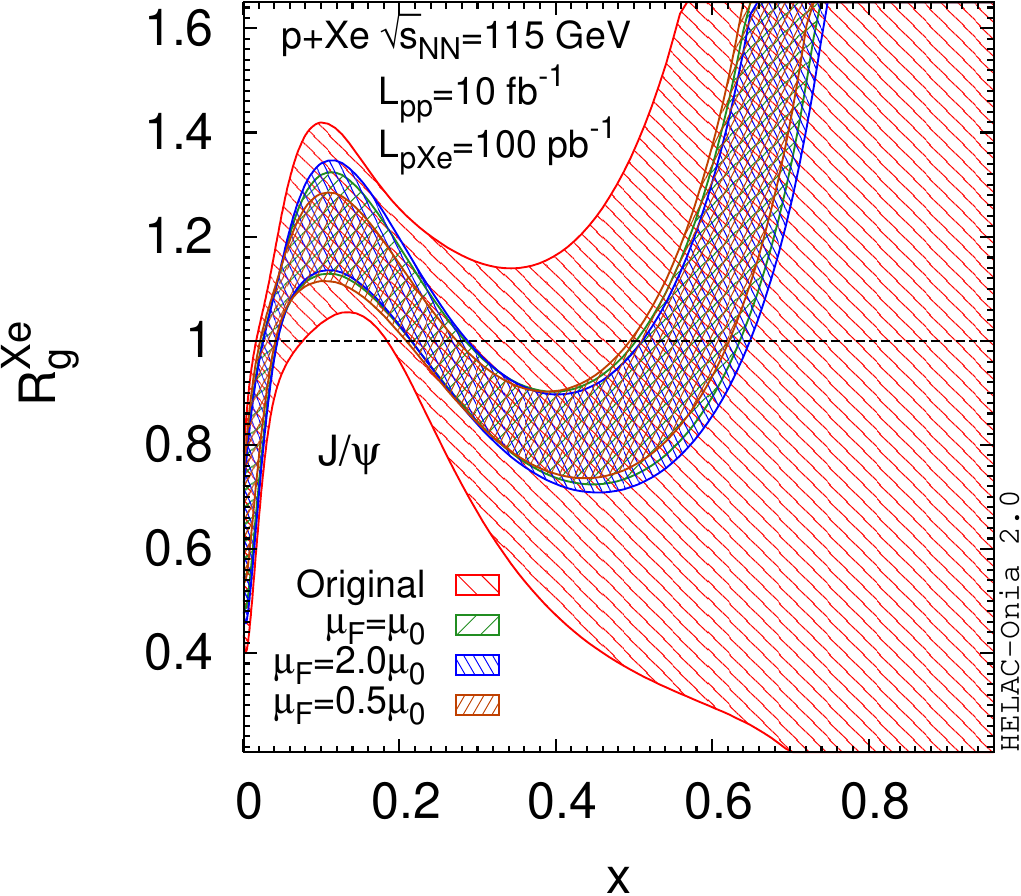}}\\
\subfigure[~]{\includegraphics[width=0.4\textwidth]{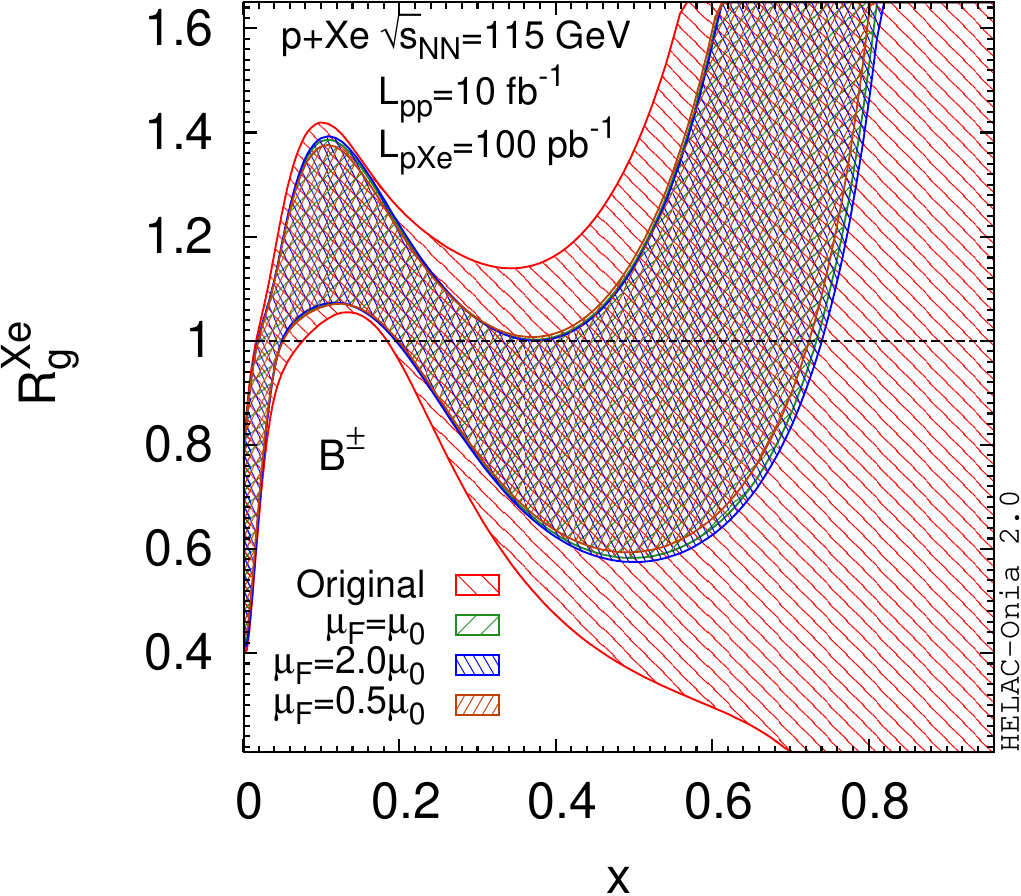}}
\subfigure[~]{\includegraphics[width=0.4\textwidth]{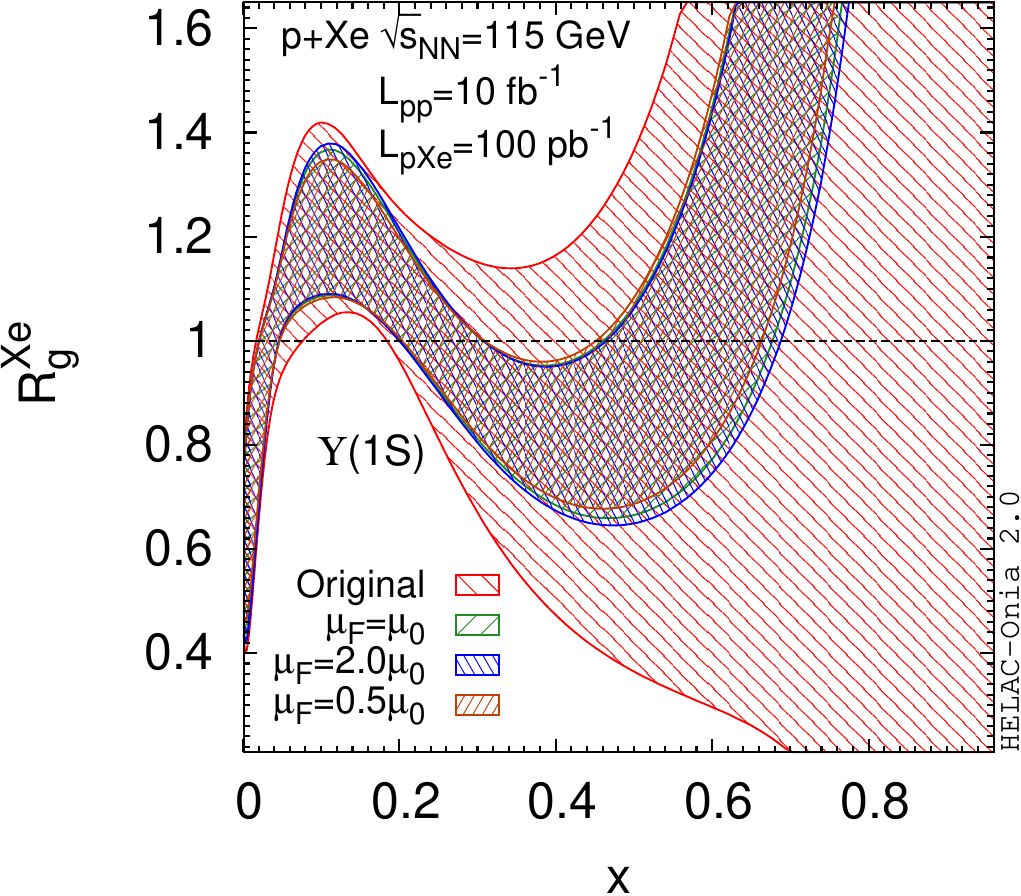}}
\caption{Same as \cf{fig:gluon_from_RpA_vs_y_HF} using a linear $x$ axis in order to highlight
the high-$x$ region.}
\label{fig:gluon_from_RpA_vs_y_HF_lin}
\end{figure}

In \cf{fig:npdf_rew} we display the nPDFs before and after the reweighting
using the \AFTER\ $R_{p\text{Xe}}$ pseudo-data.
We can see a significant decrease of the errors for up and down quark distributions
showing the potential of the \AFTER\ to constrain nPDFs.
In practice, due to the limited amount of data, the current nPDF errors are considerably
underestimated and the actual importance of these data can not be fully demonstrated
in this kind of study. However, \cf{fig:large_x_DY_pA} clearly shows how complementary 
the kinematical coverage of \AFTER\ will be compared to the current DY data for the nPDF determination.
Similarly to the proton case, the $W^{\pm}$ data could be used for a determination of the
high-$x$ nPDF in particular the light quark sea distributions. 

Additionally, we have also investigated what would be the impact of such DY
measurements on nPDFs in case less data would be collected. For this purpose
we assumed 10 times reduced luminosities for both $pp$ and $p$Xe samples.
The results for such scenario are presented in Fig.~\ref{fig:RpA_vs_y_DY_Lred10}
which shows that even in this case one can obtains a substantial reduction of
uncertainties for up and down distributions.

\AFTER\ is also able to constrain the high-$x$ nuclear gluon distribution, which is
the least known nPDF. A prime example we show here is to use heavy flavour production
at \AFTER, where the gluon shadowing effect on $J/\psi$ and $\Upsilon$ production
in \pPb\ collisions at \AFTER\ energies has been studied in Ref.~\cite{Vogt:2015dva}.
We now discuss the potential of both open and hidden heavy flavour mesons
($D^0,J/\psi,B^{+},\Upsilon(1S)$) production in \pXe\ collision at $\sqrt{s_{NN}}=115$ GeV
to pin down the high-$x$ gluon density in nPDF by performing a Bayesian-reweighting analysis.
A similar study in the LHC energies has been carried out in Ref~\cite{Kusina:2017gkz}.
We used the data-driven approach proposed in Ref.~\cite{Lansberg:2016deg} to fit matrix
elements of the heavy flavour hadrons, and then folded them with proton CT14 PDFs and nCTEQ15~\cite{Kovarik:2015cma} nPDFs to get the yields and the nuclear modification factors\footnote{The calculations are carried out using the framework of HELAC-Onia~\cite{Shao:2012iz,Shao:2015vga}.}.

\begin{figure}[!htb]
    \begin{center}
      \includegraphics[width=7.1cm]{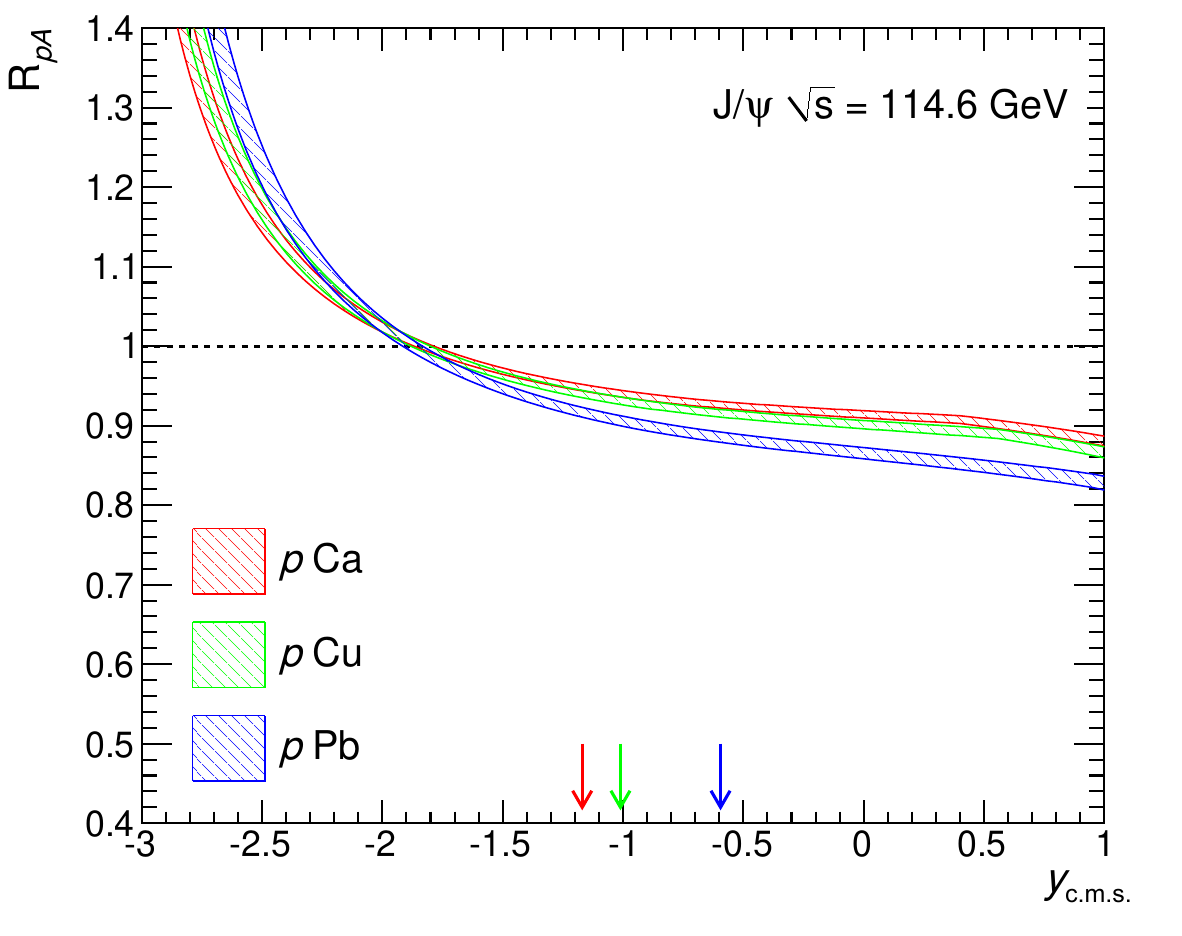}
      \includegraphics[width=7.1cm]{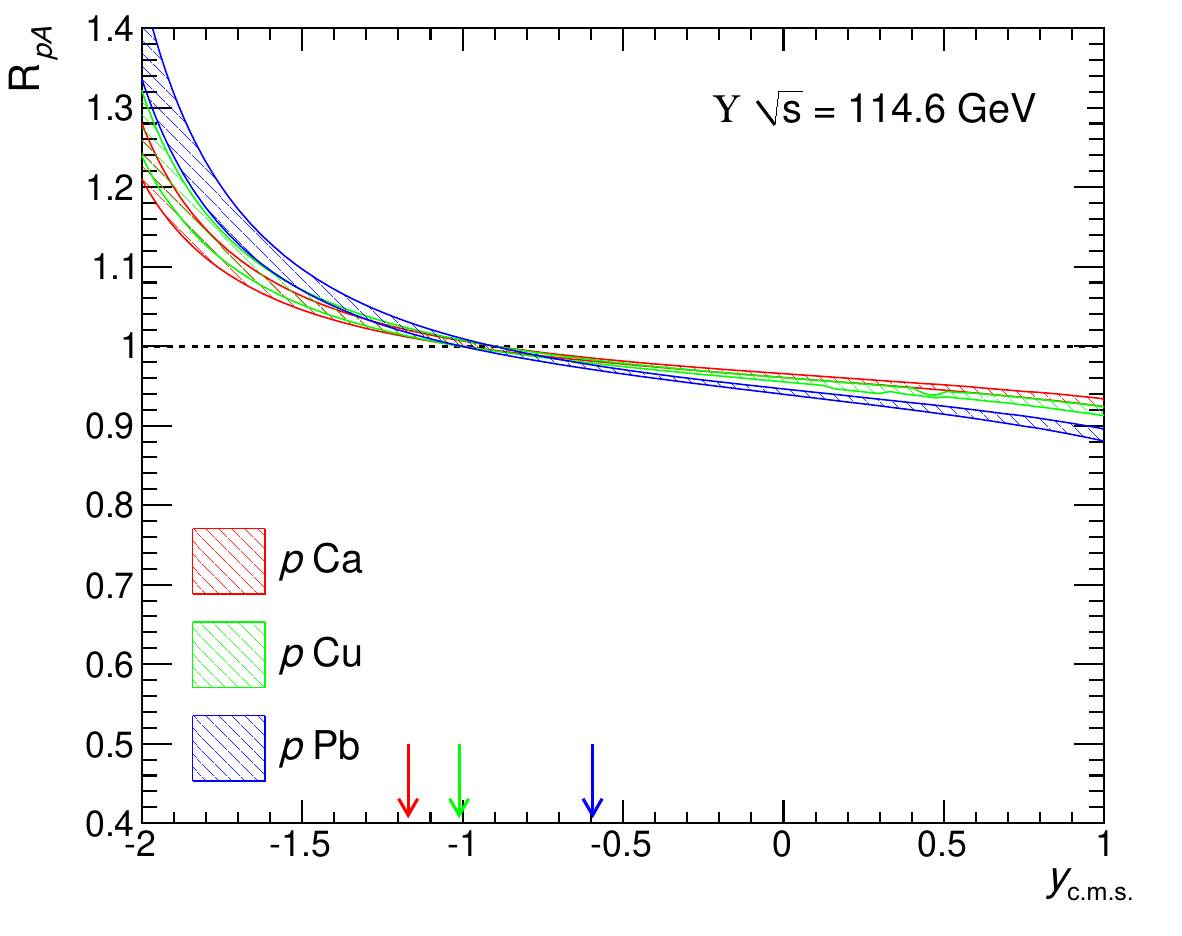}
    \end{center}
    \caption{$\jpsi$ (left) and $\Upsilon$ (right) nuclear modification factor in \pPb, \pCa, and \pCu collisions at $\sqrts=114.6$~GeV resulting from the coherent energy loss. [Adapted from~\cite{Arleo:2015lja}.]}
    \label{fig:RpPb_energyLoss}
\end{figure}

The pesudo-data are generated to match the central theoretical predictions. Their projected
statistical uncertainties are estimated by assuming the yearly integrated luminosities
$\mathcal{L}_{pp}=10\ {\rm fb}^{-1}$, $\mathcal{L}_{\pXe}=100\ {\rm pb}^{-1}$
and the reconstruction efficiency $\varepsilon=0.1$. The branching ratios of $D^0\rightarrow K\pi$,
$B^+\rightarrow K (J/\psi \rightarrow \mu^+\mu^-)$, $J/\psi\rightarrow \mu^+\mu^-$
and $\Upsilon(1S)\rightarrow \mu^+\mu^-$ have also been taken into account.
For $D^0$ and $B^+$, their charge-conjugated particles are also summed up. After considering
a $2\%$ systematic error and a $5\%$ global error, the nuclear modification factors
\RpXe\ for the four hadron productions are shown in \cf{fig:RpA_vs_y_HF} before
(red bands) and after (blue bands) reweighting, together with the variations of factorisation
scales. The uncertainty from the factorisation scale can be the dominant theoretical error
after nPDF reweighting as already pointed out in Ref.~\cite{Kusina:2017gkz}. The impact
of these pesudo-data can be transfered into nPDFs as illustrated in \cf{fig:gluon_from_RpA_vs_y_HF}
and~\ref{fig:gluon_from_RpA_vs_y_HF_lin}.
We can see that even accounting for the scale uncertainty, a substaintial reduction
of nPDF uncertainty at high-$x$ values ($x\gtrsim 0.3$) is still achieved. It clearly demonstrates
the uniqueness of \AFTER\ programme in exploring gluon densities in nuclei especially at high-$x$.

It is important to note here that the above projections for the constraints on the 
gluon nPDF were obtained assuming only the modification of nPDFs and the absence of other cold nuclear
matter effects, or that such other  effects can be subtracted. At LHC collider energies, 
this kind of leading-twist-factorisation approach was applied with success to a large
class of existing data~\cite{Kusina:2017gkz}. At lower energies, especially in the backward region, 
quarkonium break up will likely play a role and should be separated out. For that matter, the extensive access
of \AFTER\ to quarkonium excited-state studies will be crucial.
Another example of an effect that can matter when gluons are involved  is the coherent energy loss. It was recently studied 
in the context of \AFTER~\cite{Arleo:2015lja} and results in a modification of the $p$A 
cross-sections compared to the $pp$ one which is depicted in \cf{fig:RpPb_energyLoss}
for $\jpsi$ and $\Upsilon$ in terms of $R_{p\mathrm{A}}$ .
Combining all the heavy-flavour related measurement possible at \AFTER will certainly allow one to disentangle all these effects
in order to perform a reliable fit of the gluon nPDFs using these data.

%% file: physics-high-x/physics-high-x_astro.tex
\subsubsection{Astroparticle physics}

Recently, measurements of cosmic rays (CRs) with very high energies, ranging from about tens of MeV up to hundreds of TeV, became possible for many particle species ($e^\pm$ \cite{Aguilar:2014fea,Ambrosi:2017wek}, $\gamma$ \cite{Atwood:2009ez,TheFermi-LAT:2015kwa}, $\nu$ \cite{Aartsen:2016ngq,Aartsen:2017nbu}, $p$ \cite{Aguilar:2015ooa}, $\bar p$ \cite{Aguilar:2016kjl}, $A$ \cite{Aguilar:2015ctt,Aguilar:2017hno,Aguilar:2018njt}) and attracted much attention.
The mechanism responsible for the generation of such Ultra High-Energy CRs (UHECRs) is still under 
intense discussion, with two main scenarios: (i) the acceleration of particles 
due to astrophysical phenomena and (ii) dark matter decay/annihilation.
The mechanism generating CRs can only be determined if we can identify characteristic shapes of the spectrum such as sharp cutoffs which will indicate the decay of massive dark matter particles.
In such precision tests of CRs, the spectrum has to be accurately determined, thus naturally requiring precise investigations of other sources acting as backgrounds. 
Here we present two cases where the \AFTER programme can play a critical role.

\paragraph{UHECR neutrinos and the proton charm content}

The terrestrial observation of UHE neutrinos lately became 
possible thanks to IceCube, with the highest energy recorded on the order of 
PeV~\cite{Aartsen:2016ngq,Aartsen:2017nbu}. {\it Atmospheric}
neutrinos, generated by the weak decays of final-state particles of the 
collisions between CRs and atmospheric nuclei, are however an important background
to these ground observations of cosmic neutrinos.
The major source of these atmospheric neutrinos is from the weak decay of hadrons. Those originating from 
the decay of long-lived mesons, like $\pi$ and $K$ mesons, dominate 
the energy spectrum below $10^5$ GeV. Those with energy above $10^5$ GeV are 
from charm-hadron decays. Indeed charmed hadrons loose significantly less 
energies in the atmosphere than $\pi$ and $K$ because they decay almost 
instantaneously.

The yield of neutrinos from charm naturally follows that of charmed hadrons produced in
 the collisions of UHECRs and atmospheric nuclei. An accurate
 evaluation of the charm hadroproduction cross section is therefore crucial
to assess their importance. The charm production from $pp$ scattering was first evaluated in perturbative
 approaches, which considered charm quark-antiquark pairs virtually created by
 the gluon splitting $g\to c \bar c$. 
This contribution is mostly relevant at low-$x$ in the target PDF~\cite{Enberg:2008te,Bhattacharya:2015jpa} which
can also be shadowed.

The relevance of a nonperturbative charm content (or IC) 
in the projectile proton was also recently considered~\cite{Laha:2016dri}.
Just as most global analyses of PDFs rely on the assumption that the charm and bottom PDFs are
generated perturbatively by gluon splitting, most of the 
studies of neutrino fluxes were based on the same assumption. 
The intrinsic charm however has a harder distribution~\cite{Brodsky:1984nx,Brodsky:1980pb} (for a recent review, see \cite{Brodsky:2015fna}) and tends 
to predict charmed hadrons at higher energies, which would result into higher 
energy neutrinos.

As we discussed in section~\ref{subsec:nucleon}, the \AFTER programme offers a novel playground to study
the excess of charm at high-$x$.
Not only a detector such as that of LHCb used in the fixed-target mode can access
the high negative $x_F$, it also has a wide coverage close to $40\%$~\cite{private:Riehn} of the weighted charm cross section, $x_F^2 d\sigma/dx_F$
which is quasi proportional to the neutrino flux. As a comparison, in the collider mode, 
it covers less than 10\% of the same weighted cross section~\cite{Ulrich:2015uwb}. 
In other words, the charmed hadron measurements of the AFTER@LHC programme can provide the most
decisive constraints on the non-pertubative charm content in the proton in the region relevant to understand
the neutrino-from-charm flux. \cf{fig:charm_neutrinos} shows the uncertainties on the neutrino flux due 
to that on the IC in the projectile proton~\cite{Giannini:2018utr} (see also~\cite{Laha:2016dri}).

\begin{figure}[!hbt]
\centering
{\includegraphics[width=0.6\textwidth]{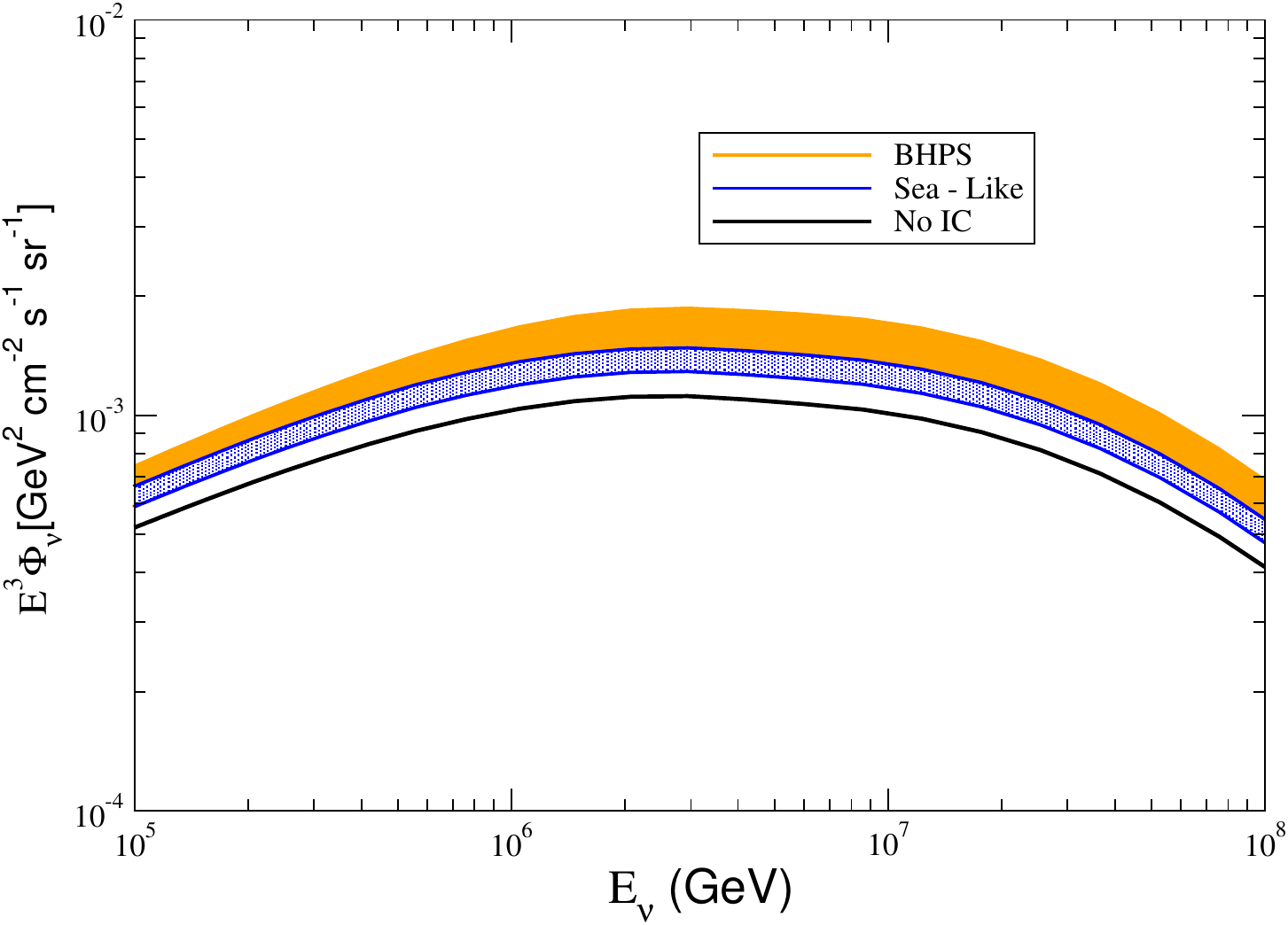}}
\caption{
Impact of the uncertainties on the charm content of the proton on the neutrino flux [Courtesy of V. Gon\c calves, based on \protect\cite{Giannini:2018utr}]. The uncertainties on the $D$ yield correspond to \protect	\cf{fig:large_x_2} with the use of the BHPS and sea-like IC.
}
\label{fig:charm_neutrinos}
\end{figure}

\paragraph{Antiproton cross section and UHECR Monte-Carlo tuning}

Among the cosmic rays, the antiprotons ($\bar p$) are the object of a specific attention.
In the current understanding, $\bar p$ are almost always of secondary origin, \ie\ created through high-energy scatterings between the interstellar matter and primary cosmic rays, which were mainly generated by acceleration in supernova remnants \cite{Grenier:2015egx,Blasi:2013rva,Strong:2007nh}. 
The cosmic $\bar p$ were measured by the AMS-02 experiment in the range of 1 GeV to 400 GeV with a typical accuracy of 5\% \cite{Aguilar:2016kjl}.
On the theory side, the spectrum of secondary $\bar p$ can be predicted using diffusion equations with the $\bar p$ production cross section as input \cite{Moskalenko:1997gh,Vladimirov:2010aq,Orlando:2017tde}.
The determination of the discrepancy between the above two can open a new window on the indirect detection of dark matter or unknown astrophysical mechanisms of CR acceleration.

The accurate evaluation of the cosmic $\bar p$ spectrum requires a precise knowledge of their production cross section for several nuclear channels.
The nuclear primary CRs are composed of $p$, $^4$He, $^{12}$C, $^{14}$N, $^{16}$O.  Other nuclei need not to be considered
due to their small contribution to $\bar p$ production.
Moreover, the interstellar matter, which acts as a fixed target for the cosmic $\bar p$ production, is composed of $p$ and $^4$He.  Other nuclei acting as targets are negligible.
The estimation of the contribution of $\bar p$-production cross sections for each nuclear channel to the $\bar p$ spectrum is roughly 50\% for $p+p \to \bar p + X$, 30-40\% for $p+^4$He$ \to \bar p + X$ ($p$, $^4$He can either be the target or the projectile), 5\% for $^4$He $+^4$He$ \to \bar p + X$, and 5\% for $p+A \to \bar p + X$ with $A=$ $^{12}$C, $^{14}$N, $^{16}$O \cite{Korsmeier:2018gcy}.
The other channels contribute in all less than 5\%.

Here we note that, in unveiling new astrophysical phenomena, the most energetic region of the cosmic $\bar p$ spectrum is the most interesting, since the determination of the high-energy CR source (whether it is due to dark matter decays/annihilations or astrophysical accelerations) relies on the shape of the CR spectrum in the high-energy region.
It is thus required to quantify the high-energy part of the production cross section of secondary $\bar p$.
At the partonic level, the generation of highly energetic $\bar p$ is dominantly due to the collision of two gluons with highly asymmetric momentum fractions, and the $\bar p$ production cross section off two nuclei is sensitive to the gluon content of the nuclei at high momentum fraction, which is not very well known.
Precise measurements of $\bar p$ production off nuclei are therefore important to improve quantitative predictions of the cosmic $\bar p$ spectrum.

Currently, the production cross sections of $pp \to \bar p +X$ and $p + ^4$He$ \to \bar p +X$ ($^4$He as target) were measured~\cite{Aaij:2018svt,Aduszkiewicz:2017sei} in the energy range of AMS-02~\cite{diMauro:2014zea,Kappl:2014hha,Donato:2001ms}. 
In particular, the LHCb experiment with SMOG measured the production of $\bar p$ from the scattering of $p$ beam off $^4$He target in the range of $\bar p$ momentum 12 GeV - 110 GeV \cite{Aaij:2018svt,Korsmeier:2018gcy} and contributed to the improvement of the prediction of the secondary cosmic $\bar p$ spectrum.
In this context, we note that \AFTERALICECB (\ie\ the ALICE Central Barrel (CB) used in the fixed target mode) can measure very slow $\bar p$, with almost zero momentum. The production of such slow $\bar p$ (which are apparently outside the range measurable by AMS-02) with the LHC $p$ beam  corresponds to the highest possible energies in the inverse kinematics, where the nuclear target (C, N, O, He) travels at TeV energies, hit an interstellar $p$ at rest and produces a $\bar p$ in the limit of $x_F \to 1$. 

Having this observation in mind and considering any arbitrary stable nucleus as target, let us discuss more precisely the cases which can be studied with \AFTERALICECB and \AFTERLHCb assuming the additional possibility of $^{16}$O beam, as envisioned at LHC for a short period in Run 3~\cite{Jebramcik:2019eot}. 
Using a gas target, the $p + ^4$He $\to \bar p +X$ process (with $p$ as a projectile) with slow emerging $\bar p$ falls within
the acceptance of ALICE. \cf{fig:antiproton} shows the $\bar p$ kinematical reach\footnote{The purpose of this plot is to show the kinematical range and the statistics available in the fixed-target set-up. We do not expect to obtain different results by using another Monte-Carlo code such as EPOS for instance.} for the ALICE CB for $pp$ collisions for which the ALICE CB performances are similar as for $p$He collisions.  As expected, 
$\bar p$ with momenta as low as a few hundred MeV (which correspond to a rapidity difference between the He target
and the $\bar p$, $\Delta y_{\mathrm{He}\bar p}$, as low as 0.4) can easily be detected. Such $\bar p$ with small $\Delta y_{\mathrm{He}\bar p}$ 
for $^4$He $+p\to \bar p +X$ correspond to the high-energy tail of the $^4$He $+p \to \bar p +X$ process ($^4$He as projectile), which is one of the leading process in the cosmic $\bar p$ spectrum. Similarly, using C, N or O targets, one can study the high-energy $\bar p$ tail for (C,N,O)$+p\to \bar p +X$. To the best of our knowledge no other experimental set-up can cover this high-energy limit.

\begin{figure}[hbt!]
\begin{center}
\subfigure[$y_{\rm Lab.}$ coverage]{\includegraphics[width=0.48\textwidth]{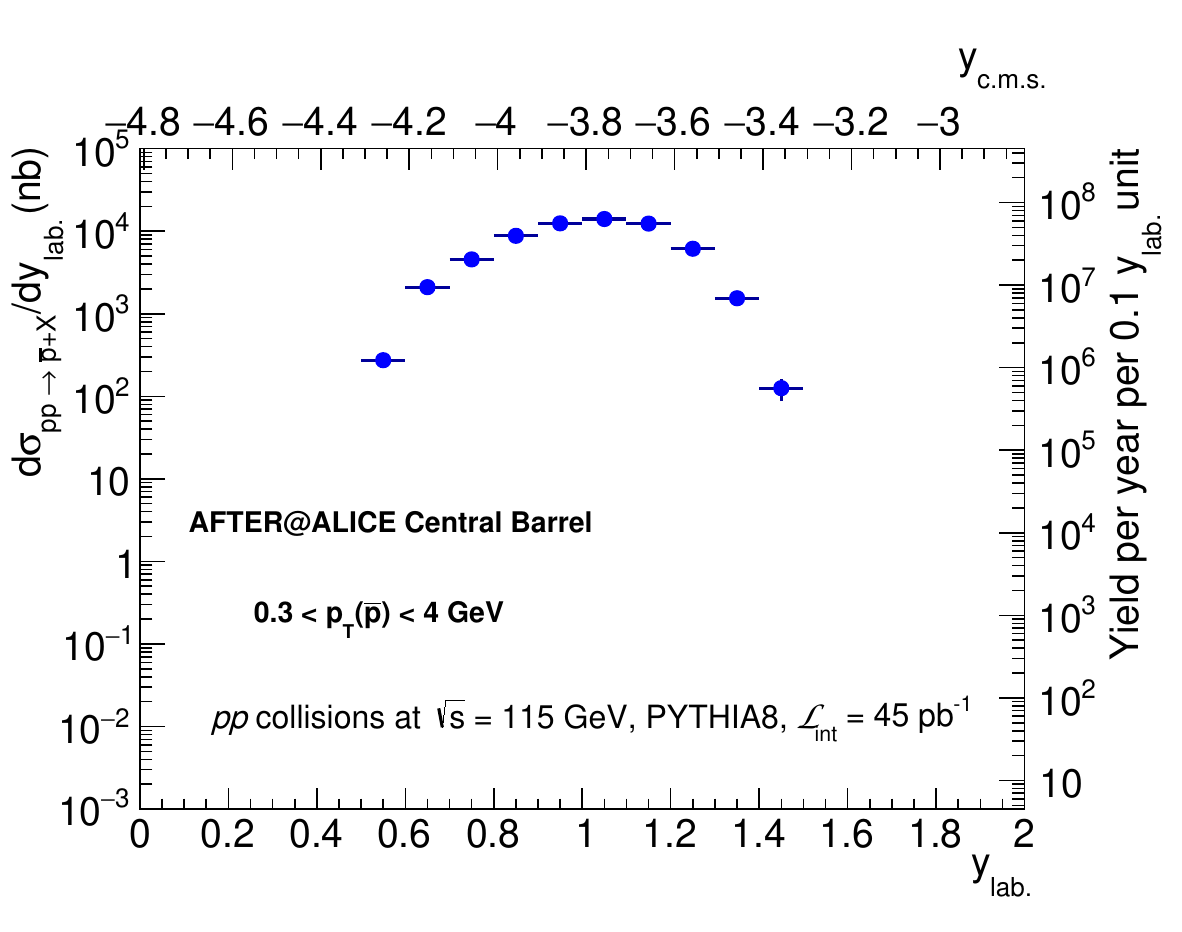}}
\subfigure[$E$ coverage]{\includegraphics[width=0.48\textwidth]{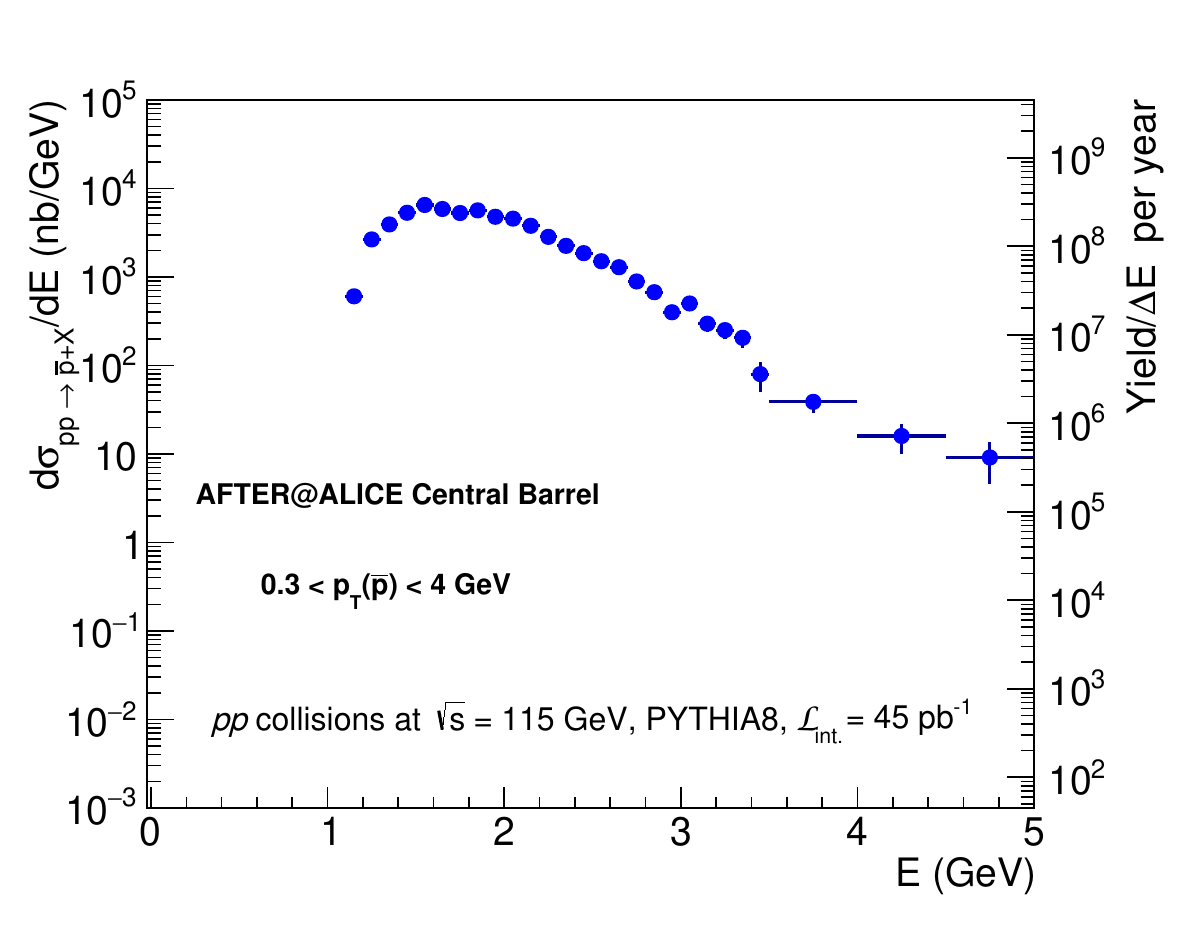}}
\caption{
The expected kinematical coverage for $\bar p$ production obtained with {\sc Pythia8}. 
 for \AFTERALICECB with a displaced vertex
at $z_{{\rm target}}=-4.7$~m from the nominal ALICE IP. Results are obtained assuming reduced track length in TPC, which gives the TPC acceptance of 1.35 $< \eta < $ 2.50 and the TOF acceptance of 0.27 $< \eta < $ 1.58. 
\label{fig:antiproton}
}
\end{center}
\end{figure}

With an oxygen beam, the study of the low-energy $\bar p$ in the laboratory frame from $^{16}$O $+p \to \bar p +X$ as well as from $^{16}$O + $^4$He $\to \bar p +X$ processes would become possible; this corresponds to the actual astrophysical situation. In such a case, \AFTERALICECB would be able to access the very low energy domain, whereas the LHCb coverage would be similar as for previous He studies, \ie\ from 10 to 100 GeV.
$^{16}$O is the most abundant nucleus in our universe after H and He, such that such experimental measurements will likely reduce the uncertainty of the cosmic $\bar p$ spectrum.

%% file: physics-spin/physics-spin.tex
 \subsection{Spin physics}
\label{section:Spin-Physics}

This section focuses on the case for spin physics at \AFTER, which is very competitive in the worldwide context. 
It can be divided into two parts. 
The first relies on the use of a polarised target to carry out many SSA measurements of common probes with a high precision, but also to measure --for the first time-- 
SSA on rare perturbative probes which would remain unaccessible otherwise\footnote{Let us recall that in this c.m.s. energy range RHIC offers significantly lower luminosities with a limited access to the high $x$ in the polarised nucleons.}. 
As such, \AFTER opens a novel domain of investigation of the Sivers-like effects for several years, in order to eventually access and (indirectly) constrain the OAM of the gluons and quarks in the proton, but also in the neutron and the deuteron.
Such a quest of measuring the parton OAM goes along with a complete tridimensional tomography of the momenta of the partons confined in hadrons.

The second part of the spin physics case bears on the very large luminosities to be accumulated thanks to the fixed-target mode, combined with an acceptance towards low transverse momenta at high $x$ in the target. 
This allows one to probe in a systematic way a class of azimuthal asymmetries related to the violation of the Lam-Tung relation~\cite{Lam:1980uc} in DY production, and potentially~\cite{Lambertsen:2016wgj} to the Boer-Mulders effect~\cite{Boer:1997nt}. 
\AFTER also allows one to study its counterpart in the gluon sector, with probably the first systematic measurement of the distribution of the linearly-polarised gluons in unpolarised nucleons at large momentum fractions.

SSAs can be described by nonperturbative twist-2 TMD matrix elements and nonperturbative twist-3 collinear matrix elements (CT3) in different kinematic regions: the TMD picture holds when there are two separate scales (such as the transverse momentum $q_T$ and the invariant mass $M$ of the lepton pair in DY production with $\lqcd\leq q_T<M$), whereas the CT3 picture holds when one single hard scale is present (such as the transverse momentum $p_T$ of a pion produced in hadronic collisions with $\lqcd< p_T$). 
Both nonperturbative descriptions can in general be related in the overlapping kinematic region in $q_T$ by means of coefficients which are perturbatively calculable (see e.g.~\cite{Bacchetta:2008xw,Bacchetta:2019qkv} for some caveats in this respect). 
Both twist-2 TMD and twist-3 collinear matrix element contain essential information on the spin structure of the nucleon and their knowledge is intertwined.
\AFTER will test with high precision whether these two formalisms offer the right description of SSAs.

We first discuss the (target) STSAs (\ie\ $A_N\equiv A_{UT}$), both for quark-induced and gluon-induced processes, which give access to both quark and gluon Sivers functions and several 3-parton correlation functions.
Then we present the prospects for the measurement of spin-related azimuthal asymmetries in unpolarised hadron collisions, which also probe either the quark or the gluon content of the nucleon.
Besides the discussion of spin and azimuthal asymmetries, we also elaborate on the relation between the TMDs that can be constrained by \AFTER\ and the OAM.
Next we discuss the use of ultraperipheral collisions to access GPDs through exclusive photoproduction processes and quarkonium production.
Finally we discuss the possibility to constrain the strange-quark-helicity distribution at large $x$ through $\Lambda$ production.

\subsubsection{Quark Sivers effect}
\label{ss:quark_sivers}

\input{physics-spin/physics-spin_1-quark-sivers.tex}

\subsubsection{Gluon Sivers effect}
\label{ss:gluon_sivers}

\input{physics-spin/physics-spin_2-gluon-sivers.tex}

\subsubsection{Quark-induced azimuthal asymmetries}
\label{ss:quark_BM}

\input{physics-spin/physics-spin_3-quark-BM.tex}

\subsubsection{Gluon-induced azimuthal asymmetries}
\label{ss:gluon_BM}

\input{physics-spin/physics-spin_4-gluon-BM.tex}

\subsubsection{From TMD PDFs to the partonic orbital angular momentum}

\input{physics-spin/physics-spin_5-OAM.tex}

\subsubsection{Ultraperipheral collisions}

\input{physics-spin/physics-spin_6-UPC.tex}

\subsubsection{Accessing the strange quark-helicity densities at high $x$}

\input{physics-spin/physics-spin_7-StrangePDF.tex}

%% file: physics-spin/physics-spin_1-quark-sivers.tex
The STSA $\An$ is schematically defined experimentally as~\footnote{Notice that another singled-out direction is needed, as the transverse momentum of the produced particles. See e.g.~\cite{Anselmino:2009st} for more details.}
\begin{equation}
A_{N} = \frac{1}{{\cal P}_{\text{eff}}} \frac{\sigma^{\uparrow} - \sigma^{\downarrow}}{\sigma^{\uparrow} + \sigma^{\downarrow}}
\,,
\end{equation}
where $\sigma^{\uparrow\,(\downarrow)}$ is a differential-production cross section of particles produced with the target spin polarised upwards (downwards), and ${\cal P}_{\text{eff}}$ is the effective polarisation. 
$\An$ is of particular interest because it was predicted to be small ($\An \propto m_q/p_T \sim O(10^{-4})$ in the collinear pQCD approach at the leading twist, while the measured $\An$ reaches $0.1$ at high $x_F$ ($=x_1-x_2$) in polarised collisions over a broad range of energy~\cite{Bonner:1988rv,Adams:1991rw,Adare:2013ekj}).
 
As introduced in section~\ref{sec:motivations}, $\An$ can be addressed either using the TMD formalism through the Sivers function, or using the CT3 formalism through 3-parton correlation functions.
One of the most important predictions, shared by both approaches, is the sign change of this asymmetry between SIDIS and DY processes.
We explain below how \AFTER can contribute to precisely measure this sign change.

Moreover, the accurate measurements to be performed by \AFTER will help to constrain the non-perturbative input that enters the TMD evolution kernel~\cite{Collins:2011zzd,Echevarria:2012pw,Rogers:2015sqa,Collins:2017oxh,Bacchetta:2017gcc,Scimemi:2017etj,Scimemi:2018xaf}, which has an important effect on the STSA (see e.g. \cite{Echevarria:2014xaa,Sun:2013dya}).

\paragraph{Drell-Yan production}

DY lepton-pair production is a unique tool to study the Sivers effect, because it is theoretically very well understood and the Sivers function $f_{1T}^{\perp q}(x,k_T^2)$ for quarks (which represents the difference of number densities of unpolarised quarks with transverse momentum $\kT$ and collinear momentum fraction $x$ for a given two opposite configurations of the transverse spin of the proton) is predicted to have an opposite sign for DY and SIDIS processes:
\begin{equation}
f_{1T}^{\perp q}(x,k_T^2)_{\rm DY} = - f_{1T}^{\perp q}(x,k_T^2)_{\rm SIDIS}
\,.
\end{equation}

Within the TMD formalism, and up to angular integrations, $A_N$ in $pp^\uparrow$ collisions can be schematically written as
\begin{align}
A_N \sim
\frac{f_1^q(x_1,k_{T1}^2)\otimes f_{1T}^{\perp \bar q}(x_2,k_{T2}^2)}{f_1^q(x_1,k_{T1}^2)\otimes f_1^{\bar q}(x_2,k_{T2}^2)}
\,,
\end{align}
where $f_1^q$ stands for the unpolarised quark TMD PDF, and $\otimes$ represents a convolution in momentum space and a sum over quark and anti-quark flavours.

The verification of the sign change of the Sivers function is the main physics case of the DY COMPASS programme~\cite{Quintans:2011zz}, which recently performed the first measurement of the asymmetry in DY production~\cite{Aghasyan:2017jop}, and the experiments E1039~\cite{Klein:zoa} and E1027~\cite{Isenhower:2012vh} at Fermilab.
The \AFTER programme will allow one to further investigate the quark Sivers effect by measuring DY STSA~\cite{Liu:2012vn,Anselmino:2015eoa} over a wide range of $x^\uparrow$ ($=x_2$) and masses.
With the high precision that \AFTER will be able to achieve, one will accurately measure the Sivers function, if the sign change happens to be already established by the mentioned experiments.
In case the asymmetry turns out to be small and these experiments cannot get to a clear answer, then \AFTER will be able to confirm/falsify the sign change.
Table~\ref{tab:DY-SSA-projects} shows a compilation of the relevant parameters of future or planned polarised DY experiments. 
As can be seen, the \AFTER\ program offer the possibility to measure the DY \An\ in a broad kinematic range with an exceptional precision. 

\flushfootnote

\begin{table}[htb!]
\begin{center} \setlength{\arrayrulewidth}{.8pt}  \renewcommand{\arraystretch}{1.2}\small
\begin{tabularx}{\textwidth}{p{4.6cm}|p{1.3cm}|p{1.2cm}|p{0.8cm}|p{1.6cm}|p{1.4cm}|p{0.8cm}|p{1.5cm}}
Experiment  & Colliding systems & Beam energy {\small [GeV]} &  $\sqrt{s}$ {\small [GeV]} &$x^\uparrow$  & $\cal L$ {\small[cm$^{-2}$s$^{-1}$]} & ${\cal P}_{\text{eff}} $  &  ${\cal F}$ [cm$^{-2}$s$^{-1}$] \\ 
\hline  \hline
{\small \AFTERLHCb: $z=0$}     		& \multirow{3}{*}{$p$H$^{\uparrow}$} 				& \multirow{3}{*}{7000}     & \multirow{3}{*}{115} & $0.05\div 0.95$ & \multirow{3}{*}{$9.2\times 10^{32}$}   &  \multirow{3}{*}{80\%} & \multirow{3}{*}{5.9 $\times$ 10$^{32}$}  \\
{\small \AFTERLHCb: $z=-0.4$ m }    		& 				&      &  & $0.02\div 0.95$ &    &   \\
{\small \AFTERLHCb: $z=-1.5$ m}     		& 				&      &  & $0.01\div 0.15$ &    &   \\
{\small \AFTERLHCb: $z=0$     }		& \multirow{3}{*}{$p^{3}$He$^{\uparrow}$ }		& \multirow{3}{*}{7000}     & \multirow{3}{*}{115} & $0.05\div 0.95$ & \multirow{3}{*}{$1.3\times 10^{33}$}    	&  \multirow{3}{*}{23\%} & \multirow{3}{*}{2.1 $\times$ 10$^{32}$}  \\       
{\small \AFTERLHCb: $z=-0.4$ m     }		& 		&   &  & $0.02\div 0.95$ &   	&    \\       
{\small \AFTERLHCb: $z=-1.5$ m     }		& 		&   &  & $0.01\div 0.15$ &   	&    \\  

{\small \AFTERLHCb: $z=0$     }		& \multirow{3}{*}{$pD^{\uparrow}$ }		& \multirow{3}{*}{7000}     & \multirow{3}{*}{115} & $0.05\div 0.95$ & \multirow{3}{*}{$5.6\times 10^{32}$}    	&  \multirow{3}{*}{78\%} & \multirow{3}{*}{6.8 $\times$ 10$^{32}$}  \\       
{\small \AFTERLHCb: $z=-0.4$ m     }		& 		&   &  & $0.02\div 0.95$ &   	&    \\       
{\small \AFTERLHCb: $z=-1.5$ m     }		& 		&   &  & $0.01\div 0.15$ &   	&    \\

\hline     
{\small \AFTERALICEMU: $z=0$ }     		& \multirow{3}{*}{$p$H$^{\uparrow}$} 				& \multirow{3}{*}{7000 }    & \multirow{3}{*}{115} & $0.10\div 0.70$ & \multirow{3}{*}{$2.6\times 10^{31}$}   &  \multirow{3}{*}{80\%} & \multirow{3}{*}{1.7 $\times$ 10$^{31}$}  \\
{\small \AFTERALICEMU: $z=-4.7$ m}     		&  				&      &  & $0.08\div 0.35$ &    &   \\
{\small \AFTERALICECB: $z=-4.7$ m}     		&  				&      &  & $0.40\div 0.95$ &    &   \\

\hline   
\multirow{2}{*}{COMPASS (CERN)\cite{Gautheron:2010wva}}  	& $\pi^{-}$NH$_3^{\uparrow}$  	& \multirow{2}{*}{190}      & \multirow{2}{*}{19}  & \multirow{2}{*}{$0.05\div 0.55$}  & $2.0\times 10^{32}$ & 16\% & 8.7 $\times$ 10$^{31}$
   \\   
  	& $\pi^{-}$ $^6$LiD  	&       &   &   & $8.2\times 10^{32}$ & 22\% & 3.2 $\times$ 10$^{32}$
   \\              
PHENIX/STAR (RHIC)~\cite{RHIC-DY}	& $p^{\uparrow}p^{\uparrow}$   & collider & 510 & $0.05\div 0.10$ & $2.0\times 10^{32}$ & 50\% &  5.0 $\times$ 10$^{31}$\\
E1039 (FNAL)~\cite{Brown:2014sea}		& $p$NH$_3^{\uparrow}$       		& 120	   & 15  & $0.10\div 0.45$ & $3.9\times 10^{34}$ & 15\%  & 1.5 $\times$ 10$^{34}$ \\
E1027 (FNAL)~\cite{Isenhower:2012vh}		& $p^{\uparrow}$H$_2$       		& 120	   & 15  & $0.35\div 0.90$ & $1.0\times 10^{35}$ & 60\% & 7.2 $\times$ 10$^{34}$\\
NICA  (JINR)~\cite{nica}		& $p^{\uparrow}p$             	& collider & 26  & $0.10\div 0.80$  & $1.0\times 10^{32}$ & 70\%  & 4.9 $\times$ 10$^{31}$ \\
fsPHENIX (RHIC)~\cite{fsphenix}	& $p^{\uparrow}p^{\uparrow}$	& collider & 200 & $0.10\div 0.50$  & $8.0\times 10^{31}$ & 60\%  & 2.9 $\times$ 10$^{31}$ \\
fsPHENIX (RHIC)~\cite{fsphenix} 			& $p^{\uparrow}p^{\uparrow}$	& collider & 510 & $0.05\div 0.60$ & $6.0\times 10^{32}$ & 50\% & 1.5 $\times$ 10$^{32}$ \\
PANDA (GSI)~\cite{PANDA}  		& $\bar{p}H^{\uparrow}$       	&  15      & 5.5 & $0.20\div 0.40$  & $2.0\times 10^{32}$ & 20\% & 8.0 $\times$ 10$^{30}$ \\ \hline
\end{tabularx}
\caption{
Compilation inspired by \cite{Brodsky:2012vg,Barone:2010zz} of the relevant parameters for the future or planned polarised DY experiments. The effective polarisation (${\cal P}_{\text{eff}} $) is a beam polarisation (where relevant) or an average polarisation times a (possible) average dilution factor  $\langle f\rangle $ (for a gas target, similar to the one developed for HERMES~\cite{0034-4885-66-11-R02,Steffens:2015kvp}) or a target polarisation times an average dilution factor $\langle f\rangle $ (for the targets used by COMPASS and E1039). For the \AFTERLHCb, \AFTERALICECB and \AFTERALICEMU lines, the numbers correspond to a gas target with a storage cell (see ~\ct{tab_lumi_comp_alice} and ~\ct{tab_lumi_comp_lhcb}) and $4 < M_{\ell\ell} < 9$~GeV (for the $x^\uparrow$ range).  $\cal{F}$ is the (instantaneous) spin figure of merit of the set-up defined as ${\cal F} ={\cal L} \hspace{0.1 true cm} {\cal P}^{2}_{\text{eff}}    \sum_i A_i$, with  ${\cal L}$ being the instantaneous luminosity.  We stress that the values of ${\cal F}$ between different set-ups should be compared with care as it does not
account for isospin and nuclear effects (via the variation of $f$ for instance) or acceptance effects neither for any energy or kinematical dependences of the DY production cross section which both alter the measured rates and the uncertainty of the asymmetry measurements. We refer to section~\ref{lhc_param} for more details.
}
\label{tab:DY-SSA-projects}
\end{center}
\end{table}

\begin{figure}[hbt!]
\centering
\subfigure[~]{\includegraphics[width=0.48\textwidth]{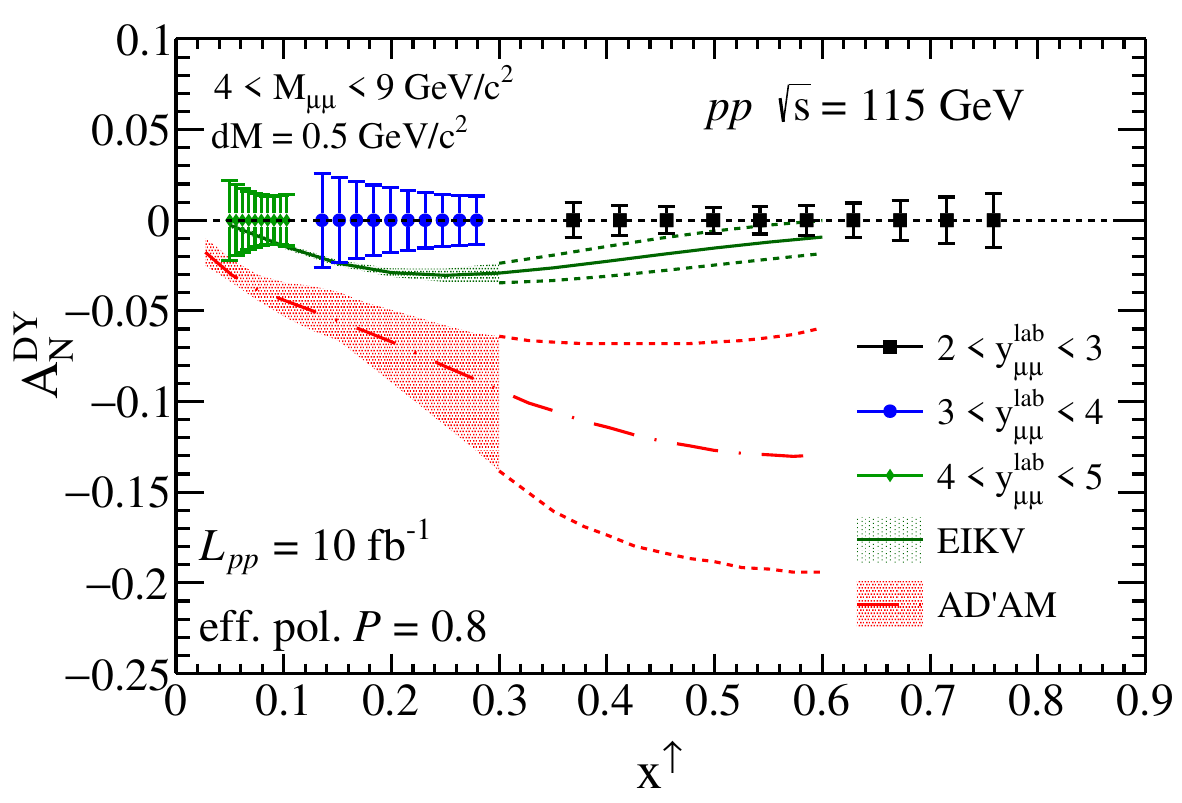}\label{fig:An:DY-pp}} 
\subfigure[~]{\includegraphics[width=0.48\textwidth]{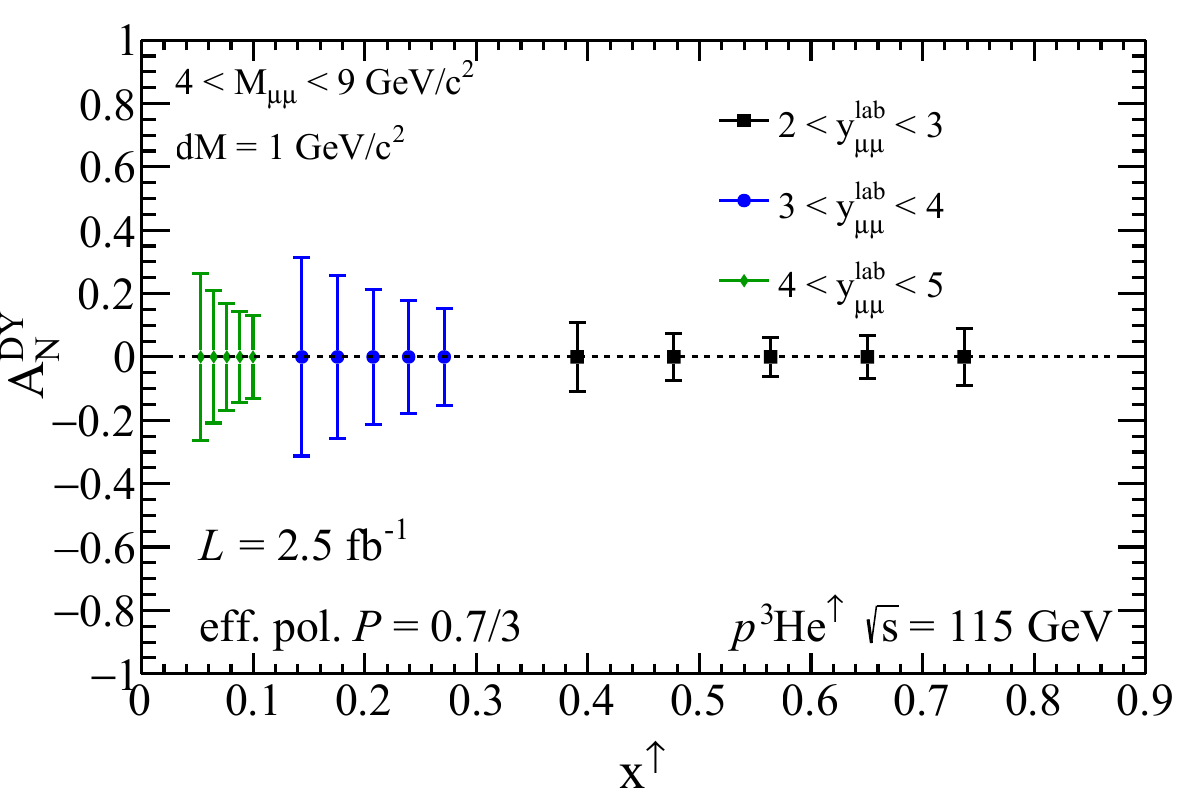}\label{fig:An:DY-pHe}} 
\caption[DY predictions]{(a) Two predictions (denoted AD'AM~\cite{Anselmino:2015eoa} and EIKV~\cite{Echevarria:2014xaa}) of the DY $\An$ as a function of $x^{\uparrow}$ at \AFTER, compared to the projected precision of the measurement~\cite{Kikola:2017hnp}. 
The bands are filled in the region where the fits use existing SIDIS data, i.e. for $x^\uparrow\lesssim 0.3$, and hollow where they are extrapolations.
(b) Similar projections for the DY $\An$ as a function of $x^{\uparrow}$ in $p+^{3}$He$^{\uparrow}$ collisions at $\sqrt{s}=115$~GeV~\cite{Kikola:2017hnp}. [In both cases, the bars show the statistical uncertainties for the quoted luminosisities accounting for the background subtraction and polarisation-dilution effects].}
\label{fig:An:DY}
\end{figure}

The DY measurement is the key to validate/falsify the Sivers effect for quarks. 
At \AFTER, the target-rapidity range corresponds to a negative \xF\ where the $\An$ asymmetry is predicted to be large (\cf{fig:An:DY}) with large theoretical uncertainties.
\cf{fig:An:DY-pp} shows the expected precision for DY $\An$ measurement at \AFTER for $\mathcal{L}  = 10~\rm{fb}^{-1}$ (which corresponds to one year of running)~\footnote{The statistical uncertainty $\delta$ on $\An$ is calculated as $\delta_{\An} = \frac{2}{{\cal P}_{\text{eff}}\,({\sigma^{\downarrow} + \sigma^{\uparrow}})^2} \sqrt{(\delta_{\sigma^{\uparrow}}\sigma^{\downarrow})^2 + (\delta_{\sigma^{\downarrow}}\sigma^{\uparrow})^2}$, where $\delta_{\sigma} = \sqrt{\sigma + 2B}$, $\sigma$ is the cross section for a given configuration and $B$ is the background in that measurement. The yields are calculated at fixed $y_{\mu\mu}^{\rm Lab.}=[2.5,3.5,4.5]$, fixed $M_{\mu\mu}=[4.5,5.5,6.5,7.5,8.5]~\gev$ and integrating over the transverse momentum of the lepton pair.}, compared to two different theoretical predictions: AD'AM~\cite{Anselmino:2015eoa} and EIKV~\cite{Echevarria:2014xaa}.
These two works performed fits of \An in SIDIS data, available for $x^\uparrow\lesssim 0.3$, using two different theoretical setups.
The uncertainty band of AD'AM curve represents the statistical uncertainty of their fitted parameters after performing a variation of the total $\chi^2$ of about 20, while the one of EIKV is obtained by using the replica method (see e.g. Ref.~\cite{Bacchetta:2017gcc}) with an effective variation of the total $\chi^2$ of about 1; this explains the difference of width among the curves. 
Thus the DY data at \AFTER will put strict constraints on the Sivers effect for quarks, help to discriminate among different approaches, and accurately test one of the most important predictions of the TMD factorisation formalism, \ie\ its sign change w.r.t. SIDIS.
In addition, given that this effect can be framed as well within the CT3 approach, \AFTER will obtain very useful data to constrain also the 3-parton correlation functions.

\AFTER\ with a gas target offers also a unique opportunity for studies of STSA in polarised $p+^{3}$He$^{\uparrow}$ collisions.
These have been studied at JLab Hall-A by several DIS/SIDIS experiments in the last two decades (see e.g.~\cite{Qian:2011py,Huang:2011bc}), which the \AFTER program could then complement. 
Such reactions give access to polarised neutrons and thus to the Sivers functions in a neutron which can in principle shed some light on its isospin dependence. 
Let us also note that a polarised deuterium target can also provide a proxy to a polarised neutron target.
\cf{fig:An:DY-pHe} shows the statistical-uncertainty predictions for DY measurements. In the case of $^3$He$^\uparrow$, a polarisation of P = 70\% can be achieved~\cite{0034-4885-66-11-R02}. 
However, the effective polarisation, ${\cal P}_{\rm eff}$, is diluted by a factor of 3 since only the neutron is polarised in the $^3$He$^\uparrow$. 
The projections for $^3$He$^\uparrow$ are prepared based on simulations for \pp\ collisions and applying corrections to account for change in signal and background yields. The combinatorial background is proportional to the number of binary nucleon-nucleon collisions $N_{coll}$, thus the background increases by a factor $N_{coll} \approx \sqrt{3}$ compared to \pp.  
An additional isospin factor of $9/6$ for DY studies is included. 
The available integrated luminosity of 2.5~fb$^{-1}$ will allow for an exploratory measurement for DY production and precision study for quarkonium production (see section~\ref{ss:gluon_sivers}).

In addition, DY production with an unpolarised fixed-target will be extremely valuable to study the simplest TMD function at large $x$, namely the unpolarised TMD PDF~\cite{Signori:2013mda,Anselmino:2013lza,
DAlesio:2014mrz,Angeles-Martinez:2015sea,Bacchetta:2017gcc,Scimemi:2017etj}. 
A good knowledge of unpolarised TMDs is of fundamental importance in order to validate our understanding of their scale evolution and to reliably study azimuthal and spin asymmetries, as they always enter the denominators of these quantities.

\paragraph{Pion and kaon production}

Pion and kaon STSAs have been extensively studied in the last three decades at Fermilab and BNL with hadron
beams and at Jefferson Lab, CERN (COMPASS) and DESY (HERMES) with lepton beams (see e.g. \cite{Adams:1991cs,Arsene:2008aa,Abelev:2008af,Airapetian:2004tw,Airapetian:2013bim,Adolph:2012sp,Alexakhin:2005iw,Qian:2011py}), observing large asymmetries in the valence region at large $x^\uparrow$, which motivated the introduction of the Sivers effect.
As for now, similar studies have not been carried out with hadron beams on $^3$He, thus on a polarised neutron target, which however could give us original insights on the flavour symmetries of the correlation between the partonic transverse momentum and the nucleon spin.
Along these lines, the \AFTER programme relying on the LHCb and/or ALICE detectors, can play a crucial role.

Indeed, as shown in \cf{fig:An-pion-TMD-CT3}, the predicted $\An$ for pion production with a neutron (a-b) and proton (c-d) target, based on the generalised parton model (GPM) approach (which is an extension of the parton model to include the transverse-momentum dependence)\footnote{The isospin symmetry is used to implement the neutron Sivers functions by using the extracted proton ones in Ref.~\cite{Anselmino:2015eoa}: $f_{1T}^{\perp u/neutron}=f_{1T}^{\perp d/proton}$ and $f_{1T}^{\perp d/neutron}=f_{1T}^{\perp u/proton}$. This is supported by the STSA on deuteron target measured by COMPASS~\cite{Alexakhin:2005iw,Ageev:2006da}, well compatible with zero. These predictions include both Sivers and Collins contributions, which are added together in the estimates of the central values as well as of the statistical uncertainty bands.} and the CT3 approach\footnote{The isospin symmetry is used again to implement the neutron twist-3 matrix elements by using the extracted proton ones in Ref.~\cite{Gamberg:2017gle}.}, both indicate a ``flavour'' sign change.

From a pQCD point of view, pion and kaon production at \AFTER\ can thus improve the current knowledge of the CT3 matrix elements involved in the production and fragmentation mechanisms~\cite{Kanazawa:2015fia}, and help to clarify if the pion and kaon STSAs are really mainly generated by the Collins mechanism, as recently suggested~\cite{Gamberg:2017gle}.

\begin{figure}[hbt!]
\centering
\subfigure[~]{\includegraphics[width=0.48\textwidth,draft=false]{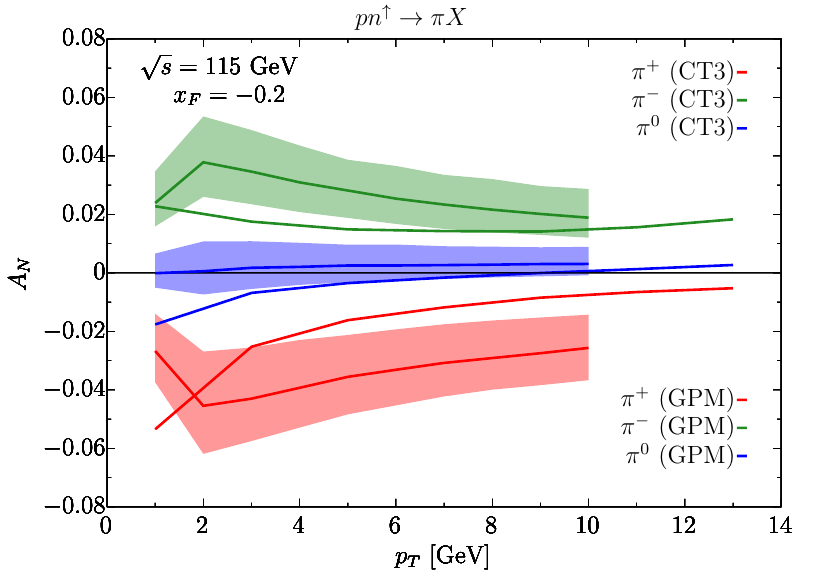}}
\subfigure[~]{\includegraphics[width=0.48\textwidth,draft=false]{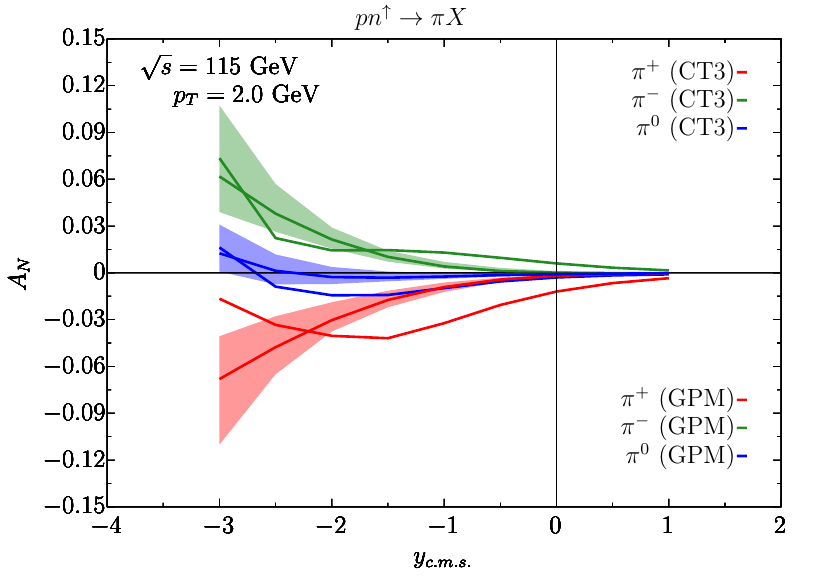}} \\
\subfigure[~]{\includegraphics[width=0.48\textwidth,draft=false]{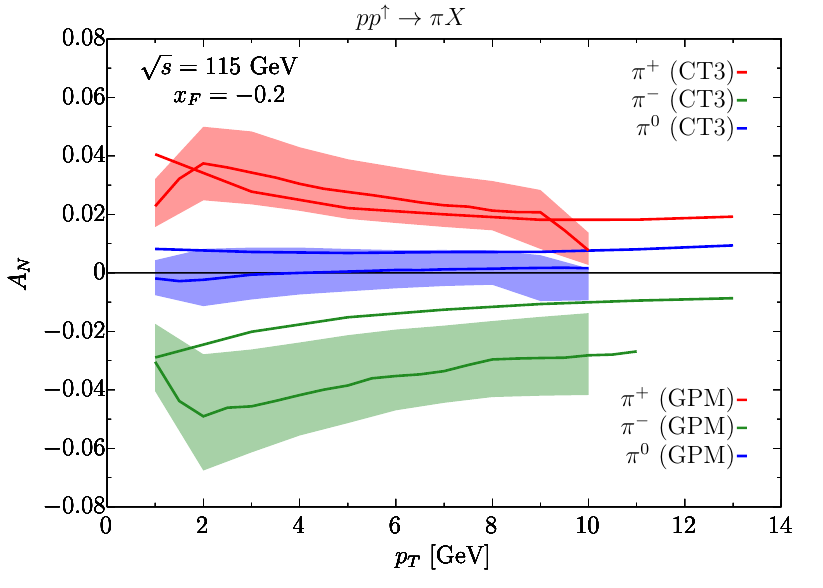}}
\subfigure[~]{\includegraphics[width=0.48\textwidth,draft=false]{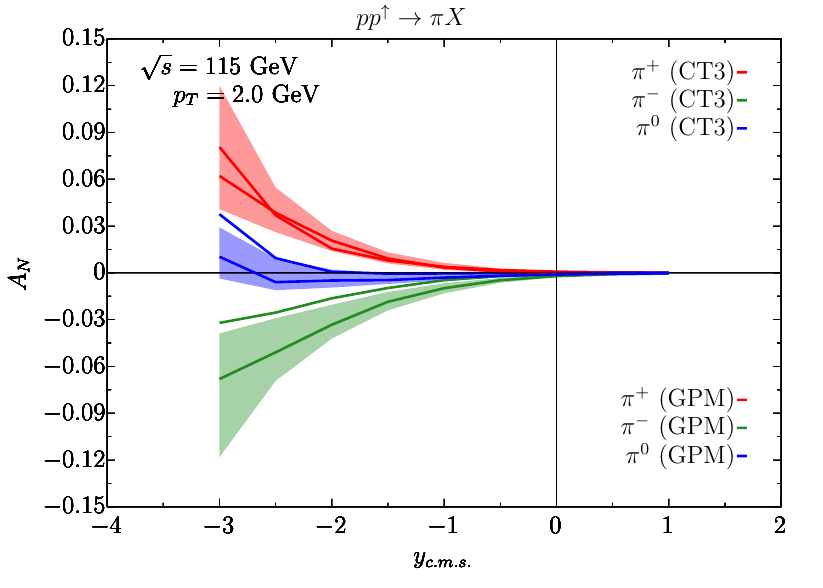}}
\caption[Neutron/proton Sivers predictions in GPM (quasiTMD) and CT3]{Predicted $\An$ for pion production as a function of (a) transverse momentum $p_T$ and (b) rapidity $y$, in the \AFTER kinematics for a neutron target (accessible with $^3$He$^\uparrow$), based on both GPM and CT3 formalisms.
(c-d): same as (a-b) but for a proton target.}
\label{fig:An-pion-TMD-CT3}
\end{figure}

\paragraph{W$^\pm$ boson production}

So far, only the valence $u$ and $d$ quark Sivers functions have been constrained, while sea-quark Sivers functions remain largely unknown \cite{Echevarria:2014xaa}.
In this sense, STSAs for vector-boson production offer a complementary tool to STSAs for DY production, giving access to the flavour dependence of the Sivers function.
Moreover, they can also serve to test the sign change of the Sivers effect with respect to SIDIS.

As explained in section \ref{subsec:nucleon}, roughly 250 (off-shell) $W^+$ and 60 (off-shell) $W^-$ per year are expected to be collected with a luminosity of 10~fb$^{-1}$.
These yields would allow one to achieve a statistical uncertainty for $A_N$ of roughly $0.1-0.2$ \footnote{The uncertainty is calculated in the same way as for DY $A_N$, and we assumed that the background is negligible.}.
This uncertainty is comparable to the precision of the $A_N$ measurement for 
$W^{\pm} \rightarrow l^{\pm}\nu$ in $p^\uparrow+p$ production at RHIC~\cite{Adamczyk:2015gyk}. In 2017 the STAR experiment carried out a measurement of the transverse single-spin asymmetry in $p^\uparrow+p \to W^{\pm}/Z^0$ aiming at a statistical precision on the level of 5\%~\cite{Aschenauer:2016our}. The \AFTER $W^{\pm}$ measurement  will cover different range in $x^\uparrow$. As such, measurements at \AFTER and RHIC will be complementary.

In general, both transverse and longitudinal polarization of a gas target are feasible, as it was demonstrated by the HERMES experiment. In that case, a setup with longitudinal target polarization was used in 1996-2000 and then changed to a transverse one in 2001~\cite{Airapetian:2004yf}. Therefore, we finally comment on the possibility to constrain the sea-quark-helicity distributions at \AFTER by measuring the spin asymmetries for a longitudinally polarised target.
At the leading order in $\alpha_s$, the asymmetries for $W^+$ and $W^-$ bosons are simply related to the (helicity) PDFs as
\begin{align}
A_{UL}^{W^+} &\propto 
\frac{g_1^u(x_1) f_1^{\bar d}(x_2) - g_1^{\bar d}(x_1) f_1^{u}(x_2)}
{f_1^u(x_1) f_1^{\bar d}(x_2) + f_1^{\bar d}(x_1) f_1^{u}(x_2)}
\,,\quad
A_{UL}^{W^-} \propto 
\frac{g_1^d(x_1) f_1^{\bar u}(x_2) - g_1^{\bar u}(x_1) f_1^{d}(x_2)}
{f_1^d(x_1) f_1^{\bar u}(x_2) + f_1^{\bar u}(x_1) f_1^{d}(x_2)}
\,,
\end{align}
where $g_1^q$ is the helicity PDF of the quark $q$. 
Dedicated simulations are however needed to quantify such constraints.

\paragraph{Direct-photon production}

The quark Sivers effect can also be studied via direct-photon production STSAs. \cf{fig:An:DirectGamma} shows the expected 
$A_N$ as a function of the photon $\pT$ for $x_F=-0.2$ (\ie\ in a range accessible by LHCb) for both aforementioned approaches. 
Contrary to the DY and $\pi$ production cases, the predicted signs of $A_N$ differ. This is related to the sign ``mismatch'' issue (see \eg~\cite{Gamberg:2012iq}). 
Even though the magnitude of $A_N$ is the largest at low $\pT$ ($<1$ GeV) where the background is probably very challenging and the application of pQCD questionable, measurements with a precision on the order of 1\% for $x_F=-0.2$ should however be sufficient to discriminante between the predictions of both approaches for $\pT$ above 5 GeV.

\begin{figure}[hbt!]
\centering
\includegraphics[width=0.48\textwidth]{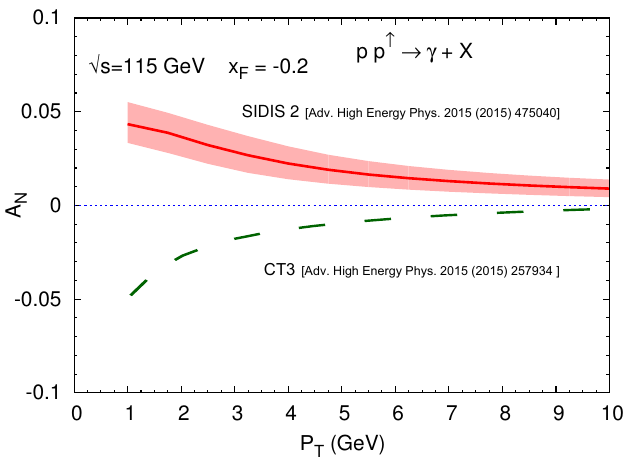}
\caption[Direct gamma predictions]{Prediction for the direct-photon $\An$ as a function of $p_T$~\cite{Anselmino:2015eoa,Kanazawa:2015fia} for $x_F=-0.2$. No theoretical uncertainties are shown for the CT3 prediction.}
\label{fig:An:DirectGamma}
\end{figure}

By measuring $\An$ for quark-induced processes which do not involve fragmentation, in particular prompt photons (compared to charged and neutral pions and kaons), the \AFTER programme will allow one to distinguish the Sivers and Collins mechanisms~\cite{Barone:2010zz,DAlesio:2007bjf}, and in turn constrain the quark contribution to the transverse spin of the proton, along with the constraints from DY production data.

%% file: physics-spin/physics-spin_2-gluon-sivers.tex
DY production is the golden process to access the intrinsic transverse motion of quarks in a nucleon. 
However, there is no analogous process to probe the gluon content, which is both experimentally clean and theoretically well-controlled.
For instance, $H^0$ boson production serve as such a process, since it is a gluon-gluon fusion process into an observed colour singlet, but it is experimentally demanding in terms of luminosity and energy.

Currently, one of the best tools at our disposal is the production of quarkonia.
In fact, a major strength of \AFTER is the large production rates for open heavy-flavour mesons and quarkonium states.
The expected rates for a single year of data taking are roughly $10^6$ $\Upsilon$ and $10^9$ $\jpsi$.
In addition, these processes are mainly sensitive to the gluon content of the colliding hadrons. 
They are very useful probes to precisely access and constrain the gluon Sivers effect, which is essentially unknown~\cite{Kurek:2016nqt,Boer:2015vso}

Moreover, due to the inherent process dependence of the TMDs, different processes will probe different gluon TMD functions~\cite{Pisano:2013cya,Buffing:2013kca}.
In fact, the generalised universality of gluon TMDs is more involved than that of the quark TMDs, due to the richer gauge-link structure in their operator definition.
In particular, in the case of the gluon Sivers function, it is known that all the functions that can be probed in different processes can be reduced to only two independent ones~\cite{Buffing:2013kca}.
The \AFTER programme offers a unique possibility to test all these theoretical predictions, either confirming them or quantifying the potential discrepancies, if any.

\paragraph{Open heavy-flavour production}

The $\An$ for open heavy-flavour production gives access to the gluon Sivers effect (see e.g. \cite{Godbole:2016tvq,DAlesio:2017rzj}).
It also offers the possibility to study the process dependence of $\An$, by measuring charm quarks and anti-quarks separately~\cite{Kang:2008ih}, being a unique probe of the $C$-parity odd twist-3 tri-gluon correlator~\cite{Ji:1992eu,Beppu:2010qn}.
 
\begin{figure}[hbt!]
\centering
\subfigure[]{
\includegraphics[width=0.48\textwidth]{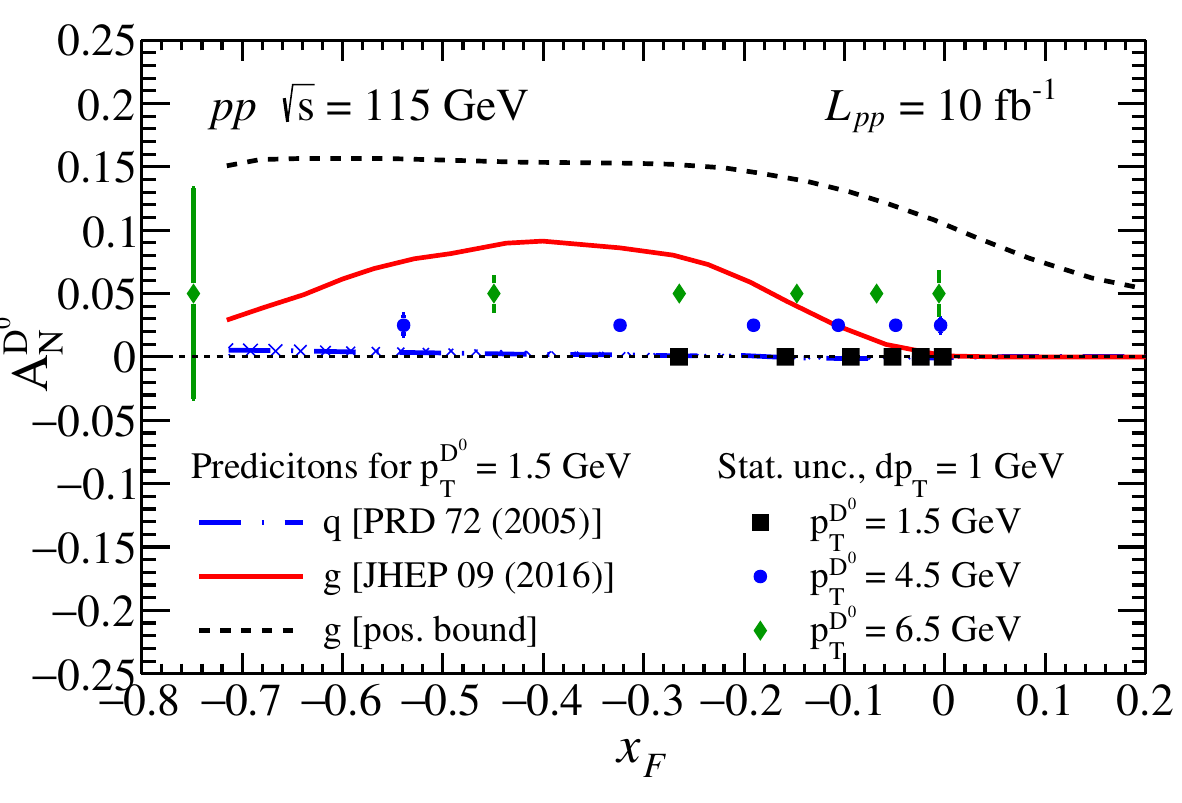} }
\subfigure[]{
\includegraphics[width=0.48\textwidth]{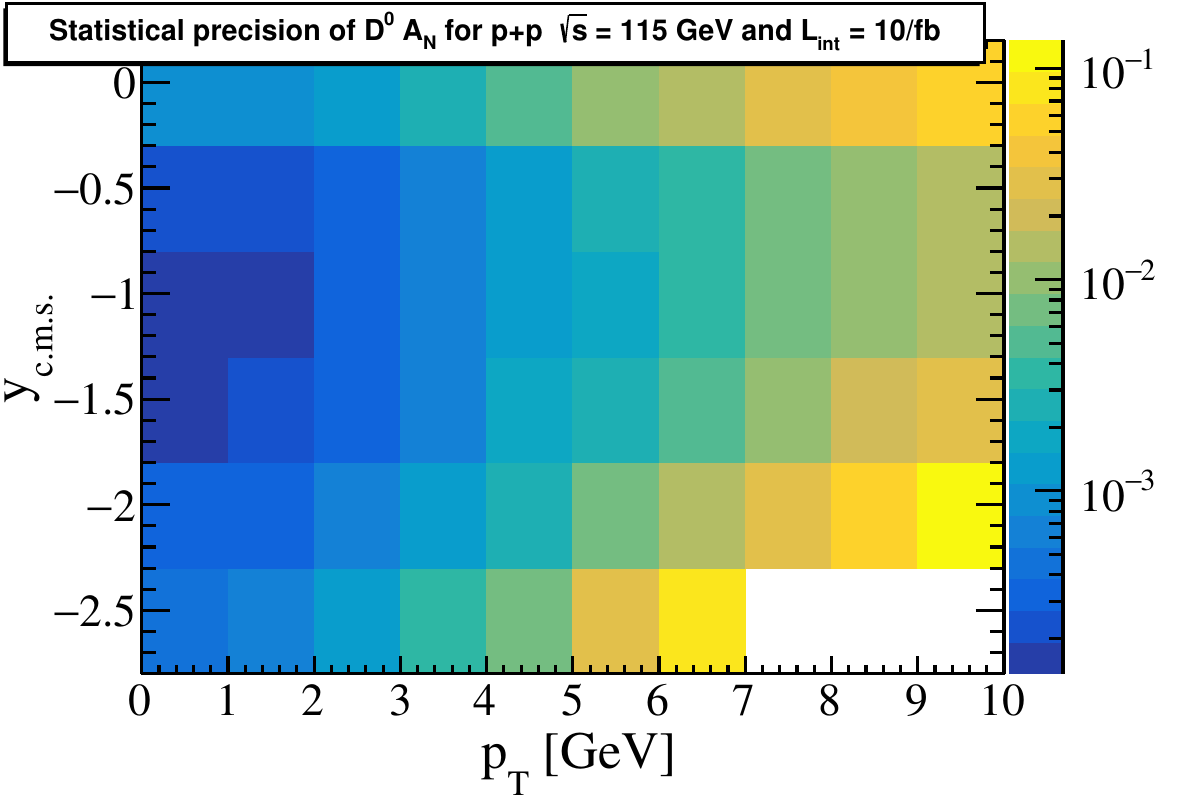}}
\caption[D0An]{Projections for the statistical precision of open charm $\An$ as a function of (a) $x_F$ at fixed $\pT$ and of (b) $\ycms$ \& $\pT$, at $\sqrtsNN=115$~GeV and 10 fb$^{-1}$ of luminosity using a LHCb-like detector. 
The $D^0$ yields are taken from FONLL \cite{Cacciari:1998it,Cacciari:2001td} calculations within the LHCb acceptance; the efficiency and the S/B ratio are extrapolated from \cite{LHCb:2017qap}. For most of the points, statistical uncertainties are smaller than the marker size. Estimates in (a) for different $\pT$ bins were made assuming bin width $d\pT = 1\;\gev$, and were displaced vertically for a better visibility. The theory curves represent the SIDIS predictions for quarks and gluons, along with the positivity upper bound of the gluon Sivers effect~\cite{Anselmino:2004nk, DAlesio:2015fwo}. 
}
\label{fig:An:D0}
\end{figure}

\begin{figure}[hbt!]
\centering
\includegraphics[width=0.48\textwidth]{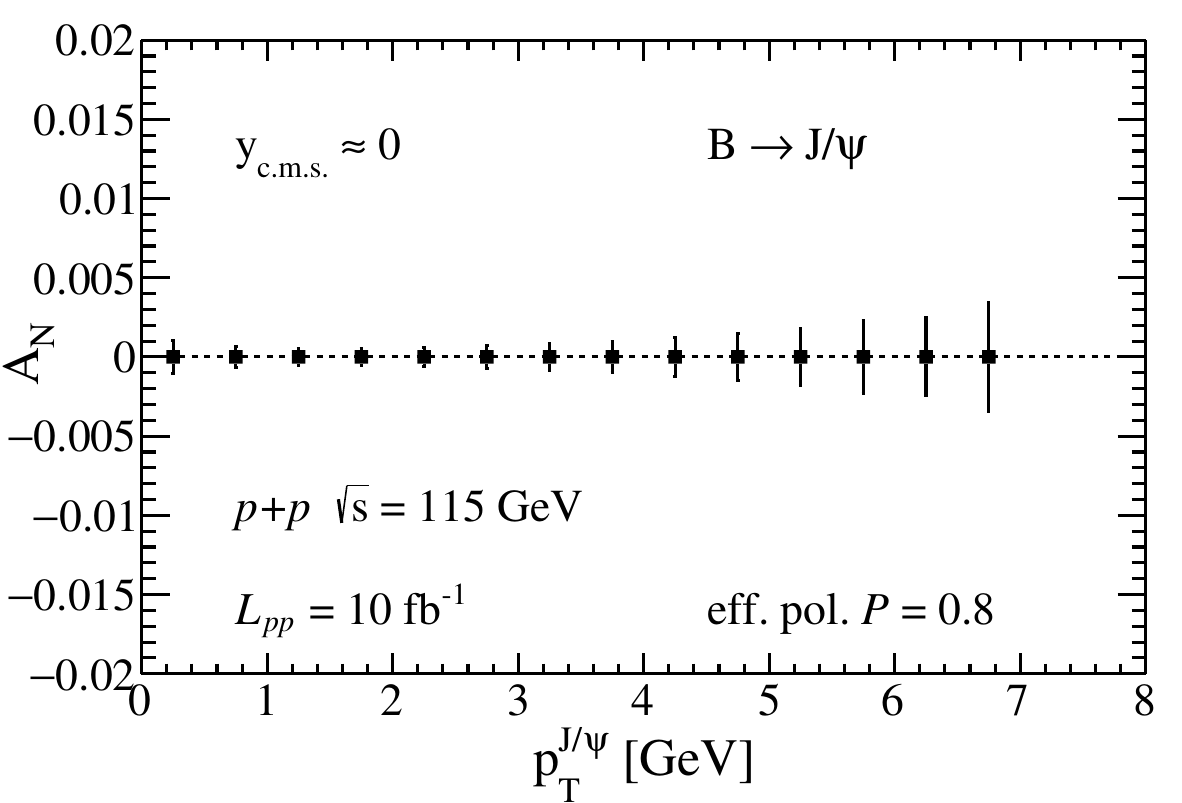} 
\caption[BAn]{Projections for the statistical precision of open bottom $\An$ (measured via $B \rightarrow J/\psi$) as a function of $\pT$ at $y_{\cms}\approx 0$ and $\sqrtsNN=115$ GeV measured with a LHCb-like detector with integrated luminosity of 10 fb$^{-1}$. 
The $B \rightarrow J/\psi$ yields are calculated based on the FONLL predictions for the bottom-quark yields, and the $B \rightarrow J/\psi$ fragmentation is calculated using PYTHIA. 
The calculations include the acceptance and reconstruction efficiency of a LHCb-like detector. 
See~\cite{Kikola:2015lka} for more details.
}
\label{fig:An:B}
\end{figure}

\cf{fig:An:D0} shows the estimated statistical precision for charmed-meson $\An$ at backward and mid-rapidity (in the \cms).
As can be seen, even in the case of a moderate target polarisation we would expect a very precise measurement for $\pT\lesssim 5 \ \gev$. 
Such measurements can only be performed at \AFTER .
It can definitely be achieved in the $K\pi$ decay channel, and possibly in the lepton-decay channel, despite of the presence of many more sources for the background (see e.g. \cite{Aidala:2017pum}).

\cf{fig:An:B} shows the estimated statistical precision for open bottom $\An$, measured via non-prompt \jpsi, as a function of $\pT$ at mid-rapidity, which is similar to that of the prompt charmonium at the sub-percent level.
It is thus clear that such a measurement would open a new era of precision studies of STSAs for open heavy-flavour production.

\paragraph{Vector quarkonium production}

\cf{fig:An:Upsi-pp} shows the estimated uncertainties at \AFTER for the $\Upsilon$ $\An$ as a function of $\xF$ in \pp\ collisions at $\sqrt{s} = 115$~GeV, while \cf{fig:An:psi-pp} shows the projected uncertainties for \jpsi compared to the already existing measurements at $\sqrt{s} = 200$~GeV~\cite{Adare:2010bd,Aidala:2018gmp}. 
With the expected charmonium yields~\cite{Massacrier:2015qba} these results are only limited by the systematic uncertainties, and can thus give a very precise access to the gluon content of the proton over a much wider $x$ range than at RHIC.  
As for now, the experimental measurements are compatible with zero, which could be due to the following reasons:
first, the gluon Sivers function might be zero;
second, the effect of the Sivers function might disappear when integrated over $p_T$, in case it has a node;
third, the dominant \jpsi production mechanism might be via colour-octet transitions, whose STSA in $pp$ collisions is thought to vanish as compared to the STSA for colour-singlet transitions~\cite{Yuan:2008vn}.
However, we note that the arguments in support of the latter reason are derived at leading-twist, and thus deviations on the order of $10\div 20\%$ of the measured asymmetry should certainly not be excluded.
On the other hand, the generation of such STSA through a Collins-like fragmentation mechanism has not been discussed in the literature.
In any case, the precise measurements performed within the \AFTER programme can give us a handle to discriminate the discussed scenarios.

\begin{figure}[hbt!]
\centering
\begin{tabular}{cc}
\subfigure[~]{\includegraphics[width=0.48\textwidth]{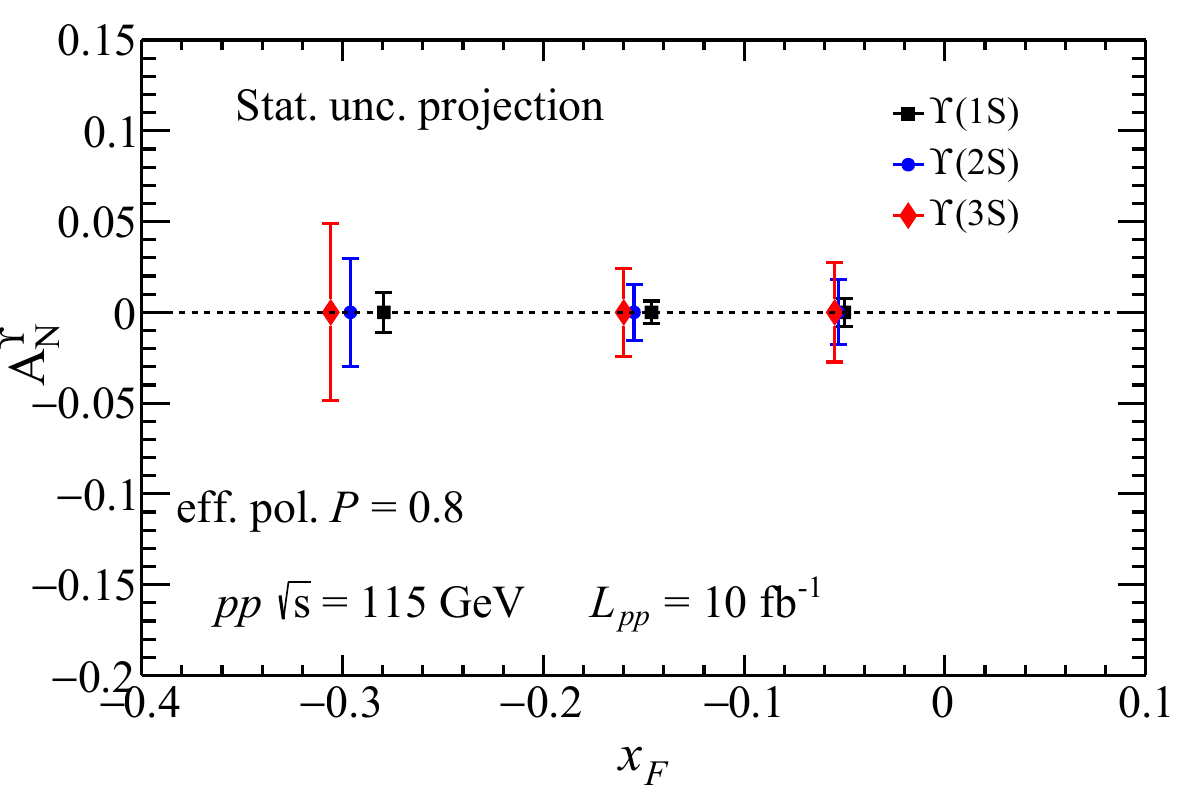}\label{fig:An:Upsi-pp}} &
\subfigure[~]{\includegraphics[width=0.48\textwidth]{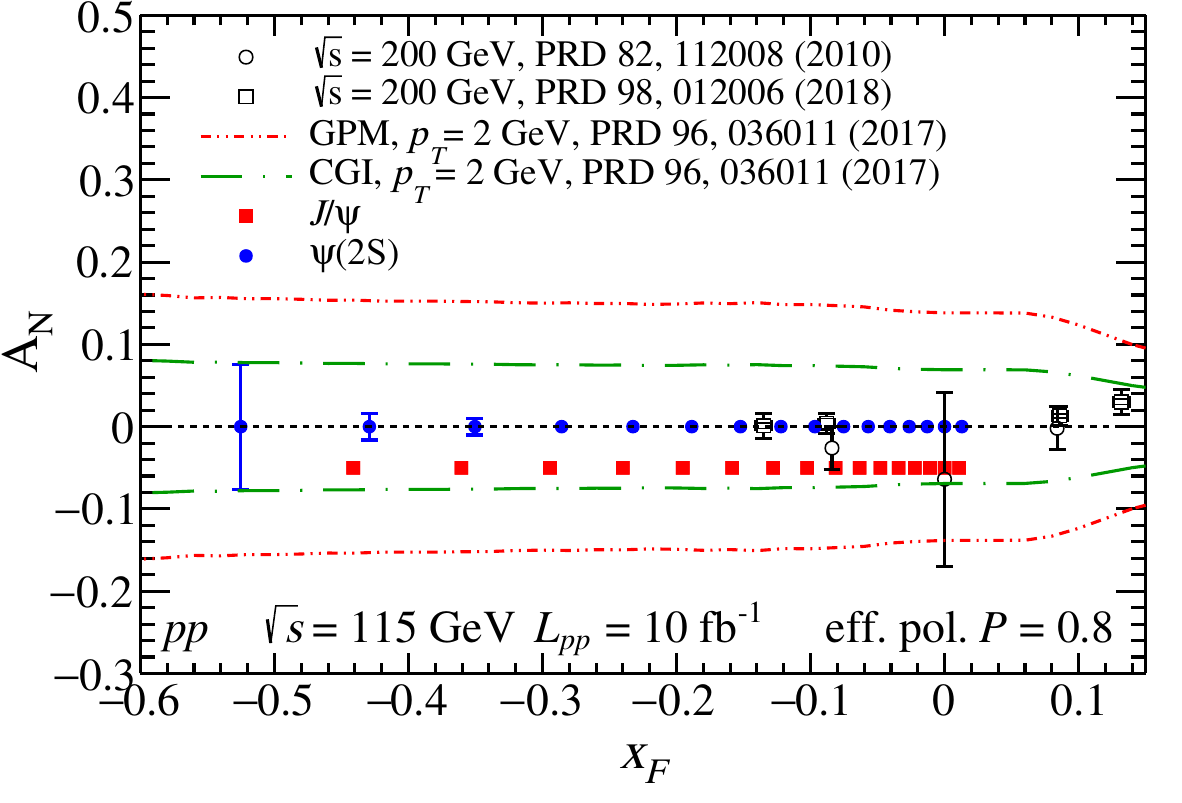}\label{fig:An:psi-pp}} \\
\end{tabular}
\caption{Statistical-precision projections for (a) $\Upsilon(nS)$ and (b) $\psi(nS)$ $\An$ as a function of $\xF$ compared to the existing measurements~\cite{Adare:2010bd,Aidala:2018gmp} and predictions by the generalised parton model (GPM) and the colour gauge invariant (CGI) version of  the GPM model~\cite{DAlesio:2017rzj}. The quarkonium states are assumed to be measured in a di-muon channel with a LHCb-like detector. 
The signal and the background are calculated in realistic simulations that take into account the performance of the LHCb detector~\cite{Massacrier:2015qba,Kikola:2017hnp}.}
\label{fig:An:VectorOnium}
\end{figure}

\begin{figure}[hbt!]
\centering
\includegraphics[width=0.48\textwidth]{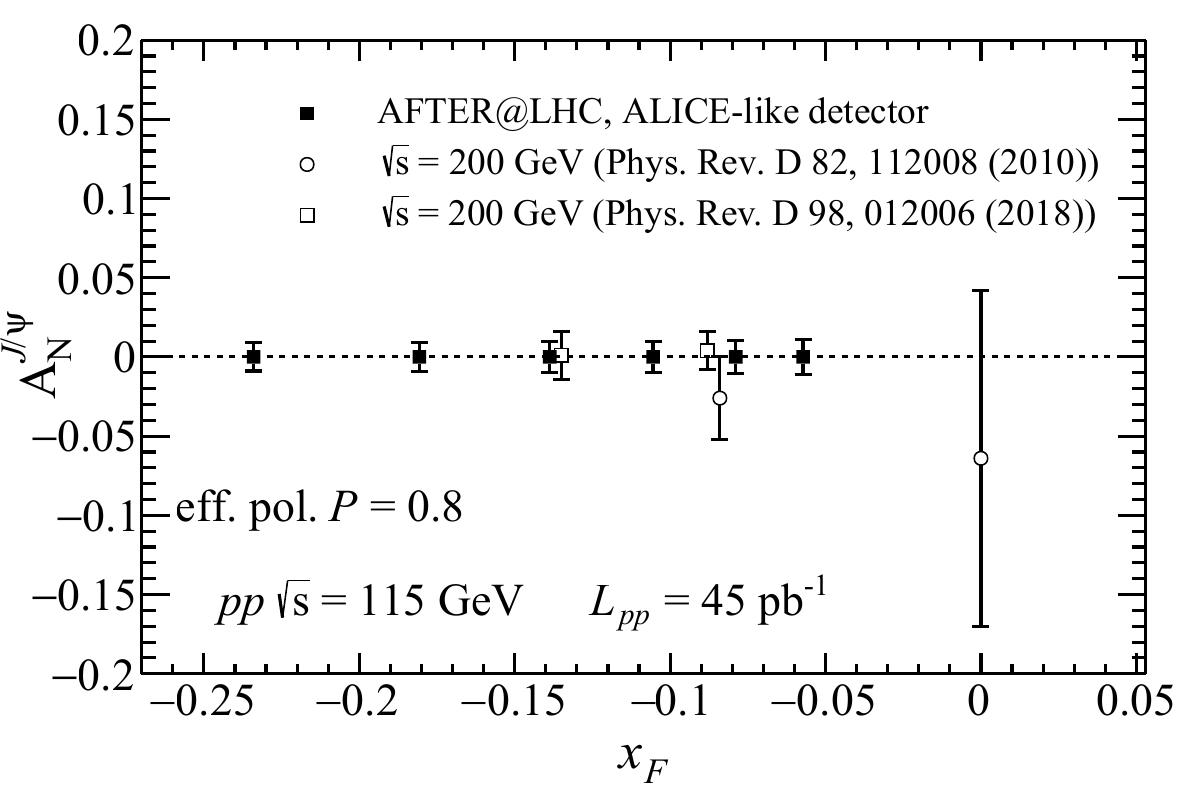}
\caption{Statistical-precision projections for $J/\psi$ $\An$ as a function of $\xF$ compared to the existing measurements~\cite{Adare:2010bd,Aidala:2018gmp} for AFTER@ALICE$_{\mu}$ with target located at the nominal IP ($z_{\rm{target}}\approx 0)$.
The \jpsi\ di-muon spectrum is assumed to be measured in the Muon Spectrometer of the ALICE detector. 
The signal and the background are extrapolated at $\sqrtsNN=115$~GeV from the ALICE measurements in~\cite{Adam:2015rta}. }
\label{fig:An:VectorOnium:ALICE}
\end{figure}

\cf{fig:An:VectorOnium:ALICE} shows the expected statistical precision for \jpsi \An with an ALICE-like detector for $pp^{\uparrow}$ collisions with 45 pb$^{-1}$ of luminosity. 
The expected \jpsi and the background yields  were extrapolated from the \jpsi-rapidity spectrum and the signal-to-background ratios of \cite{Adam:2015rta} with the procedure described in \cite{Kikola:2017hnp}. 
The signal-to-background ratio at 115 GeV is 1.2 and an efficiency of 13\% was assumed~\cite{Abelev:2014qha}.

\paragraph{$C$-even quarkonium production}

The production of $C$-even states can fruitfully be investigated~\cite{Boer:2012bt,Boer:2014tka,Echevarria:2015uaa,Signori:2016jwo}.
With a LHCb-like detector\footnote{That is, assuming a detector that has a good momentum and energy resolution for muons, a decent energy resolution for photons, PID for protons and antiprotons and excellent vertexing capabilities.}, STSA measurements for $\chi_c$, $\chi_b$ and $\eta_c$ are possible as demonstrated by studies of various $\chi_c$ states~\cite{LHCb:2012af,LHCb:2012ac} in the busier collider environment down to a transverse momentum $\pt$ as low as 2 GeV.
LHCb allowed for the first study of inclusive $\eta_c$ production \cite{Aaij:2014bga} above $\pt=6$~GeV as well as non-prompt $\eta_c(2S)$~\cite{Aaij:2016kxn}. Prompt studies are definitely within the LHCb reach~\cite{Lansberg:2017ozx}.
At lower energies, the reduced combinatorial background will definitely allow one to access lower $\pt$ with very reasonable statistics. Indeed, the cross section for  pseudoscalar-charmonium production should be similar to that of the vector ones. The main differences in the expected yields come from likely smaller branchings (with the notable exception of the  $KK\pi$ decay) and detection efficiencies. 
Measuring the STSA of $\eta_c$ also gives  a clean access to tri-gluon correlation functions~\cite{Schafer:2013wca} but also, if one can measure its $\pt$ dependence, to the transverse momentum dependence of the gluon Sivers function relevant for hadron-induced reactions.

\paragraph{Associated production}

\begin{figure}[hbt!]
\centering
\begin{tabular}{cc}
\subfigure[~]{\includegraphics[width=0.48\textwidth]{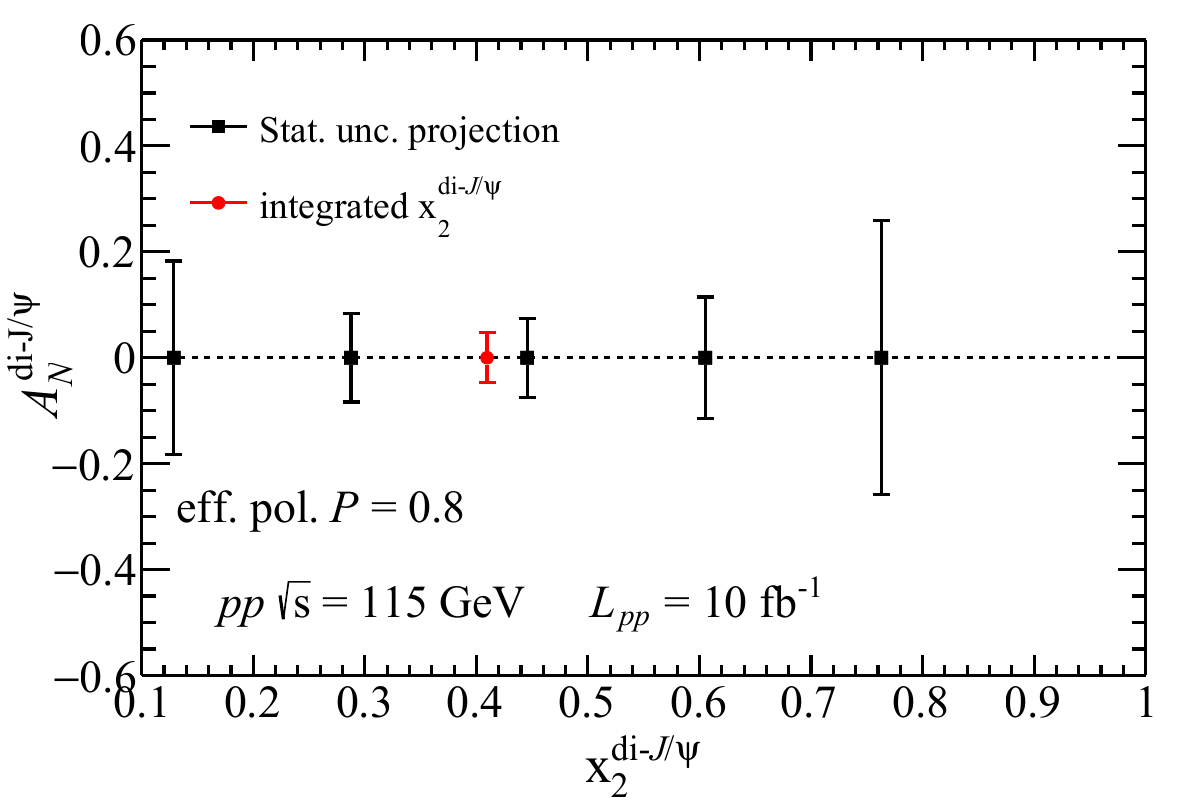}\label{fig:An:diJpsi1}} &
\subfigure[~]{\includegraphics[width=0.48\textwidth]{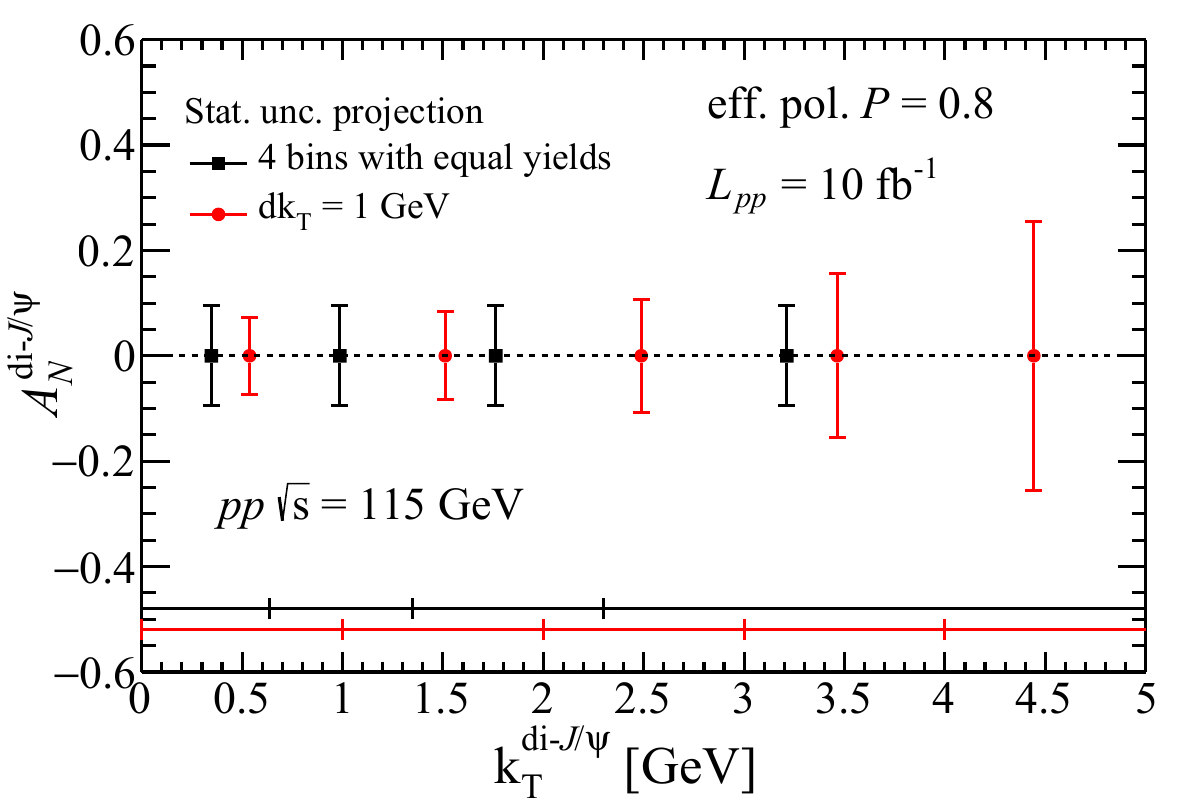}\label{fig:An:diJpsi2}} \\
\end{tabular}
\caption{Statistical-precision projections for di-$\jpsi$ $\An$ as a function of (a) the pair $x_2$ and (b) the pair $k_T$ with a LHCb-like detector. The horizontal lines in (b) denote the width of the $k_T$ bins used for the calculations.}
\label{fig:An:diJpsi}
\end{figure}

Associated-production channels~\cite{Qiu:2011ai,Dunnen:2014eta,Boer:2014lka,Lansberg:2015hla,Signori:2016jwo,Signori:2016lvd,Boer:2016bfj,Scarpa:2019fol} are fundamental tools to access the gluon Sivers effect and also probe the gluon TMD sector in general. 
Even if the production rates are lower than for single production, they would allow one to probe the TMD evolution mechanism~\cite{Echevarria:2012pw,Echevarria:2015uaa} by tuning the mass of the final state.

A few different processes are potentially interesting in this context, for instance $\jpsi-\jpsi$ , $\jpsi-\gamma, \gamma-\gamma$, $\ups-\gamma$. 
The $\jpsi-\jpsi$ production seems to be the most practical one since the yields are not too small~\cite{Lansberg:2015lva} and the measurement is relatively straightforward (compared, for instance, to direct $\gamma$ studies). 
\cf{fig:An:diJpsi} shows the \An\ for double \jpsi\ production as a function of the transverse momentum of the pair, \kT, and the corresponding $x_2$. 
We consider two scenarios for the analysis of $\An$ as a function of \kT: bins with a fixed width of 1~\gev\ ($dk_T = 1\, \gev$, red points) and bins with equal number of yields.
Here, we model the \kT\ dependence as a Gaussian distribution with the width $\sigma = 2 \, \gev$. 
The $x_2$-integrated \An\ will allow for the determination of the STSA with a few percent precision, and the $A_N(k_T)$ gives access --for the first time-- to the $k_T$ dependence of the gluon Sivers TMD up to $\kT \approx 4 \, \gev$.

%% file: physics-spin/physics-spin_3-quark-BM.tex
In section \ref{ss:quark_sivers} we discussed the extraction of the Sivers asymmetry from the DY production cross-section.
However this process can also give valuable information on other asymmetries, and thus on other TMDs.
In fact, the cross-section for a transversely polarised target (and an unpolarised beam) can be schematically written in terms of the following structure functions~\cite{Arnold:2008kf}:
\begin{align}
A_{UU}^{cos2\phi} &\sim
\frac{h_1^{\perp q}(x_1,k_{1T}^2)\otimes h_1^{\perp \bar q}(x_2,k_{2T}^2)}
{f_1^{q}(x_1,k_{1T}^2)\otimes f_1^{\bar q}(x_2,k_{2T}^2)}
\,,
\\
A_{UT}^{sin\phi_S}  &\sim
\frac{f_1^{q}(x_1,k_{1T}^2)\otimes f_{1T}^{\perp \bar q}(x_2,k_{2T}^2)}
{f_1^{q}(x_1,k_{1T}^2)\otimes f_1^{\bar q}(x_2,k_{2T}^2)}
\,,
\\
A_{UT}^{sin(2\phi+\phi_S)}  &\sim
\frac{h_1^{\perp q}(x_1,k_{1T}^2)\otimes h_{1T}^{\perp \bar q}(x_2,k_{2T}^2)}
{f_1^{q}(x_1,k_{1T}^2)\otimes f_1^{\bar q}(x_2,k_{2T}^2)}
\,,
\\
A_{UT}^{sin(2\phi-\phi_S)}  &\sim
\frac{h_1^{\perp q}(x_1,k_{1T}^2)\otimes h_1^{\bar q}(x_2,k_{2T}^2)}
{f_1^{q}(x_1,k_{1T}^2)\otimes f_1^{\bar q}(x_2,k_{2T}^2)}
\,,
\end{align}
where $h_{1}^q$ is the transversity, $h_{1}^{\perp q}$ the Boer-Mulders function and $h_{1T}^{\perp q}$ the pretzelosity
($f_1^q$ and $f_{1T}^{\perp q}$ are the already introduced unpolarised TMD PDF and the Sivers function, respectively).
Again $\otimes$ stands for a convolution in momentum space, and a sum over parton flavours is understood.
The superscript on the $A$'s means that we weight the cross-section with that angular term to single out the corresponding angular modulation.

Let us focus on the Boer-Mulders function $h_1^{\perp}$, which encodes the correlation between the quark transverse spin and its transverse momentum, namely it represents a spin-orbit effect for the quark inside an unpolarised proton.
This function, like the quark Sivers function, is \emph{naive} time-reversal odd (T-odd), and thus it changes sign under time-reversal transformations~\footnote{\emph{Naive} time reversal stands for time reversal but without the interchange of initial and final states \cite{Collins:2011zzd}.}.
In particular, a sign change is predicted for $h_1^{\perp}$ probed in SIDIS and DY production.
Moreover, it might help explain~\cite{Lambertsen:2016wgj} the violation of the Lam-Tung relation in unpolarised DY reaction~\cite{Lam:1980uc}.
Hints about the transverse momentum dependence of the Boer-Mulders function $h_1^{\perp}$ have been extracted from SIDIS data in~\cite{Barone:2009hw}. 
\AFTER will contribute to the study of the Boer-Mulders function in DY production, shedding light on its process dependence and on the TMD formalism in general.

\begin{figure}[hbt!]
\centering
\subfigure[~]{\includegraphics[width=0.32\textwidth]{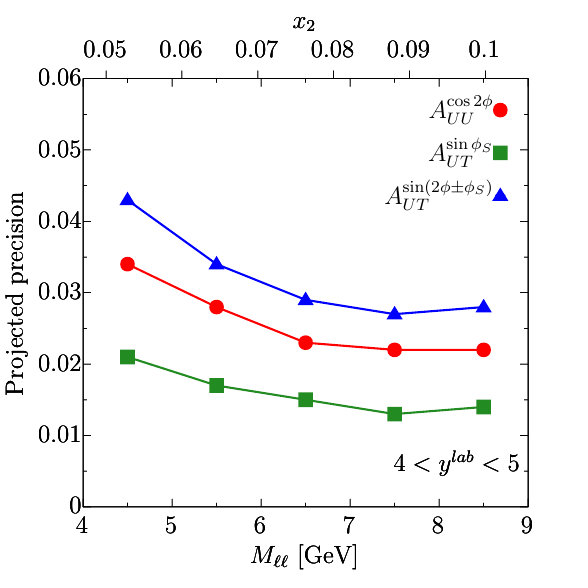}\label{fig:AN_unc_y4p5}} 
\subfigure[~]{\includegraphics[width=0.32\textwidth]{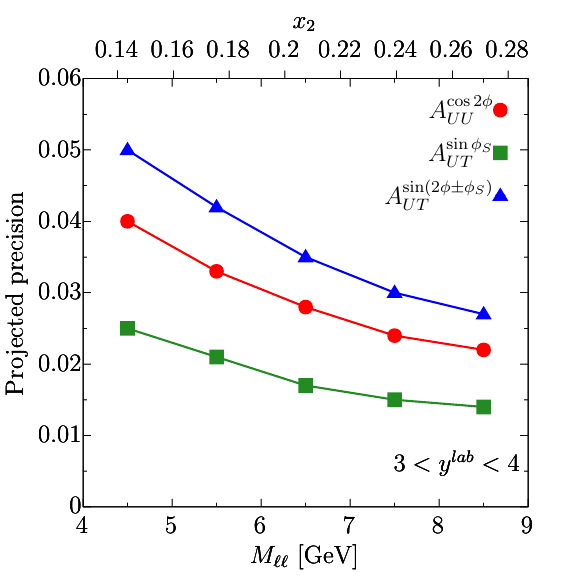}\label{fig:AN_unc_y3p5}} 
\subfigure[~]{\includegraphics[width=0.32\textwidth]{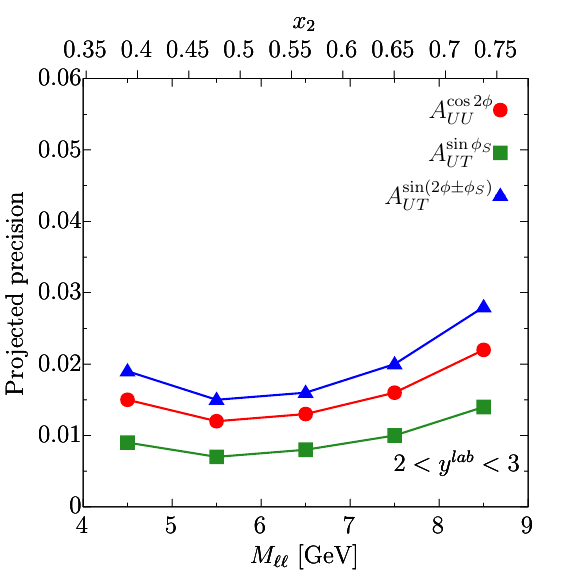}\label{fig:AN_unc_y2p5}} 
\caption{Expected statistical uncertainty 
on asymmetries in DY production at \AFTERLHCb, computed all for ${\cal L}_{pp}=10$~fb$^{-1}$ and ${\cal P}_{\rm eff.}=0.8$. The rapidity has been integrated over the bins specified in the plots, as well as the mass in bins of $dM=1$~GeV.
[The statistical uncertainties are calculated using following expressions: 
$\delta(A_{UT}^{\sin \phi_S}) = {1}/{{\cal P}_{\rm eff.}} \times {\sqrt{2}}/{\sqrt{S+2B}}$, $\delta(A_{UU}^{\cos 2\phi_S}) = {2\sqrt{2}}/{\sqrt{S+2B}}$ and $\delta(A_{UT}^{\sin(2\phi \pm \phi_S)}) = {2}/{{\cal P}_{\rm eff.}} \times {\sqrt{2}}/{\sqrt{S+2B}}$, 
where $S$ is the signal yield, $B$ is the background yield and ${\cal P}_{\rm eff.}$ is the effective polarisation in a given measurement.] 
}
\label{fig:AN_unc}
\end{figure}

In \cf{fig:AN_unc} we show the expected precision achievable at \AFTER for different angular modulations of the DY production cross-section in different kinematic regions (rapidity, invariant mass, momentum fraction in the (un)polarised target nucleon). We note that $A_{UU}^{\cos 2 \phi}$ could be measured without a polarised target and that asymmetries with faster modulations are usually determined with a poorer precision.

%% file: physics-spin/physics-spin_4-gluon-BM.tex
In the quark case, there are two leading-twist TMDs, as we have discussed, the unpolarised $f_1^q(x,k_T^2)$ and the Boer-Mulders $h_1^{\perp q}(x,k_T^2)$ functions. 
For a gluon in an unpolarised proton, the relevant functions are the unpolarised distribution $f_1^g(x,k_T^2)$ and the distribution of linearly polarised gluons $h_1^{\perp g}(x,k_T^2)$~\cite{Mulders:2000sh,Boer:2016xqr}.
 
The phenomenology of $h_1^{\perp g}$ is potentially easier than that for the Boer-Mulders function in the quark case, because it is T-even and matched onto the twist-2 unpolarised collinear distributions $f_1^{g,q}$, whereas $h_1^{\perp q}$ is matched onto the twist-3 collinear matrix elements, which are so far unknown. 
However, no experimental extractions of $h_1^{\perp g}$ have been performed yet.
Recently, it has been proposed to access both $f_1^g$ and $h_1^{\perp g}$ in di-$J/\psi$ and $\Upsilon$ production in hadronic collisions \cite{Lansberg:2017dzg,Scarpa:2019fol}, for which data with sensitivity to transverse momenta have been collected at the LHC. 
It is expected that $h_1^{\perp g}$ reaches its maximal size in the small-$x$ regime \cite{Boer:2016xqr,Meissner:2007rx,Metz:2011wb,Dominguez:2011br}.
Its role in different $x$-regions has yet to be explored. Factorisation proofs have recently been provided for $\eta_{c,b}$ production~~\cite{Echevarria:2019ynx,Fleming:2019pzj}.
It is also expected to be constrained from azimuthal-asymmetry measurements at the future EIC and the LHeC~\cite{Boer:2010zf,Pisano:2013cya}, and also possibly from measurements at RHIC and the LHC~\cite{Qiu:2011ai}.

The impact of linearly polarised gluons in $H^0$ production has been addressed \eg\ in \cite{Catani:2010pd,Boer:2011kf,Boer:2013fca,Echevarria:2015uaa}.
Their effect has been predicted for gluon fusion into two photons in~\cite{Nadolsky:2007ba,Qiu:2011ai}, for (pseudo)scalar quarkonium production in~\cite{Boer:2012bt,Signori:2016jwo}, for vector quarkonium production in~\cite{Mukherjee:2015smo,Mukherjee:2016cjw} and for $H^0$ plus jet production in \cite{Boer:2014lka}. 
Associated production of quarkonium and $Z$ boson has been investigated in \cite{Gong:2012ah}.
Associated production of quarkonium plus one photon~\cite{Dunnen:2014eta} is also promising, due to the possibility of producing final states with different invariant masses, suited thus to be analysed using TMD factorisation and to test TMD evolution.
This process, together with $\eta_{b,c}$ production~\cite{Ma:2012hh,Boer:2012bt,Signori:2016jwo} and double $J/\psi$ production~\cite{Lansberg:2015lva}, can be investigated within the \AFTER programme.

\begin{table}[hbt!]
\begin{center}\renewcommand{\arraystretch}{1.2}
\begin{tabular}{c|c|c|c|c}
Process& Expected yield & $x_2$ range & $M$ [GeV] & $q_T$ modulation \\
\hline
\hline
$\eta_{c}$~\cite{Boer:2012bt,Signori:2016jwo} & ${\cal O}(10^6)$ & $0.02\div 0.5$ & ${\mathcal O}(3)$ & $0\div 80 \%$ \\
\cline{1-5}
$\chi_{c0}(1P)$~\cite{Boer:2012bt} & ${\cal O}(10^4)$ & $0.02\div 0.5$ & ${\mathcal O}(3)$ & $0\div 80 \%$  \\
\cline{1-5}
$\chi_{c2}(1P)$~\cite{Boer:2012bt} & ${\cal O}(10^6)$ & $0.02\div 0.5$ & ${\mathcal O}(3)$ & $<1\%$ \\
\cline{1-5}
$\chi_{b0}(nP)$~\cite{Boer:2012bt} & ${\cal O}(10^2)$ & $0.1\div 1$ & ${\mathcal O}(10)$ & $0\div 60 \%$ \\
\cline{1-5}
$\chi_{b2}(nP)$~\cite{Boer:2012bt} & ${\cal O}(10^3)$ & $0.1\div 1$ & ${\mathcal O}(10)$ & $<1\%$ \\ \hline
\end{tabular}
\caption{
Expected $q_T$ modulations generated by $h_{1}^{\perp g}$ for a selection of quarkonium-production observables, along with the expected yields and $x_2$ ranges derived from $x_2=M e^{y_{\cms}}/\sqrt{s}$ for a rapidity coverage $-2.8<y_{\cms}<0.2$ and $\sqrt{s}=115$ GeV. The modulation expectations are meant to approximately account for TMD evolution effects~\cite{Signori:2016jwo}.
}	 
\label{t:processes1}
\end{center}
\end{table}

\begin{table}[hbt!]
\begin{center}\renewcommand{\arraystretch}{1.2}
\begin{tabular}{c|c|c|c|c|c}
Process & Expected yield & $x_2$ range & $M$ [GeV] & $\cos2\phi$ modulation& $\cos4\phi$ modulation\\
\hline
\hline
$J/\psi + \gamma$~\cite{Dunnen:2014eta} & $1000\div 2000$ & $0.1\div 0.6$ & ${\mathcal O}(10)$ & $0\div 5\%$ & $0\div 2\%$ \\
\cline{1-6}
$J/\psi + J/\psi$~\cite{Lansberg:2017dzg} & $300\div 1500$ & $0.1\div 0.8$ & $8\div 12$ & $0\div 8\%$ & $0\div 20\%$ \\ \hline
\end{tabular}
\caption{
Expected azimuthal asymmetries generated by $h_{1}^{\perp g}$ for a selection of quarkonium-associated-production observables, along with the expected yields and $x_2$ ranges derived from $x_2=M e^{Y_{\cms}}/\sqrt{s}$ for a rapidity coverage $-2.8<Y_{\cms}<0.2$ and $\sqrt{s}=115$ GeV ($Y_{\cms}$ refers to the rapidity of the observed 2-particle system). The modulation expectations are meant to approximately account for TMD evolution effects~\cite{Scarpa:2019fol}. 
}	 
\label{t:processes2}
\end{center}
\end{table}

Several processes can be measured at the proposed \AFTER programme in order to constrain $h_1^{\perp g}$ in yet unexplored kinematic regions.
In \ct{t:processes1} we show those in which the effect of the presence of $h_1^{\perp g}$ is the modulation of the transverse-momentum spectrum,  referred to as ``$q_T$ modulation", while in \ct{t:processes2} we show those for which $h_1^{\perp g}$ creates an azimuthal modulations of the spectrum, referred to as ``$\cos n\phi$ modulation".
We notice that in all the mentioned processes the same $h_1^{\perp g}$ function is probed, since the gauge-link structure is the same.
As can be seen, overall the \AFTER programme offers a great opportunity to constrain $h_1^{\perp g}$ through all these processes.

At \AFTER, it will be possible to study the potential TMD factorisation breaking effects~\cite{Ma:2014oha} in the production of 
$\chi_{c0}$ and  $\chi_{c2}$~\cite{Boer:2012bt}.  
Moreover, $\eta_c$ production at low transverse momentum~\cite{Echevarria:2019ynx} will be accessed, complementing the high transverse momentum region measured by LHCb and going beyond RHIC's capabilities. 

As already mentioned for the case of quark TMDs, the \AFTER\ program can be useful also to better constrain the simplest of the gluon TMDs, \ie\ the unpolarised gluon TMD PDF, which remains so far unconstrained (see however a first attempt in \cite{Lansberg:2017dzg}). 
Like its quark counterpart, it enters the denominators of all spin and azimuthal asymmetries, and thus its knowledge is fundamental in order to reliably study any TMD-related asymmetry.

%% file: physics-spin/physics-spin_5-OAM.tex
\newcommand{\ud}{d}
\newcommand{\uvec}[1]{\boldsymbol{#1}}

On top of providing a handle on the intrinsic spin of partons, the TMD formalism intuitively connects with the orbital angular momentum of the quarks through the correlations proportional to the partonic transverse momenta.
In fact, conservation of total angular momentum imposes that off-diagonal TMDs (i.e. those with $\Delta\lambda=(\Lambda'-\Lambda)-(\lambda'-\lambda)\neq 0$, where $\Lambda^{(\prime)}$ and $\lambda^{(\prime)}$ are the initial (final) target and parton light-front helicities) would vanish in absence of orbital angular momentum (OAM). 
Some of these off-diagonal TMDs appear to be experimentally sizeable \cite{Airapetian:2004tw,Alexakhin:2005iw,Boglione:2015zyc}, confirming henceforth the presence of a significant amount of OAM inside the nucleon. 
This rises the question as to whether TMDs can be used to quantify more precisely the OAM.

It has been observed within many effective quark models that the expectation value of the \emph{canonical} quark OAM can be expressed in terms of some TMDs~\cite{Ma:1998ar,She:2009jq,Avakian:2010br,Efremov:2010cy}

\begin{equation}\label{LzTMD}
\begin{aligned}
\langle L_{\text{can}}^q \rangle
&=\int\ud x\,\ud^2k_\perp\left[h^q_1(x,\uvec k^2_\perp)-g^q_{1L}(x,\uvec k^2_\perp)\right],\\
&=-\int\ud x\,\ud^2k_\perp\,\tfrac{\uvec k_\perp^2}{2M^2}\,h_{1T}^{\perp q}(x,\uvec k^2_\perp).
\end{aligned}
\end{equation}

Unfortunately, it has also been shown that the validity of these relations cannot be extended to QCD~\cite{Lorce:2011dv,Lorce:2011kn,Lorce:2011zta}. 
Although not exact, they remain phenomenologically interesting as they provide at least some indication about the sign and the magnitude of the canonical quark OAM. Note that the results can be quite different~\cite{Burkardt:2008ua,Lorce:2011kd} from the \emph{kinetic} quark OAM derived from the Generalised Parton Distributions (GPDs) through the Ji relation~\cite{Ji:1996ek} which contains also quark-gluon interactions.

Interestingly, Burkardt~\cite{Burkardt:2002ks,Burkardt:2003uw} suggested that the quark Sivers TMD $f_{1T}^{\perp q}(x,\uvec k^2_\perp)$ and the quark GPD $E^q(x,\xi,t)$ could be related by
a chromodynamic lensing mechanism 
\begin{equation}
\begin{aligned}
&\int\ud^2k_\perp\,\tfrac{\uvec k^2_\perp}{2M^2}\,f_{1T}^{\perp q}(x,\uvec k^2_\perp)
\propto
\int \ud^2b_\perp \,\bar{\mathcal I}(\uvec b_\perp)
\,(\uvec S_T\times\uvec \partial_{b_\perp})_z\, {\cal E}^q(x, \uvec b_\perp^2),
\end{aligned}
\end{equation}
where $\bar{\mathcal I}(\boldsymbol b_\perp)$ is called the lensing function and ${\cal E}^q(x,\boldsymbol{b}_\perp^2) =\int \frac{d^2 \Delta_\perp}{(2\pi)^2} \, e^{-i\boldsymbol{b}_\perp \cdot \boldsymbol{\Delta}_\perp}\, E^q(x,0,-\boldsymbol{\Delta}_\perp^2)$. 
$\uvec S_T$ is the transverse spin of the proton, while $\xi$ and $t=-\boldsymbol{\Delta}_\perp^2$ are the longitudinal and transverse components of the momentum transfer.
The Sivers function could then be used to constrain the GPD $E^q$ and hence the kinetic OAM via the Ji relation. 
Despite some support from model calculations~\cite{Burkardt:2003uw,Burkardt:2003je,Meissner:2007rx,Gamberg:2009uk}, such a relation can hardly be put on a firmer theoretical grounds. 
A variation of it has however been used by Bacchetta and Radici~\cite{Bacchetta:2011gx} to fit SIDIS data for the Sivers effect with the integral constrained by the anomalous magnetic moments, leading to a new estimate of the total angular momentum $\langle J^{q}_\text{kin}\rangle$, in good agreement with most common GPD extractions~\cite{Guidal:2004nd,Diehl:2004cx,Ahmad:2006gn,Goloskokov:2008ib,Diehl:2013xca}. 
A similar relation may a priori hold in the gluon sector, but has never been investigated so far.

In fact, one should not expect any \emph{direct} quantitave relation between TMDs and OAM since the latter requires some information about the correlation between the position and the momentum of the parton, information which is integrated out at the TMD level. 
However, TMDs provide essential information about several angular-momentum correlations~\cite{Lorce:2015sqe} and can be used to constrain, to some extent, the nucleon wave function, \emph{indirectly} providing us  with valuable information about its OAM content.
Thus the \AFTER\ programme, with its unique capabilities of measuring quark and gluon TMDs, and in particular Sivers functions, can shed light on the partonic OAM by giving us a handle to constrain it.

%% file: physics-spin/physics-spin_6-UPC.tex
Ultraperipheral collisions (UPCs) provide a unique way to study photoproduction processes in hadron-hadron interactions \cite{Baltz:2007kq}, also in the fixed target experiments \cite{Lansberg:2015kha}. 
Such processes are conveniently described in the Equivalent Photon Approximation, where the relation between the hadron-hadron cross section, $(d)\sigma^{h_Ah_B}$,
and the (differential) cross section for a photo-hadron scattering ($h_A$ or $h_B$), $(d)\sigma^{\gamma h_{A,B}}$, is naturally given by the following convolution in the photon momentum, $k_\gamma$, of the photon fluxes $dn/dk_\gamma$  from each hadron and the aforementioned photo-hadron cross section:
\begin{eqnarray}
\label{eq:intWW}
d\sigma^{h_Ah_B}=
\int dk_\gamma \Big[ \frac{dn^{h_A}}{dk_\gamma} \, d\sigma^{\gamma h_B}(k_\gamma)+
 \frac{dn^{h_B}}{dk_\gamma} \, d\sigma^{\gamma h_A}(k_\gamma)\Big]\quad.
\end{eqnarray}
The relevant parameters of such photon beams, for various projectiles and targets at AFTER@LHC, can be found in Table 1 of \cite{Lansberg:2015kha}.

Exclusive photoproduction processes which can be studied in the UPCs allow one to probe the internal structure of hadrons in terms of GPDs \cite{Diehl:2003ny, Belitsky:2005qn}, which through the Ji's sum rule are directly related to the total OAM carried by quarks and gluons. They also allow one to explore the tri-dimensional ``tomography'' of hadrons~\cite{Burkardt:2000za}.

\begin{table}[htpb]
\begin{center}\renewcommand{\arraystretch}{1.3}
\begin{tabular}{c|c|c|c}         
& $\sigma_{BH} (\textrm{pb})$   &Luminosity& Events year $^{-1}$  	\\ \hline \hline
 $p$ on Pb & $1940 ~\textrm{pb}$   & 0.16 fb$^{-1}$year $^{-1}$	 & $3 \times 10^5$	\\ \hline
 $p$ on H  & $7.1 ~\textrm{pb}$    & 20 fb$^{-1}$ year$^{-1}$		 & $1.4\times 10^5$	\\ \hline
 Pb on H & $5500 ~\textrm{pb}$	 & 11 nb$^{-1}$ year$^{-1}$		 & $6\times 10^3$ \\ \hline
\end{tabular}
\end{center}
\caption{The Bethe-Heitler cross section and number of events predicted~\cite{Lansberg:2015kha} for three different modes of operation for \AFTER [Note that the assumed luminosity in~\cite{Lansberg:2015kha} may not correspond to yearly
luminosities].}
\label{tab:BH}
\end{table}

One of such exclusive processes, that has not yet been measured, is Timelike Compton Scattering (TCS)~\cite{Berger:2001xd}, contributing to  exclusive lepton-pair photoproduction.  
Although the purely electromagnetic Bethe-Heitler (BH) amplitude gives much larger contributions than the TCS one, it is possible to study the interference term between TCS and BH processes, which may be projected out through the analysis of the angular distribution of the produced leptons. Such an interference then allows one to extract information on the GPDs.
The ratio of these two contributions for the kinematics relevant to \AFTER\ was found to be of the order of 10 \% \cite{Lansberg:2015kha}.

\ct{tab:BH} gathers the BH cross section, the luminosity and the yearly expected lepton-pair yields in 3 operation modes\footnote{The Pb target should be considered as an illustrative case for a heavy nuclear target.} in the kinematical region where the TCS-signal extraction is possible.
With a magnitude of 10 \% for the interference term, the azimuthal modulation should be observable in the 3 cases.

We also note that the cross sections for $\eta_c$ production by photon-pair fusion has been derived in \cite{Lansberg:2015kha}. 
This process, in particular in this energy range, is sensitive to the method used to compute the photon flux in the \pp\ case. We found out that $10^4$ $\eta_c$ can be produced per year in $pp$ UPCs with \AFTER. 
The same paper also discussed possible competing hadronic processes via pomeron or odderon exchanges. These could be separated out by a careful analysis of the transverse-momentum dependence of the produced particles.
$\eta_c$ production from $\gamma\gamma$ fusion in UPCs at \AFTER\ was also discussed in~\cite{Goncalves:2015hra}.

\begin{figure}[!hbt]
\centering
\subfigure[~]{\includegraphics[width=0.48\textwidth,clip]{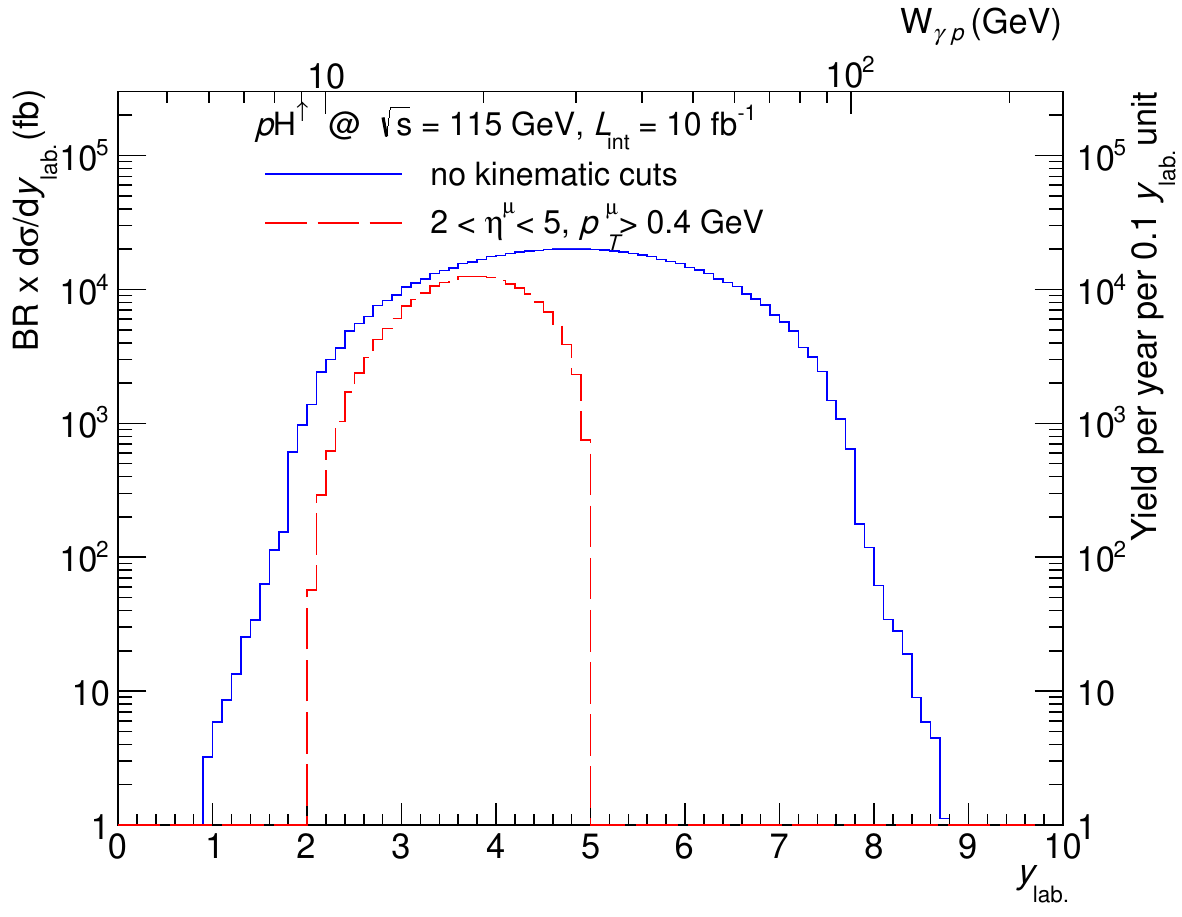}} 
\subfigure[~]{\includegraphics[width=0.48\textwidth,clip]{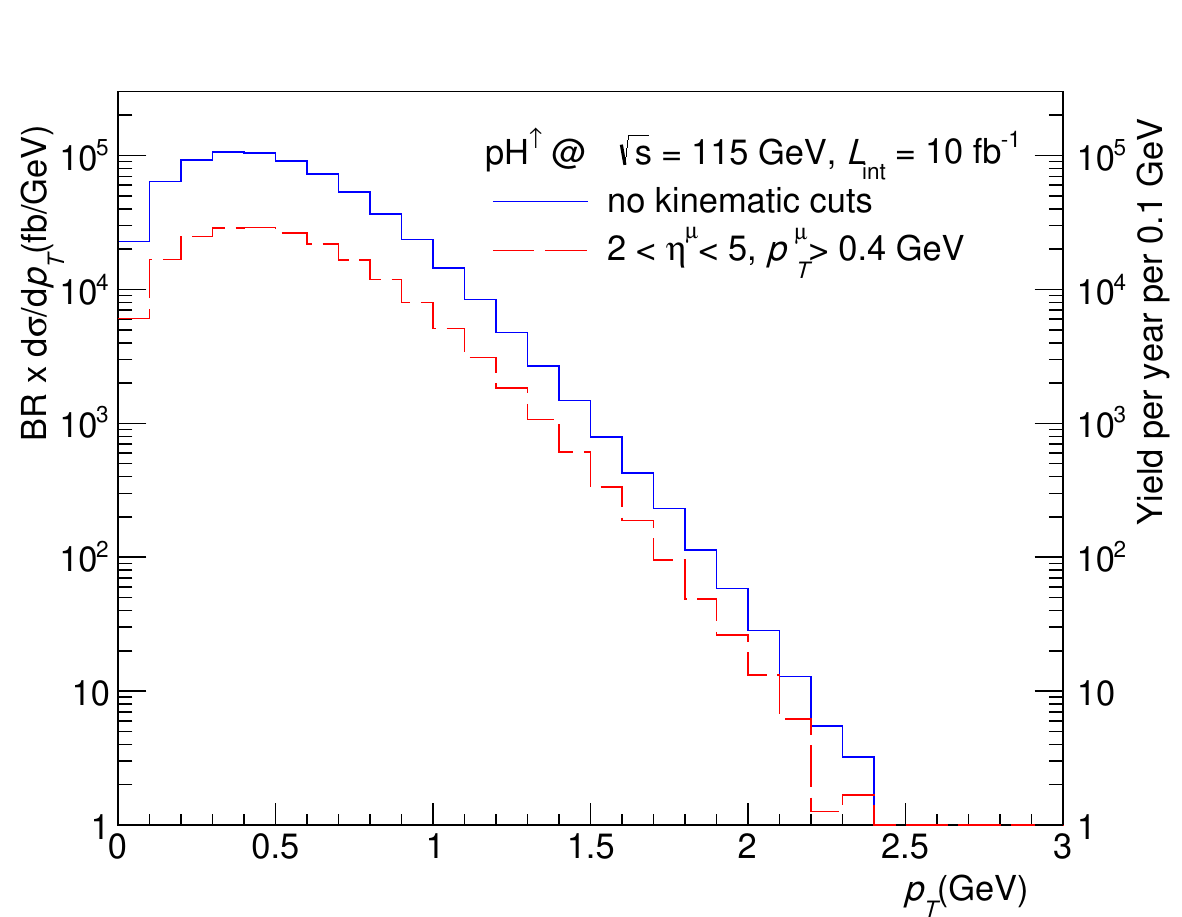}} 
\caption{Rapidity-differential (a) and $\pT$-differential (b) cross sections of the photoproduced \jpsi\ in the laboratory frame for $p$H with 7 TeV proton beam, from the STARLIGHT generator~\cite{Klein:2016yzr}. The blue curves have been produced without applying kinematical cut, while the red curves are produced by applying cuts on the two daughters of the \jpsi\ (2 $< \eta^{\mu} <$ 5 and $\pT^{\mu} >$~0.4~\gevc) which correspond to a LHCb-like detector. For (a), the upper $x$ axis indicates the corresponding invariant mass of the initial $\gamma p$ system, $W_{\gamma p}$, which is equal to that of the final $J/\psi p$ system. [Adapted from \cite{Lansberg:2018fsy}.]}
\label{pH_AFTER}       %
\end{figure}

\begin{figure}[!hbt]
\centering
\subfigure[~]{\includegraphics[width=0.48\textwidth,clip]{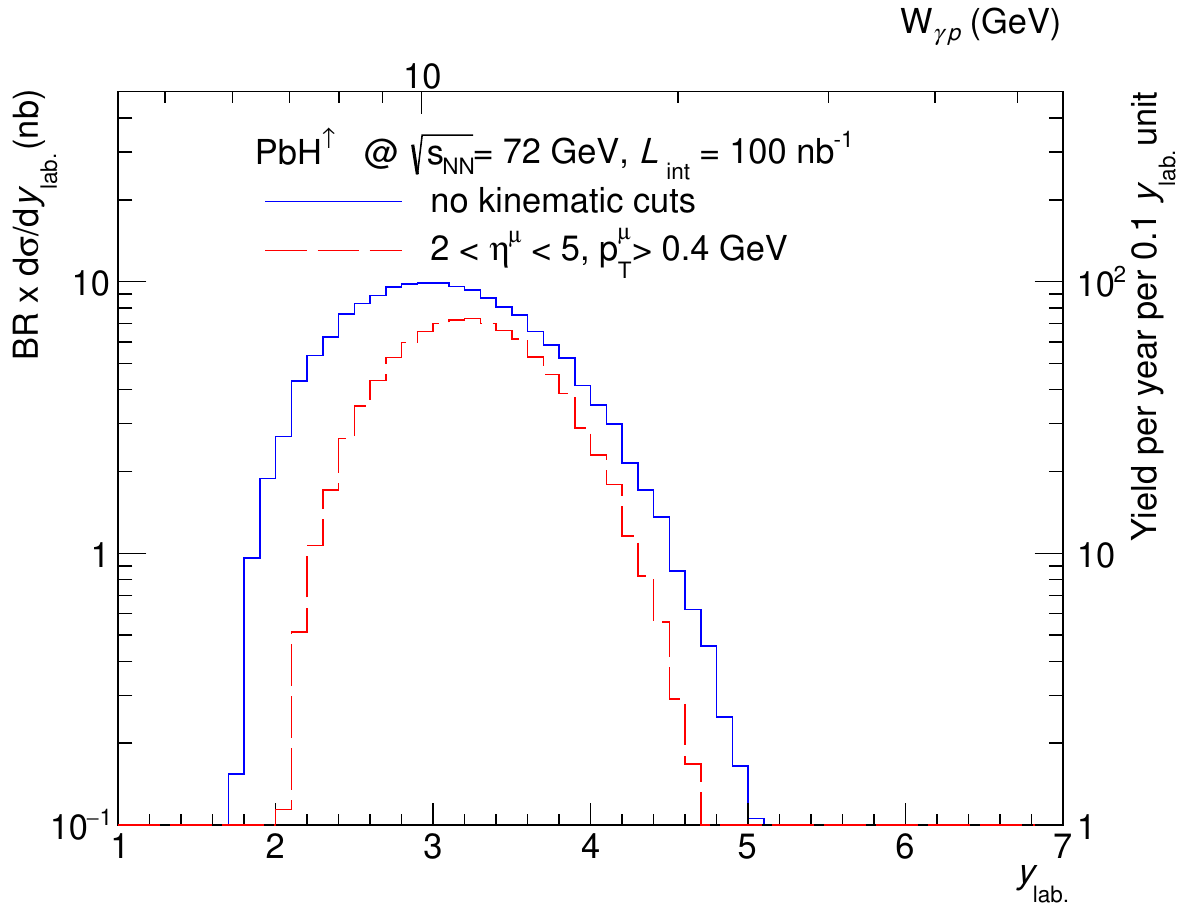}} 
\subfigure[~]{\includegraphics[width=0.48\textwidth,clip]{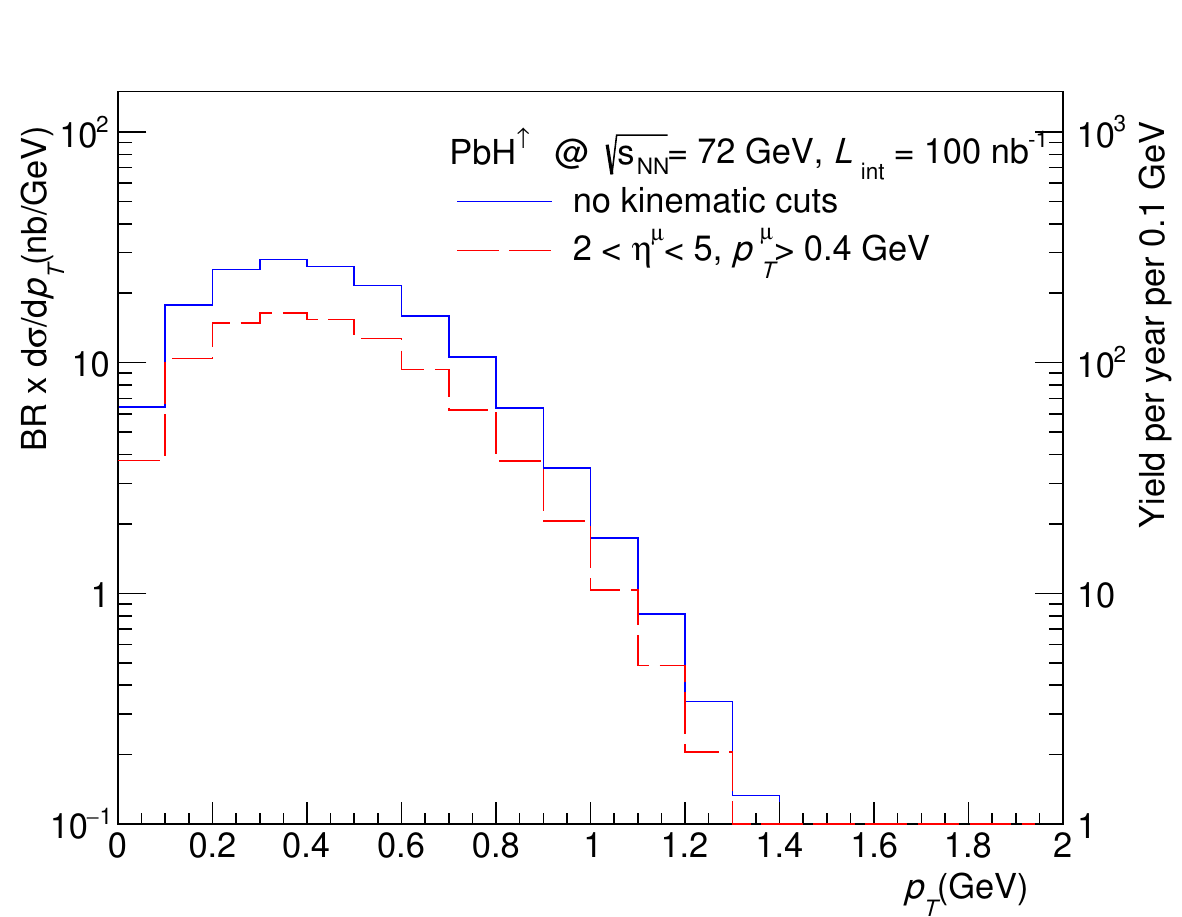}}
\caption{Rapidity-differential (a) and $\pT$-differential (b) cross sections of the photoproduced \jpsi\ in the laboratory frame for PbH with 2.76 A TeV Pb beam, from the STARLIGHT generator~\cite{Klein:2016yzr}. The blue curves have been produced without applying kinematical cut, while the red curves are produced by applying cuts on the two daughters of the \jpsi\ (2 $< \eta^{\mu} <$ 5 and $\pT ^{\mu} >$~0.4~\gevc) which correspond to a LHCb-like detector. [Adapted from \cite{Lansberg:2018fsy}.]}
\label{PbH_AFTER}       %
\end{figure}

\begin{figure}[!hbt]
\centering
\subfigure[~$p$H with 7 TeV proton]{\includegraphics[width=8.cm,clip]{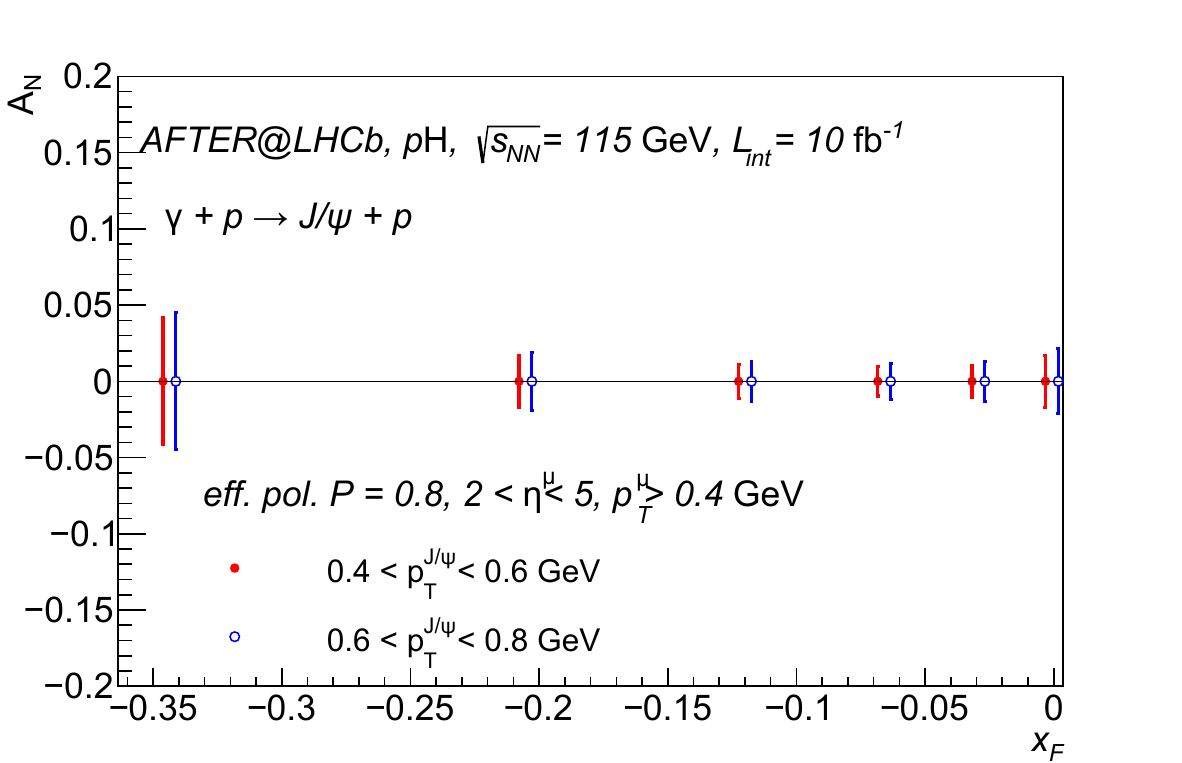}}
\subfigure[~PbH with 2.76 TeV Pb]{\includegraphics[width=8.cm,clip]{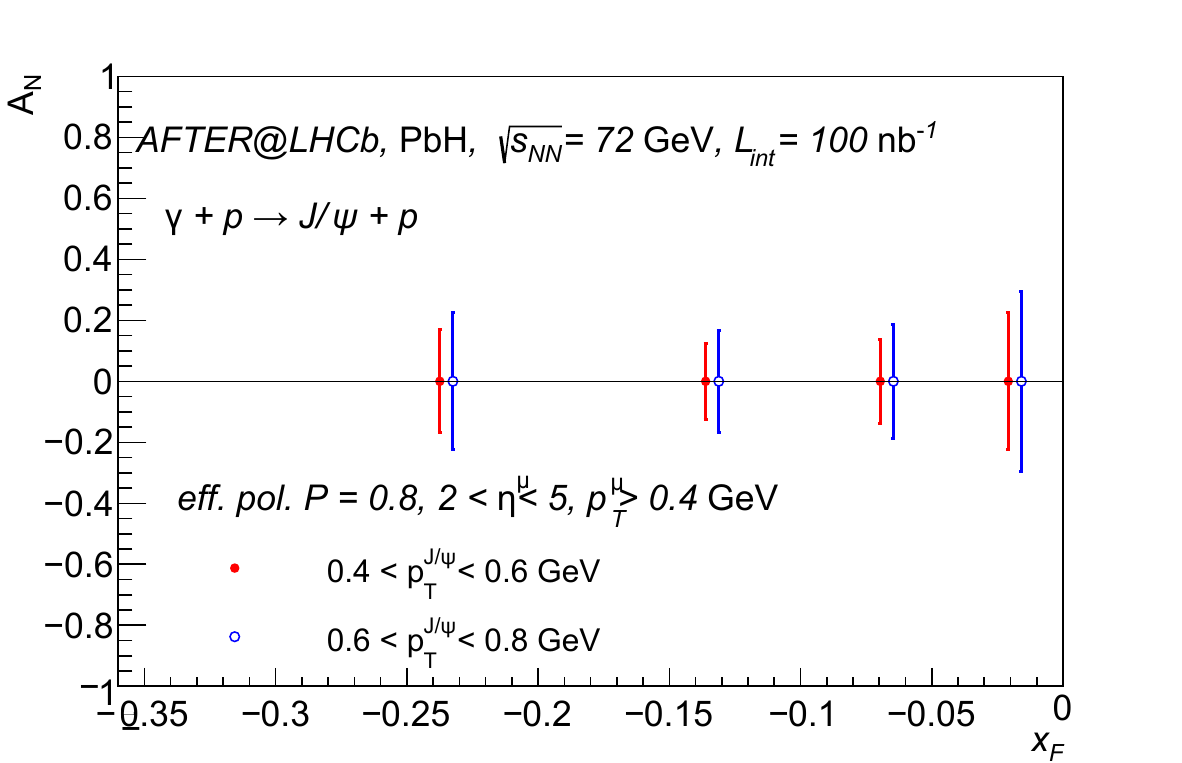}}
\caption{Statistical uncertainty projections for the STSA in exclusive $J/\psi$ photoproduction in UPCs of a proton beam (a) and a lead beam (b) on a hydrogen target. [Adapted from \cite{Lansberg:2018fsy}.]}
\label{AN_pH}       %
\end{figure}

Exclusive $J/\psi$ production \cite{Ivanov:2004vd} draws a lot of attention due to the fact that at the leading order it is only sensitive to gluon GPDs. 
It has already been measured in the ultraperipheral collisions at LHCb, ALICE and CMS in the collider mode. 
However, the \AFTER\ programme would create a unique possibility to study STSAs in such a process~\cite{Massacrier:2017lib}, which are sensitive to the yet unknown GPD $E_g$~\cite{Koempel:2011rc}, an important piece of the spin sum rule. An analogue UPC process~\cite{Goncalves:2017fkt}, where a $J/\psi$ is (semi-){\it inclusively} photoproduced, would however give access to the gluon Sivers function.

In~\cite{Massacrier:2017lib,Lansberg:2018fsy} two LHC fixed-target operation modes were studied: proton-hydrogen and lead-hydrogen collisions.
The $y$ distribution and $p_T$ distribution for those cases are shown in  the \cf{pH_AFTER} and \cf{PbH_AFTER}. 
The statistical uncertainties in the bins relevant for the GPD extraction are presented in the \cf{AN_pH}, and indicate that \AFTER\ is able to perform the first determination of $E_g$.
The exclusive vector-meson photoproduction in UPCs at \AFTER\ was also studied in~\cite{Goncalves:2018htp}, including that of the light $\rho$ and $\omega$ mesons.

%% file: physics-spin/physics-spin_7-StrangePDF.tex
The distributions of longitudinally polarised (anti)quarks inside a nucleon are far from well known. 
For instance, there are still sizeable uncertainties in the case of antiquarks (see e.g.~\cite{Nocera:2014gqa}). 
Specifically, the understanding of the distribution of polarised strange and antistrange quark distributions ($\Delta s$ and $\Delta \bar{s}$, respectively), and their possible asymmetry is one of the most intriguing open quests in hadronic physics. 
Besides of being a key information on the structure of matter, it is also an essential ingredient of theoretical calculations in astrophysics ($\Delta s$ enters the hydrodynamical modelling of core-collapse supernova explosions~\cite{Melson:2015spa,Hobbs:2016xlg}). 
As of today, the precise value of $\Delta s$ remains unknown, although the experimental data of COMPASS \cite{Adolph:2015saz} and several lattice calculations \cite{QCDSF:2011aa,Alexandrou:2017oeh,Yamanaka:2018uud,Liang:2018pis,Lin:2018obj} are suggesting a small negative value
while, at large $x$, HERMES data hint at a slightly positive value~\cite{Airapetian:2008qf}.
The region where the quarks carry the majority of the proton momentum (high $x$, $x\sim 1$) is of special importance. 
It is pivotal to reveal the content of nucleons observed in nature: both the quark structure and the total spin of a nucleon are in fact arising from the (valence) quark PDFs at high $x$. 
Currently, insufficient data exist in this kinematic range, which leads to unacceptably large uncertainties in the extracted polarised quark PDFs at $x \rightarrow1$. 
For example, the $d/u$ quark ratio at high $x$ remains a puzzle (see section~\ref{subsec:nucleon}). 
In general, any additional data at high $x$ can reduce the uncertainties in the determination of (un)polarised quark distributions.
The \AFTER programme provides opportunities for such studies. 
The ALICE CB detector used in the fixed-target mode covers extremely backward rapidities in the \cms , corresponding to $x \to 1$ in the target (see section~\ref{section:detector:alice}).

Moreover, the ALICE CB detector excels in particle identification and is capable of measuring identified hadrons (for example $\pi^{\pm}$, $K^{\pm}$, $K^{0}_S$ mesons and $\Lambda$ baryons). 
It can therefore provide data to study quark and antiquark densities at $x \rightarrow 1$. 
Specifically, it gives access to the strange quark helicity densities at high $x$ via the measurement of the longitudinal spin transfer $D_{LL}$ from a longitudinal polarised target to $\Lambda$ and $\bar{\Lambda}$ hyperons. 
So far, only limited set of experimental $D_{LL}$ results exist~\cite{Abelev:2009xg, Alekseev:2009ab,Astier:2001ve,Airapetian:2006ee} and their precision is far from being satisfactory.

The $\Lambda$ (and $\bar{\Lambda}$) baryons, which contain a strange (anti-strange) quark, are popular tools in studies of spin effects in high-energy collisions because of their self-spin analysing decays in $p\pi^-$ ($\bar{p}\pi^+$). 
In this decay, a proton is preferably emitted along the spin direction of a parent baryon, which gives a convenient access to the spin orientation of the latter.  
In practice, $D_{LL}$ is defined as the ratio of the difference of the inclusive cross sections with a positive or negative polarisation to their sum for a given target polarisation:
\begin{equation}
D^\Lambda_{LL} \equiv \frac{\sigma_{pp^\rightarrow \rightarrow \Lambda^\rightarrow} - \sigma_{pp^\rightarrow \rightarrow \Lambda^\leftarrow}}{\sigma_{pp^\rightarrow \rightarrow \Lambda^\rightarrow} + \sigma_{pp^\rightarrow \rightarrow \Lambda^\leftarrow}},
\end{equation}
where the $\rightarrow$ and $\leftarrow$ signs denote positive or negative helicities. 
Within perturbative QCD, $D_{LL}$ is sensitive to both polarised quark densities and polarised fragmentation functions. 
However, the interpretation of the experimental data depends on the assumed theoretical model of a fragmentation function. 
Besides the strange quark, the spin transfer from $u$ and $d$ quarks could contribute to $D^\Lambda_{LL}$ (see for example \cite{Burkardt:1993zh}); $D^{\bar{\Lambda}}_{LL}$  provides a cleaner probe of polarised anti-strange quark density. In what follows, we will omit the $\Lambda$ subscript of $D^\Lambda_{LL}$.

For the evaluation of the expected statistical precision of $D_{LL}$ measurements, we assumed that one will use a similar technique as in \cite{Abelev:2009xg}. 
We estimated the statistical uncertainty on $D_{LL}$ taking $\sigma({D}_{LL}) = \frac{1}{\alpha_{\Lambda} P} \frac{1}{\sqrt{N} }$, where $P$ is an effective target polarisation, $\alpha_{\Lambda} = 0.750 \pm 0.010$~\cite{Zyla:2020zbs} is the weak decay parameter and $N$ is the overall $\Lambda$ yield registered in the ALICE detector. 
The $\Lambda$ production is simulated with the PYTHIA8 event generator, and our estimation takes into account the geometrical acceptance of the ALICE CB detector for the $\Lambda$ daughters and the overall $\Lambda$ reconstruction efficiency \cite{Aamodt:2011zza}.
\cf{fig:LambdaDll} shows the expected statistical precision of $D_{LL}$ as a function of (a) the transverse momentum and (b) $x^{\rightarrow}$ for a target located in front of the ALICE TPC detector along the beam direction ($z_{\rm target}$ = 4700 cm).
The assumed integrated luminosity reflects the expected performance of the ALICE TPC detector and the target polarisation corresponds to a polarised hydrogen gas-jet target (see section \ref{subsubsec:gas_jet} for details).

\begin{figure}[hbt!]
\centering
\subfigure[~]{\includegraphics[width=0.48\textwidth]{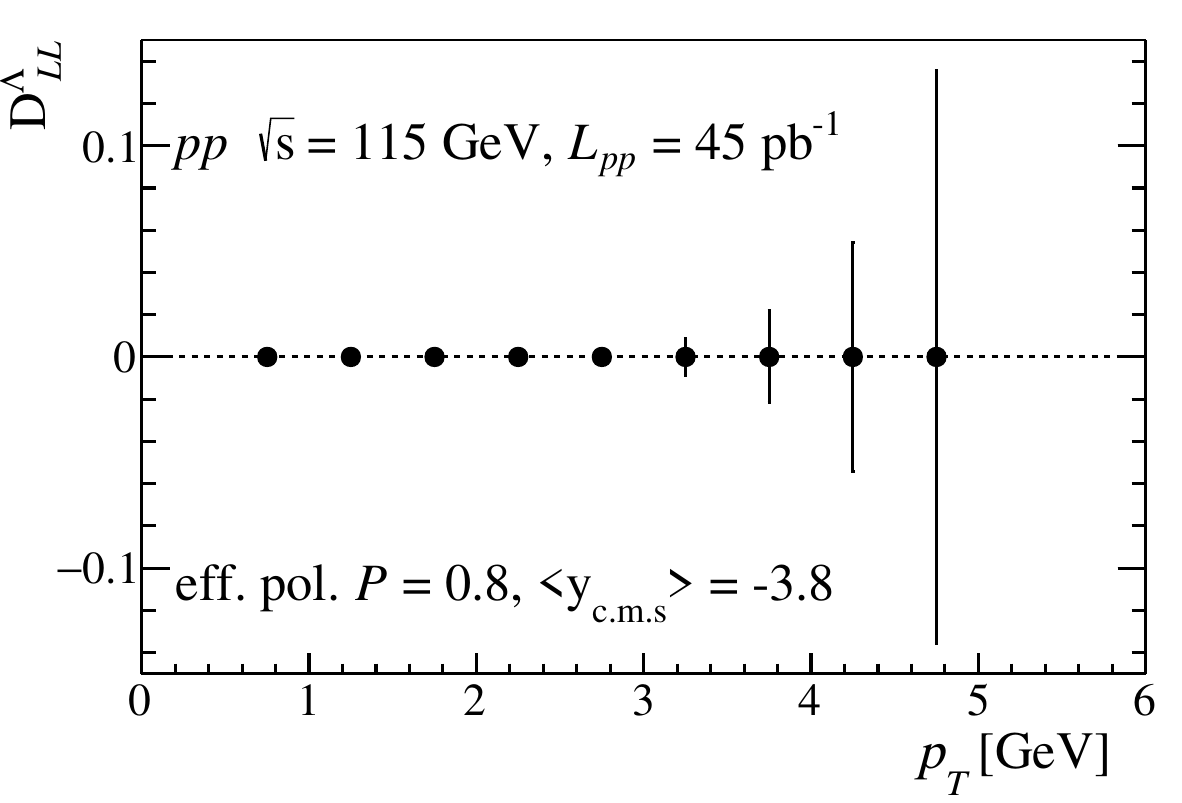}}
\subfigure[~]{\includegraphics[width=0.48\textwidth]{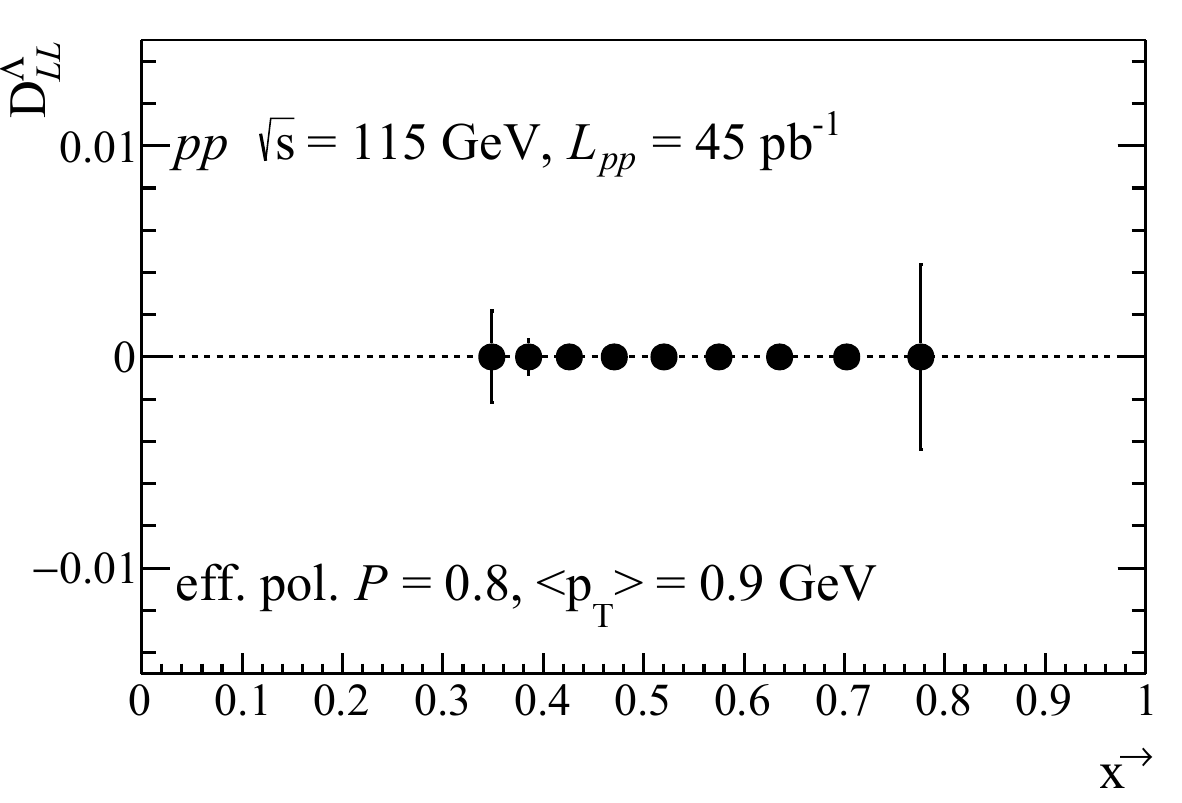}}
\caption{The statistical precision expected for the longitudinal spin transfer $D_{LL}$ to $\Lambda$ hyperons with the ALICE detector and a target located ahead from the ALICE TPC at $z_{\rm target}$ = -4.7~m. 
The $\Lambda$ yields are calculated using PYTHIA, taking into account the acceptance of the ALICE TPC detector and a realistic reconstruction efficiency of $\Lambda$ baryons in the ALICE experiment~\cite{Aamodt:2011zza}}
\label{fig:LambdaDll}
\end{figure}

With a single year of data taking it will be possible to measure $D_{LL}$ with sub-percent precision for $x^{\rightarrow} \rightarrow 1$; however, the $\pT$ reach is limited due to the proximity of the end of the phase space. 
Therefore, these results could be a challenge for theoretical calculations, given the relevance of perturbative QCD calculations in these predictions.
Anyhow, such data could give a unique opportunity to study polarised strange and anti-strange quark distributions at $x^{\rightarrow} \rightarrow 1$.

As a final remark, we note that similar studies with a LHCb-like detector remain to be explored.

\paragraph{Strange-quark-transversity distribution}

The integrated quark transversity, also called the nucleon tensor charge, is a useful input in the search for new physics beyond the standard model, especially in the context of the electric dipole moment \cite{Engel:2013lsa,Yamanaka:2017mef} or the neutron beta decay \cite{Gonzalez-Alonso:2018omy}.
The quark transversity of light quarks has recently been extracted through a TMD-evolution analysis of the Collins azimuthal asymmetries in $e^+ e^-$ annihilation and SIDIS processes \cite{Kang:2015msa}, as well as through a global analysis of $ep$ DIS and $pp$ collisions \cite{Radici:2018iag}.
The tensor charges deduced from the above analyses are consistent with each other, but are in conflict with recent lattice QCD results \cite{Bali:2014nma,Bhattacharya:2015esa,Alexandrou:2017qyt,Yamanaka:2018uud}.
The extraction of the nucleon tensor charge from experimental data is now under intense debate, and an improvement of the analysis of DIS experimental data considering hadrons in jet final states was recently proposed \cite{Accardi:2017pmi,Accardi:2019luo}.

Complementary measurements of the transverse-spin transfer $D_{TT}$ to hyperons can be carried out within the \AFTER\ programme with a similar precision as that for $D_{LL}$, discussed above.
$D_{TT}$ is sensitive to both the transversity distribution of $s$ and $\bar s$ and transversely-polarised fragmentation functions. Thus, $D_{TT}$ give insight into transversity distribution of quarks which has never been extracted from experimental data (as well as the transversely-polarised fragmentation functions). However, the extraction of the nucleon tensor charge would require a reliable model of the fragmentation functions. 

Recent lattice calculations suggest a small value of $\delta s \sim O(10^{-3})$ for the strange quark contribution to the nucleon tensor charge \cite{Alexandrou:2017qyt,Gupta:2018lvp}. Recently the STAR experiment reported $D_{TT}$ in polarised proton-proton collisions at \sqrts = 200 GeV at mid-rapidity (that is, in the low-$x$ regime), and indeed $D_{TT}$ is compatible with zero within uncertainties~\cite{Adam:2018wce}. The sea-quark distributions (at large $x$)  will hardly be measurable at any other experimental facility, and thus the \AFTER programme might provide a very useful handle to further constrain the nucleon tensor charge and the transversely polarised fragmentation functions.

%% file: physics-heavy-ion-collisions/physics-heavy-ion-collisions.tex
\subsection{Heavy-ion physics}
\label{section:heavy Ion Physics}

Despite considerable progress achieved in the last three decades at AGS, SPS, RHIC and LHC in understanding the properties of the hadronic matter at extreme conditions \cite{Andronic:2015wma,Brambilla:2014jmp,Adams:2005dq,Adcox:2004mh,Arsene:2004fa} produced in high-energy proton-nucleus and nucleus-nucleus collisions, crucial aspects of the resulting system remain obscure. Important open questions regarding the properties of a new phase of matter, assumed to be a Quark Gluon Plasma (QGP), remain to be addressed:
\begin{enumerate}
\item the nature of the phase transition between the hadronic matter and the deconfined phase of quark and gluons;
\item the transport properties of this medium, including its specific shear viscosity; 
\item the interaction of hard partons with this medium and their energy loss via collisional and radiative processes; 
\item the flavour dependence of the energy loss in the hot medium;
\item the thermodynamic properties of this hot medium.
\end{enumerate}

In the following text, we focus on two types of observables used in QGP studies: modification of hadron yields in heavy-ion (or proton-nucleus) collisions compared to proton-proton reactions, and anisotropy of the azimuthal-angle distribution of hadrons in the final state. We quantify the medium effects on hadron yields using the nuclear modification factor, $R_{AA}$ (or $\RpA$) which is a ratio of invariant yields in \AA\  (\pA) interactions, normalised by the respective numbers of binary collisions, and yields in \pp\ collisions. If heavy-ion reactions are a simple superposition of \pp\ interactions, then $R_{AA} \sim 1$. The azimuthal anisotropy is conventionally quantified using a Fourier series of the particle azimuthal angle $\phi$ distribution with respect to the reaction plane $\Psi$ (a plane defined by the beam and a vector connecting the centres of colliding nuclei): 
\begin{equation}
\frac{d^2 N }{dp_T d\phi } \propto \sum_{n=1}^{\infty} 2 v_n (p_T ) \cos ( n (\phi - \Psi) )
\end{equation}
where $v_1$ is called the directed flow, $v_2$, the elliptic flow and $v_3$, the triangular flow, \etc. The anisotropy parameters $v_n$ provide information about the early dynamics of the created system and the collective behaviour of hadrons under study. They can shed more light on the equation of state and the thermodynamic properties, like the shear-viscosity $\eta$ to entropy-density $s$\footnote{Not to be confused with the center-of-mass energy \sqrts.} ratio $\eta/s$, of the QGP. Since $v_n$ are measured using final-state particles, they can  contain signals from the QGP phase, fluctuations from the initial conditions as well as some final-state effects such as correlations of the decay products from short-lived hadron decays or jet correlations. There are several different experimental approaches to determine the $v_n$ parameters, for example using two- or multi-particle correlations. In general, the latter are the most sensitive to the collective behaviour of these particles.

The \AFTER\ heavy-ion program will take place at the \cms\ energy of 72 GeV with a 2760 GeV Pb beam. With lighter species the \cms\ energy is only slightly larger \footnote{For 2890 GeV Xe beam, $\sqrtsNN=73.7$ GeV.}.  The Beam Energy Scan (BES) programme at RHIC has shown that in AuAu collisions at \sqrtsNN = 62 GeV the produced hadrons have a large elliptic flow~\cite{Adamczyk:2015fum,Adamczyk:2013gw,Adamczyk:2012ku} and jet-quenching effects were observed~\cite{Adamczyk:2017nof}. These results suggest that quark and gluons are probably deconfined in this energy range, which implies that \AFTER\ will be capable of studying both the properties of this deconfined medium and the phase transition. The large kinematic coverage of the available detectors together with different colliding systems, and high luminosities, will allow \AFTER\ to deliver data that can provide definitive answers to aforementioned questions. The purpose of this section is to demonstrate how  \AFTER\ will be able to address them.

\subsubsection{Precise quarkonium studies in a new rapidity and energy domain}

Since more than thirty years, studies of the production of various quarkonium states in heavy-ion collisions have been performed in order to provide insights into the thermodynamic properties of a possible deconfined matter, via the observation of a sequential melting of quarkonia. However, global analyses including SPS, RHIC and LHC data clearly show that such endeavour is much more complex than initially thought~\cite{Rapp:2008tf,Andronic:2015wma,Brambilla:2010cs}. Many facts support this viewpoint: the complexity of both charmonium and bottomonium feed-down combined with the absence of measurements of direct yields; the competition with conventional nuclear effects (see sections  \ref{sec:high_x_nucl_structure} for discussion of one of them, the modification of the PDF in a nucleus), the intrinsic complexity of modelling such a new state of matter and the fate of these bound states when they cross it and finally the smaller cross section of these hard probes compared to light flavoured hadrons.

Indeed, while quarkonium production in \AA\ collisions was predicted to be suppressed (relative to the \pp\ case) due to the (Debye-like) screening of the $Q \bar Q$ potential in the deconfined medium, where the coloured charges are mobile, it is now becoming clear that dynamical effects beyond such a static screening should be taken into account. These are due to the Landau damping following from the inelastic scatterings of the pair with the constituents of the deconfined medium, or to colour rotations of the colour singlet $Q \bar Q$ pair leading to its dissociation. Conversely, these colour rotations may also lead to quarkonium regeneration. At high energies, the charmonium  production can even be more complex with the possibility of the recombination of uncorrelated charm and anticharm quarks produced in the same collisions.

The bottomonium case seems to provide an easier path to (partially) achieve the goal of using quarkonium sequential suppression as a thermometer. Indeed, the three states are observable\footnote{Despite the aforementioned caveats related to the oversimplified picture of the thermometer based on the uniqueness of the Debye screening, it is in any case mandatory to measure more than 2 states to calibrate it and then to ``measure'' a temperature.} in the dimuon channel 
and $b\bar b$ recombination is far less likely. This is even more true in the energy range of \AFTER. This will be the object of the first quarkonium section. 

In what concerns the charmonium family, in spite of significantly larger cross sections, the situation is much more intricate and unquestionably calls for measurements which have never been done. Merely improving the precision of past studies is bound to be insufficient. Indeed, the $\psi(2S)$ state is likely too fragile to fit in any -- idealised -- thermometer picture. In addition, the access to information about a
third state, in this case the $\chi_c$ triplet via feed-down, has been shown to be close to impossible since there is no consensus, after 20 years of data, on whether they indicate that the  $\chi_c$ suppression is closer to that of the $\psi(2S)$ or of the $\jpsi$. In this context, \AFTER can play a crucial role by providing completely novel observations ranging from direct $\chi_c$ suppression measurements  to new correlations studies. This will be the object of the second quarkonium section.

\paragraph{Measurements of the 3 $\Upsilon(nS)$ states in \pp, \pA\ and \AA\ collisions}
\label{heavy-ion-collisions:upsilon}

\input{physics-heavy-ion-collisions/physics-heavy-ion-collisions-upsilon.tex}

\paragraph{Advanced charmonium studies in heavy-ion collisions}
\label{heavy-ion-collisions:quarkonium}

\input{physics-heavy-ion-collisions/physics-heavy-ion-collisions-charmonium.tex}

\subsubsection{Study of the heavy-quark energy loss and their interaction with the surrounding nuclear matter}
\label{heavy-ion-collisions:charm}
\input{physics-heavy-ion-collisions/physics-heavy-ion-collisions-charm.tex}

\subsubsection{Soft probes at large rapidities -- a precise tool to study the bulk properties of the nuclear matter}
\label{heavy-ion-collisions:flow}
\input{physics-heavy-ion-collisions/physics-heavy-ion-collisions-flow.tex}

\subsubsection{Search for a collective dynamics of partons in small systems at low energies}
\label{heavy-ion-collisions:small-system}
\input{physics-heavy-ion-collisions/physics-heavy-ion-collisions-small-system.tex}

\subsubsection{Test of the factorisation of the initial-state nuclear effects in \AA\ collisions with Drell-Yan lepton-pair production}
\label{heavy-ion-collisions:DY}
\input{physics-heavy-ion-collisions/physics-heavy-ion-collisions-DY.tex}

%% file: physics-heavy-ion-collisions/physics-heavy-ion-collisions-upsilon.tex
In this context, we find it useful to start by discussing the unique reach of \AFTER\ for $\Upsilon(nS)$ production in \pp, \pA\ and \AA\ collisions as a function of the system size (for various colliding systems or vs. collision centrality), \pt\ and rapidity, in \PbA\ collisions at $\sqrt{s_{NN}} =$ 72 GeV. This lies nearly half way between the SPS and RHIC energy ranges and significantly lower than the only existing studies by CMS at the (collider mode) LHC~\cite{Sirunyan:2017lzi,Khachatryan:2016xxp,Chatrchyan:2012lxa}.
Although the relative suppression of the $\Upsilon(nS)$ cannot  readily be used as a thermometer it will bring crucial new inputs for our understanding of the nature of the hot medium created in this energy range as opposed to that presently available at the LHC.

\begin{figure}[hbt!]
\centering
\begin{tabular}{ll}
\subfigure[~]{\includegraphics[width=0.45\textwidth]{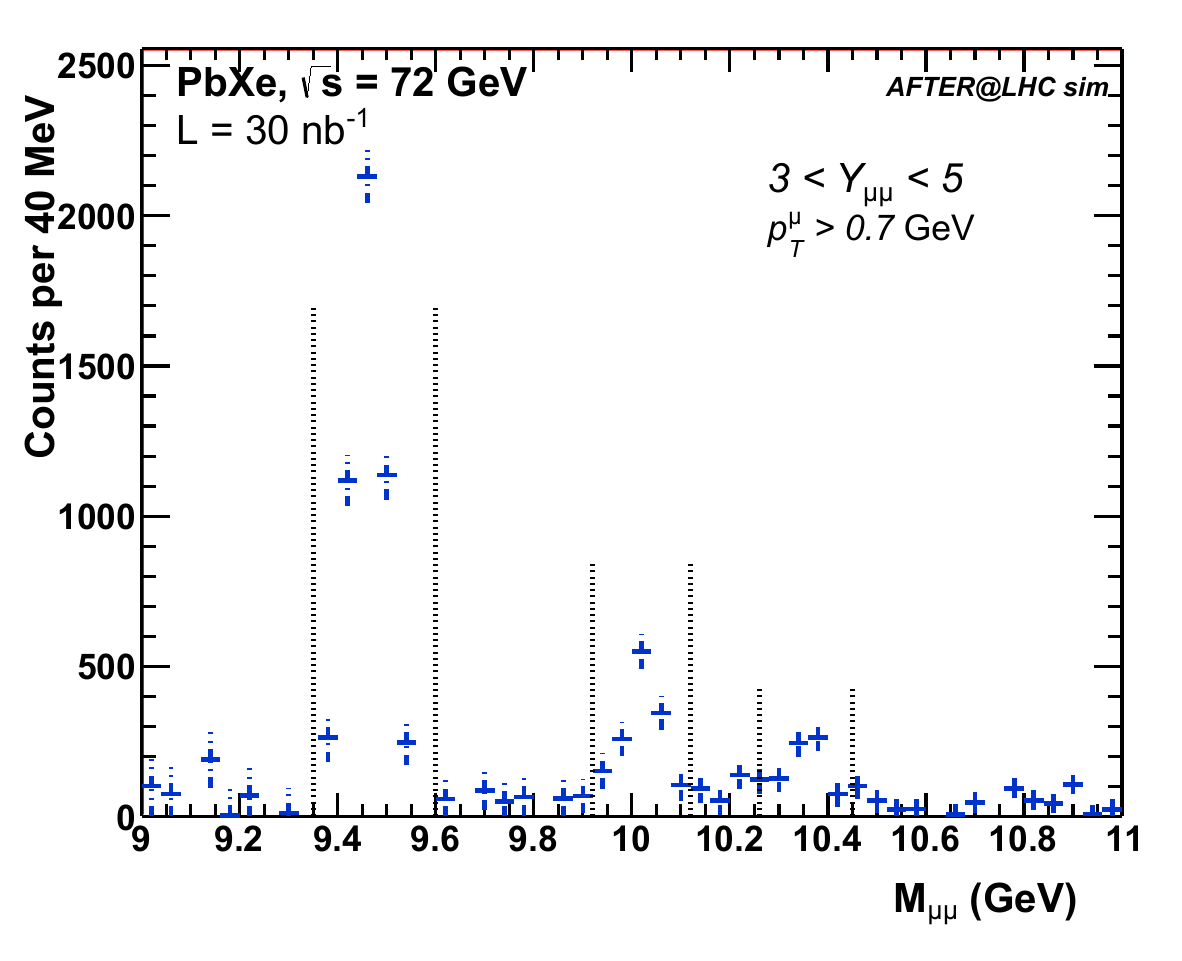}\label{fig:Upsilon_HI:mass}} &
\subfigure[~]{\includegraphics[width=0.53\textwidth]{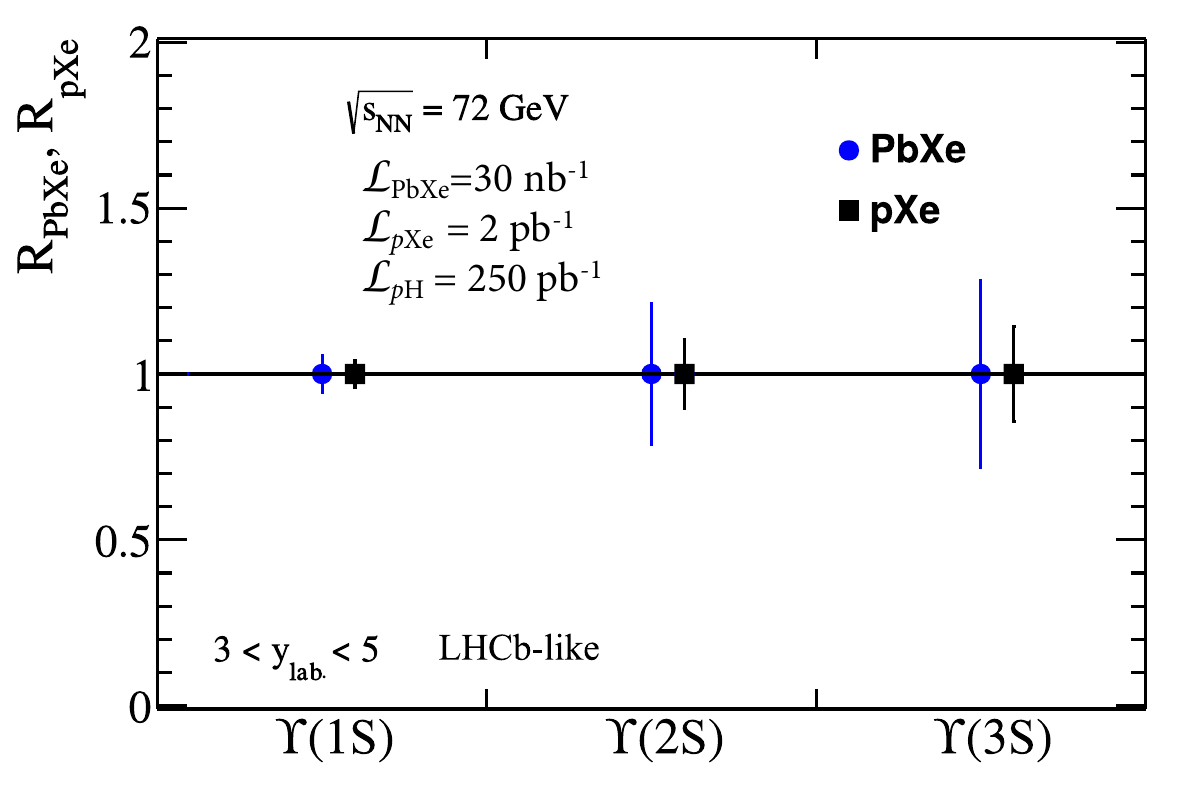}\label{fig:Upsilon_HI:RAA}} \\
\end{tabular}
	\caption{(a) The $\Upsilon(nS)$ signal after the (like-sign) uncorrelated-background subtraction with the expected statistical uncertainties in \PbXe collisions at $\sqrt{s_{NN}} =$ 72 GeV in the $\Upsilon$ acceptance range of 3 $< Y^{\rm lab}_{\mu\mu} < $ 5, for a LHCb-like detector. No nuclear modifications are assumed, $\mathcal L_{\rm PbXe}$ = 30 nb$^{-1}$. (b) Statistical-uncertainty projections for measurements of the nuclear modification factors \RPbXe\ and \RpXe~\cite{Trzeciak:2017csa}. The uncertainties are calculated using the yields given in \ct{tab:Upsilon_yields}.}
\label{fig:Upsilon_HI}
\end{figure}

\begin{table}[!htb]
\begin{center}
{\renewcommand{\arraystretch}{1.5}
\subtable{\begin{tabular}{c|c|c}
 $\Upsilon(1S)$ & $S$ & $S/B$ \\ \hline \hline
\pp & 1.33 $\times 10^{3}$ & 29.0 \\ 
\pXe & 1.39 $\times 10^{3}$ & 7.8 \\
 \PbXe & 4.33 $\times 10^{3}$ & 1.8 $\times 10^{-1}$ \\ \hline
\end{tabular}}
\subtable{\begin{tabular}{c|c|c}
$\Upsilon(2S)$  & $S$ & $S/B$ \\ \hline \hline
\pp & 2.92 $\times 10^{2}$ & 8.2 \\ %
\pXe & 3.06 $\times 10^{2}$ & 2.2 \\
\PbXe & 9.56 $\times 10^{2}$ & 5.0 $\times 10^{-2}$ \\ \hline
\end{tabular}}
\subtable{\begin{tabular}{c|c|c}
$\Upsilon(3S)$  & $S$ & $S/B$ \\ \hline \hline
\pp & 1.37 $\times 10^{2}$ & 10.3 \\ %
\pXe & 1.44 $\times 10^{2}$ & 2.8 \\
\PbXe & 4.49 $\times 10^{2}$ & 6.2 $\times 10^{-2}$ \\ \hline
\end{tabular}}}
\end{center}
\caption{$\Upsilon(nS)$ signal yields ($S$) and signal-over-combinatorial-background ratios ($S/B$) for \pp, \pXe and \PbXe collisions at $\sqrt{s_{NN}} = 72$~GeV and $3 < y_{\rm Lab.} <  5$ assuming LHCb-like performances~\cite{Trzeciak:2017csa} with $\int \mathcal L_{pp} = 250\, {\rm pb}^{-1}$, $\int \mathcal L_{p{\rm Xe}} = 2\, {\rm pb}^{-1}$ and $\int \mathcal L_{\rm PbXe} = 30\, {\rm nb}^{-1}$.}
\label{tab:Upsilon_yields}
\end{table}

\cf{fig:Upsilon_HI:mass} shows the di-muon invariant-mass distribution in the $\Upsilon(nS)$ mass range for a single year of  data taking with a LHCb-like setup.
The expected \ups\ yields are clearly large enough with the excellent resolution of LHCb to clearly distinguish each $\Upsilon(nS)$ state
in an energy domain where \ups\ studies are extremely demanding. The yields together with the signal over background ratios in \pp, \pXe\ and \PbXe\ collisions are also gathered in Table~\ref{tab:Upsilon_yields}. Projections of the statistical precision of the nuclear modification factors for the $\Upsilon(nS)$ states measured in \pXe\ and \PbXe\ collisions are presented in \cf{fig:Upsilon_HI:RAA}. The statistical uncertainties take into account the background subtraction procedure using the like-sign method. If needed, this can further be improved by using the mixed-event technique. The predictions do not include potential modifications of the $\Upsilon(nS)$ yields due to the nuclear effects. 

\begin{figure}[hbt!]
	\centering
	\includegraphics[width=0.6\textwidth]{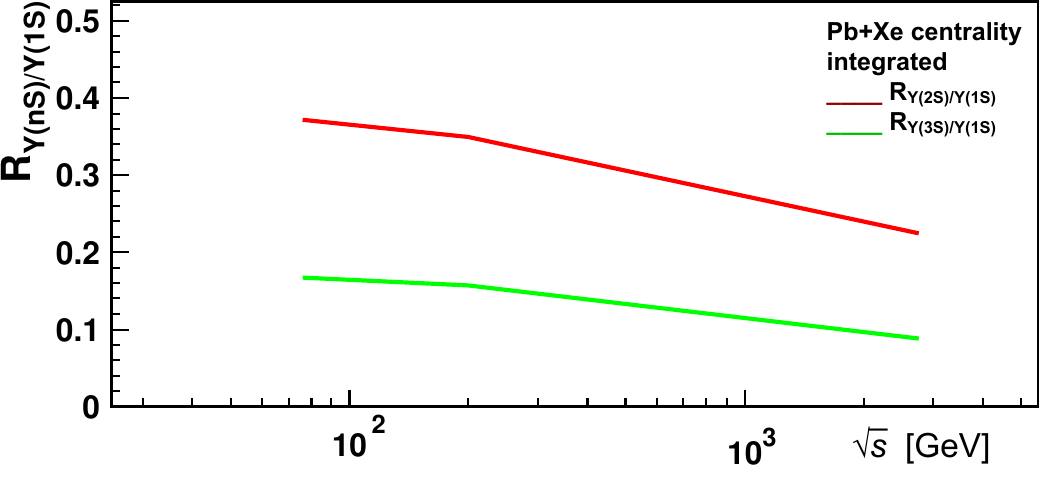}
	\caption{The relative suppression of the excited $\Upsilon(nS)$ states to the $\Upsilon(1S)$ as a function of the energy in iCIM~\cite{Ferreiro:2018wbd}.}
	\label{fig:R:Upsilon_comovers}
\end{figure}

\cf{fig:R:Upsilon_comovers} shows the expected relative suppression $R_{\Upsilon(nS)/\Upsilon(1S)}$\footnote{
The relative suppression R is defined as the double ratio of excited $\Upsilon$ states to the $\Upsilon(1S)$ in \AA\ and \pp\ collisions, $R_{\Upsilon(nS)/\Upsilon(1S)} = [\Upsilon(nS)/\Upsilon(1S)]_{AA}/[\Upsilon(nS)/\Upsilon(1S)]_{pp}$~\cite{Chatrchyan:2011pe,Sirunyan:2017lzi}.} in the improved Comover Interaction Model (iCIM) recently applied to describe the $\Upsilon(nS)$ suppression at the LHC~\cite{Ferreiro:2018wbd}. It can be seen as an effective approach to deal with quarkonium suppression, accounting for the Landau damping -- the pair gets broken by a scattering with a gluon -- and the colour rotation -- a scattering with a gluon turns the pair into a colour-octet state which cannot hadronise any more. Predictions in other approaches are expected to yield similar suppressions. Given the foreseen accuracy of $\Upsilon(nS)$ measurements, the \AFTER\ program will allow one to verify such predictions in a completely new energy domain.

%% file: physics-heavy-ion-collisions/physics-heavy-ion-collisions-charmonium.tex
In addition to studies of the $\Upsilon(nS)$ states, the \AFTER\ programme will explore an array of new charmonium observables that are virtually not accessible elsewhere. As compared to the SPS experiments, the higher energies at \AFTER\ allow for quarkonium-correlation studies. None have been carried out so far in \pA\ and \AA\ collisions. The use of a detector like LHCb (or maybe the joint usage of the ALICE CB (to detect a photon and a muon) along with the muon arm (to detect the second muon) like in~\cite{ALICE-UPC-CB-MUON})  without absorber enables $\chi_c$ studies at backward rapidities where the multiplicities are reduced. As compared to RHIC and LHC experiments, which have to cope with a large combinatorial background, studies of 
the $\eta_c$ suppression should be within reach in \pA\ collisions and, possibly, in the most backward part of the acceptance, in semi-central \AA\ collisions. These studies would bear on the natural access toward negative rapidities in the \cms, on large luminosities typical of the fixed-target mode and on more modern detectors as compared to those used in the 90's at the SPS and, to a lesser extent, to the ageing RHIC detectors. 

\begin{figure}[hbt!]
\centering
\begin{tabular}{ll}
\subfigure[~]{\includegraphics[width=0.45\textwidth]{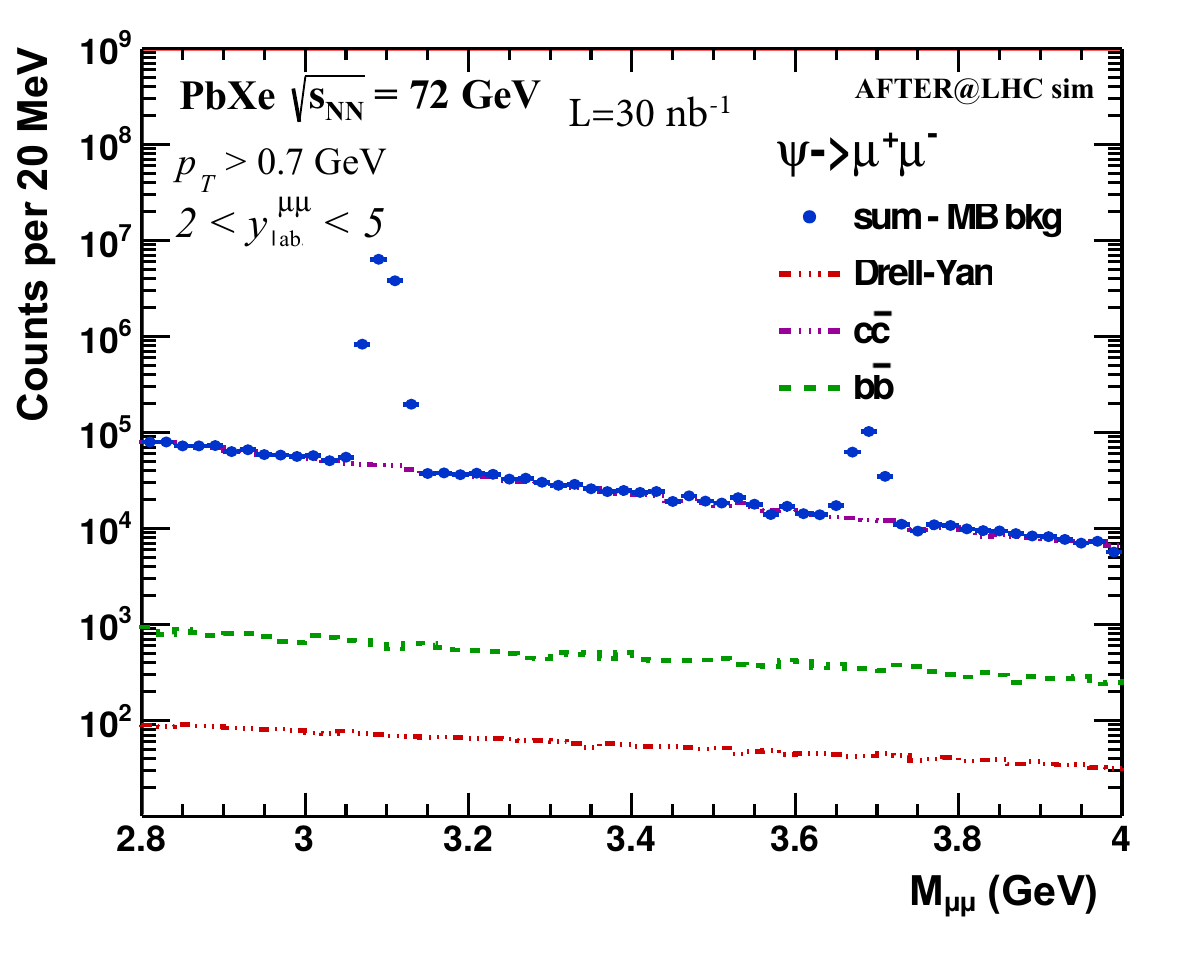}\label{fig:psi_HI-mass}}
\subfigure[~]{\includegraphics[width=0.45\textwidth]{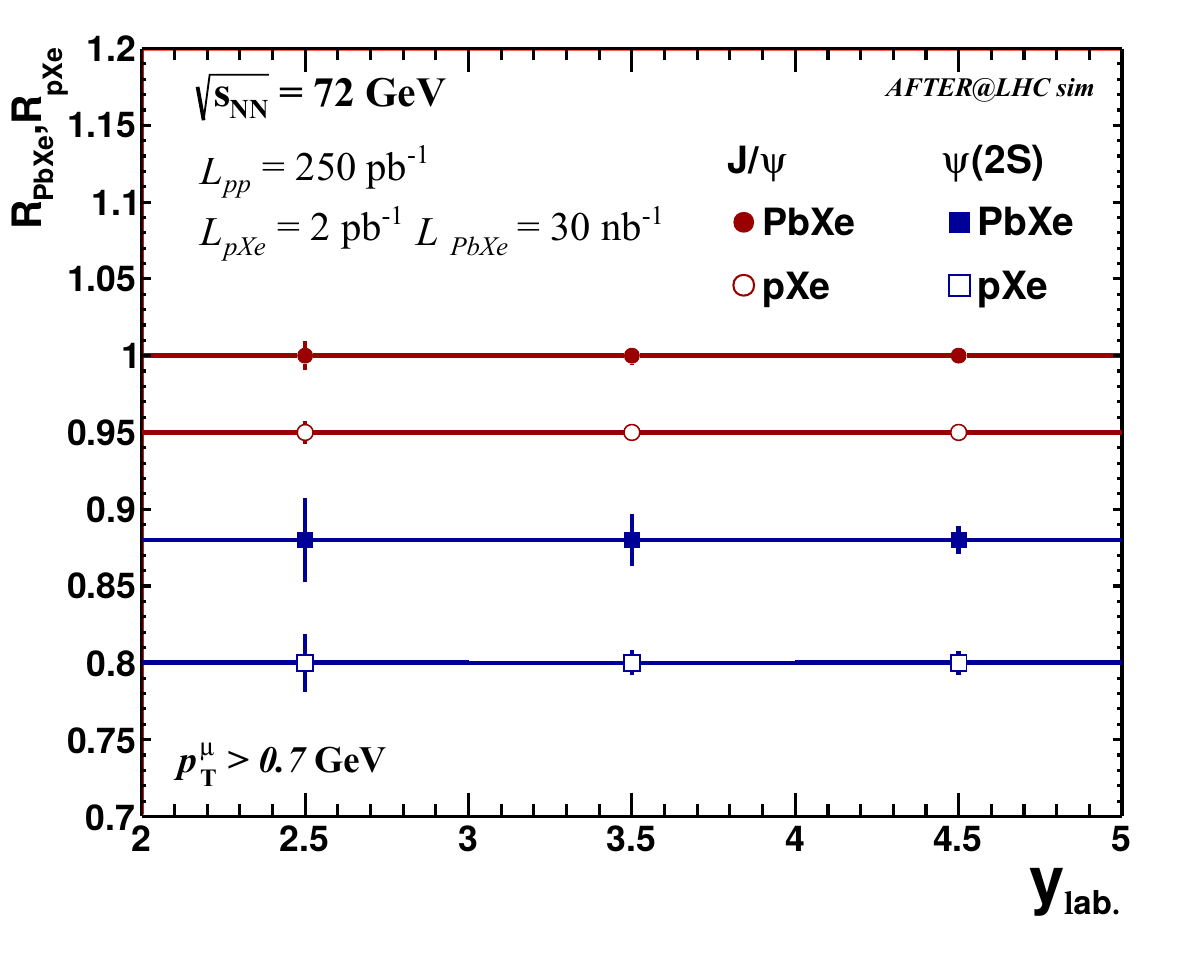}\label{fig:psi_HI-RAA}}
\end{tabular}
\caption{(a) \Jpsi and \psip signal after the (like-sign) uncorrelated-background subtraction with the expected statistical uncertainties for \PbXe collisions at $\sqrt{s_{NN}} =$ 72 GeV in 2 $< y_\cms < $ 5 for $\int \mathcal L_{\rm PbXe}$ = 30 nb$^{-1}$. No nuclear modifications are assumed.  (b) Projected statistical precision of the nuclear modification factor as a function of the laboratory rapidity in \PbXe (\RPbXe) and \pXe (\RpXe) at $\sqrt{s_{NN}} =$ 72 GeV, assuming the uncorrelated-background subtraction with the like-sign technique using $\int \mathcal L_{pp}$ = 250 pb$^{-1}$, $\int \mathcal L_{pXe}$ = 2 pb$^{-1}$, $\int \mathcal L_{\rm PbXe}$ = 30 nb$^{-1}$ which fits in one month of Pb ion run. Calculations were done for a LHCb-like detector performance. [Adapted from~\cite{Trzeciak:2017csa}.]}
\label{fig:psi_HI}
\end{figure}

The particularly large \pA\ rates (orders of magnitude larger than those reachable at the collider LHC and RHIC) with a wide rapidity coverage are also crucial to
disentangle between the different sources of the charmonium suppression. In that regard, the interpretation of such studies, in an energy range where the charm recombination
cannot be relevant~\cite{Adamczyk:2016srz}, will also be much easier. This asset should not be underestimated.

Overall, one expects the measurements of \psip, $\chi_c$, $\Jpsi+\Jpsi$ and $\Jpsi+D$ productions and correlations to be possible, each of them with enough precision to bring in constraints to the charmonium-suppression puzzle.
As an illustration, \cf{fig:psi_HI-mass} shows a di-muon invariant mass distribution for \PbXe collisions at $\sqrt{s_{NN}}=$ 72 GeV in the \Jpsi and \psip mass ranges. It is clear that the background is well under control yielding a very precise determination of the charmonium rates. With such a background for the \jpsi, we find legitimate to highlight the possibility for the aforementioned more demanding studies.

Correspondingly, the statistical projections of the nuclear modification factors in \pXe\ and \PbXe\ collisions as a function of the rapidity  presented on \cf{fig:psi_HI-RAA}\footnote{Statistical projections were calculated with the same assumptions as in the case of $\Upsilon(nS)$ predictions.} allows us to expect a precision at the per cent level for the $\jpsi$ and $\psip$ cases, depending on the rapidity. Without a doubt, more differential studies as well as elliptic flow $v_2$ measurements will also be possible. Projections for the other charmonium-like observables remain to be done. However, we anticipate that they should show a precision around the five per cent level, which would be a clear breakthrough in the field since none of them are within the reach of any other experiment at present. Finally, let us stress that the present discussion only bore on a single target, Xe. However, almost independently of the target implementation (see section \ref{sec:implementation}), a few other species could be used -- even during a single year. This will allow for systematic studies of the $A$ dependence of the nuclear effects to be complemented with that of the centrality.

%% file: physics-heavy-ion-collisions/physics-heavy-ion-collisions-charm.tex
Heavy quarks (charm and bottom) are unique tools to study and characterise the QGP properties. They are produced in hard scatterings at the early stage of the nuclear collisions and loose their energy while traversing the hot and dense medium. The energy loss and elliptic flow of open heavy-flavour hadrons are sensitive to the dynamics of the medium: such measurements could be used to determine the fundamental properties of the QGP, such as the transport coefficients, including $\hat{q}$ (which characterises the (squared) momentum transfer per mean free path of the fast partons) and the charm quark diffusion coefficients. Precision measurements of the elliptic flow of heavy quarks can give insights into the degree of thermalisation of the created nuclear matter (\ie\ whether it is in a local thermal equilibrium or not) and can help discriminate between different models of the heavy-quark interactions with the QGP~\cite{Andronic:2015wma}. 

A significant suppression of open heavy-flavour production at high \pt and a significant elliptic flow of the heavy quarks were observed at the top RHIC energy~\cite{Adamczyk:2017xur,Adamczyk:2014uip,Adamczyk:2014yew,Adare:2010de}. These experimental data can be described assuming two main effects: a medium-induced gluon radiation (radiative energy loss, $dE/dx_{\rm rad}$) and a collisional energy loss, $dE/dx_{\rm coll}$, due to binary interactions of the quark with other objects in the QGP. 
A major difficulty in modelling the heavy-quark energy loss is that the relative contributions of $dE/dx_{\rm coll}$ and $dE/dx_{\rm rad}$ are still not precisely known~\cite{Andronic:2015wma} and need to be constrained using experimental data. To better understand the interplay between both mechanisms, precise measurements of the individual suppression of charm and bottom quark yields are necessary. \cf{fig:D0:Raa:PbPb} shows calculations of the central-to-peripheral nuclear modification factor $R_{CP}$\footnote{$R_{CP}$ is the ratio of the yields in central and peripheral collisions, normalised by the respective numbers of binary collisions.} for $D$ mesons as a function of \pt\ in two rapidity ranges assuming $dE/dx_{\rm coll}$ or $dE/dx_{\rm rad}$ as the sole source of suppression. Since both mechanisms have a different \pt\ and rapidity dependence, the high quality $D^0$ data to be provided by \AFTER\ in \PbA\ collisions at \sqrtsNN=72~GeV in different rapidity ranges can help  pin down $dE/dx_{\rm rad}$ and $dE/dx_{\rm coll}$.

\begin{figure}[!hbt]
        \centering
                \subfigure[~]{\includegraphics[width=0.48\textwidth]{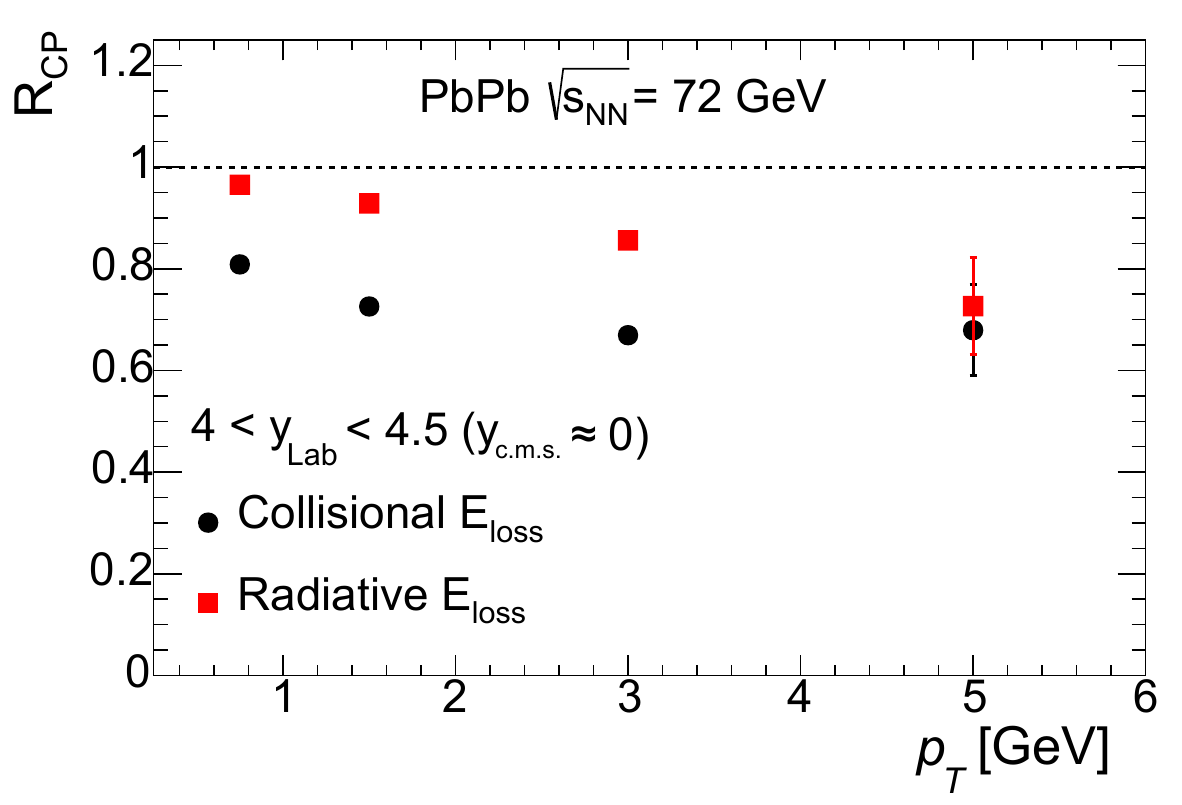}} 
                \subfigure[~]{\includegraphics[width=0.48\textwidth]{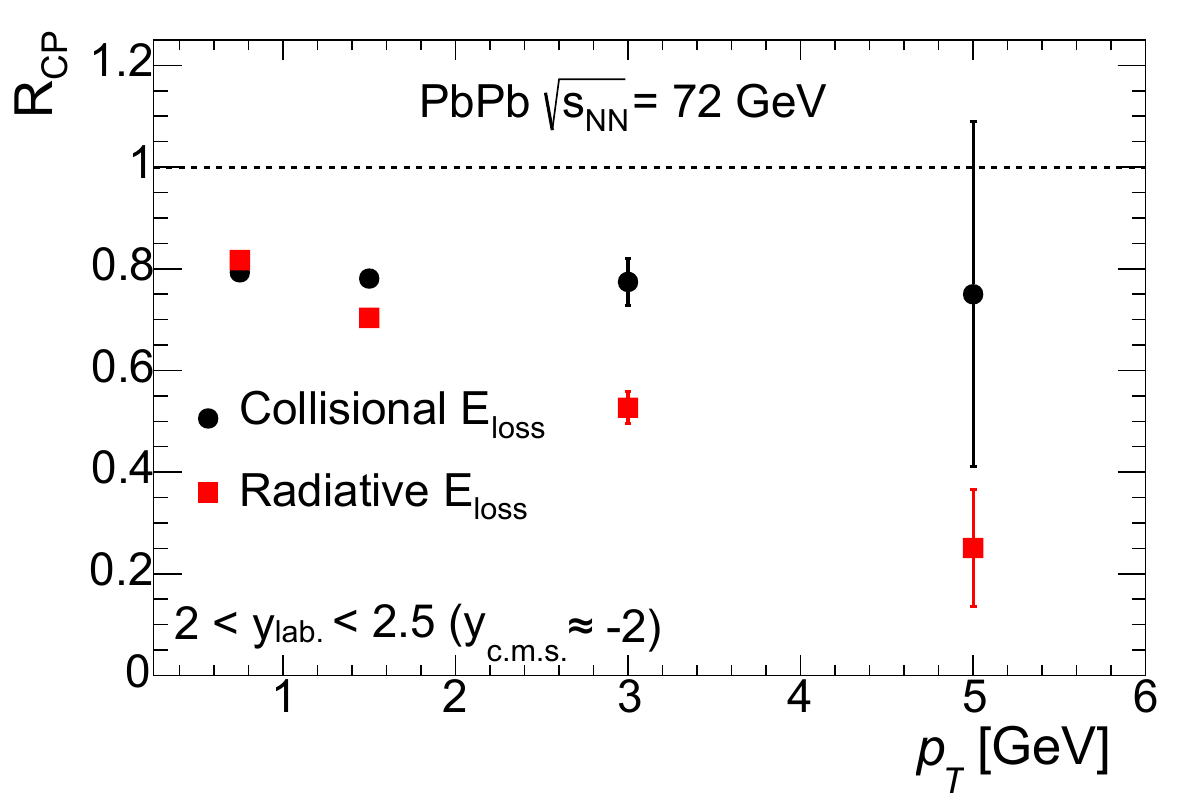}}
        \caption{Expected statistical uncertainties of $R_{CP}$ for $D^0$ mesons measured by a LHCb-like detector in \PbPb\ collisions at \sqrtsNN=72~GeV for $\int \mathcal L_{\rm PbPb}$ = 7 nb$^{-1}$ [Adapted from~\cite{Kikola:2015lka}].}
        \label{fig:D0:Raa:PbPb}
\end{figure}

With the high-statistics data accessible at \AFTER, heavy-flavour azimuthal correlations ($D-D, J/\psi - D$) and heavy-flavoured jets can also be studied to further understand the heavy-quark in-medium interactions. Simultaneous precise measurements of the $D$-meson elliptic flow and the nuclear modification factor will improve the determination of the QGP transport properties. In addition, correlation measurements of heavy-flavour pairs will provide new means to disentangle collisional from radiative interactions and to test Langevin against Boltzmann transport approaches~\cite{Rapp:2018qla}. 
In general, the temperature and mass dependences of the transport coefficients can be extracted with precise $R_{AA}$ and $v_{2}$ measurements for $B$ and $D$ at various beam energies. \AFTER\ will clearly extend such studies towards the low-energy domain.

Furthermore, the study of $D$ mesons is a natural continuation of the investigation of the \Jpsi and \psip formation and dissociation in nucleus-nucleus collisions. Precise measurements of the $D$ yields in $pA$ and $AA$ collisions are necessary to constrain the initial-state cold nuclear matter effects, that is the modification of the $c\bar{c}$ production due to shadowing/anti-shadowing (see section~\ref{sec:high_x_nucl_structure}). In that sense, $D$ mesons are part of the useful measurements to establish a baseline to understand how the QGP affects quarkonium production. As we also discussed earlier, the nuclear effects on bottom-quark production can be also studied via $B \rightarrow J/\psi$ in \pA\ collisions with a good precision (see also~\cite{Kikola:2015lka}).

%% file: physics-heavy-ion-collisions/physics-heavy-ion-collisions-flow.tex
One of the unique assets of the \AFTER\ programme stems from an expected large rapidity coverage. The combined acceptance of the LHCb detector and the central barrel of the ALICE experiment covers the range of $-5.2 \lesssim y^{\cms} \lesssim 0.7$ in the case of the 2.76 TeV Pb beam - such a wide kinematic acceptance is not accessible in any collider experiment. Both detectors also offer  excellent particle identification abilities, which altogether provides a large lever arm for studies related to the longitudinal expansion of the nuclear matter in heavy-ion collisions. In contrast, most of the heavy-ion experiments were designed to study the transverse dynamics of such a system at mid-rapidity, thus the longitudinal evolution is hardly accessible there. 

The longitudinal dynamics of a system created in nuclear collisions is a topic of intensive experimental \cite{Aad:2014fla, ATLAS:2018jht,Aaboud:2017tql} and theoretical~\cite{Schenke:2010rr,Bozek:2010vz,Jia:2014ysa,Xiao:2012uw} studies via azimuthal flow, flow correlations and decorrelation measurements. The two- and multi-particle pseudorapidity correlations allow for an examination of the long-range collective phenomena, \eg\ the ridge, as well as the initial-state fluctuations. 
The studies of the flow (de)correlation vs. rapidity offer an independent test of theoretical calculations that assume collective dynamics (collective flow). Most of them involve a hydrodynamic phase of the system evolution. Since the flow-decorrelation effect increases with the decreasing \cms\ energy~\cite{Wu:2018cpc}, \AFTER\ will provide an excellent setting for such an analysis thanks to the aforementioned large rapidity coverage. Moreover, a measurement of the directed flow $v_1$ of charmed mesons as a function of rapidity was proposed recently to map out the three-dimensional distribution of the nuclear matter produced in heavy-ion collisions~\cite{Chatterjee:2017ahy}. The large $D$ meson yields expected in 
\AFTER\  will facilitate such a study with an unparalleled precision. 
In the following paragraphs, we discuss a few examples of other studies involving the ``soft'' (low $\pt$) probes that are pivotal for our understanding of the QGP properties and the QCD phase diagram.

\paragraph{Determination of the temperature dependence of the ratio of the shear viscosity to the entropy density: $\eta/s$}
The shear viscosity $\eta$ (or the ratio of the viscosity to the entropy $s$, $\eta/s$), is one of the most fundamental properties of the the QGP. Yet, our understanding of $\eta/s$ is far from satisfactory. Indeed, $\eta/s$ cannot directly be measured : it must be derived from a comparison of the experimental data and theoretical calculations (for example hydrodynamic models). After decades of developments, these models reached a high predictive power for the QGP macroscopic behaviour at RHIC and LHC energies. Until recently, such calculations focused on the transverse dynamics of the QGP at mid-rapidity. Intense efforts have been made to include the medium longitudinal expansion in hydrodynamic models (see \eg\ \cite{Schenke:2010rr}). These calculations indicate that particle yields and the azimuthal anisotropy coefficients $v_n$ measured at large rapidities are powerful tools to study the medium  $\eta/s$ and its temperature dependence (see Fig.~\ref{fig:vn_rapidity})~\cite{Denicol:2015nhu}.

\begin{figure}[hbt!]
\begin{tabular}{cc}	
\centering
\subfigure[]{	\includegraphics[width=0.47\textwidth]{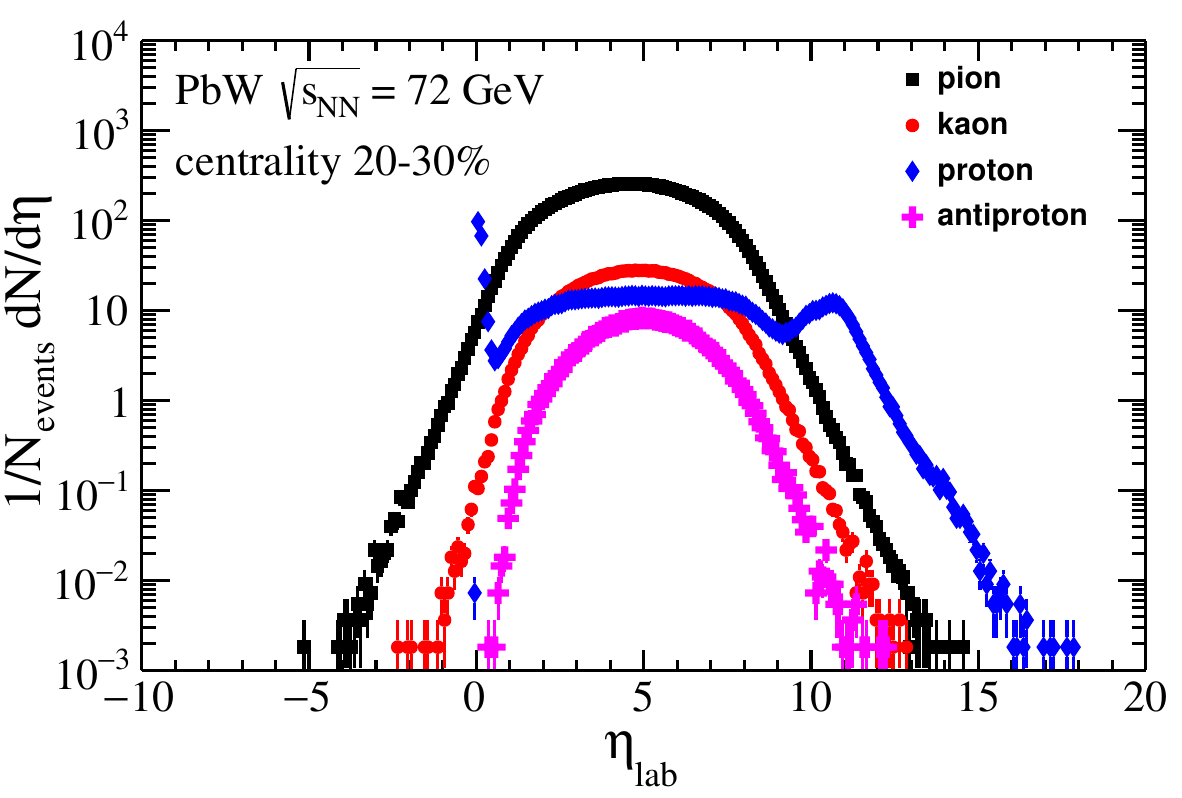}\label{fig:PbW:yields}} &
\subfigure[~]{	\includegraphics[width=0.47\textwidth]{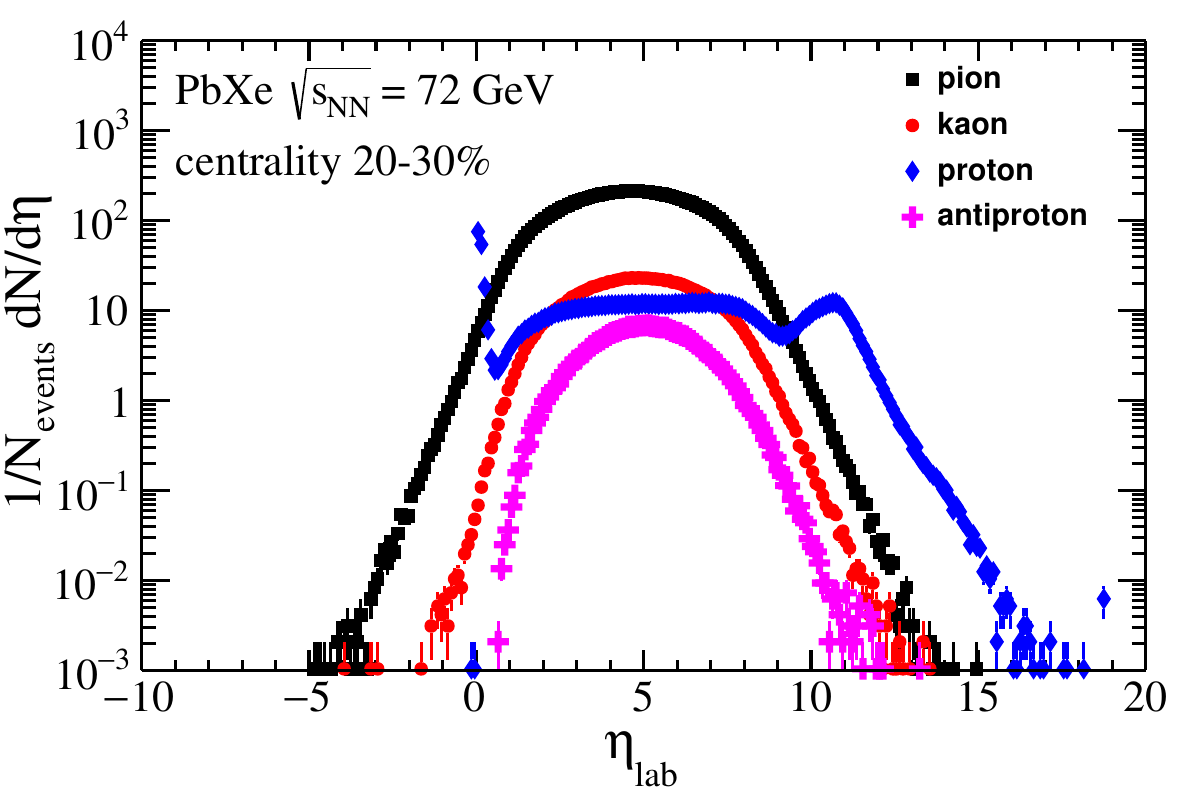}\label{fig:PbXe:yields}} \\
\subfigure[]{	\includegraphics[width=0.47\textwidth]{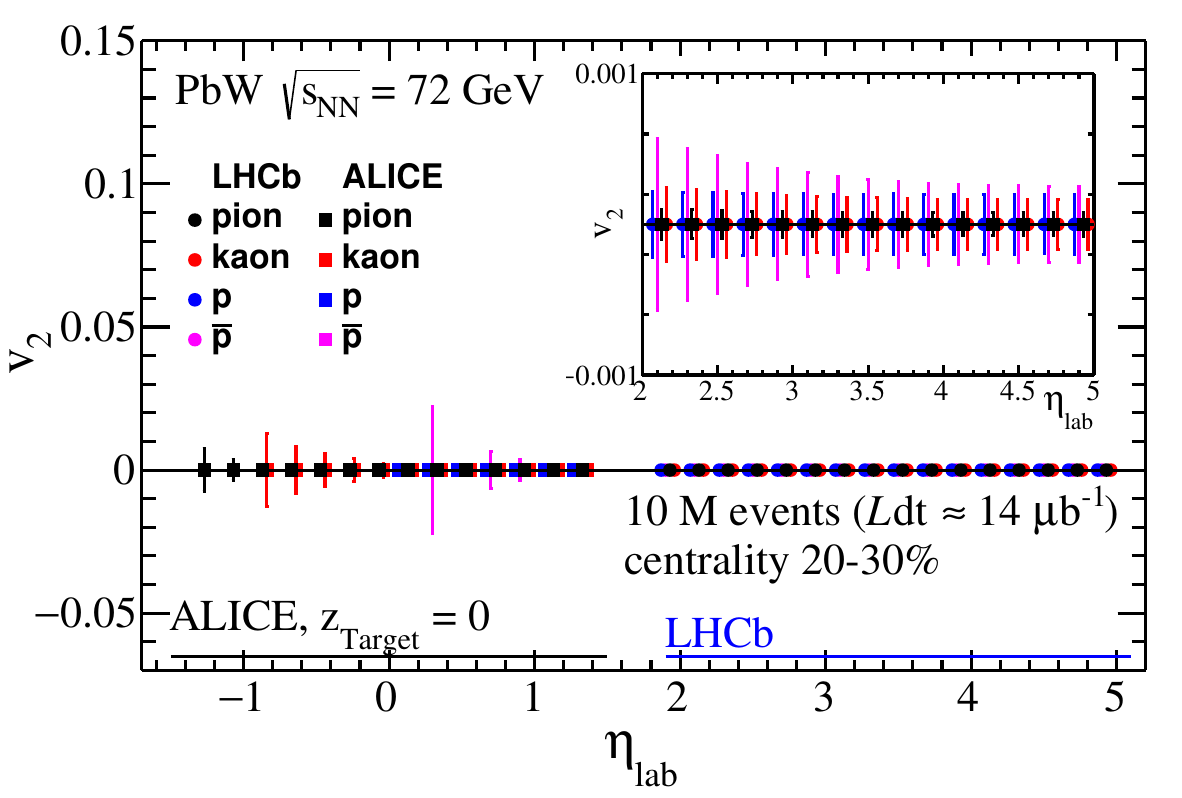}\label{fig:PbW:Z0:flow}} &
\subfigure[~]{	\includegraphics[width=0.47\textwidth]{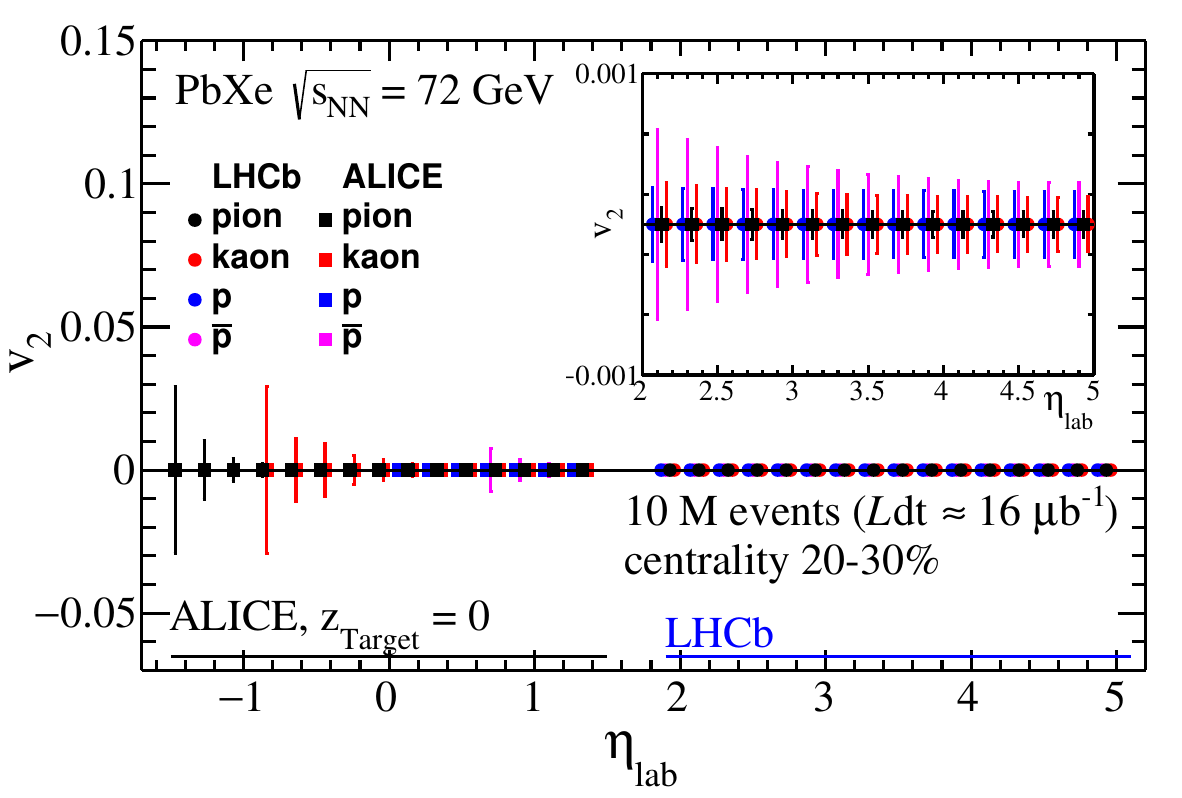}\label{fig:PbXe:Z0:flow}} \\
\subfigure[]{	\includegraphics[width=0.47\textwidth]{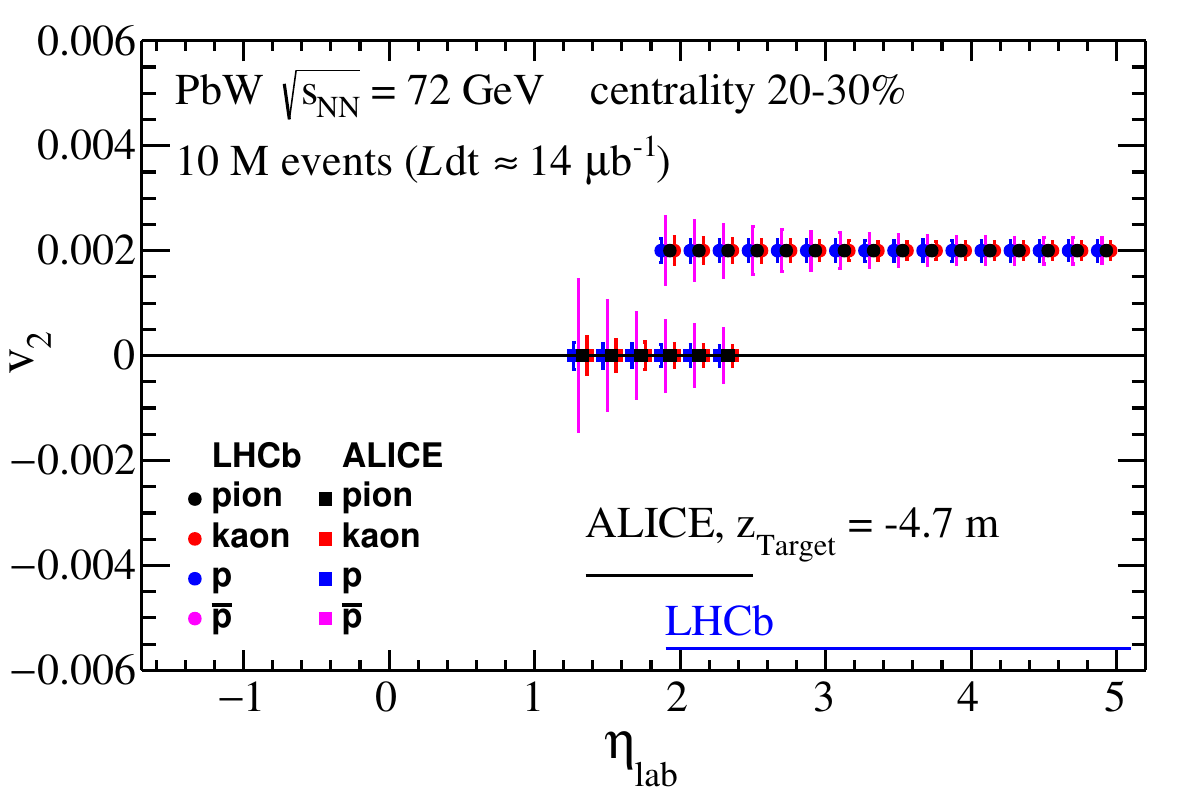}\label{fig:PbW:Z47m:flow}} &
\subfigure[~]{	\includegraphics[width=0.47\textwidth]{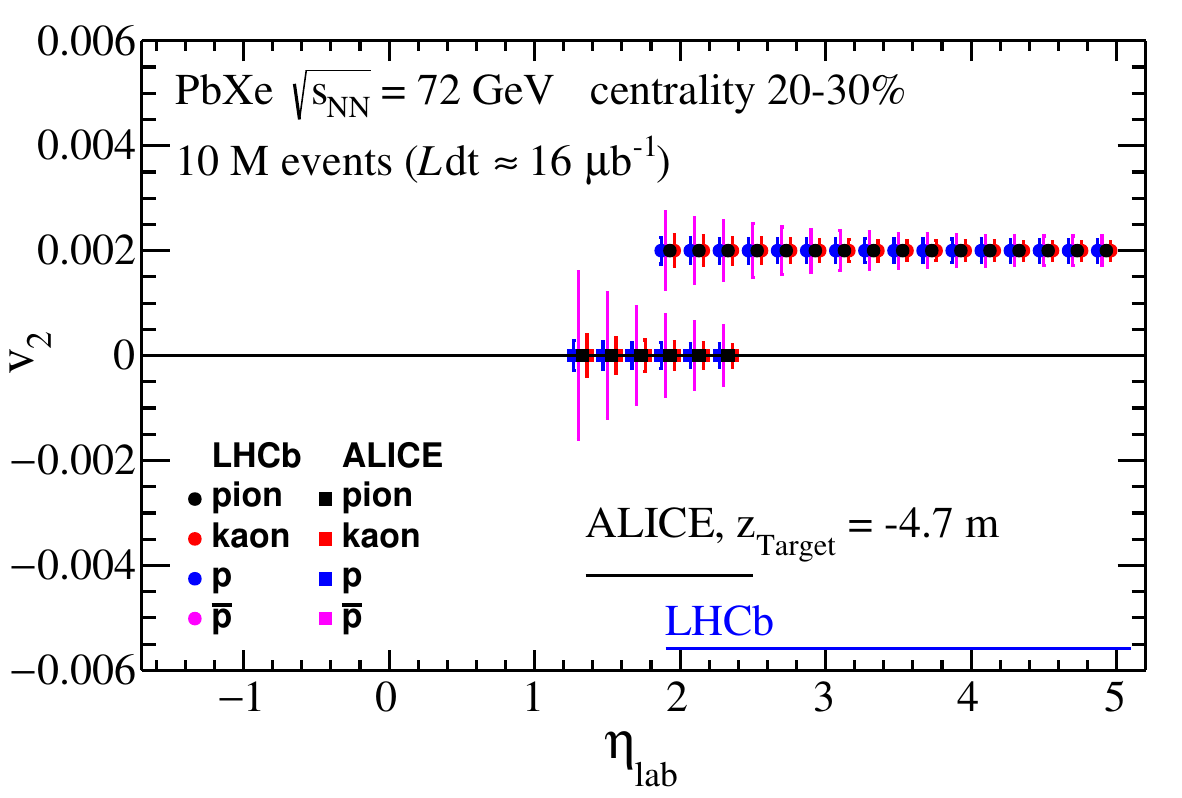}\label{fig:PbXe:Z47m:flow}} \\
\end{tabular}	 
\caption{Identified particle multiplicities per an event in a simulation of (a) 5500 and (b) 9600 events of mid-central \PbW\ and \PbXe\ collisions at \sqrtsNN = 72 GeV and centrality 20-30\% from EPOS~\cite{Pierog:2013ria,Werner:2005jf}. (c) -(f) Projections of the statistical uncertainty on the measurement of the $p_{T}$-integrated ($\pt>0.2$ GeV for an ALICE-like setup and $\pt>0.5$ GeV for a LHCb-like detector) elliptic flow of identified hadrons as a function of the pseudorapidity in the laboratory frame for two target locations for ALICE. The points for a LHCb-like detector in (e) and (f) were shifted vertically for clarity. Projections in (c) and (e) correspond to 10 million mid-central \PbW\ collisions and the integrated luminosity $\mathcal{L}_{\PbW} = 14 \mub^{-1}$, and the results in (d) and (f) represent 10 million mid-central \PbXe\ events and $\mathcal{L}_{\PbXe} = 16\, \mub^{-1}$. }
\label{fig:v2HydroPrediction}
\end{figure}

The \AFTER\ program is well suited for these new frontiers of the QGP hydrodynamic studies, providing a large rapidity coverage to measure several particle azimuthal asymmetries and the possibility to obtain large statistics for different targets. \cf{fig:v2HydroPrediction}~(a)-(b) shows particle yields as a function of pseudorapidity ($\eta_{\rm lab.}$) in mid-central (20-30\%) \PbW\ and \PbXe\  collisions at \sqrtsNN =72 GeV obtained using  EPOS. 
The expected statistical uncertainties on the elliptic flow ($v_{2}$) measurement for identified hadrons with 10 million 20-30\% central  \PbW\ and \PbXe\ events (\ie\ 100 million minimum-bias events) are presented in~\cf{fig:v2HydroPrediction}~(c)-(f). Even this small data sample, corresponding to an integrated luminosity of 14 $\mu$b$^{-1}$ for \PbW and 16 $\mu$b$^{-1}$ for \PbXe\ collisions, will allow for a precision study of $v_n$ over a very broad rapidity range. Such high-quality data will allow for an accurate determination of the temperature dependence of $\eta/s$. 

We believe that the \AFTER\ program is essential to advance our knowledge of the QGP macroscopic properties. This program will complement and continue the existing studies of $v_n$ with high precision thus providing a detailed account of the system evolution (including transverse and longitudinal dynamics) between SPS and the top RHIC energy. As such, \AFTER\ will be a perfect place to study collective effects in the system produced in \pA to \PbA\ collisions. It will provide high precision studies of the QCD matter, complementary to the ones performed at the RHIC BES program.

\paragraph{The rapidity scan: a new tool to study the QCD phase diagram}

Thermal model calculations indicate that the baryonic chemical potential $\mu_{B}$ and the temperature T depend on the rapidity~\cite{Becattini:2007qr}. The recent calculations using the Hadron Resonance Gas (HRG) model~\cite{Begun:2018efg} and a viscous hydro+cascade model vHLLE+UrQMD~\cite{Karpenko:2018xam} show that in \PbPb\ collisions at \AFTER\ $\mu_{B}$ strongly varies  with the rapidity: $0 < \mu_{B} < 250$~MeV (as vHLLE+UrQMD predicts, see~\cf{fig:muB_vs_y}) or even 80~MeV$ < \mu_{B} < 400$~MeV (given by the HRG model,~\cf{fig:muB-y-HRG}). These $\mu_B$ values cover a large fraction of the $\mu_B$ range accessible at the RHIC BES program~\cite{Keane:2017kdq,Adamczyk:2017iwn}, as illustrated by~\cf{fig:muB-T-HRG}. The $\mu_{B}$ vs. $y_{\cms}$ dependence suggests that one can perform a ``rapidity scan'' of the QCD diagram~\cite{Brewer:2018abr,Begun:2018efg}, complementary to the BES programs at RHIC and SPS. Measurements of correlations and fluctuations of the conserved quantities (electric charge, baryon and strangeness number, \etc) in small rapidity windows could provide a new approach to search for the QCD critical point and possible evidences of the first-order phase transition. However, the interpretation of such a data could be a challenge since the origin of the $\mu_{B}$ vs. rapidity dependence is an open question. For example, it might result from a superposition of different rapidity spectra of particles produced in a single, homogeneous fireball; or a product of small subdomains with different $\mu_B$ and T values, moving with different longitudinal velocities. In the latter case, one could indeed perform phase-diagram studies by varying the
particle rapidity.
Nevertheless, the large rapidity coverage of the \AFTER\ project, combined with an excellent particle identification capabilities of the ALICE and the LHCb detectors, makes it a perfect place for such a ``rapidity scan'' of the QCD phase diagram.

\begin{figure}[hbt!]
\centering
\subfigure[
]{ \includegraphics[width=0.4\textwidth]{physics-heavy-ion-collisions/muB_vs_y_HRG_72GeV.pdf}\label{fig:muB-T-HRG}}\quad 
\subfigure[
]{\includegraphics[width=0.4\textwidth]{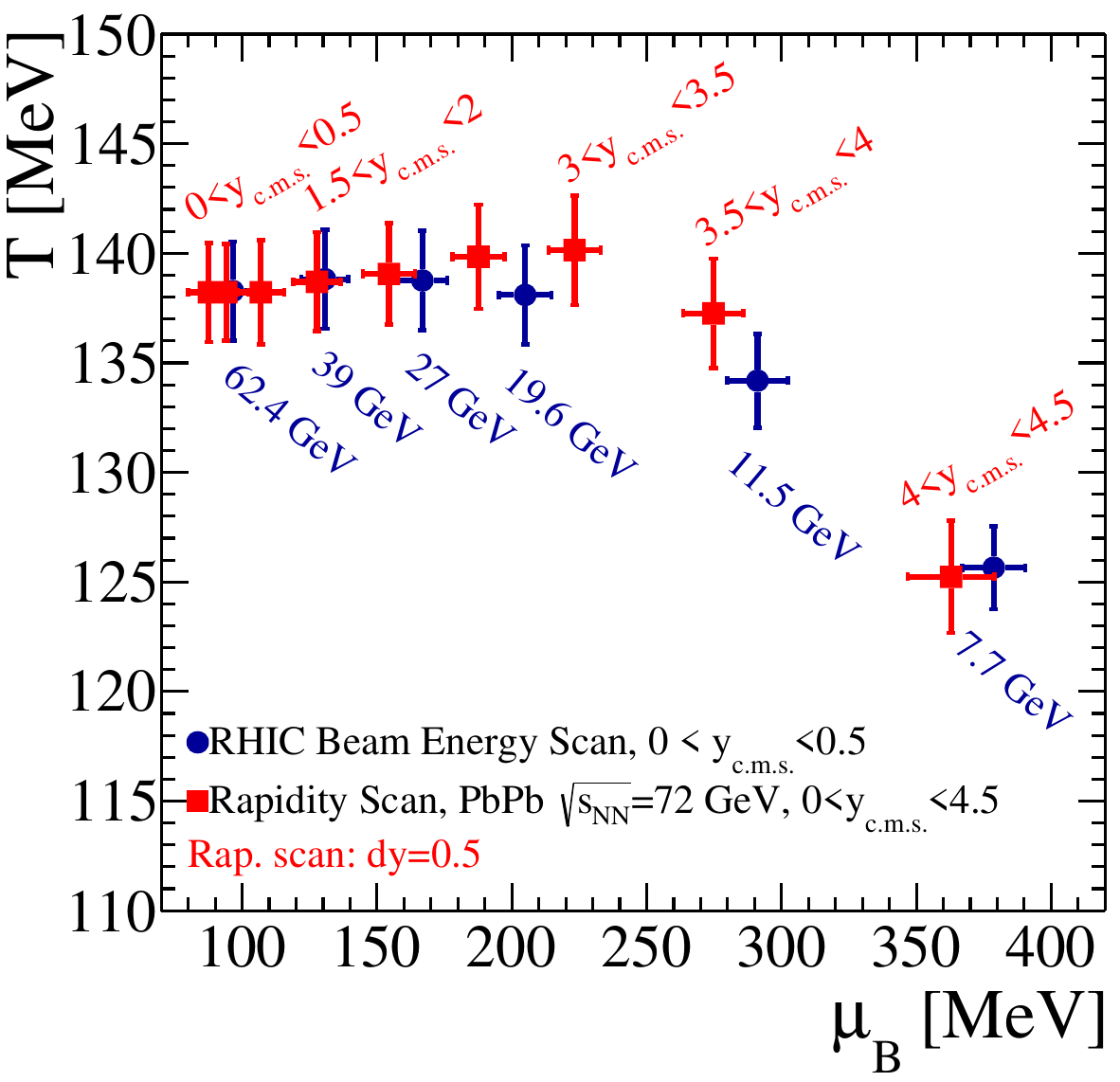}\label{fig:muB-y-HRG}}
\caption{(a) The baryonic chemical potential $\mu_B$ as a function of \ycms and (b) the temperature T as a function of $\mu_B$ in  0-10\% most central \PbPb\ collisions at \sqrtsNN = 72~GeV from the Hadron Resonance Gas model calculations [Adapted from~\cite{Begun:2018efg}]. The two series of results represent calculations that use two sets of particle densities as the input (with or without the $\Xi$ and $\Omega$ baryons). The uncertainties on the  points arise from an assumed relative uncertainty of 10\% on the particle yields measured in \AFTER.}
\label{fig:muBvsRapidityPbPb72GeV}
\end{figure}

%% file: physics-heavy-ion-collisions/physics-heavy-ion-collisions-small-system.tex
One of the recently most discussed topics in the heavy-ion community is a possibility of collective motion of partons and formation of QGP droplets in so-called small systems such as $d$Au, \pA\ and even in \pp\ collisions. These considerations were triggered by an observation of long-range angular correlations of hadrons in high-multiplicity \pp\ events at the LHC~\cite{Khachatryan:2010gv}, and later in $p$Pb collision at the LHC \cite{CMS:2012qk} and $d$Au at RHIC \cite{Adare:2013piz}. At first, it was a surprise since such correlations were earlier observed only in heavy-ion collisions. Another unexpected result was a large asymmetry of azimuthal angle distributions of hadrons produced in high multiplicity $d$Au~\cite{Adare:2013piz} and \pPb\ collisions \cite{Chatrchyan:2013nka}. A large, positive $v_2$ seen in the heavy-ion collisions is usually interpreted as an indication of collective interactions on the partonic level, thus of a possible QGP formation. Surprisingly, the values of $v_n$  observed in small systems are comparable to the ones observed in AuAu and PbPb collisions for the similar multiplicity events.  Positive flow parameters were registered in essentially all kinds of hadronic collisions with energies as low as \sqrtsNN=19.6 GeV \cite{Aidala:2017ajz}. Moreover, there is a significant positive elliptic flow of heavy-flavour particles and strange baryons in the \pA\ collisions at the LHC~\cite{CMS:2018xac,Sirunyan:2018toe}, although its magnitude is lower than the $v_2$ of light hadrons. There are also hints of non-zero $v_2$ of heavy quarks in $d$Au collisions at \sqrtsNN = 200 GeV at RHIC. 

In addition, other intriguing phenomena were observed in high-multiplicity \pp\ interactions. The production of multi-strange hadrons in such \pp\ collisions is enhanced~\cite{ALICE:2017jyt}, on a level remarkably similar to the results seen in \PbPb collisions. Moreover, heavy-flavour-particle yields (both the open-heavy flavour and charmonium) increase fast with the number of charged particles produced in a \pp\ collision~\cite{Adam:2018jmp,Abelev:2012rz,Adam:2015ota}. A few different models are able to qualitatively reproduce such a behaviour, for example, the string percolation approach~\cite{Ferreiro:2012fb} or EPOS~\cite{Werner:2013tya}. While these approaches significantly differ, they all assume some sort of collective interactions on the parton level.
	
Several explanations for the collective behavior in small systems were suggested: formation of QGP droplets \cite{Yan:2013laa,Bozek:2013uha,Weller:2017tsr,Mantysaari:2017cni}, parton escape with kinetic transport \cite{He:2015hfa}, coherent effects driven by a color glass condensate \cite{Dusling:2012iga}, as well as more basic QCD derivations of string formation \cite{Christiansen:2015yqa}. 
\AFTER\ is in the optimal position to address this issue and discriminate between the proposed scenarios. The luminosity available with  \AFTER\ is orders of magnitude larger than that at RHIC. The differential measurement of $D$ and \jpsi\ production and azimuthal asymmetries $v_n$ will be possible in \pp\ and \pA\ collisions. \cf{fig:mult:pp} shows the charged-particle-density distribution in \pp\ collisions at 115 GeV from EPOS. \AFTER\ will be able to probe very rare events, with a multiplicity of the order of 15 times larger than the average, and larger. The precision studies as a function of rapidity, transverse momentum and event multiplicity will shed new light on the problem at hand. As an example,~\cf{fig:D0:v2:pPb} shows a statistical precision of the measurement of  the elliptic flow $v_2$ of $D^0$ in \pPb\ collisions with \AFTER\ in two rapidity ranges -- the statistical uncertainty is at a sub-percent level for $\pT < 3$~GeV. While very limited number of model predictions exist for charmed hadron $v_2$ in small-system collisions (for example the POWLANG~\cite{Citron:2018lsq} and Colour-Glass-Condensate (CGC) predictions~\cite{CMS:2019isc} for \pPb\ collisions at \sqrtsNN = 8 TeV), one may consult relative variation of predictions for $v_2$ of $D^0$ in heavy-ion collisions at RHIC~\cite{Adamczyk:2017xur} to assess the discrimination power of the proposed measurement at \AFTER. At transverse momentum $\pT \approx 2$ GeV the difference between theory results is significantly larger than the estimated statistical uncertainty on $D^0$ $v_2$ in~\cf{fig:D0:v2:pPb}. Thus, together with the nuclear modification factor $R_{\pPb}$, these data will provide means to precisely determine a possible collective behaviour of heavy quarks and quantify the cold-nuclear-matter effects. Moreover, high-luminosity data samples available at \AFTER\ will allow $v_n$ to be measured for identified hadrons with multi-particle-correlation methods over a wide rapidity range. This gives a useful handle on the collective effects in small systems. Consequently, \AFTER\ offers novel opportunities to study these effects at low energies with multiple probes and over a broad kinematic range.

\begin{figure}[htp]
	\centering
	\includegraphics[width=0.6\textwidth]{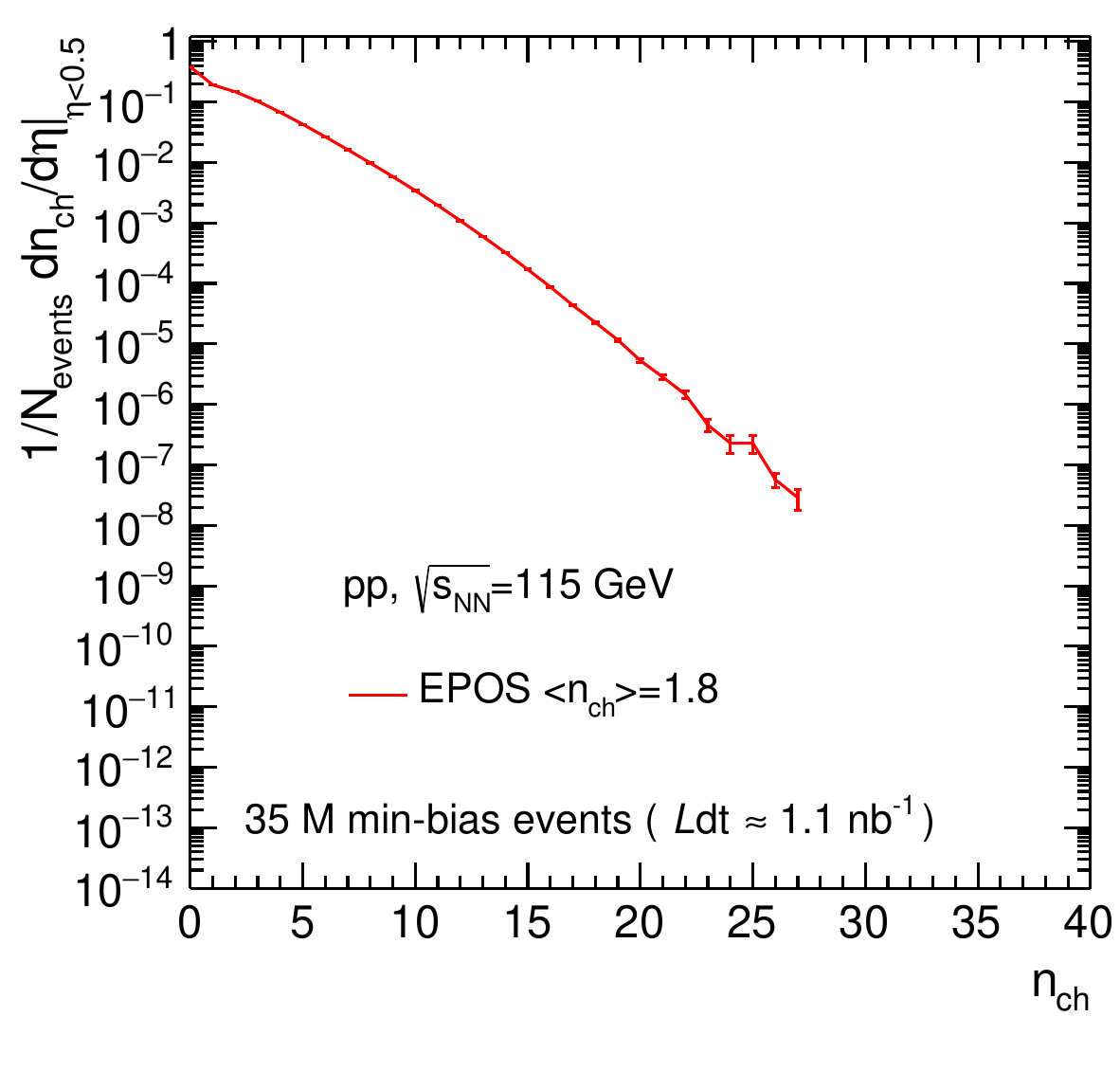} 
	\caption{Distribution of the charged-particle density in \pp\ collisios at \sqrts=115~GeV at mid-rapidity from the EPOS model for 35 million minimum-bias events, which corresponds to an integrated luminosity of 1.1 nb$^{-1}$.}
	\label{fig:mult:pp}
\end{figure}

\begin{figure}[htp]
\centering
  \includegraphics[width=0.6\textwidth]{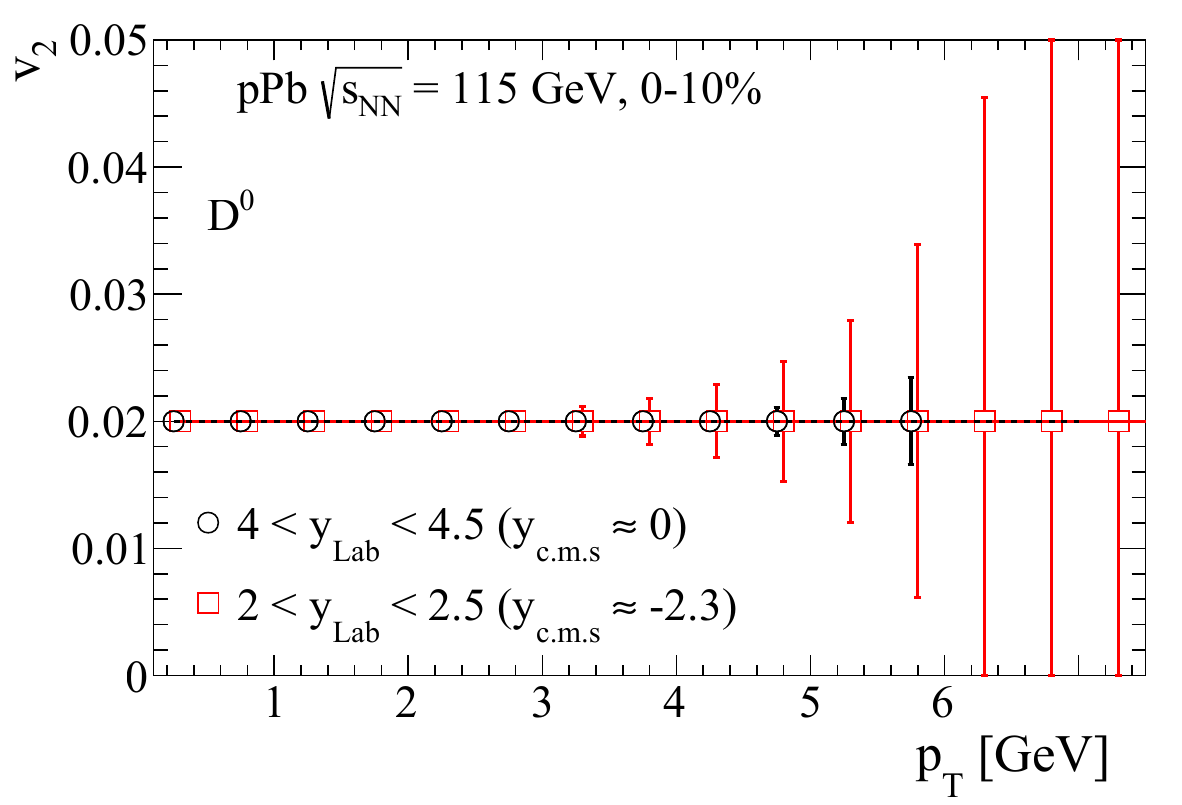} 
\caption{Expected statistical uncertainties of $D^0$ elliptic flow in $p$Pb collisions at \sqrtsNN=115~GeV at mid-rapidity and backward rapidity measured with a LHCb-like detector for $\int \mathcal L_{p\rm Pb}$ = 160 pb$^{-1}$.  Calculations include the acceptance and reconstruction efficiency of a LHCb-like detector. The results indicate that $v_2$ will be measured with sub-percent precision over a broad \pt\ and rapidity range. [Adapted from~\cite{Kikola:2015lka}].}
\label{fig:D0:v2:pPb}
\end{figure}

%% file: physics-heavy-ion-collisions/physics-heavy-ion-collisions-DY.tex
\begin{figure}[!htb]
\centering
  \subfigure[~]{\includegraphics[width=0.45\textwidth]{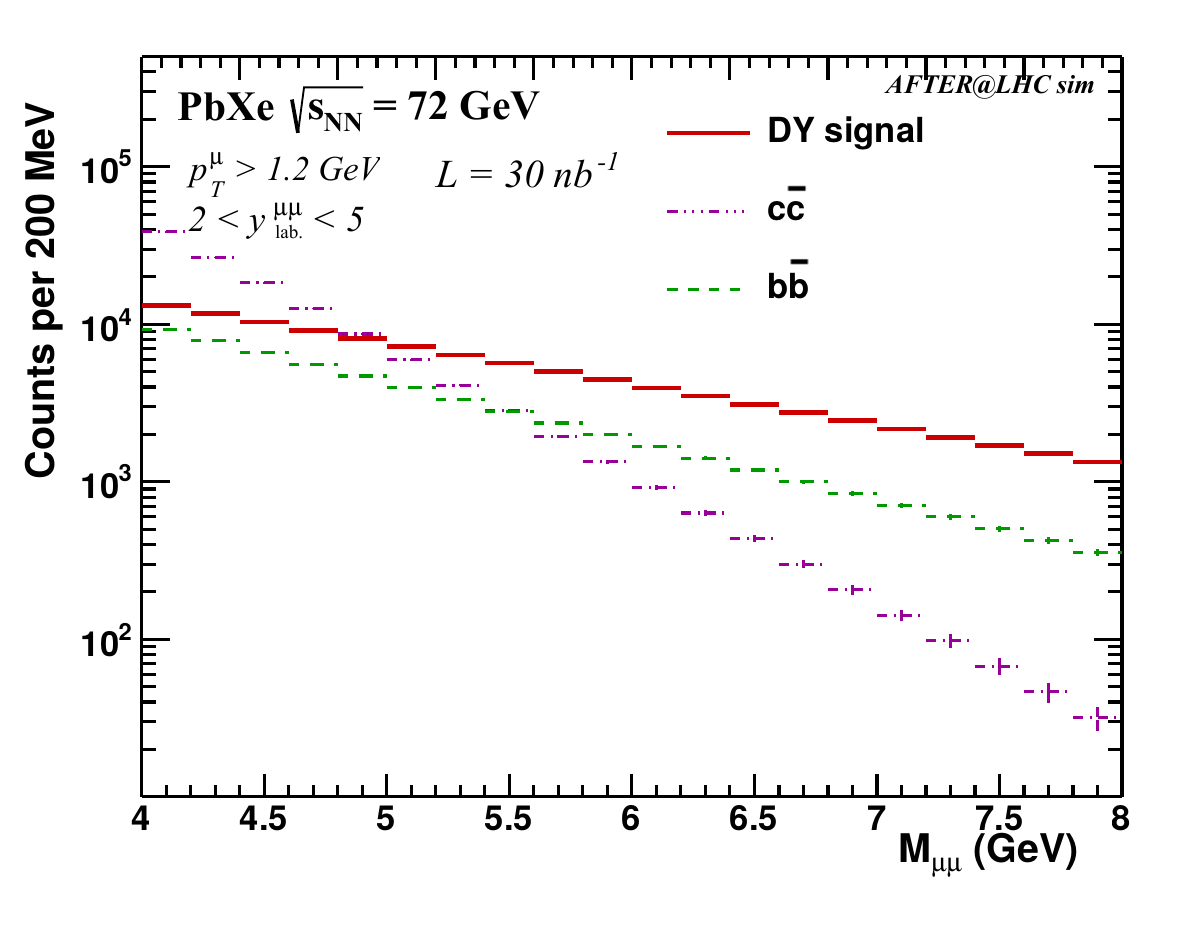}}
  \subfigure[~]{\includegraphics[width=0.45\textwidth]{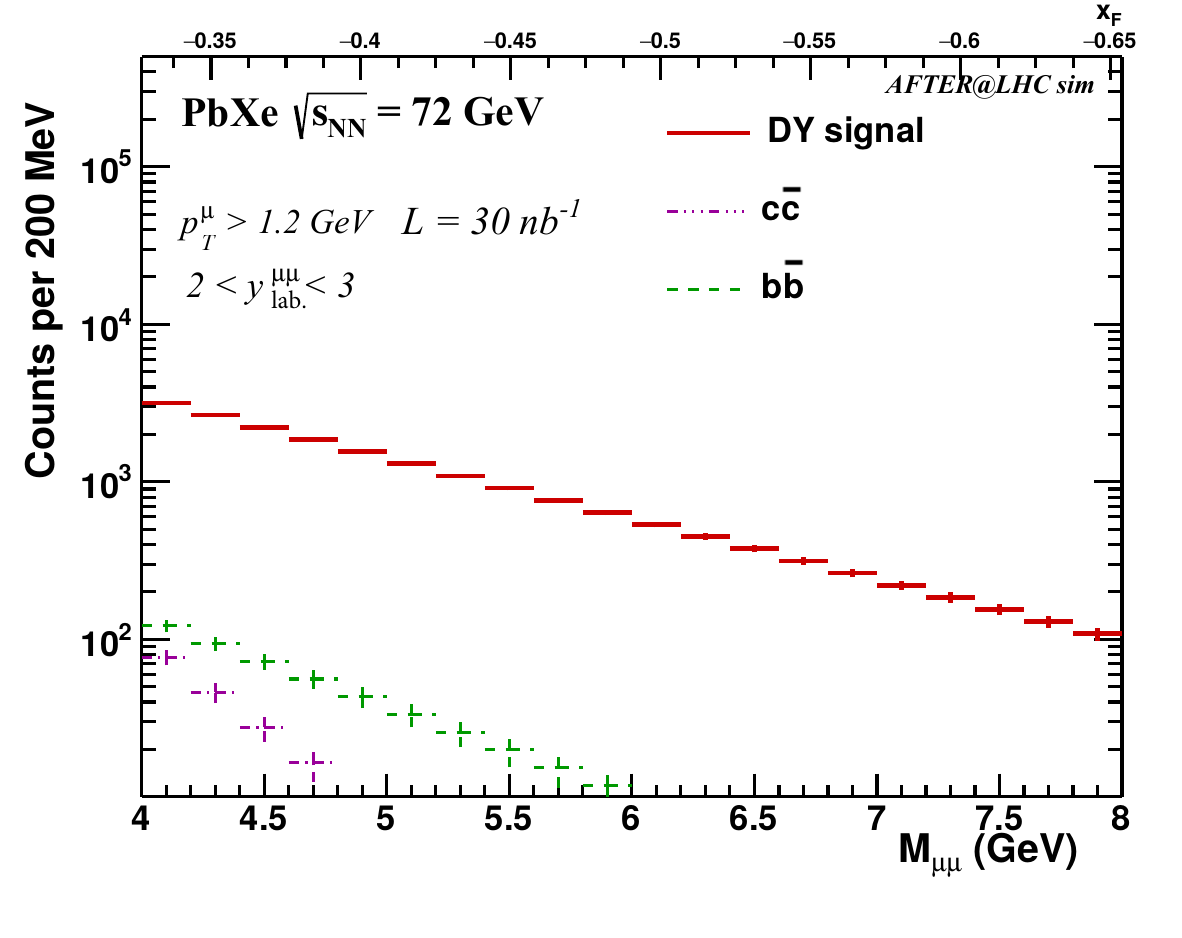}}\\
  \subfigure[~]{\includegraphics[width=0.45\textwidth]{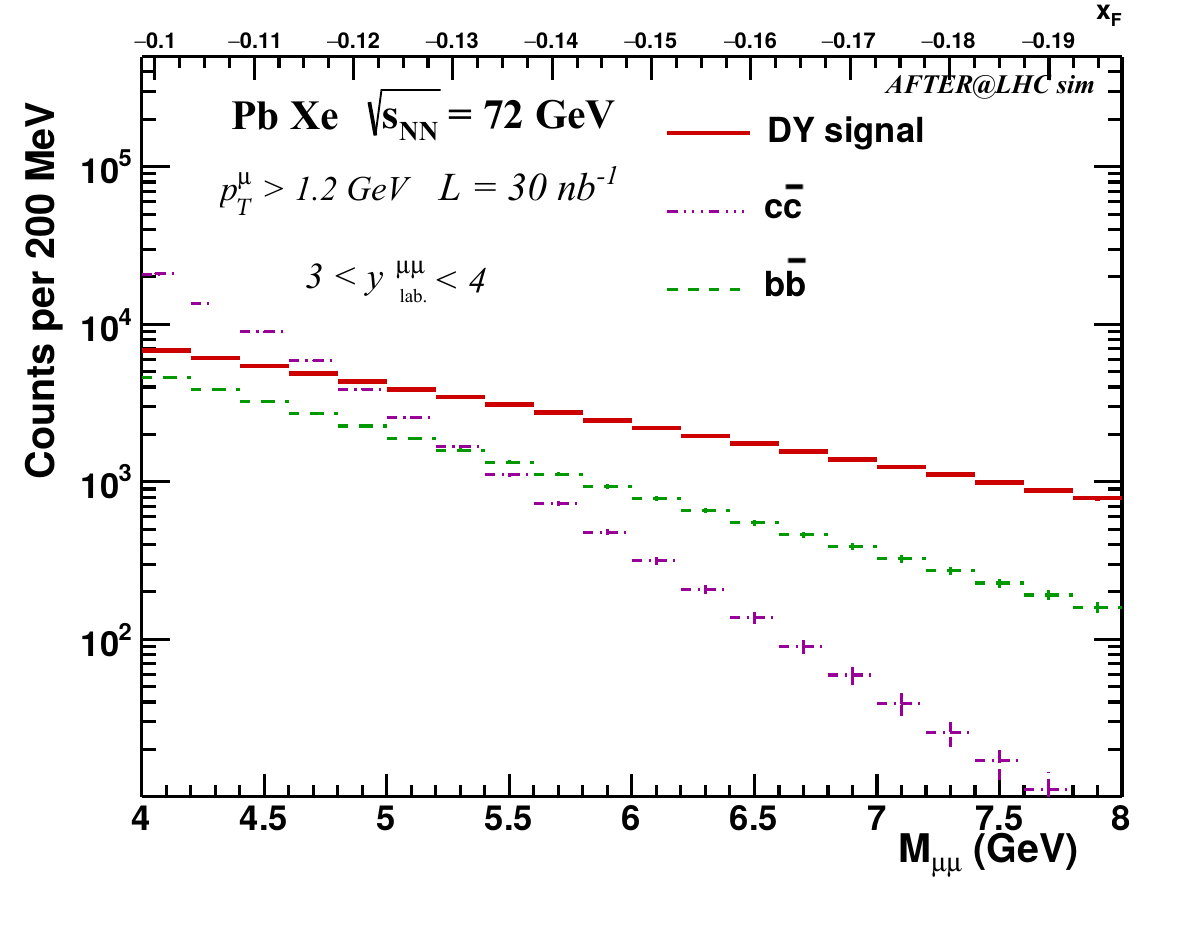}}
  \subfigure[~]{\includegraphics[width=0.45\textwidth]{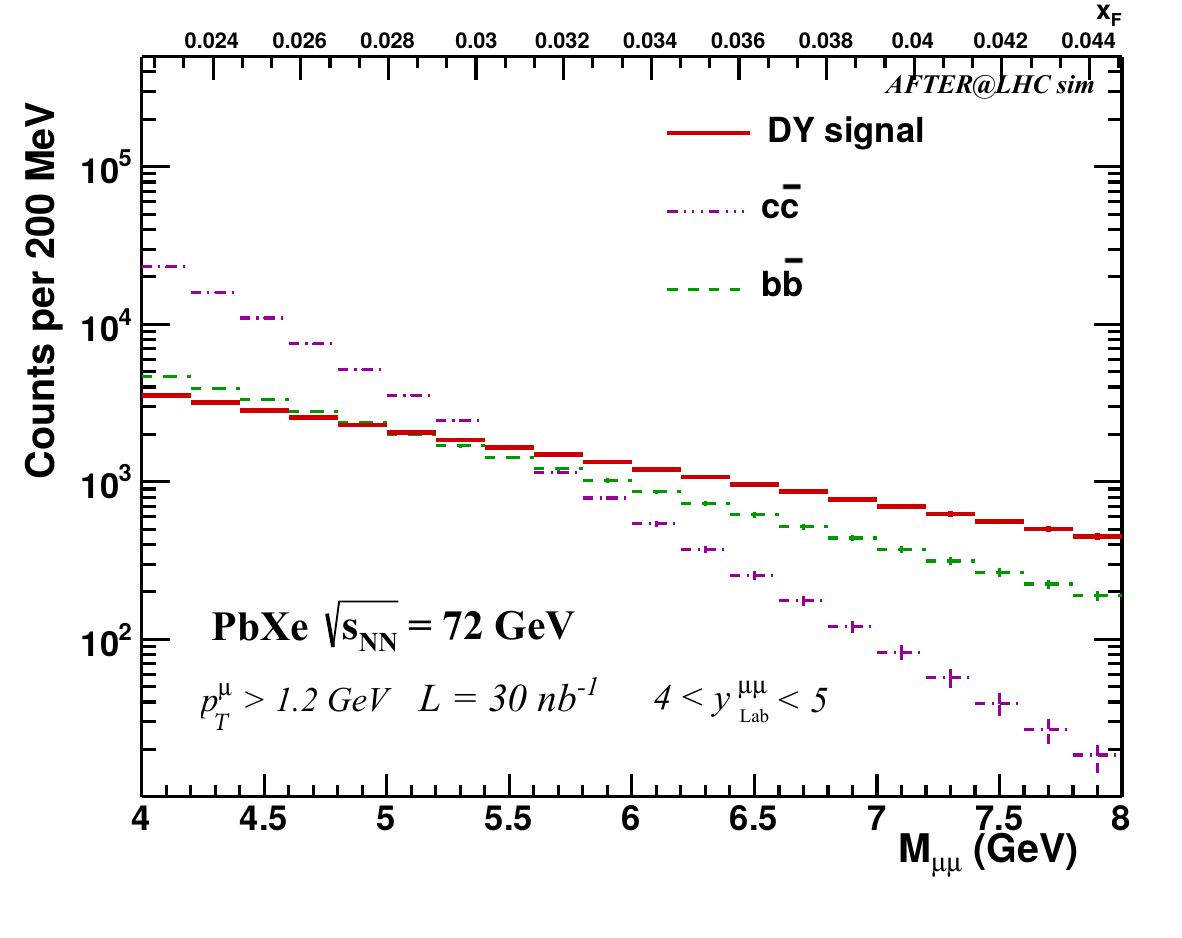}}
\caption{Di-muon invariant-mass distributions ($4 < M_{\mu^+\mu^-} < 8$~GeV) from DY, $c\overline{c}$ and $b\overline{b}$ productions registered with a LHCb-like detector, for $\PbXe$ collisions at $\sqrt{s_{NN}} = 72$~GeV with $\int \mathcal L_{\rm PbXe} = 30 {\rm nb}^{-1}$, assuming $R_{AA}=1$, in the integrated rapidity range of $2< y^{\mu\mu}{\rm lab.} <  5$ (a) and divided into the following ranges: $2 < y^{\mu\mu}_{\rm lab.} <  3$ (b), $3 < y^{\mu\mu}_{\rm lab.} <  4$ (c) and $4 < y^{\mu\mu}_{\rm lab.} <  5$ (d). For (b-d), the upper $x$-axis represents the corresponding $x_{F}$ values in a given rapidity range and invariant-mass bin. The combinatorial background is not presented and systematic uncertainties resulting from the background subtraction with with the event-mixing technique are not included.}
\label{fig:DYmass_HI}
\end{figure}

Initial-state effects observed in \pA\ collisions are currently extrapolated to \AA\ collisions assuming that the effects factorise linearly, that is, that the effects associated to initial-state sources are independent in each nucleon-nucleon binary collision. This na\"ive assumption can be tested using electromagnetic probes: high-$p_T$ isolated photons (inverse Compton process), DY, $W$ and $Z$ bosons. These probes are produced from initial-state partons and do not interact with the nuclear medium. The nuclear modifications observed for isolated photons, $W$ and $Z$ bosons measured by CMS at mid-rapidity~\cite{Chatrchyan2012256,Chatrchyan201266,Chatrchyan:2014csa} are smaller than $\sim$20\% in \PbPb\ collisions. These results rule out a scenario of a large suppression due to initial-state effects in heavy-ion collisions. However, they cannot test whether processes observed in \pA collisions are magnified in \AA\ collisions given the small nuclear modifications observed in \pA collisions to start with. Theoretical work such as that performed in \cite{Kopeliovich2011333} indicates that initial-state effects can be modified in \AA\ collisions with respect to \pA\ since the natural scale of a given process is boosted and thus affected by initial-state effects with a different magnitude. A conclusive experimental test of such a phenomenon from \pA\ to \AA\ collisions needs to be performed in a broad $x$ range where significant variation of the nuclear modifications is expected in \pA collisions (see the \pA physics section, Sec. \ref{sec:high_x_nucl_structure}).

\begin{figure}[!htb]
\centering
\begin{tabular}{ll}
  \includegraphics[width=0.5\textwidth]{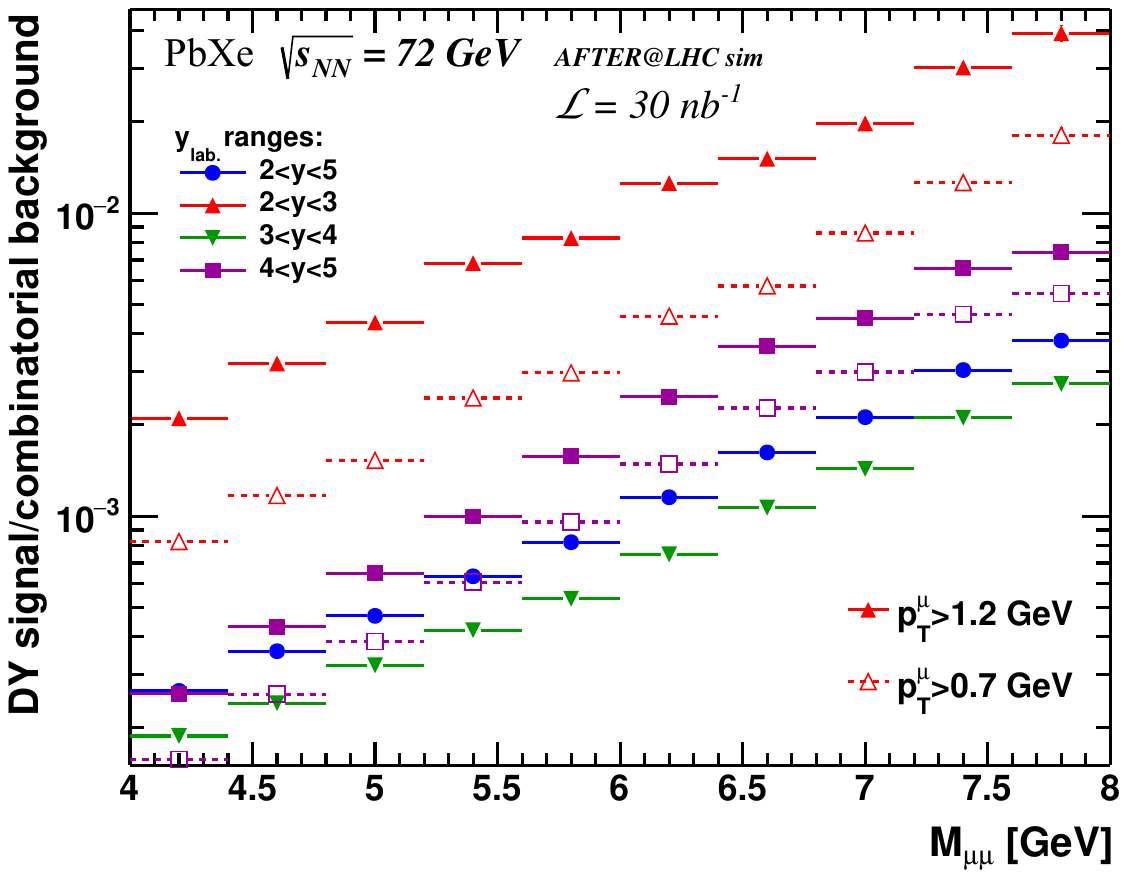}
\end{tabular}
	\caption{DY signal-over-combinatorial-background ratio ($S/B$) for a LHCb-like detector as a function of the di-muon invariant mass, for  \PbXe collisions at $\sqrt{s_{NN}} =$ 72 GeV with $\int \mathcal L_{\rm PbXe}$ = 30 ${\rm nb}^{-1}$ and for 2 $< y_{\rm lab.} < $ 5, 2 $< y_{\rm lab.} < $ 3, 3 $< y_{\rm lab.} < $ 4 and 4 $< y_{\rm lab.} < $ 5. The solid lines represent the $S/B$ with the selection condition for DY single-muon $p_{T}^{\mu} >$ 1.2 GeV, and dashed lines represent the $S/B$ with $p_{T}^{\mu} >$ 0.7 GeV.  No nuclear modifications assumed.}
\label{fig:DYSB}
\end{figure}

\begin{table}[!htb]
\small
\begin{center}
\begin{tabular}{c|c|c|c|c|c|c}
\multirow{2}{1.5cm}{} & \multicolumn{3}{c|}{$M_{\mu\mu}$: 4-5 GeV} & \multicolumn{3}{c}{$M_{\mu\mu}$: 5-6 GeV} \\
\cline{2-7}\cline{2-7}
   & signal ($\times 10^{3}$) & bkg. ($\times 10^{6}$)  & S/B ($\times 10^{-3}$)
    & ~~~~signal~~~~~ & ~~~~~bkg.~~~~~  & ~~~~S/B~~~~ \\
\hline \hline
\ylab: 2-3 & 11.48 	& 4.38 	& 2.6	 & 4.77	& 0.76 &	 6.3   \\ \hline
\ylab: 3-4 & 26.85 	& 121.90 & 0.2 & 15.58 & 36.6	& 0.4 \\ \hline
\ylab: 4-5 & 15.33  	& 45.08 	& 0.3 & 8.28	& 8.05 & 1.0  \\ \hline
\multicolumn{7}{c} ~\\
\multirow{2}{1.5cm}{} & \multicolumn{3}{c|}{$M_{\mu\mu}$: 6-7 GeV} & \multicolumn{3}{c}{$M_{\mu\mu}$: 7-8 GeV} \\
\cline{2-7}
 & signal  & bkg. & S/B & signal & bkg. & S/B  \\
\hline \hline
\ylab: 2-3 & 1.92 & 0.14 & 13.9 & 0.82 & 0.03 & 28.1  \\ \hline
\ylab: 3-4 & 8.80 & 9.68 & 0.9 & 5.11 & 2.42 & 2.1 \\ \hline
\ylab: 4-5 & 5.09 & 1.68 & 3.0 & 2.78 & 0.44 & 6.3 \\ \hline
\end{tabular}
\end{center}
\caption{\label{tab:DY_yields}DY yields ($\times 10^{3}$), uncorrelated background yields ($\times 10^{6}$) and DY over the uncorrelated backgroud ratios ($\times 10^{-3}$) for PbXe collisions at $\sqrt{s_{NN}} =$ 72 GeV in 4 di-muon invariant-mass ranges between 4 and 8 GeV and three rapidity ranges between 2 and 5, with single $\mu$ satisfying $p_{T} >$ 1.2 GeV. The results hold for LHCb-like performances and $\mathcal L_{\rm PbXe}$ = 30 nb$^{-1}.$ }
\end{table}

The physics program of \AFTER\ includes the precise measurements of the DY process which can probe initial-state effects on quarks in several \AA\ collision species. The greatest challenge in the DY measurements in colliders is the large correlated background from $b+\bar{b} \rightarrow B^++B^- \rightarrow l^+l^-$ and $c$+$\bar{c} \rightarrow D^++D^- \rightarrow l^+l^-$. This background is much smaller at the \AFTER\ \cms\ energy in \PbA\ collisions. The large combinatorial background typically expected in \PbA\ collisions can precisely be determined with the mixed-event technique and the large expected amount of like-sign di-leptons in order to reduce uncertainties in the combinatorial-background normalisation. \cf{fig:DYmass_HI} shows the invariant mass distributions of the di-muon pairs in different rapidity intervals, and \cf{fig:DYSB} shows the signal-to-background ratios in \PbXe\ collisions at $\sqrt{s_{NN}} =$ 72 GeV. Table~\ref{tab:DY_yields} shows the DY signal and background yields and signal-to-background (S/B) ratios in these reactions. Overall, the background is significant in the low-mass range ($M_{\mu\mu}$: 4-5 GeV), but it will be suppressed by imposing a stringent cut on the muon transverse momentum. Figure~\ref{fig:DYSB} indicates, that the S/B increases significantly when a single muon cut of $p_{T}^{\mu} >$ 1.2 GeV is applied, compared to the $p_{T}^{\mu} >$ 0.7 GeV case. 
Nonetheless, while the measurement could be challenging in some kinematic ranges, the expected yields will allow for a definitive test of factorisation of the initial-state effects from \pA\ to \AA\ collisions.

%% file: summary/summary.tex
\section{Conclusions}
\label{sec:conclusion}

Unlike the Fermilab-Tevatron and DESY-HERA colliders (with proton beams in the TeV range), no fixed-target program was planned for the LHC. In this review, we have put forward a strong physics case for such a program both for the multi-TeV proton and ion LHC beams. Such a physics case relies on extensive theory work and projection studies which have been performed with LHCb and ALICE-like detectors allowing for high precision studies in the backward hemisphere of $pp$, $pA$ and $AA$ collisions. 

These projections cover the 3 main research axes of the physics case, namely that of the nucleon and nucleus structure at high momentum fractions, that of the nucleon-spin decomposition in terms of the partonic degrees of freedom and that of the properties of the nuclear matter at extreme conditions such as those resulting from ultra-relativistic heavy-ion collisions. 

They are relevant for different possible implementations which we have reviewed including the state-of-the-art solutions provided by modern polarised gas targets or by splitting the beam with a bent crystal. For each of the possible implementations, we have also detailed the expected luminosities compatible with the LHCb and ALICE detector capabilities. For a number of studies where such projections are not yet available, we have collected the existing theory predictions for \eg\ cross sections, spin and azimuthal asymmetries or nuclear modification factors. 

Overall, we believe that the present review constitutes a very solid basis for the elaboration of a rich and fruitful LHC fixed-target program starting as early as 2020.

%% file: acknowledgements.tex
\section*{Acknowledgements}

We thank R. Arnaldi, V. Chambert, F. Fleuret, B. Genolini, V. Kartvelishvili, A. Nass, R. Mikkelsen, S.~Platchkov, F. Rathmann, P. Rosier, M. Schmelling, E. Scomparin, E. Steffens, U. Uggerh\o j, R. Ulrich, for their involvement at the early stage of this project.

We are particularly thankful to C. Barschel, U. D'Alesio, N. Doshita, M. Ferro-Luzzi,  V. Gon\c calves, T~.Pierog, D.~Pitonyak, K. Pressard, M. Schlegel, M. Siddikov, and H. Spiesberger for providing us with material for this review.
We thank D. d'Enterria and R. Milner for useful and constructive comments on the manuscript.

We thank F.~Arleo, N.~Armesto, E.~Aschenauer, S.~Barsuk, D.~Boer, F.~Bradamante, M.~Calviani, M.~Chiosso, Z.~Conesa del Valle, J.R.~Cudell, J.~Cugnon, T.~Dahms, A.~Dainese, O.~Denisov, M.~Diehl, B.~Espagnon, Y.~Gao,  M. Gazdzicki, S.~Glazov, G.~Graziani, A.~Holzner, P.~Jacobs, J.~Jowett, T.~Kasemets, M.J.~Kim, L.~Kluberg, S.~Klein, B.~Kopeliovich, P.~Lenisa, G.~Martinez, S.~Montesano, P.~Mulders, F.~Olness,  J.C.~Peng, B.~Pire, C.~Pisano, M.~Ploskon, A.~Poblaguev, J.W.~Qiu, B.~Saghai, H.~Satz, G.~Schnell,  D.~Sivers, T.~Stavreva, A.~Stocchi, C.~Suire, L.~Szymanowski, M.~Ubiali, T.~Ullrich, C.~Vall\'ee, R. Vogt, S.~Wallon, M.~Winn, C.~Yin Vallgren, P.~Zurita for useful comments and suggestions.

We thank D. Henry for artworks.

This project has received funding from the European Union’s Horizon 2020 research and innovation programme under grant agreement STRONG-2020 No 824093. MGE is supported by the European Research Council (ERC) under the European Union's Horizon 2020 research and innovation program (grant agreement No. 647981, 3DSPIN). AS acknowledges support from U.S. Department of Energy contract DE-AC05-06OR23177, under which Jefferson Science Associates, LLC, manages and operates Jefferson Lab, for his work on this review [JLAB-THY-18-2756]. AS also acknowledges support from the European Commission through the Marie Sk\l{}odowska-Curie Action SQuHadron (grant agreement ID: 795475). S.J.B. is supported by the Department of Energy, contract DE–AC02–76SF00515 [SLAC-PUB-17291]. GC acknowledges support from the ERC Ideas Consolidator Grant CRYSBEAM G.A. n. 615089. AK, NT, LM and CH acknowledge support by the RFBR/CNRS grant 18-52-15007 and PRC-1980. ZY is supported by NSFC under the grant number 11575094.  This work is supported by Grant No. 2017/26/M/ST2/01074 of the National Science Centre, Poland.
The work of JPL, CH and LM  was partly supported by the French CNRS via the COPIN-IN2P3 agreement, the IN2P3 project ``TMD@NLO" and ``GLUE@NLO" , the  Franco-Spanish PICS ``Excitonium", the project Quarkonium4AFTER of the Franco-Chinese LIA FCPPL, by the P2IO Labex via the ``Gluodynamics'' project. and by the Paris-Saclay U. via the P2I Department, that of HSS by the ILP Labex (ANR-11-IDEX-0004-02, ANR-10-LABX-63), that of JS by Funda\c c\~ao para a Ci\^encia e a Tecnologia under contract CERN/FIS-PAR/0015/2017, that of EGF by that of EGF by Ministerio de Ciencia e Innovacion of Spain under project FPA2017-83814-P, Unidad de Excelencia Maria de Maetzu under project MDM-2016-0692 and the Paris-Saclay U., that of NY by JSPS Postdoctoral Fellowships for Research Abroad. The work of FD is supported by the ``Departments of Excellence 2018 - 2022" Grant awarded by the Italian Ministry of Education, University and Research (MIUR) (L.232/2016).

%% file: Appendix_Implementation_B.tex
\subsection{Schematic view of the H-jet system used at the BNL-RHIC collider}
\label{appendix:Gas_jet}

\begin{figure}[hbt!]
\centering
\includegraphics[scale=0.35]{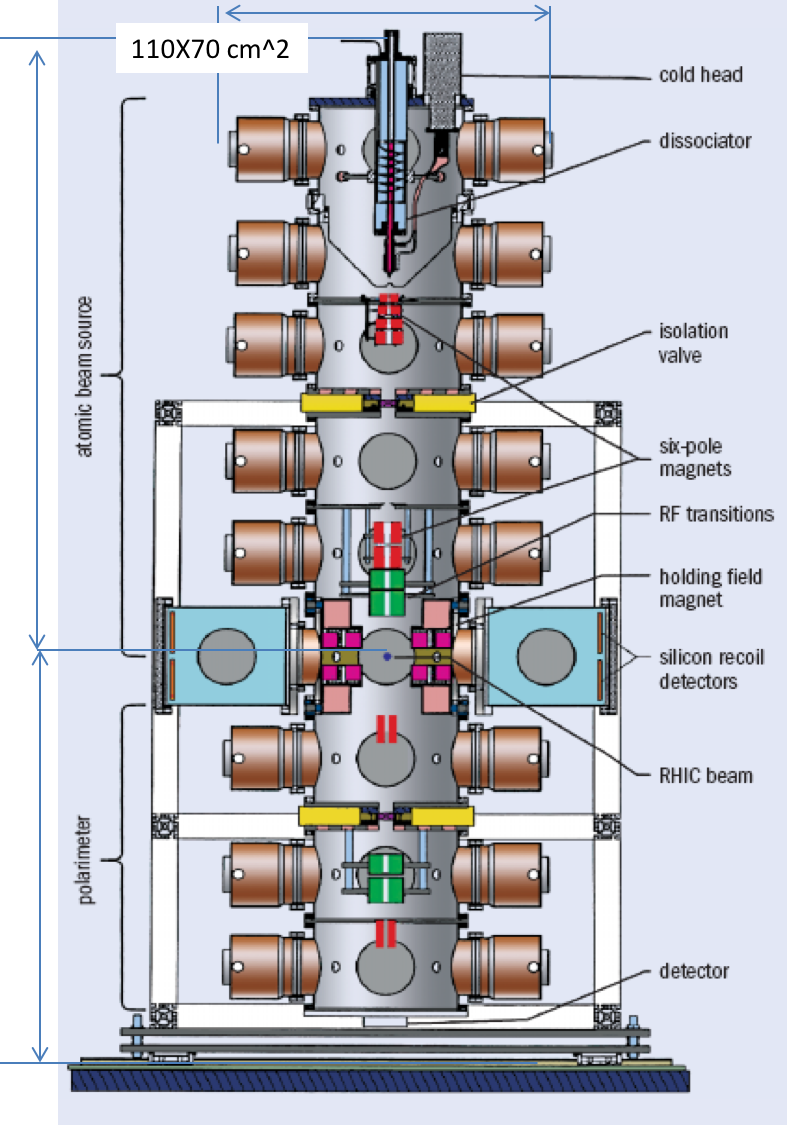}
\caption{Schematic view of the polarised H-jet system used at BNL-RHIC collider~\cite{Zelenski:2005mz} that consists of an Atomic Beam Source, a turbo-molecular pumping system and a Breit-Rabi polarimeter.}
\label{Gas-jet_figure}
\end{figure}

%% file: Appendix_Implementation_Abis.tex
\subsection{Possible setup of the beam-splitted option upstream of LHCb}
\label{appendix:UA9}

\begin{figure}[hbt!]
\centering
\includegraphics[scale=0.25]{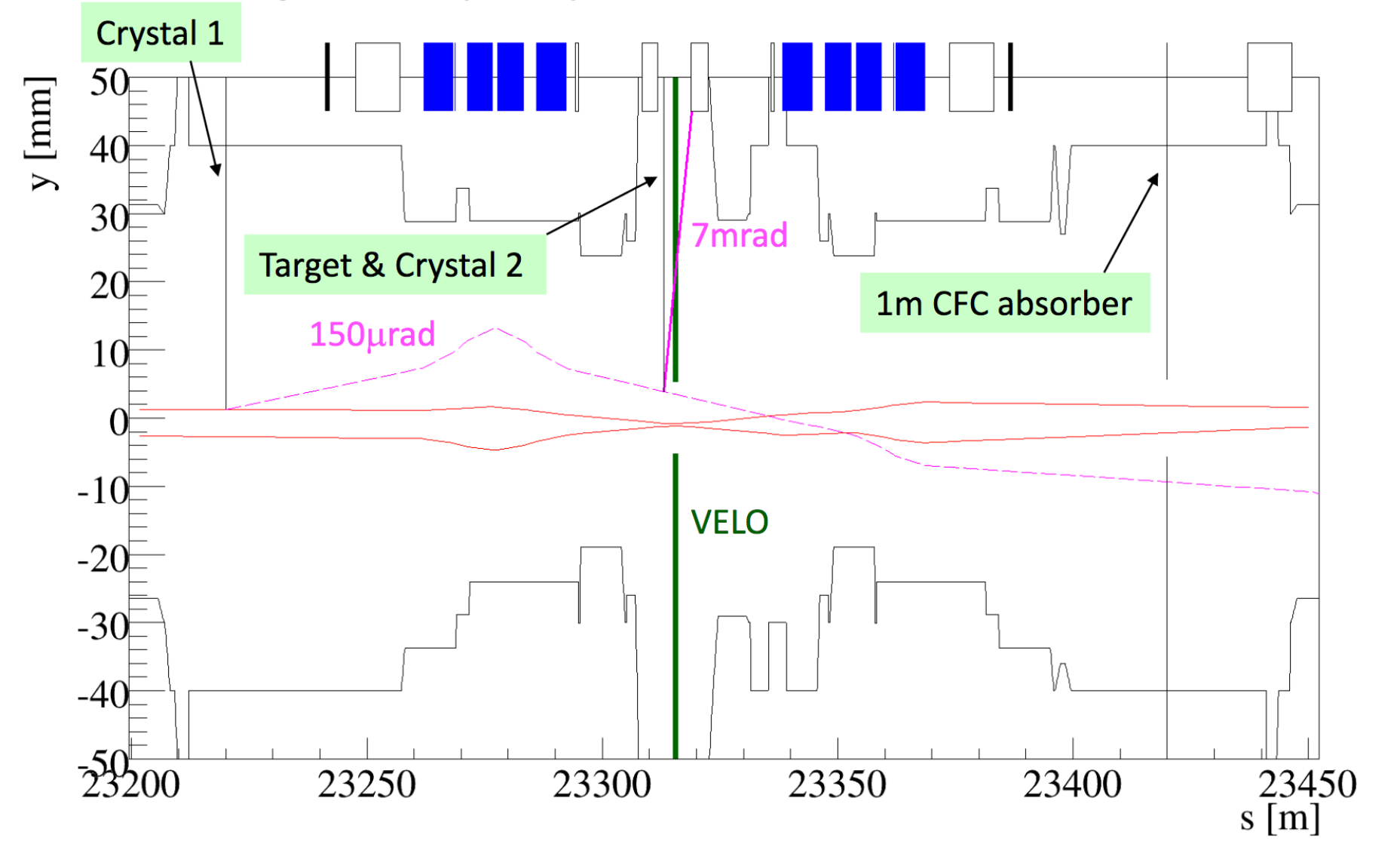}
\caption{Possible setup of the beam-split option upstream of LHCb [From~\cite{Scandale:2017MarchPBC}]: a first bent crystal deflects the halo particles upstream the LHCb IP at 5$\sigma$ from the centre line, where $\sigma$ is the beam width. The red full lines represent the main beam envelope. The deflected beam, represented in pink dashed line, is highly focused and its envelope is not visible in the figure. A target inserted in the pipe located in front of the detectors intercepts the deflected halo. For studying electric and magnetic dipole moment of charm charged baryons, a second crystal is necessary. An absorber located downstream LHCb intercept the halo particles non-interacting with the target.}
\label{E1039_target}
\end{figure}

%% file: Appendix_Implementation_C.tex
\subsection{Schematic view of the E1039 target}
\label{appendix:E1039_tg}

\begin{figure}[H]
\centering
\includegraphics[scale=0.35]{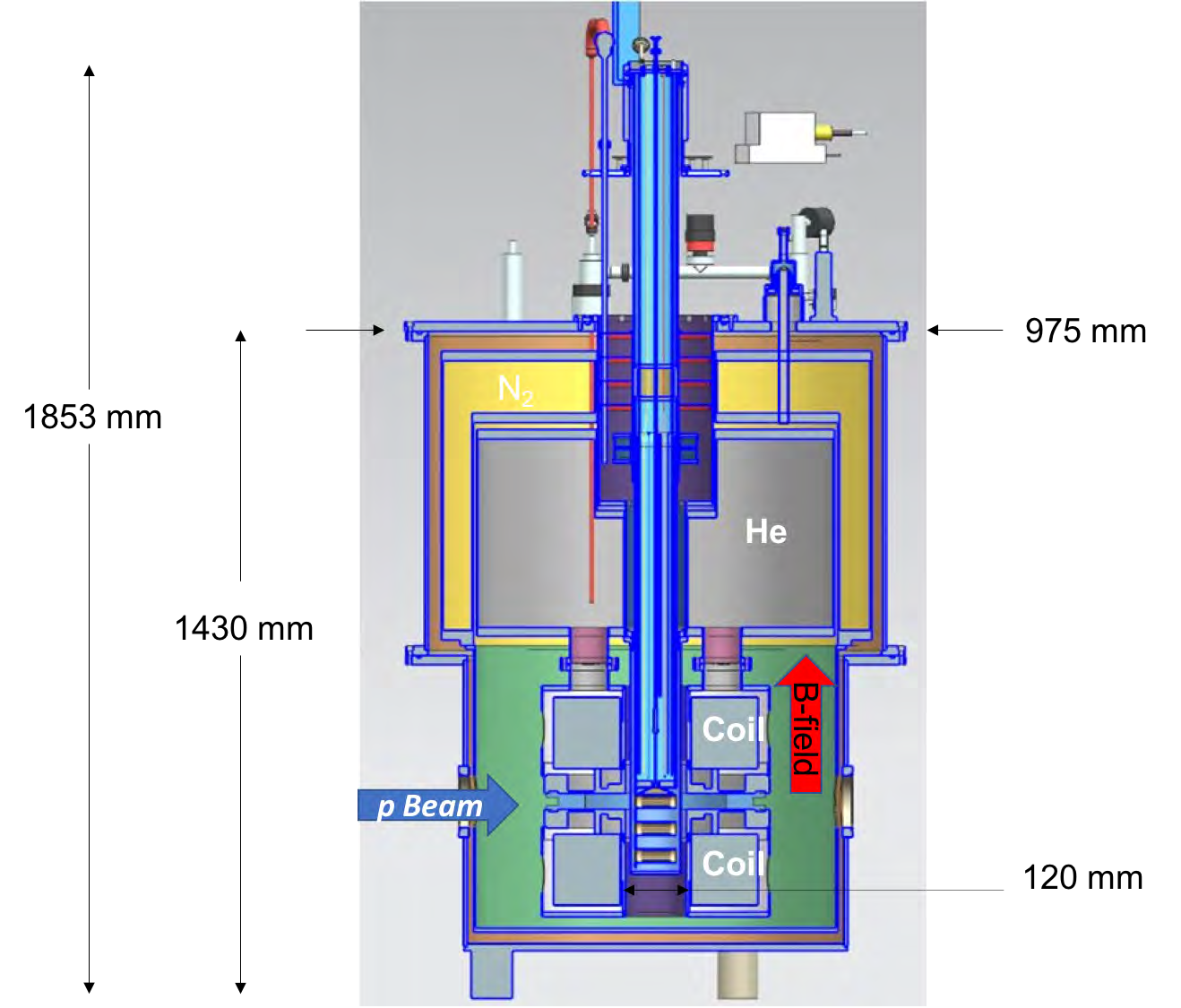}
\caption{Schematic view of the E1039 polarised target~\cite{Klein:zoa} (cylindrical geometry). It consists of a split coil superconducting magnet, of a refrigerator inside the magnet and in the centre of the system, a target cell, the microwave horn and the NMR coils. }
\label{E1039_target_tg}
\end{figure}

%% file: Appendix_Implementation_D.tex
\subsection{Schematic view of the COMPASS target}
\label{appendix:compass_tg}

\begin{figure}[hbt!]
\centering
\includegraphics[scale = 0.35]{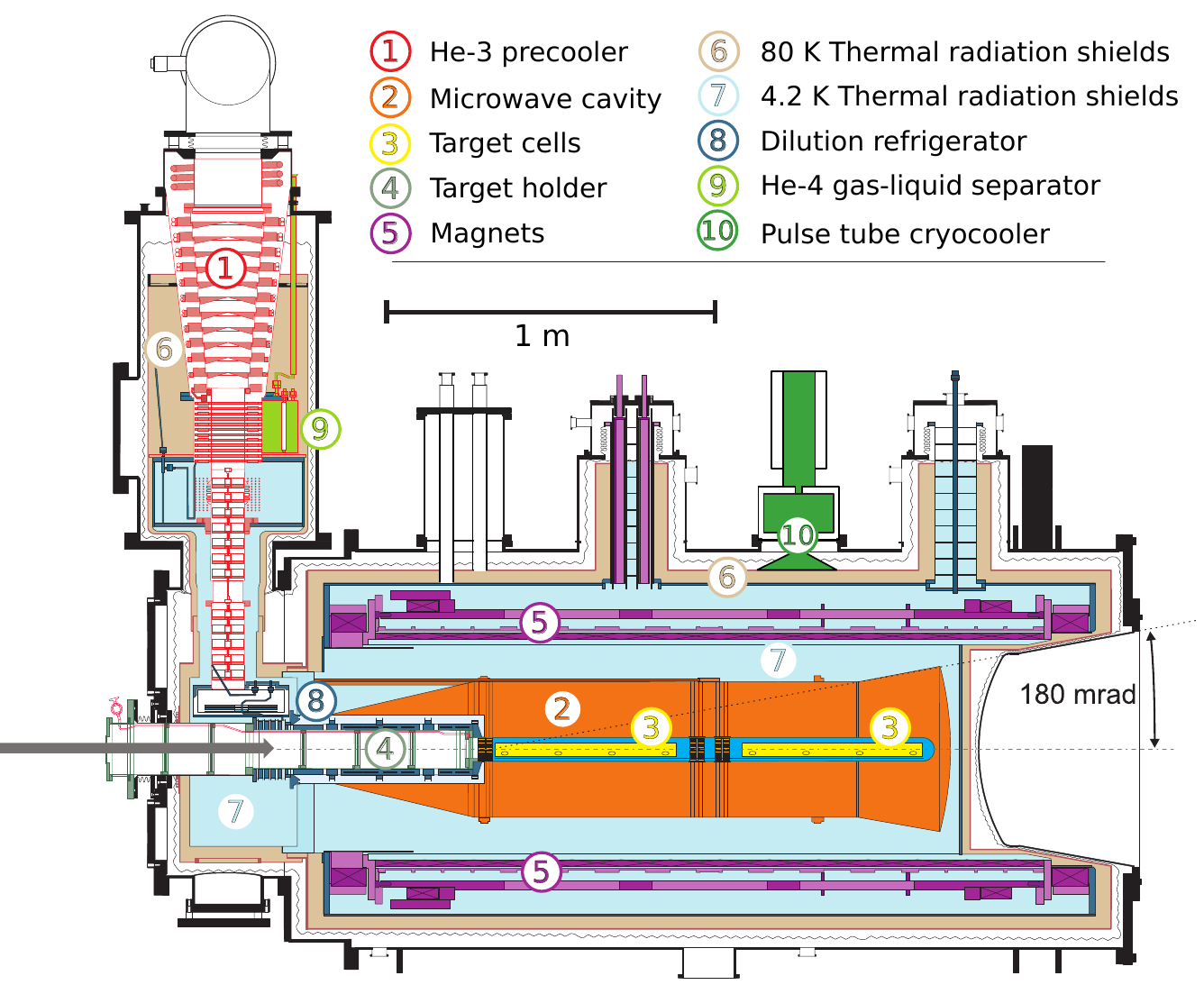}
\caption{Schematic side view of the COMPASS polarised target~\cite{Pesek:2014uua,Doshita:2004ee,Ball:2003vb}. We would like to highlight the two sets of 55 cm long target cells (denoted (3)) and the 2.5T solenoid and 0.6T dipole magnets (denoted (5)). The direction of the beam is represented by the grey arrow.}
\label{target_COMPASS}
\end{figure}

%% file: Appendix_Detector_ALICE.tex
\subsection{Schematic view of the ALICE detectors}
\label{appendix:alice_det}

\begin{figure}[H]
\centering
\includegraphics[width=0.8\textwidth]{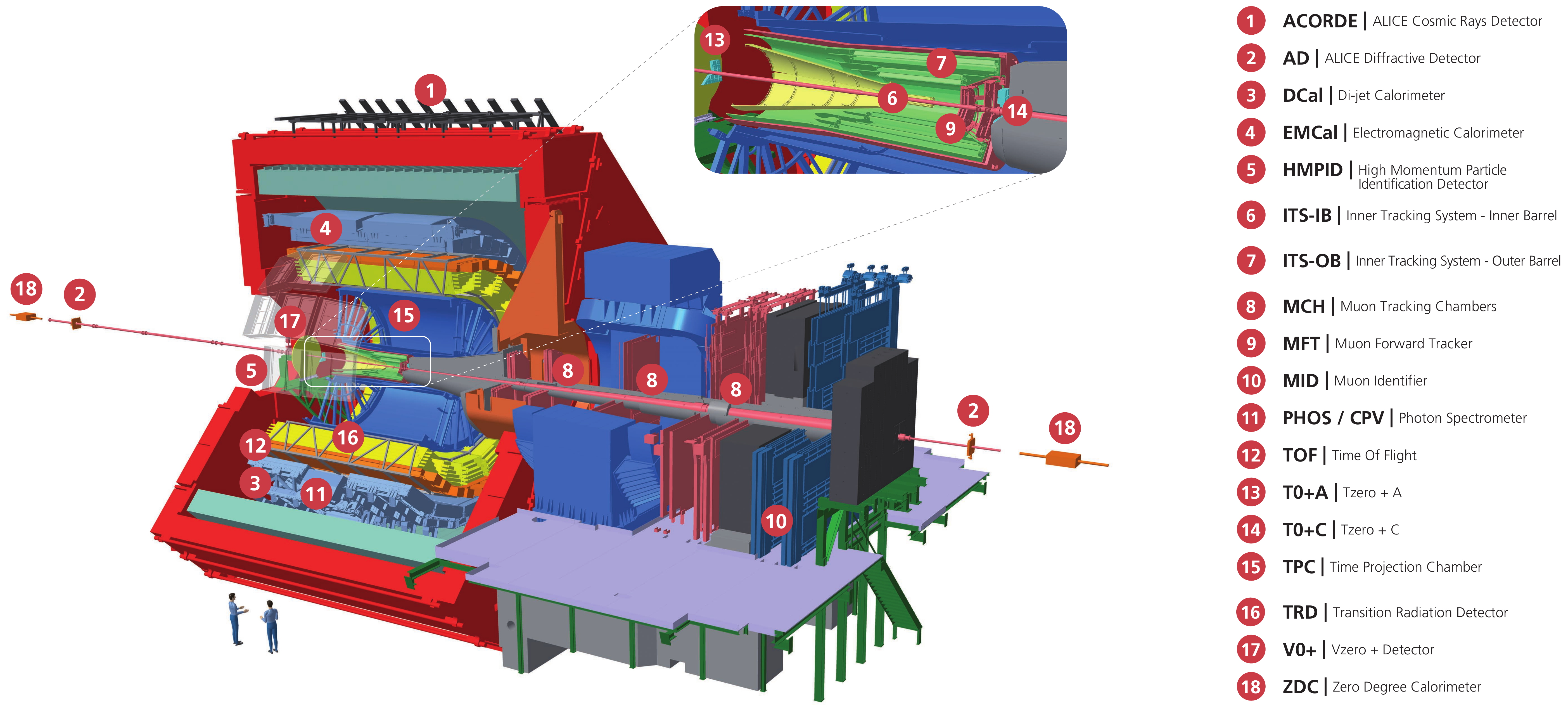}
\caption{Schematic view of the ALICE detectors for the LHC Run 3, after the upgrades~\cite{Abelev_et_al_2014}. Figure courtesy of \href{http://cds.cern.ch/record/2263642}{CERN}.}
\label{fig:Alice:Det:Run3}
\end{figure}

%% file: Appendix_Detector_LHCB.tex
\subsection{Schematic view of the LHCb detectors}
\label{appendix:lhcb_det}

\begin{figure}[hbt!]
\centering
\includegraphics[height=0.7\textwidth,angle=-90]{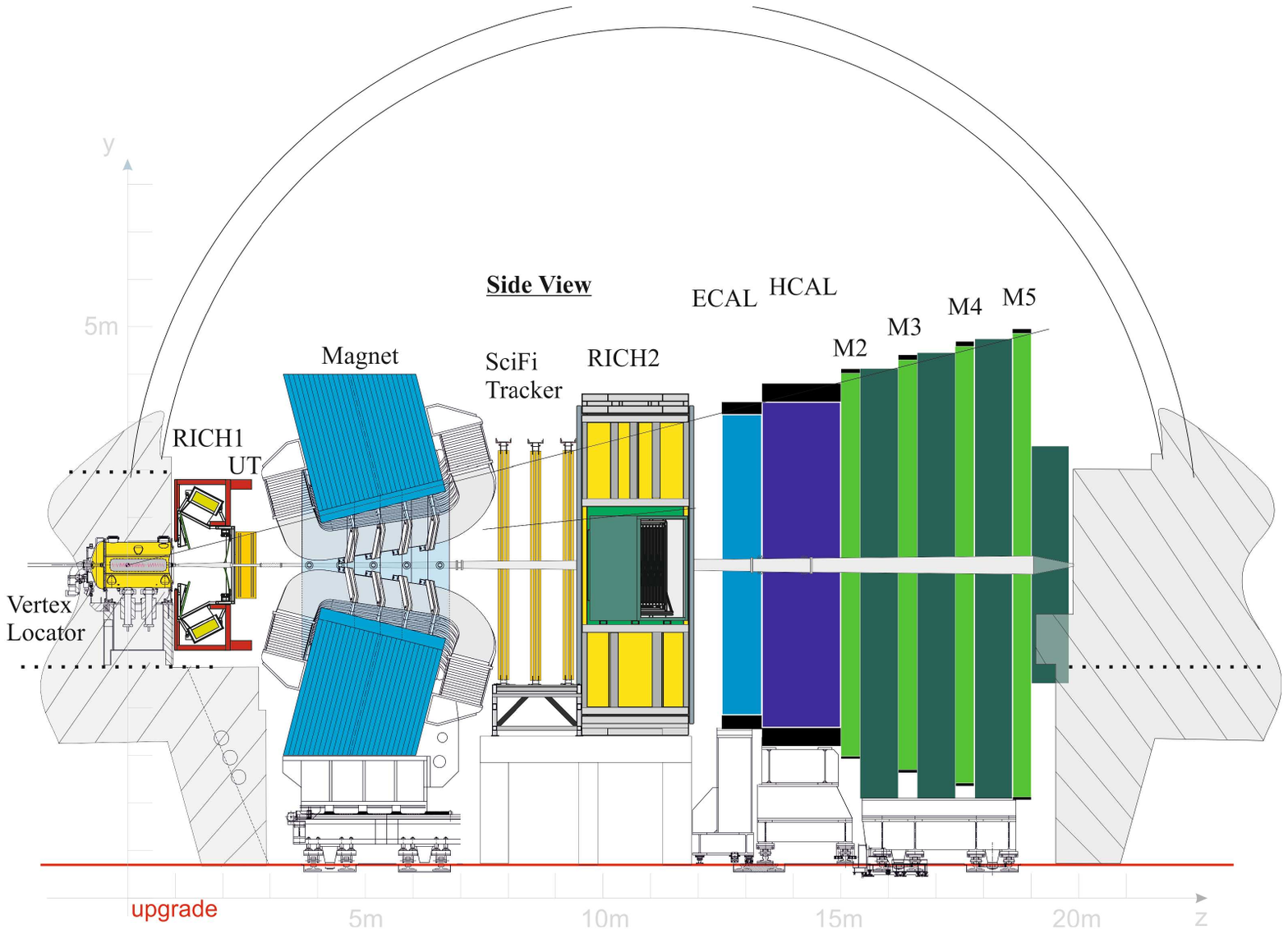}
\caption{Schematic view of the LHCb detectors for the LHC Run 3~\cite{Collaboration:1624070,Collaboration:1647400}. The main subsystems include the Vertex Locator, the Silicon Micro-strip Upstream Tracker (UT), the Scintillating Fiber (SciFi) Tracker, the Muon Chambers (M2 - M5), the Hadron Calorimeter (HCAL), the Electromagnetic Calorimeter (ECAL) and the RICH detectors. Figure courtesy of \href{http://cds.cern.ch/record/1087860}{CERN}.}
\label{fig:LHCb:det}
\end{figure}